\documentclass[12pt]{iopart}

\newcommand{\collop}[1][f_s]{\ensuremath{{C}\hspace{-0.5mm}\left[#1\right]}}
\newcommand{\vth}[1][s]{\ensuremath{v_{\mathrm{th}_#1}}}

\newcommand{\pd}[2]{\ensuremath{ \frac{\partial #1} {\partial #2} } }
\newcommand{\inpd}[2]{\ensuremath{ \infrac{\partial #1} {\partial #2} } }
\newcommand{\gyror}[1]{\ensuremath{ {\left< #1 \right>}_{\bm{r}}}}
\newcommand{\ensav}[1]{\ensuremath{ {\left< #1 \right>}_{\mathrm{turb}}}}  % Ensemble averaging
\newcommand{\fav}[1]{\ensuremath{\left< #1 \right>_{\psi}}} % Flux surface averaging
\newcommand{\gyroR}[1]{\ensuremath{{\left< #1 \right>}_{\bm{R}}}}

\providecommand{\eqref}[1]{Eq.\ (\ref{#1})}

\newcommand{\Figref}[1]{Fig.\ \ref{#1}}
\providecommand{\Tref}[1]{Table.\ \ref{#1}}

\providecommand{\Or}[1]{\mathcal{O}#1}
\providecommand{\Eref}[1]{Equation\ (\ref{#1})}
\providecommand{\eref}[1]{(\ref{#1})}

\newcommand{\dg}{\cdot\nabla}      % dg = dot grad
\newcommand{\dv}{\nabla\cdot}
\newcommand{\curl}{\nabla\times}
\newcommand{\tor}{\phi} % toroidal angle
\newcommand{\gyr}{\vartheta} % gyroangle
\newcommand{\pot}{\varphi} % electrostatic potential
\newcommand{\fpot}{\Phi} % Flow potential
\newcommand{\pol}{\theta} % Poloidal angle (straight field)
\newcommand{\gkeps}{\epsilon} % Rho over L, gyrokinetic epsilon
\newcommand{\energy}[1][s]{{{\varepsilon}_#1}} % Energy Variable
\newcommand{\denergy}[1][s]{{\dot{\varepsilon}}_{#1}}
\newcommand{\source}[1][s]{\ensuremath{{{{S}}_{#1}}}} % General Source
\newcommand{\gkupot}{\delta \pot'}
\newcommand{\gkpot}{\chi}
\newcommand{\psistar}[1][s]{{\psi^*_#1}}
\newcommand{\magmom}[1][s]{\mu_#1}
\newcommand{\dmu}[1][s]{\dot{\mu}_#1}

\newcommand{\tu}{\hat{u}_{\parallel s}}

\newcommand{\ddR}[2][s]{\pd{#2}{\bm{R}_{#1}}}
\newcommand{\dgR}[2][s]{\cdot\ddR[#1]{#2}}

\providecommand{\tensor}[1]{ {\bm{\mathsf{#1}}}}
\newcommand{\viscosity}[1][s]{\tensor{\Pi}_#1}
\newcommand{\idmat}{\tensor{I}}
\newcommand{\ParticleFlux}[1][s]{\Gamma_{#1}}
\newcommand{\HeatFlux}[1][s]{q_{#1}}
\newcommand{\MomentumFlux}[1][s]{\pi^{(\psi\tor)}_{#1}}
\newcommand{\TotMomFlux}{{\pi}^{(\psi\tor)}_{\mathrm{tot}}}
\newcommand{\inertia}{J}
\newcommand{\accel}[1][s]{\bm{a}_{#1}}
\newcommand{\daccel}[1][s]{\delta \bm{a}_{#1}}
\newcommand{\ddtpsi}{\left.\pd{}{t}\right|_{\psi}}
\newcommand{\oxford}{
Rudolf Peierls Centre for Theoretical Physics, University of Oxford, Oxford OX1 3NP, UK
}

\newcommand{\culham}{
EURATOM/CCFE Fusion Association, Culham Science Centre, Abingdon OX14 3DB, UK
}

\newcommand{\imperial}{
Blackett Laboratory, Imperial College, London SW7 2AZ, UK
}

\newcommand{\maryland}{
Department of Physics, University of Maryland, College Park, MD 20742-4111, USA
}

\newcommand{\llnl}{
Lawrence Livermore National Laboratory, Livermore, CA 94550, USA
}

\newcommand{\merton}{
Merton College, Oxford, OX1 4JD, UK
}

\newcommand{\mitpsfc}{
Plasma Science and Fusion Center, Massachusetts Institute of Technology, Cambridge, MA 02139, USA
}

\newcommand{\ippgreifs}{
Max-Planck-Institut f\"ur Plasmaphysik, 17491 Greifswald, Germany
}

\newcommand{\wint}{\int\hspace{-1.25mm} d^3 \bm{w}}
\newcommand{\Fneo}[1][s]{F^{\mathrm{(nc)}}_{#1}}
\newcommand{\FE}{F^{\mathrm{(E)}}_{s}}
\newcommand{\Fnct}{\tilde{F}^{\mathrm{(nc)}}_{s}}
\newcommand{\Fstar}[1][s]{F^{*}_{1#1}}
\newcommand{\Esource}[1][s]{S^{({E})}_{#1}}
\newcommand{\Psource}[1][s]{S^{({n})}_{#1}}
\newcommand{\Msource}{S^{({\omega})}}
\newcommand{\FreeEnergy}{W}
\newcommand{\perpav}[1]{\left<#1\right>_\perp}
\newcommand{\timeav}[1]{\left<#1\right>_T}

\newcommand{\entropy}[1][]{\widetilde{H}_{#1}}
\newcommand{\MeanEntropy}[1][]{H_{#1}}

\newcommand{\infrac}[2]{ {#1}/{#2} }
\newcommand{\binfrac}[2]{\left(\infrac{#1}{#2}\right)}
\newcommand{\CollEnergy}[1][s]{C^{{(E)}}_#1}
\newcommand{\ViscousHeat}[1][s]{P^{\mathrm{visc}}_#1}
\newcommand{\JouleHeat}[1][s]{P^{\mathrm{Ohm}}_#1}
\newcommand{\PotEng}[1][s]{P^{\mathrm{pot}}_#1}
\newcommand{\CompHeat}[1][s]{P^{\mathrm{comp}}_#1}
\newcommand{\TurbInj}[1][s]{P^{\mathrm{drive}}_#1}
\newcommand{\TurbColl}[1][s]{P^{\mathrm{diss}}_#1}
\newcommand{\TurbPow}[1][s]{P^{\mathrm{turb}}_#1}
\newcommand{\EMViscosity}{\pi_{\mathrm{EM}}^{(\psi\tor)}}
\newcommand{\NeoMomFlux}[1][s]{\pi^{\mathrm{(nc)}}_{#1}}
\newcommand{\ClassMomFlux}[1][s]{\pi^{\mathrm{(cl)}}_{#1}}
\newcommand{\vchi}{\bm{V}_\chi}
\newcommand{\vchiR}{\gyroR{\vchi}}
\newcommand{\vdrift}[1][s]{\bm{V}_{\mathrm{D}#1}}

\newcommand{\EProd}{C^{\mathrm{(H)}}}
\newcommand{\Rentropyflux}{\Gamma^{({H})}}
\newcommand{\JUflux}{\Gamma^{{(U)}}}
\newcommand{\CollEntropy}{C^{{(H)}}}
\newcommand{\EntropySource}{S^{{(H)}}}
\newcommand{\DebyeLength}{\lambda_{\mathrm{De}}}
\newcommand{\NotN}[1][s]{N_{#1}}

\newcommand{\delB}{{\ensuremath{\delta\bm{B}}}}
\newcommand{\delBp}{{\ensuremath{\delta B_\parallel}}}

\newcommand{\delAp}{{\ensuremath{\delta A_\parallel}}}

\newcommand{\delE}{{\ensuremath{\delta\bm{E}}}}
\newcommand{\delA}{{\ensuremath{\delta\bm{A}}}}
\newcommand{\delj}{{\ensuremath{\delta\bm{j}}}}
\newcommand{\delpot}{{\delta\pot}}
\newcommand{\MagDelB}{{\delta B}} 

\newcommand{\MeanB}{{\bm{B}}}
\newcommand{\MeanE}{{\bm{E}}}
\newcommand{\MeanMagB}{{B}}
\newcommand{\Meanb}{{\bm{b}}}
\newcommand{\Meanj}{{\bm{j}}}

\newcommand{\Efield}{{\widetilde{\bm{E}}}}
\newcommand{\Bfield}{{\widetilde{\bm{B}}}}

\newcommand{\current}{{\widetilde{\bm{j}}}}
\newcommand{\chargedens}{{\widetilde{\varrho}}}
\newcommand{\Apot}{{\widetilde{\bm{A}}}}
\newcommand{\Epot}{{\widetilde{\pot}}}
\newcommand{\MeanA}{{\bm{A}}}

\newcommand{\FHat}[1][s]{\widehat{F}_{1#1}}
\newcommand{\angvel}{\omega}
\newcommand{\cycfreq}[1][s]{\Omega_{#1}}

\newcommand{\chempot}[1][s]{\Upsilon_{#1}}
\newcommand{\quantconc}[1][s]{n_{\mathrm{Q}#1}}

\newcommand{\vpsi}{\bm{V}_\psi}

\newcommand{\delEnt}[1][s]{{\Delta S_{#1}}}

\newcommand{\lincol}[1][\cdot]{\ensuremath{{C}_{L}\hspace{-0.5mm}\left[#1\right]}}
\newcommand{\npol}{\ensuremath{n^{\mathrm{pol}}_s}}
\newcommand{\poldrift}{\ensuremath{\bm{V}^{\mathrm{pol}}_s}}

\usepackage{iopams}
\usepackage{setstack}
\usepackage{floatflt}
\usepackage{graphicx}
\usepackage{bm,ulem,color}
\usepackage{xr}
\usepackage{xr-hyper}
\usepackage[colorlinks=true]{hyperref}
\begin{document}

\date{\today}

\title[Multiscale Gyrokinetics for Rotating Tokamak Plasmas]{Multiscale Gyrokinetics for Rotating Tokamak Plasmas: Fluctuations, Transport and Energy Flows}

\author{I.~G.~Abel,$^{1,2,3}$ G.~G.~Plunk,$^{4}$ E.~Wang,$^5$ M.~Barnes,$^{6}$ S.~C.~Cowley,$^{2,7}$ W.~Dorland,${^8}$ and A.~A.~Schekochihin$^{1,3}$}
\ead{i.abel1@physics.ox.ac.uk}
\address{$^1$\oxford}
\address{$^2$\culham}
\address{$^3$\merton}
\address{$^4$\ippgreifs}
\address{$^5$\llnl}
\address{$^6$\mitpsfc}
\address{$^7$\imperial}
\address{$^8$\maryland}

\begin{abstract}
This paper presents a complete theoretical framework for studying turbulence and transport in rapidly-rotating tokamak plasmas.
The fundamental scale separations present in plasma turbulence are codified as an asymptotic expansion in the ratio $\gkeps = \infrac{\rho_i}{a}$ of the gyroradius to the 
equilibrium scale length. Proceeding order-by-order in this expansion, a set of coupled multiscale equations is developed. They describe an instantaneous equilibrium, the fluctuations driven by gradients in
the equilibrium quantities, and the transport-timescale evolution of mean profiles of these quantities driven by the interplay between the equilibrium and the fluctuations. The equilibrium distribution functions are 
local Maxwellians with each flux surface rotating toroidally as a rigid body. 
The magnetic equillibrium is obtained from the generalized Grad-Shafranov equation for a rotating plasma, determining the magnetic flux function from the mean pressure and velocity profiles of the plasma.
The slow (resistive-timescale) evolution of the magnetic field is given by an evolution equation for the safety factor~$q$.
Large-scale deviations of the distribution function from a Maxwellian are given by neoclassical theory. 
The fluctuations are determined by the ``high-flow'' gyrokinetic equation, from which we derive the governing principle for gyrokinetic turbulence in tokamaks: the conservation and {\it local} (in space) cascade of the free energy of the fluctuations (i.e., there is no turbulence spreading).
Transport equations for the evolution of the mean density, temperature and flow velocity profiles are derived. These transport equations show how the neoclassical and fluctuating corrections to the equilibrium Maxwellian act back upon the mean profiles through fluxes and heating.
The energy and entropy conservation laws for the mean profiles are derived from the transport equations.
Total energy, thermal, kinetic, and magnetic, is conserved and there is no net turbulent heating. Entropy is produced by the action of fluxes flattening gradients, Ohmic heating, and the equilibration of interspecies temperature differences. This equilibration is found to include both turbulent and collisional contributions.
Finally, this framework is condensed, in the low-Mach-number limit, to a more concise set of equations suitable for numerical implementation.
\end{abstract}
\maketitle

\section{Introduction}
\label{sec_intro}
Plasma turbulence in fusion devices is a fundamentally multiscale problem both in space and in time. The turbulent fluctuations driven by background gradients typically occur at scales associated with the ion Larmor radius, $\rho_i$ (or smaller), whilst the mean temperature, density, and bulk velocity profiles vary smoothly over the system scale (in tokamaks, the minor radius $a$). Similarly, the fluctuation frequency $\omega$ is much larger than the rate at which the mean profiles evolve, $\sim\tau_E^{-1}$, where $\tau_E$ is the energy confinement time. 
\Tref{TableInfo} gives approximate values for these space and time scales in some large tokamaks -- it is manifest that the separation of scales is very strong in such plasmas.
Because of this scale separation, it is possible to average over the time and space scales associated with the turbulent fluctuations and consider mean fields that slowly evolve due to turbulent and collisional transport, with the turbulence in turn driven by gradients in the same mean fields~\cite{sugama1998neg,sugama1997tpe,sugama1996tpa,gabethesis,ericthesis,michaelthesis}.

There are also scale separations associated with the turbulence itself. Turbulence in a strongly magnetized plasma occurs at frequencies $\omega$ which are much smaller than the cyclotron frequency of the ions, $\cycfreq[i]$ (see \Tref{TableInfo}).
The turbulence is also strongly anisotropic, viz., correlation lengths along the mean magnetic field are much longer than correlation lengths across the field; particles can stream rapidly along field lines but only drift slowly across them. These two properties of the turbulence are the foundation of the gyrokinetic theory~\cite{taylor1968sgp,rutherford1968dig,catto1978lgk,antonsenjr2006kel,frieman1982nge}, in which the fast cyclotron time scales are averaged out and the full 6D kinetics reduced to a simpler 5D formulation, the kinetics of charged rings.

In this paper, we unify this hierarchy of timescales and spatial scales in one formulation. We use the physical scale separations inherent in plasma dynamics to determine how the mean fields influence the evolution of the small-scale turbulence, and how the turbulent fluctuations react back upon the mean fields.
To quantify this back reaction, we derive the transport equations, in which it becomes manifest by what physical mechanisms the turbulence can affect the mean distribution functions and fields. 

The structure of this paper is as follows.
We start by recapitulating the fundamental equations of plasma physics in \Sref{kinetic-eq} and splitting all quantities into mean and fluctuating parts in \Sref{ordering}. After formalising our assumptions about scale separation, the practical validity of such assumptions is discussed in \Sref{Sscale}. We then impose order on our
multiple small parameters and scale separations by introducing the fundamental gyrokinetic ordering in \Sref{gkord}. This allows us to formulate the entire problem as a systematic expansion in the small ratio $\gkeps = \rho_i / a$. In subsequent sections, we proceed to expand the equations of \Sref{kinetic-eq} order by order in $\gkeps$,
zeroth order in \Sref{s0ord}, first in \Sref{s1ord}, second in \Sref{s2ord} and finally, third (transport order) in \Sref{transport}; pausing in \Sref{rotvar} to introduce rotating gyrokinetic variables, which are convenient for handling the turbulent kinetics in toroidally rotating plasmas.

Sections \ref{kinetic-eq} to \ref{transport} are, necessarily, quite technical.  Readers who believe themselves to be already familiar with the formalism presented in these sections may wish to skip directly to \Sref{thermo}.
Sections~\ref{thermo}~and~\ref{Sentropy} attempt to explain the physical content of multiscale gyrokinetics by
combining all our earlier results and examining how energy and entropy flow between the various constituent parts of our multiscale system. In \Sref{energyNentropy}, we prove that our transport equations conserve the total energy, and that the fluctuations can do no net work on the mean fields.
Explaining this result leads us in \Sref{SFreeEnergy} to consider the balance of free energy of the fluctuations, as this is the fundamental conserved quantity of kinetic turbulence~\cite{Krommes2,hallatschek,Tome,schekcrete}. The equation derived for the conservation of free energy clearly demonstrates a local (to a given flux surface) turbulent cascade that takes the energy injected by instabilities and dissipates it via collisions. In \Sref{Sentropy}, we link the free-energy cascade and the mean-field transport through the evolution of the mean entropy of the system.
\begin{table}[h]
\centering
\begin{tabular}{|c||c|c|c|c|c|}
	\hline
Parameter & JET & D-IIID & K-STAR & TFTR & ITER \\
&&&&&(projected)\\
\hline
\hline

$B$, T & 3.5 & 2 & 3.5 & 5-6 & 5 \\
$n_e$, cm$^{-3}$ & $5\times 10^{13}$ & $5\times 10^{13}$ &$5\times 10^{13}$ & $10^{14}$ & $10^{14}$ \\
$T_i$, keV & 5--15 & 7 & 5--10 & 5--32 & 25 \\
${u}$, km/s & $500$ & $200$ & $0.1$--$100$ & $100$ & $50$\\
$M = {u} / \vth[i]$ & 0.3--0.5 & 0.35 & 0.01--0.2 & 0.1--0.2 &  0.05 \\
\hline
$\DebyeLength$, cm & $1.3\times10^{-2}$ & $8.8\times10^{-3}$& $8.2\times10^{-3}$ & $4\times10^{-3}$ & $1.2\times10^{-2}$ \\
$\rho_i$, cm &  $0.051$ & $0.06$ & $0.045$ & $0.032$ & $0.032$ \\
$a$, cm & 100 & 50 & 50 & 87 & 200 \\
\hline
$\nu_{ee}$, $s^{-1}$ & $1.6\times10^3$ & $4.5\times10^3$& $7.2\times10^3$ & $1.3\times10^3$ & $1.3\times10^3$ \\
$\omega\sim\vth[i]/a$, $s^{-1}$ & $2.0\times10^4$ & $1.6\times10^5$ & $1.5\times10^5$ & $1.4\times10^5$ & $5.5\times10^4$ \\
$\cycfreq[i]$, $s^{-1}$ & $1.7\times10^8$& $9.6\times10^7$ & $1.7\times10^8$& $2.6\times10^8$ &$2.4\times10^8$\\
$\tau_E$, s & 0.5 & 0.1 & 0.3 & 0.3 & 3.5 \\
\hline
$\epsilon = \rho_i / a$  & $5\times10^{-3}$ & $1.21\times10^{-2}$ & $6.9\times10^{-3}$& $5.2\times10^{-3}$ & $2.28\times10^{-3}$\\
$\beta_i$ & 2.5\% & 3.5\% & 1--4\% & 4.5\% & 4\% \\
\hline
\end{tabular}
\caption{Typical length and time scales in selected fusion devices (approximate).
JET parameters estimated from \cite{vries2009internal}, DIII-D from \cite{diiid-aug-hybrid}, TFTR from \cite{pppl3147}, and ITER from TRANSP studies \cite{budny2002alphas,halpern:062505}.
}
\label{TableInfo}
\end{table}

Sections \ref{kinetic-eq}--\ref{Sentropy} present a pedagogical derivation all the way from the Vlasov-Landau-Maxwell system of equations to the non-equilibrium mean-field thermodynamics of the system. 
The order of presentation is the order of the asymptotic expansion, in the spirit of asymptotology~\cite{kruskal1960asymptotology}.
However, there are three conceptual strands interwoven in the derivation that deserve to be highlighted separately.

Firstly, there is the equilibrium -- the instantaneous solution for the mean fields.
First we recover the toroidally rotating Maxwellian for the mean distribution function (\Sref{maxback}) and find that the toroidal rotation is a rigid-body motion of nested flux surfaces (\Sref{constraints} and \Sref{maxback}). This solution has an arbitrary density and temperature for each species and an arbitrary angular velocity on each flux surface. Then, in \Sref{poldens}, we find that the 
the poloidal variation of the density is determined by the balance between centrifugal forces and electrostatic fields set up within a flux surface.
With these results in hand, the poloidal flux function is determined, in \Sref{ampmag}, by a generalized Grad-Shafranov equation \eref{gradshaf}.
Finally, the first-order correction to the mean distribution function is given by the solution of the neoclassical drift-kinetic equation, which is also (re)derived within our unified formalism (\Sref{nclasse}).

Secondly, there are the fluctuations that feed on the free-energy sources (gradients) present in this (local) equilibrium.
The fluctuating distribution function splits into the Boltzmann response to the fluctuating electromagnetic fields and a distribution of charged rings (\Sref{boltzresp}).
This non-Boltzmann part of the distribution function is governed by the gyrokinetic equation (\Sref{Sgke}). This system of equations is closed by Maxwell's equations for the fluctuating electromagnetic fields in \Sref{Sfmag}.
Finally, in \Sref{SFreeEnergy}, we conclude this strand by deriving the conservation law that governs the nature of the fluctuations: the conservation of free energy.

Thirdly, there is the long-time evolution of the mean fields. This starts in \Sref{magevolve}, where we determine the evolution of the mean magnetic field. This turns out to be completely independent of the fluctuations.
The back-reaction of the small-scale turbulence on the mean profiles is found in the transport equations of \Sref{transport}.
 Examining particle, momentum and energy conservation, we find \eref{dndt}, \eref{domegadt}, and \eref{dpdt}, which determine the transport-timescale evolution of the density, angular velocity and temperature profiles in terms of fluxes and sources, which in turn are given as functions of the turbulent and neoclassical distribution functions and fields.
 This strand concludes with the results of \Sref{energyNentropy} and \Sref{Sentropy}: that the total energy is conserved on the transport timescale and that the increase of mean entropy can be written in the usual way as a combination of heating terms and the product of fluxes and thermodynamic gradients.

This paper is written in an entirely self-contained way and so presents both rederivations of many known results, cast in forms suitable for our unified framework, and a number of new results -- we have, throughout the exposition, striven to give credit where credit is due without attempting to provide a fully exhaustive literature review. To guide a reader aiming to learn gyrokinetic theory from this paper, it is perhaps useful to put our approach into the context of other, alternative, approaches. Our exposition falls within what might be termed ``traditional gyrokinetics,'' in which the theory is viewed as an order-by-order asymptotic expansion. We have made no attempt to adjust our equations to achieve a Hamiltonian structure or exact energy conservation {\textit{within}} each asymptotic order (see \ref{parallelNote} for further comments on this subject; for an exposition of the Hamiltonian approach, see \cite{brizard2007} and references therein) -- indeed, the salient point of \Sref{thermo} is that energy in a multiscale system flows between fluctuations and mean fields and so ``between different orders'' of the asymptotic expansion. The expansion terminates at the transport order (\Sref{transport}) in the sense that the equations are closed and energy is conserved overall. Similarly, we make no attempt to formulate ``global'' equations that simultaneously describe the long and short scales -- in fact, we take scale separation to be a virtue and consistently enforce it within our formalism. This said, alternative approaches (sometimes termed ``modern gyrokinetics'') have many virtues of their own to recommend them -- not least some fascinating mathematics -- and we refer the curious reader to recent reviews \cite{brizard2007,garbet2010gyrokinetic} where they are presented.

In \Sref{LowMach}, we present a low-Mach-number limit of the system of equations given in this paper. This limit removes many cumbersome technical complications and so Sections~\ref{LM-Mean}--\ref{LM-Energy} can be used as a concise summary of the basic structure of our multiscale hierarchy. It is this set of equations which is implemented in current linked-flux-tube transport codes~\cite{barnes2009trinity,candy2009tgyro}.

Finally, in \Sref{SConc}, we finish the paper by summarizing the conclusions of this work and how they fit into the broader landscape of fusion plasma physics.
\section{Fundamental Equations} 
\label{kinetic-eq}

As our starting point we take the Fokker-Planck kinetic equation for $f_s$, the distribution function of species $s$,
\begin{equation}
\label{vfp}
\frac{d f_s}{dt}=\pd{f_s}{t} + \bm{v}\dg f_s + \frac{Z_s e}{m_s}\left( \Efield + \frac{1}{c} \bm{v}\times\Bfield\right)\cdot\pd{f_s}{\bm{v}} = \collop[f_s] + \source,
\end{equation}
where $Z_s$ is the charge of the particles of species $s$ as a multiple of the fundamental charge $e$, $m_s$ their mass and $\bm{v}$ their velocity.
 We will work in Gaussian units throughout with $c$ the speed of light,  $\Efield$ the electric field and $\Bfield$ the magnetic field (throughout this work, tildes denote exact fields containing both the mean and the fluctuating parts; see \Sref{ordering}).
On the right hand side, $\collop[f]$ is the Landau collision operator and $\source$ an arbitrary source term that stands in for all physical processes not yet accounted for, e.g., atomic physics, fusion reactions, Bremsstrahlung, radio frequency heating and current drive. 

 The electric and magnetic fields obey Maxwell's equations:
\begin{eqnarray}
\label{max1}
\dv \Efield &=& 4\pi \chargedens,\\
		\label{maxB}
		\dv{\Bfield} &=& 0,\\
		\label{maxF}
\quad\pd{\Bfield}{t} &=& -c\curl\Efield,\\
	\label{maxN}
\curl \Bfield &=& \frac{4\pi}{c} \current + \frac{1}{c} \pd{\Efield}{t},
\end{eqnarray}
where the charge density $\chargedens$ and current $\current$ are
\begin{eqnarray}
\label{cdensDef}
\chargedens &= \sum_s Z_s e \int d^3 \bm{v} f_s,\\
\label{curDef}
\current &= \sum_s Z_s e \int d^3 \bm{v}\bm{v} f_s.
\end{eqnarray}

For the purposes of this work, we neglect Debye-scale effects and relativistic effects, viz.,
\begin{eqnarray}
\label{nodebye}
k_\perp^2 \lambda_{\mathrm{De}}^2 \ll 1,\\
\label{nonrel}
\frac{\vth[s]^2}{c^2} \ll 1,
\end{eqnarray}
where $\lambda_{\mathrm{De}} = \sqrt{T_e \left/ 4\pi n_e e^2 \right.}$ is the Debye length, $\vth[s] = \sqrt{\left. 2 T_s \right/ m_s}$ the thermal speed and $T_s$ and $n_s$ are the mean temperature and density of species $s$, where the notion of mean will be rigorously defined in \Sref{ordering}. The assumptions \eref{nodebye} and \eref{nonrel} are well satisfied in most modern fusion experiments and will be in future ones. 

We use \eref{nodebye} to replace \eref{max1} with the quasineutrality constraint
\footnote{The quasineutrality constraint implicitly defines the fluctuating electrostatic potential and the mean electric field within a flux surface but cannot, in practice, be used to determine the mean radial electric field \cite{parra2008lgt}, so we derive an equation for the radial electric field from momentum conservation in \Sref{momtrans} .}
\begin{equation}
\label{qn}
\chargedens = 0,
\end{equation}
and \eref{nonrel} to drop the displacement current in \eref{maxN}, giving Amp\`ere's Law,
\begin{equation}
\label{ampere}
\curl\Bfield = \frac{4\pi}{c} \current.
\end{equation}
We also satisfy \eref{maxB} and \eref{maxF} by introducing the scalar potential $\Epot$ and the vector potential~$\Apot$:
\begin{eqnarray}
\label{Edef}
\Efield &= -\nabla \Epot - \frac{1}{c} \pd{\Apot}{t}, \\
			  \label{Bdef}
\Bfield &= \curl \Apot.
\end{eqnarray}
We will work in the Coulomb gauge $\dv\Apot = 0$. Thus, the four Maxwell equations \eref{max1}--\eref{maxN} are replaced by \eref{qn}, \eref{ampere}, \eref{Edef} and \eref{Bdef}.

\section{Fluctuations and Mean Fields}
\label{Sfluctmean}

\subsection{Small-Scale Averaging}
\label{ordering}
In order to analyse the mean and fluctuating quantities separately, we introduce the concept of an average over fluctuations, denoted by $\ensav{\cdot}$.  Formally, we demand that it separate any arbitrary physical quantity $g$ into an averaged part $\ensav{g}$ and a fluctuating part $\delta g = g - \ensav{g}$, which by construction vanishes under the average, $\ensav{\delta g} = 0$.

\begin{figure}[h]
\centering
\includegraphics[width=.5\textwidth]{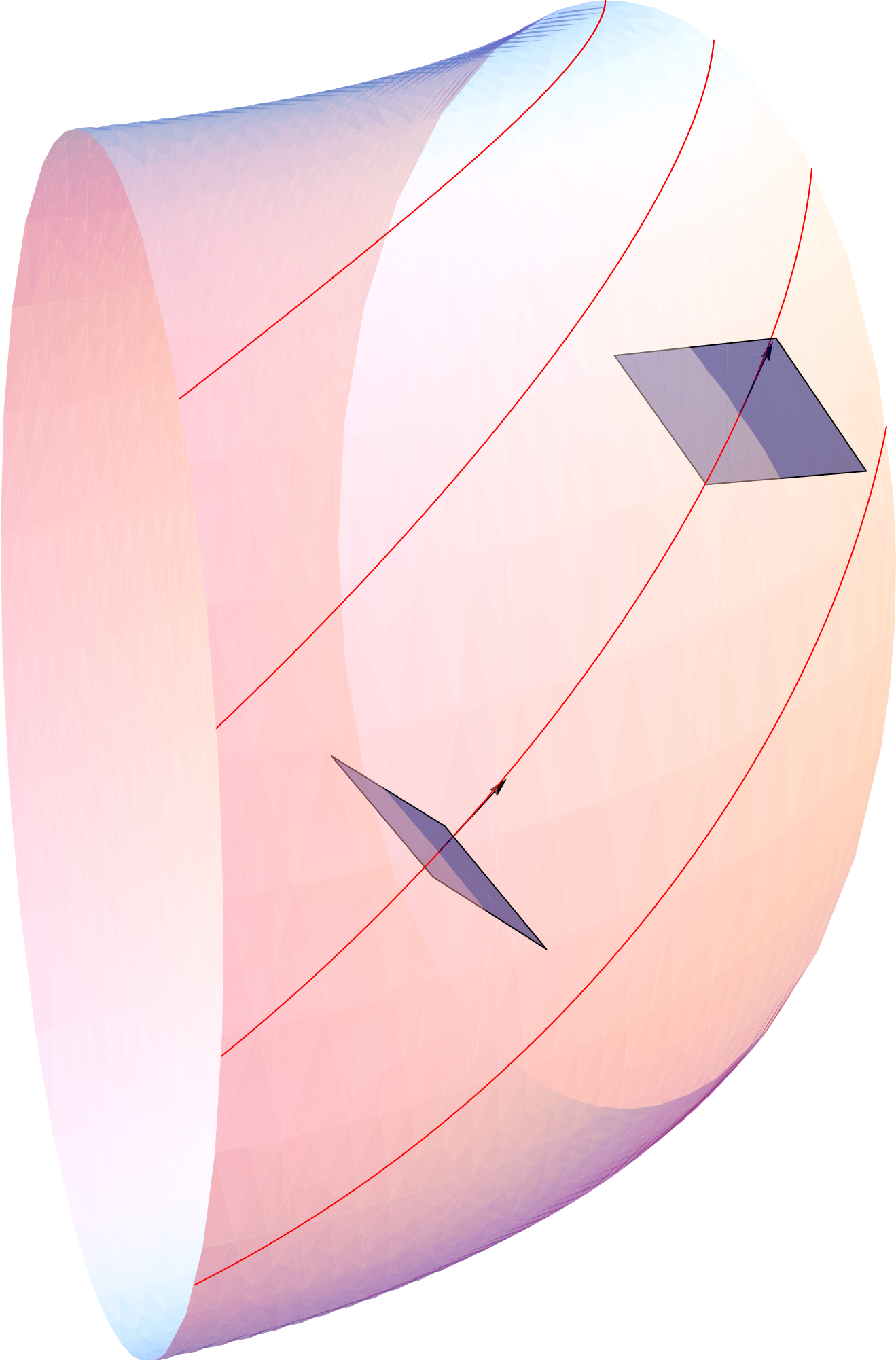}
\caption{A section of a toroidal flux surface showing field lines and, in blue, the small perpendicular patches $\lambda_\perp^2$ over which fluctuations are averaged in \eref{PerpAvDef}. Black arrows denote the normals to the patches, and are aligned with the magnetic field.}
\label{figPatchAvg}
\end{figure}

As there is a separation of scales, we can interpret this average as an average over the fluctuating temporal and spatial scales.
The equilibrium length scale, denoted $a$ (and understood to be, e.g., the tokamak minor radius), is well separated from the fluctuation length scale, taken to be the Larmor radius $\rho_s = \vth[s]\left/\cycfreq\right.$, where $\cycfreq = \left.Z_s e \MeanMagB\right/m_s c$. Therefore, we can pick an intermediate scale $\lambda$ that satisfies
\begin{equation}
a \gg \lambda \gg \rho_s
\label{LambdaScale}
\end{equation}
and define the perpendicular spatial average $\left<\cdot\right>_\perp$ of a function $g(\bm{r},\bm{v},t)$ by
\begin{equation}
\perpav{g(\bm{r},\bm{v},t)} = \left.\int\limits_{\lambda_\perp^2} d^2 \bm{r}'_\perp g(\bm{r}'_\perp,l,\bm{v},t) \right/ \int\limits_{\lambda_\perp^2} d^2\bm{r}_\perp,
\label{PerpAvDef}
\end{equation}
where $\lambda^2_\perp$ is a small surface which is everywhere normal to the magnetic field and has spatial extent of the order of $\lambda$ in both perpendicular directions. The integrals are taken at constant $\bm{v}$, $t$ and $l$, where $l$ is the distance along a given field line, or any other field-aligned coordinate -- see Fig.~\ref{figPatchAvg}.  Clearly, an averaged function cannot vary on the small length scales $ \sim \rho_s$, and any function that varies only on the equilibrium length scale $\sim a$ is unaffected by the average. 
Similarly, the typical fluctuation time scale $\omega^{-1}$ and the timescale of the evolution of the mean profiles,  taken to be the transport time $\tau_E$, are also well separated. 
Therefore, we can pick an intermediate time $T$ that satisfies
\begin{equation}
\tau_E \gg T \gg \omega^{-1}
\label{TScale}
\end{equation}
and use it to define the temporal average $\left<\cdot\right>_T$ by
\begin{equation}
\timeav{g(\bm{r},\bm{v},t)} = \frac{1}{T} \int\limits_{t-T/2}^{t+T/2} dt' g(\bm{r},\bm{v},t'),
\label{TavDef}
\end{equation}
with the integral taken at constant $\bm{r}$ and $\bm{v}$.
Once again, we see that averaged functions cannot vary on the short timescale $\omega^{-1}$, and that functions that vary on the transport timescale are unchanged.

We now define the full average over fluctuations by
\begin{equation}
\label{avdef}
\ensav{g(\bm{r},\bm{v},t)} = \timeav{\perpav{g}}.
\end{equation}
We can split the distribution function $f_s$ and all fields into mean and fluctuating parts:
\begin{eqnarray}
\label{deltaFdecomp}
f_s &= F_s + \delta f_s,\qquad F_s &= \ensav{f_s},\\
\Efield &= \MeanE + \delE, \qquad \MeanE &= \ensav{\Efield},\\
\Bfield &= \MeanB + \delB, \qquad \MeanB &= \ensav{\Bfield},\\
\Apot &= \MeanA + \delA, \qquad \MeanA &= \ensav{\Apot},\\
\Epot &= \pot + \delpot,\qquad\,\,\, \pot &= \ensav{\Epot}.
\end{eqnarray}

\subsection{Scale Separation}
\label{Sscale}
Let us use the average \eref{avdef} to restate the formal assumptions about temporal and spatial variation of the mean and fluctuating quantities ($\ensav{g}$ and $\delta g$):
\begin{eqnarray}
	\label{ord1}
\pd{}{t} \ln\ensav{g} &\sim \tau_E^{-1},\\
\pd{}{t} \ln\delta g &\sim \omega,\\
\nabla \ln \ensav{g} &\sim \Meanb\dg\ln\delta g\sim a^{-1},\\
	\label{ordN}
\nabla_\perp \ln \delta g &\sim k_\perp \sim \rho_s^{-1},
\end{eqnarray}
where $\Meanb$ is the unit vector in the direction of the averaged magnetic field, and $\perp$ and $\parallel$ denote components of vectors or operators perpendicular and parallel to the averaged magnetic field, respectively. In \Sref{gkord}, we will formally order all time scales, spatial scales and fluctuation amplitudes with respect to the small parameter $\gkeps = \rho_s \left/ a\right.$.

Equations (\ref{ord1})--(\ref{ordN}) encode the assumption of complete scale separation. Whilst we will consider {\textit{only}} plasmas in which the scales {\textit{are}} separated, there are reasons why one might question the validity or usefulness of this assumption. 
In part this is a question of how asymptotic any given experiment (numerical or physical) is, i.e., how small $\rho_i / a$ is (strictly speaking, we only determine how the plasma behaviour changes as $\rho_i / a$ decreases, but it should be a better approximation the smaller $\rho_i/a$ is). It is important to realise that,
however accurate the theory that we present in this paper might be, we cannot capture effects that vanish as $\rho_i/a \rightarrow 0$. There are, broadly-speaking, two questions one must consider with regard to this limitation.

Firstly, to what extent is it valid to assume that all fluctuations are small-scale, $k_\perp \rho_i \sim 1$, and short-timescale, $a/\vth \ll \tau_E$?
From linear studies, we know that the instabilities that drive fluctuations in tokamaks are most virulent at $k_\pol \rho_i \sim 1$, where $k_\pol$ is the poloidal wavenumber, but many linear calculations lead to a radial eigenmode structure that is extended across a finite fraction of the radius of the plasma. At first glance, this violates our assumption of scale separation. However, we are considering a {\textit{turbulent}} plasma, i.e., one in which the nonlinear processes on each flux surface occur much more rapidly than the timescales on which these radial eigenmode structures are formed. This can limit the radial correlation length (the important nonlinear length scale) to a small spatial scale, well separated from the scale of profile variation -- even if the linear mode structure is {\textit{set}} by such profile variation. So long as this condition holds, we can use the results of this paper to investigate the effect such turbulence has on the plasma as a whole -- even though we will not correctly determine the linear mode structure (effectively we are neglecting $\rho_i / a$ corrections to the growth rate and mode structure).

The second question is to what extent are the mean profiles confined to the long spatial ($a$) and slow temporal ($\tau_E$) scales?  In current experiments, some interesting phenomena arise in parameter regimes where the mean length scales can approach $\rho_i$ (or, equivalently $\rho_{\mathrm{pol}}$ the poloidalion gyroradius or banana width) and where the timescales of profile evolution can become short as, e.g., in transport barriers, L-H transitions, and heat pulse / cold pulse experiments. It is currently unclear whether the physics of these phenomena is intrinsically related to a violation of scale separation (i.e., whether the effect disappears as $\rho_i/a \rightarrow 0$ and so our theory is inapplicable) or whether the scale separation is instead merely obscured by the particular parameters of current experiments (i.e., $\rho_i/a$ is no longer numerically small, but our formalism may still qualitatively describe the behaviour of the plasma). Thus, the applicability of the theory presented in this paper to such problems is an open question.

\subsection{Axisymmetry and Magnetic Geometry}
\label{maggeom}
\begin{figure}[h]
\centering
\includegraphics[height=.3\textheight]{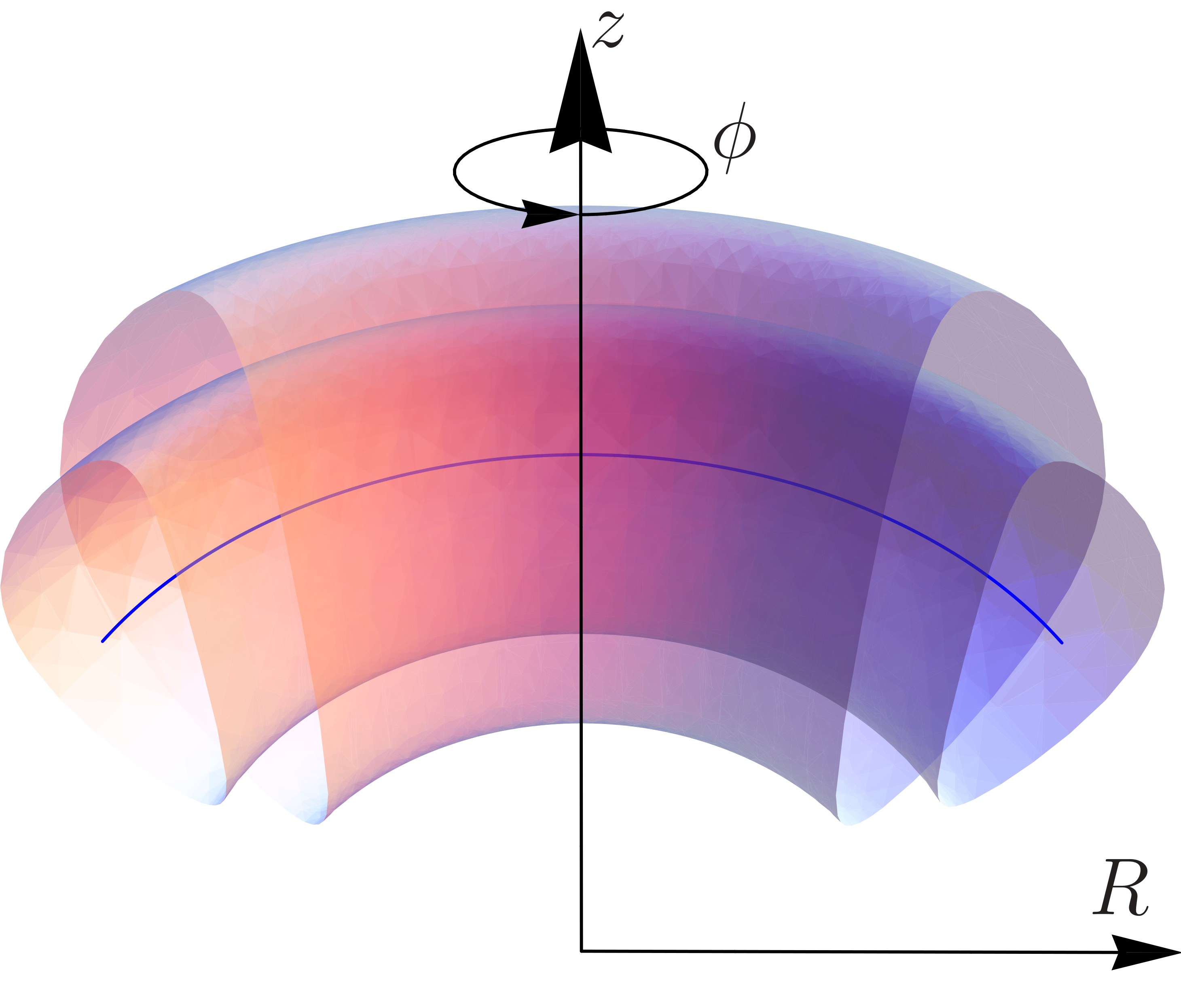}
\caption{The toroidal coordinate system ($R$,$z$,$\tor$) showing the magnetic axis and flux surfaces}
\label{coordDef}
\end{figure}

We will assume the axisymmetry of all mean quantities. In the case of modern tokamaks, the deviation from axisymmetry in the magnetic field is extremely small, much smaller than $\rho_s / a$ in such devices. 
Let us introduce the cylindrical coordinate system: major radius $R$, vertical position $z$ and toroidal angle $\tor$. We pick this system such that $(R, \tor, z)$ is a right-handed coordinate system -- see Fig.~\ref{coordDef}.
Assuming that the toroidal magnetic field variation is the same as the toroidal variation of any averaged quantity, we formally demand that for any quantity $g(\bm{r})$,
\begin{equation}
\label{axisymmetry}
\pd{\ln \ensav{g}}{\tor} = 0.
\end{equation}
At the end of \Sref{derivInt}, we will extend the notion of axisymmetry to distribution functions, which depend upon velocity space as well as the position $\bm{r}$.
The averaged form of \eref{Bdef} is
\begin{equation}
\MeanB = \curl\MeanA = \left( \pd{A_R}{z} - \pd{A_z}{R} \right) R \nabla\tor + \frac{1}{R} \pd{\left(R A_\tor\right)}{R} \nabla z - \pd{A_\tor}{z} \nabla R,
\end{equation}
where
\begin{equation}
\MeanA = A_\tor \nabla \tor + A_R \nabla R + A_z \nabla z
\end{equation}
  and we have used \eref{axisymmetry} to drop all $\tor$ derivatives.
Therefore, the magnetic field can be written in the usual toroidal decomposition~\cite{boozer2005pmc}:
\begin{eqnarray}
\label{magfield}
	\MeanB &= I \nabla\tor + \nabla\psi \times \nabla \tor,
\end{eqnarray}
where
\begin{equation}
\psi(R,z) = A_\tor = R^2\MeanA\dg\tor
\label{psiDef}
\end{equation}
is the poloidal flux function and
\begin{equation}
I(R,z) = R\left( \pd{A_R}{z} - \pd{A_z}{R} \right) = R^2\MeanB\dg\tor.
\label{IDef}
\end{equation}

The toroidal symmetry guarantees the existence of well-defined flux surfaces~\cite{white2006ttc,arnold1989mmc}. Topologically, these are nested tori. Since $\MeanB\dg\psi = 0$, these surfaces can be labelled by $\psi$. 
We will see that many mean quantities will only depend on $R$ and $z$ through $\psi(R,z)$.
%The innermost of these toroidal surfaces encloses no volume and is, by axisymmetry, a circle lying in a plane of constant $z$. This circle is called the magnetic axis (to distinguish it from the symmetry axis $R=0$).

\subsection{Flux-Surface Averaging and the Motion of Flux Surfaces}
\label{Sfluxav}
\begin{figure}[h]
\centering
\includegraphics[width=.85\textwidth]{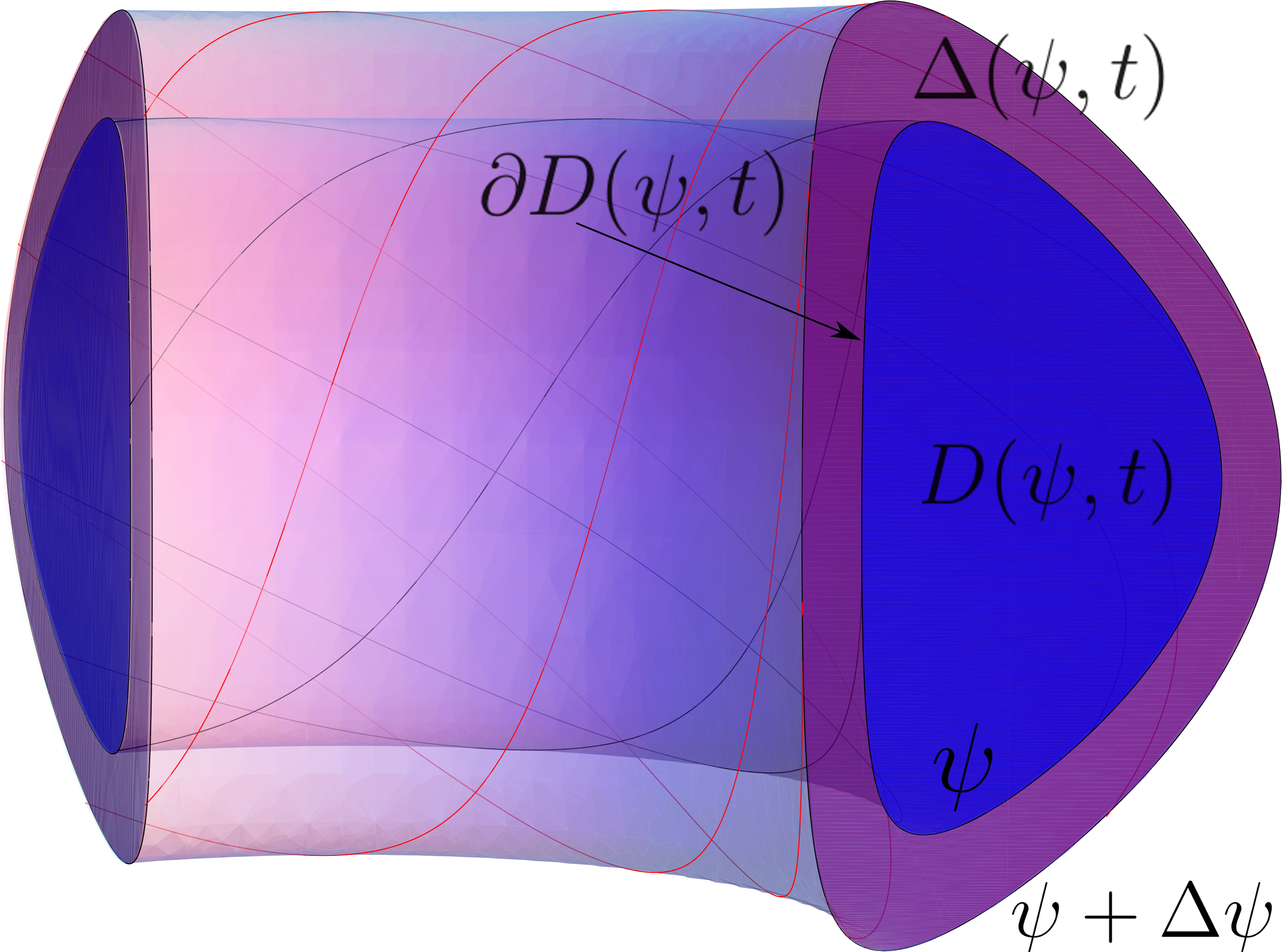}
\caption{A section of the torus showing the regions $D(\psi,t)$ and $\Delta(\psi,t)$ used in defining the flux-surface average \eref{favdef}, \eref{newfav} and the surface $\partial D$ separating these regions.}
\label{figFluxAv}
\end{figure}

It will be convenient to define an average over the surface labelled by $\psi$, which we do as follows \cite{DhaeseleerFlux,hinton1976tpt}.
For an arbitrary function $g(\bm{r})$,
\begin{equation}
\label{favdef}
\fav{g(\bm{r})}(\psi) = \lim_{\Delta \psi \rightarrow 0} \left\{ \int\limits_{\Delta(\psi)} d^3{\bm{r}} g(\bm{r}) \right/ \left. 
\int\limits_{\Delta(\psi)} d^3{\bm{r}}\right\},
\end{equation}
where the domain of integration $\Delta(\psi)$ is the annulus between the flux surface labelled by $\psi$ and that labelled by $\psi+\Delta\psi$ (see \Figref{figFluxAv}).

Note that
\begin{equation}
\fl\begin{eqalign}{
\pd{}{\psi} \int_{D(\psi,t)} d^3\bm{r} g &= \lim_{\Delta\psi\rightarrow 0} \frac{1}{\Delta\psi}\left( \int_{D(\psi+\Delta\psi,t)} d^3\bm{r} g - \int_{D(\psi,t)} d^3\bm{r} g\right) \\
	&= \lim_{\Delta\psi\rightarrow 0} \frac{1}{\Delta\psi} \int_\Delta d^3\bm{r} g = V' \fav{g},
}\end{eqalign}
\end{equation}
where ${D(\psi,t)}$ is the volume enclosed by the flux surface labelled by $\psi$ (i.e., the volume between that surface and the magnetic axis; see \Figref{figFluxAv}), $V = \int_{D(\psi,t)}d^3\bm{r}$ is the volume
of this region and 
\begin{equation}
\label{Vprimdef}
V' = \lim_{\Delta\psi \rightarrow 0} \frac{1}{\Delta\psi} \int_\Delta d^3\bm{r}= \pd{V}{\psi}.
\end{equation}
Thus, an alternate definition of the flux-surface average is
\begin{equation}
\fl
\fav{g} = \frac{1}{V'} \pd{}{\psi} \int_{D(\psi,t)} d^3\bm{r} g = \frac{1}{V'}\pd{}{\psi} \int^\psi_0d\psi'\int_{\partial D(\psi',t)}\frac{dS}{|\nabla\psi|} g =  \frac{1}{V'} \int_{\partial D(\psi,t)} \frac{dS}{|\nabla\psi|} g,
\label{newfav}
\end{equation}
where the surface integral is taken over the boundary of $D$, $\partial D(\psi,t)$,
and we have defined $\psi$ in such a way that $\psi = 0$ at the magnetic axis.\footnote{In our choice of coordinates, $\psi$ increases away from the magnetic axis if the toroidal current flows in the negative $\tor$ direction and the toroidal field is in the positive $\tor$ direction. Similarly, if the toroidal field and current are both in the positive $\tor$ direction, $\psi$ will decrease away from the magnetic axis. If the positive $\tor$ direction is fixed to be opposite to that of the toroidal current then $\psi$ will always increase outwards.}

We will discover in the subsequent sections (Sections \ref{magevolve}, \ref{transport} and \ref{thermo}) that flux-surface averaging allows us to close evolution equations for the mean fields.
Two mathematical identities will be useful in those derivations: the formula for the flux-surface average of a divergence and that for the flux-surface average of a time derivative.

First, using \eref{newfav}, we find for any vector field $\MeanA(\bm{r})$,
\begin{equation}
\begin{eqalign}{
\fav{\dv\MeanA} &= \frac{1}{V'} \pd{}{\psi} \int_{D(\psi,t)} d^3\bm{r}\dv\MeanA = \frac{1}{V'} \pd{}{\psi} \int_{\partial D(\psi,t)} \frac{dS}{|\nabla\psi|} \MeanA\dg\psi.
}\end{eqalign}
\label{favdivtmp}
\end{equation}
Therefore,
\begin{equation}
\fav{\dv\MeanA} = \frac{1}{V'}\pd{}{\psi} V'\fav{\MeanA\dg\psi},
\label{favdiv}
\end{equation}
which is the first of the two identities we will require later. This identity also implies that
\begin{equation}
\fav{\MeanB\dg g} = \fav{\dv\left(g\MeanB\right)} = \frac{1}{V'} \pd{}{\psi} V'\fav{g\MeanB\dg\psi} = 0.
\label{annhilator}
\end{equation}
Thus, the flux-surface average annhilates the operator $\MeanB\dg$~\cite{hinton1976tpt}.

Since the constant-$\psi$ surfaces change in time, flux-surface averages do not commute with time derivatives.
Let us consider
\begin{equation}
\begin{eqalign}{
\ddtpsi \int_{D(\psi,t)} d^3\bm{r}g &= \int_{D(\psi,t)} d^3\bm{r} \pd{g}{t} + \int_{\partial D(\psi,t)} \frac{d{S}}{|\nabla\psi|}g \vpsi\dg\psi\\
	&= \int_{D(\psi,t)} d^3\bm{r} \pd{g}{t} + V' \fav{ g \vpsi \dg \psi },
}\end{eqalign}
	\label{movingfluxtmp}
\end{equation}
where $\vpsi$ is the velocity with which the boundary of $D(\psi,t)$ moves. The time derivative in the left-hand side of \eref{movingfluxtmp} is taken at constant flux label $\psi$.
Taking the derivative of \eref{movingfluxtmp} with respect to $\psi$ and using \eref{newfav}, we find
\begin{equation}
\fl
\fav{\pd{g}{t}} = \frac{1}{V'} \pd{}{\psi} \int_{D(\psi,t)} d^3\bm{r}\pd{g}{t} = \frac{1}{V'} \ddtpsi V'\fav{g} - \frac{1}{V'} \pd{}{\psi}V'\fav{g\vpsi\dg\psi}.
\label{movingflux-intermed}
\end{equation}
Finally, as the boundary of $D(\psi,t)$ is defined to be a constant-$\psi$ surface, we have
\begin{equation}
\pd{\psi}{t} + \vpsi \dg\psi = 0.
\end{equation}
As $\vpsi$ is defined to be the velocity of a flux surface, we can demand that $\vpsi$ has no component in the surface and so
\begin{equation}
\label{VpsiDef}
\vpsi = -\pd{\psi}{t} \frac{\nabla\psi}{|\nabla\psi|^2}.
\end{equation}
Therefore,
\begin{equation}
\label{movingflux}
\fav{\pd{g}{t}} = \frac{1}{V'} \ddtpsi V'\fav{g} + \frac{1}{V'} \pd{}{\psi}V' \fav{g\pd{\psi}{t}},
\end{equation}
which is the second of the identities we were seeking.
\subsection{The Gyrokinetic Ordering}
\label{gkord}
To proceed further we need to impose order on the three small parameters we have, $\gkeps = \rho_s / a$, $\omega  / \cycfreq$ and the anisotropy of the turbulence
$k_\parallel / k_\perp$, as well as on the amplitudes of the fluctuations.
Then each term in the kinetic equation \eref{vfp} will have a well defined order in terms of $\gkeps$ when compared to $\cycfreq f_s$.

We postulate the following standard ordering~\cite{frieman1982nge} of the time and space scales with respect to $\gkeps$\footnote{In this paper, we use the symbol $\sim$ to mean ``is the same order as'' rather than the more usual ``is asymptotically equivalent to''.},
\begin{equation}
\frac{|\delB|}{|\MeanB|} \sim \frac{|\delE|}{|\MeanE|} \sim \frac{\delta f_s}{f_s} \sim 
\frac{k_\parallel}{k_\perp} \sim \frac{\omega}{\cycfreq} \sim \frac{\rho_s}{a} = \gkeps
\end{equation}
Note that when treating the electromagnetic fields, we order the electric field $\MeanE$ and the magnetic field $\MeanB$ as
\begin{equation}
\MeanE \sim \frac{\vth[s]}{c} \MeanB,
\end{equation}
which is equivalent to assuming that the $\MeanE\times\MeanB$ flows are at most sonic and not relativistic. 

From these assumptions, we can make a simple random-walk estimate of the turbulent thermal diffusivity $\chi_{T_s} \sim \rho_s^2 \omega$ (gyro-Bohm diffusion~\cite{dimits:969}) and then order the transport time on the basis of this estimate:
\begin{equation}
\label{TauEScale}
 \frac{1}{\tau_E} \sim \frac{\chi_T}{a^2} \sim \frac{\omega}{\cycfreq}\left(\frac{\rho_s}{a}\right)^2 \cycfreq \sim \gkeps^3 \cycfreq.
	\end{equation}

For the collision operator $\collop[f_s]$, we choose an ordering that allows the plasma to be either collisional or collisionless, namely $\collop[f_s] \sim \omega f_s$. Thus, 
all collision frequencies $\nu$ are ordered
\begin{equation}
\label{nuord}
\frac{\nu}{\cycfreq} \sim \gkeps,
\end{equation}
and any further assumption about the collisionality will be handled as a subsidiary expansion. 
Even though the collision frequency $\nu$ is often smaller than $\omega$, the collision operator $\collop[\delta f_s]$ must be retained as the turbulence will otherwise generate arbitrarily small scales in velocity space~\cite{Krommes2,Tome,schekcrete,abelcollisions,barnesVspace}
\footnote{We choose not to order velocity-space derivatives, which would allow the introduction of smaller collision frequencies whilst retaining $\collop[f_s] \sim \omega f_s$.}.
We will return to this point in \Sref{SFreeEnergy} when discussing free-energy balance.

Finally, we order the source term as $\source \sim F_s / \tau_E$ which restricts us to sources of particles, energy and momentum that do not alter the Maxwellian form of the equilibrium distribution function (found in \Sref{maxback}). In \cite{flowtome3-alphas}, this restriction is relaxed to consider sources that produce high-energy tails and other non-Maxwellian distributions.

We can now expand the mean (large-scale) distribution function $F_s$ and the fluctuating (small-scale) distribution function $\delta f_s$ as follows
\begin{eqnarray}
\label{expandF}
F_s &=& F_{0s} + F_{1s} + F_{2s} + \cdots, \\
		\label{expanddF}
\delta f_s &=&	\delta f_{1s} + \delta f_{2s} +\cdots,
\end{eqnarray}
where $F_{0s} \sim f_s$, $F_{1s} \sim \delta f_{1s} \sim \gkeps f_s$, $F_{2s} \sim \delta f_{2s} \sim \gkeps^2 f_s$, etc.

In the following sections, we insert \eref{expandF} and \eref{expanddF} into the Fokker-Planck equation \eref{vfp} and expand order by order in $\gkeps$.
\section{Zeroth Order \texorpdfstring{$\Or\left(\cycfreq f_s\right)$}{O(0)}}
\label{s0ord}

In this section, we derive the lowest-order implications of imposing the ordering of \Sref{gkord} on the equations of \Sref{kinetic-eq}. 

We start by taking the lowest-order components of \eref{qn} and \eref{ampere}. The former just states that the lowest-order mean densities $n_s$ must satisfy the quasineutrality constraint:
\begin{equation}
\label{quasineutral0}
\sum_s Z_s e n_s = 0,\qquad n_s = \int d^3 \bm{v} F_{0s}.
\end{equation}
Substituting the toroidal decomposition of the magnetic field, \eref{magfield}, in Amp\`ere's law, \eref{ampere}, we merely find that the mean current $\Meanj = \ensav{\current}$ must be zero to this order,
\begin{equation}
\label{amp0}
\Meanj = \sum_s Z_s e n_s \bm{u}_s = 0,\qquad \bm{u}_s = \frac{1}{n_s}\int d^3\bm{v} \bm{v} F_{0s}.
\end{equation}

By taking the lowest-order component of \eref{Edef}, we also learn that the mean electric field is predominantly electrostatic:
\begin{equation}
\label{electrostatic}
\MeanE = -\nabla \pot + \Or\left(\gkeps^2 \frac{\vth[s]}{c} B\right).
\end{equation}

\subsection{Sonic Flows}
\label{constraints}
We now prove that if the plasma has any perpendicular flow faster than the drift velocity then it is a purely toroidal $\MeanE\times\MeanB$ flow. This restriction on the flow has previously been found in the context of neoclassical collisional transport~\cite{connor1987trmt,catto1987ion,hintonwong1985nit, BishopAndCowley}. 

From \eref{vfp}, we find to lowest order in $\gkeps$,
\begin{equation}
\left(\MeanE + \frac{1}{c}\bm{v}\times\MeanB\right)\cdot\pd{F_{0s}}{\bm{v}} = 0.
\end{equation}
Multiplying by $\bm{v}$ and integrating over all velocities, we obtain
\begin{equation}
\label{EXBFlow}
\MeanE + \frac{1}{c}\bm{u}_s\times\MeanB = 0.
\end{equation}
This implies that the perpendicular part of $\bm{u}_s$ is species independent.
Using \eref{EXBFlow} and \eref{electrostatic} we have
\begin{equation}
\MeanE\cdot\MeanB = \MeanB\dg\pot = 0.
\end{equation}
Therefore, to lowest order, the electrostatic potential is a flux function and we can partially solve for $\pot$ as follows:
\begin{equation}
\label{poteq}
\pot = \fpot\left(\psi\right) + \pot_0,
\end{equation}
where $\pot_0 \sim \gkeps \fpot$ (we choose the subscript $0$ because $Z_s e \pot_0 / T_s \sim \gkeps^0$).
There is some arbitrariness in the definition of $\pot_0$ by \eref{poteq} as we can add any function of $\psi$ to it. We resolve this by requiring that $\pot_0$ vanish when averaged over a flux surface:
\begin{equation}
\label{potconstraint}
\fav{\pot_0} = 0,\qquad \fav{\pot} = \Phi(\psi).
\end{equation}

Solving \eref{EXBFlow} for $\bm{u}_s$ shows that any sonic flow that is present is a $\MeanE\times\MeanB$ flow plus some arbitrary parallel flow
\begin{eqnarray}
\label{udecomp}
\bm{u}_s = \frac{c}{B}\Meanb\times\nabla\pot + u_{\parallel s} \Meanb,
\end{eqnarray}
where $\Meanb = \MeanB \left/ \left|\MeanB\right|\right.$.
Using the solution \eref{poteq} for $\pot$ and the toroidal decomposition of the magnetic field, \eref{magfield}, to expand $\Meanb$, we have
\begin{equation}
\label{omdef}
\bm{u}_s = \angvel(\psi) R^2\nabla\tor+\left[u_{\parallel s} - \angvel(\psi) \frac{I}{B} \right]\Meanb, \qquad \angvel(\psi) = c \frac{d\fpot}{d\psi}.
\end{equation}
The first term in \eref{omdef} is a species-independent rigid-body toroidal rotation of each individual flux surface with an angular velocity $\angvel(\psi)$. This will be denoted by
\begin{equation}
\label{torrot}
\bm{u} = \angvel(\psi) R^2 \nabla\tor.
\end{equation}
The second term is a purely parallel flow, which in \Sref{maxback} will be shown to vanish to lowest order. This does not mean that there is no poloidal rotation, but that the poloidal rotation is at most diamagnetic-sized: $\Or(\gkeps\vth)$ rather than $\Or(\vth)$ and, therefore, smaller than the dominant toroidal rotation. 
The full expression for the $\Or(\gkeps\vth)$ flow is given by \eref{flow} and will be needed to determine the lowest-order non-zero component of the mean current in \Sref{ampmag}.
The fact that the mean flow is made up both of the toroidal flow given by \eref{torrot} and the next-order flow given by \eref{flow} can 
obscure the relationship between the lowest-order rotation rate $\angvel(\psi)$ appearing in \eref{torrot} (and in the rest of the multiscale gyrokinetic equations derived in what follows) and the real physical plasma flows that might be measured in an experiment. How to establish this relationship in an accurate way is discussed in \ref{experimentalOmega}.

\section{Rotating Gyrokinetic Variables}
\label{rotvar}
\setcounter{footnote}{1}
Before continuing to expand the kinetic equation \eref{vfp}, it will be convenient to introduce the following new variables~\cite{sugama1998neg,catto1978lgk,antonsenjr2006kel,frieman1982nge,catto1987ion,artun1994nonlinear,krommesGyroReview}:
the guiding-centre position $\bm{R}_s$, particle energy $\energy$, magnetic moment $\magmom$, gyrophase $\gyr$, and the sign of the parallel velocity $\sigma$. The variable transformation of the phase space is
	\begin{equation}
	\left(\bm{r},\bm{v}\right) \rightarrow \left(\bm{R}_s,\energy,\magmom,\gyr,\sigma\right)
	\end{equation}
and the new variables are defined by
\begin{eqnarray}
\label{Rdef}
\bm{R}_s &= \bm{r} - \frac{\Meanb\times\bm{w}}{\cycfreq},\\
			 \label{energydef}
\energy &= \frac{1}{2} m_s v^2 + Z_s e \left[ \fpot(\psi) + \pot_0 \right] - Z_s e\fpot\left(\psistar\right),\\
			  \label{mudef}
\magmom &= \frac{m_s w_\perp^2}{2\MeanMagB},\\
\sigma &= \frac{w_\parallel}{\left|w_\parallel\right|},
\end{eqnarray}
where $\bm{w}$ is the peculiar velocity of the particles with respect to the toroidal rotation:
\begin{equation}
\label{wDef}
\bm{w} = \bm{v} - \bm{u}= w_\parallel \Meanb + w_\perp \left(\cos \gyr\,\bm{e}_2 - \sin\gyr\,\bm{e}_1 \right),
\end{equation}
with the toroidal velocity determined by \eref{torrot},  $\bm{e}_1$ and $\bm{e}_2$ arbitrary orthogonal unit vectors perpendicular to the magnetic field (with $\Meanb = \bm{e}_2\times\bm{e}_1$),\footnote{There might be a concern that we might not be able to define such an $\bm{e}_1$ and $\bm{e}_2$. This is resolved in \cite{burby2012}, where it is proved that for toroidal confinement devices, one can always make a globally valid choice of these vectors.} and $\psi^*$ is a flux-like quantity proportional to the toroidal canonical angular momentum:
\begin{equation}
\label{psistardef}
\fl
\psistar(\bm{r},\bm{v}) = \psi(\bm{r})+ \frac{m_s c}{Z_s e}\left( \bm{v}\dg\tor \right) R^2 = \psi + \frac{m_s c}{Z_s e}\left( \bm{w}\dg\tor \right) R^2 + \frac{\MeanMagB R^2\angvel(\psi)}{\cycfreq}.
\end{equation}

The new velocity variables $\energy$ and $\magmom$ are closely related to conserved quantities of the particle motion in the mean electromagnetic fields.
The magnetic moment $\magmom$ is related to the first adiabatic invariant of the particle gyromotion\footnote{If $\MeanE$ and $\MeanB$ were constant in space and time, then $\magmom[s]$ would be precisely the adiabatic invariant associated with a particle's Larmor gyration. In the more general fields we consider here, $\magmom[s]$ as given by \eref{mudef} is the first term in the infinite asymptotic series for the exact invariant~\cite{kruskal1958gyration,alfven,hastie1967equilibrium}.}.
The energy variable, $\energy$, is constructed from the conserved total energy and the conserved quantity $\psistar$ so that it is the energy in the rotating frame to lowest order. This can be seen by expanding $\fpot(\psistar)$ around $\fpot(\psi)$, yielding
\begin{equation}
\label{eloword}
\energy = \frac{1}{2} m_s w^2 - \frac{1}{2} m_s \angvel^2(\psi) R^2 + Z_s e \pot_0 + \Or(\gkeps T_s).
\end{equation}

The kinetic equation \eref{vfp} can be written in these new variables as
\begin{equation}
\label{newvfp}
\frac{d f_s}{d t} = \pd{f_s}{t} + \dot{\bm{R}_s}\cdot\pd{f_s}{\bm{R}_s} + \dmu\pd{f_s}{\magmom} + \denergy\pd{f_s}{\energy} + \dot{\gyr}\pd{f_s}{\gyr} = \collop[f_s] + \source,
\end{equation}
where $\dot{g} = \left.{d g}\right/{d t}$ denotes the time derivative of $g$ along a particle orbit.
This form of the kinetic equation follows from the fact that $df_s/dt$ should be independent of the coordinate system that we use to describe the phase space. If we choose some set of variables $\bm{z}$ instead of $(\bm{r},\bm{v})$ then $f_s = f_s(\bm{z},t)$ and so 
$\infrac{df_s}{dt} = \inpd{f_s}{t} +  \dot{\bm{z}}\cdot \left( \inpd{f_s}{\bm{z}}\right)$. As this must be independent of our choice of $\bm{z}$, \eref{vfp} implies \eref{newvfp}. For more details, see the discussion surrounding equations (6) and (7) of \cite{littlejohn1984relativistic}.
Explicit expressions for $\dot{\bm{R}_s}$, $\dmu$, $\denergy$,  and $\dot{\gyr}$ are derived in \ref{apA}; see \eref{rdot}, \eref{mudot}, \eref{edot}, and \eref{gyrdot}.

\subsection{The Gyroaverage}
Transformation from the spatial coordinate $\bm{r}$ to the guiding-centre position $\bm{R}_s$ (known as the Catto transformation~\cite{catto1978lgk}) allows us to introduce gyroaveraging --- a crucial mathematical device that will enable us to close the equations for $F_s$ and $\delta f_s$. We define the gyroaverage of a quantity $g$ by
\begin{equation}
\label{gyroaverage}
\gyroR{g} = \frac{1}{2\pi} \oint d\gyr g(\bm{R}_s,\energy,\magmom,\gyr,\sigma),
\end{equation}
where the integral is performed holding $\bm{R}_s$, $\energy$, $\magmom$ constant.

\subsection{Derivatives and Integrals}
\label{derivInt}
All quantities that have velocity-space dependence will be functions of $\bm{R}_s$, $\energy$, $\magmom$, $\gyr$ and $\sigma$. The quantities that are only functions of space will be evaluated at the spatial position $\bm{r}$, unless explicitly stated otherwise, with $\bm{r}$ considered as a function of $\bm{R}_s,\magmom,\gyr$. In what follows, $\nabla$ is reserved for derivatives with respect to $\bm{r}$, while derivatives with respect to $\bm{R}_s$ will be written explicitly as $\partial\left/ \partial \bm{R}_s \right.$.

As the variables we have chosen are adapted to the particle motion, we will not transform the field equations into these variables. The fields are functions
of space but not velocity so they remain functions of $\bm{r}$ and
$t$ in accordance with the principle stated above. Therefore, the integrals over velocity in the definitions of density, \eref{cdensDef}, and current, \eref{curDef}, are to be taken at constant $\bm{r}$, 
so functions of $\bm{R}_s$, $\energy$ and $\magmom$ appearing in the integrand are to be considered as functions
of $\bm{r}$ and $\bm{v}$ via the definitions \eref{Rdef}, \eref{energydef} and \eref{mudef}. Namely, for any function $g$,
\begin{equation}
\label{wintdef}
\fl
\wint g(\bm{R}_s,\energy,\magmom,\gyr,\sigma) = \wint g(\bm{R}_s(\bm{r},\bm{w}),\energy(\bm{r},\bm{w}),\magmom(\bm{r},\bm{w}),\gyr(\bm{r},\bm{w}),\sigma(\bm{r},\bm{w})).
\end{equation}
This is how all velocity-space integrals will be performed in this paper unless explicitly stated otherwise.
With this definition, integrals over $\bm{v}$ and $\bm{w}$ at constant $\bm{r}$ are equivalent.

Finally, as promised in \Sref{maggeom}, we now give a precise mathematical formulation of the assumption of axisymmetry of mean fields as applied to the mean distribution function:
\begin{equation}
\left(\nabla\tor\right)\left.\dgR{}\right|_{\energy,\magmom,\gyr} F_s = 0,
\label{lowestaxi}
\end{equation}
which, to lowest order in $\gkeps$, is equivalent to
\begin{equation}
\left(\nabla\tor\right)\left.\dg\right|_{w_\parallel,w_\perp,\gyr} F_s = 0.
\end{equation}
However, this is not equivalent to $\left(\nabla\tor\right)\left.\dg\right|_{\bm{w}} F_s = 0$: indeed in \eref{wDef}, the basis vectors $\Meanb$, $\bm{e}_1$ and $\bm{e}_2$ possess non-zero variation in the $\tor$ direction.

\subsection{Gyrotropy of \texorpdfstring{$F_{0s}$}{F0}}
To zeroth order, the kinetic equation \eref{newvfp} turns out to be just
\begin{equation}
\label{f0indep}
\cycfreq \left.\pd{F_{0s}}{\gyr}\right|_{\bm{R}_s,\magmom,\energy} = 0.
\end{equation}
Thus, $F_{0s}$ is gyrophase independent. Note that this means that, to lowest order,
\begin{equation}
\label{noflow}
\wint \bm{w}_\perp F_{0s} = 0,
\end{equation}
confirming the earlier result that the lowest-order perpendicular velocity is 
completely contained in $\bm{u}_s$ as defined by~\eref{omdef}.

\section{First Order \texorpdfstring{$\Or\left(\gkeps\cycfreq f_s\right)$}{O(1)}: The Maxwell-Boltzmann Equilibrium}
\label{s1ord}
In this section, we start from the kinetic equation \eref{newvfp}, and expand it to first order. Using the lowest-order expressions for $\dot{\bm{R}_s}$, $\denergy$ and $\dot{\gyr}$ given by \eref{low-rdot}, \eref{low-edot}, and \eref{gyrdot} respectively we find
\begin{equation}
\fl
\label{ke:1}
\left(w_\parallel \Meanb +  \bm{u}\right)\cdot\pd{F_{0s}}{\bm{R}_s} + \dmu\pd{F_{0s}}{\magmom}- Z_s e \bm{w}_\perp \dg \gkupot \pd{F_{0s}}{\energy}  + \cycfreq \pd{}{\gyr}\left({F_{1s} + \delta f_{1s}}\right) = \collop[F_{0s}],
\end{equation}
where $\gkupot$ is the fluctuating electrostatic potential in the toroidally-rotating frame, given by
\begin{equation}
\label{etaDef}
\gkupot = \delpot - \frac{1}{c} \bm{u}\cdot\delA,
\end{equation}
and $\bm{u}$ without the species index is the toroidal-rotation component of $\bm{u}_s$ defined by~\eref{torrot}.

In the following subsections, we analyse \eref{ke:1} to discover that:
the lowest-order mean distribution function $F_{0s}$ is Maxwellian;
the bulk motion is the purely toroidal rotation of flux surfaces, i.e., the second term in \eref{omdef} vanishes and 
$\bm{u}_s = \bm{u}$;
 and $\delta f_s$ consists of the Boltzmann response to $\gkupot$ and a gyrophase-independent distribution of charged rings $h_s(\bm{R}_s,\magmom,\energy,\sigma,t)$.
Mathematically, we can consider \eref{ke:1} to be a first-order differential equation in $\gyr$ for $F_{1s} + \delta f_{1s}$. Then the results of \Sref{maxback} regarding $F_{0s}$ can be viewed as solubility constraints required for \eref{ke:1} to have single-valued solutions for the $\gyr$-dependence of $F_{1s}$ and $\delta f_{1s}$. The results of Sections \ref{SgyrF1} and \ref{boltzresp} then follow from solving \eref{ke:1} provided that the constraints are satisfied.

\subsection{Maxwellian Equilibrium}
\label{maxback}
We first prove that $F_{0s}$ is Maxwellian.
Gyroaverging \eref{ke:1} and using the fact that, to lowest order, $\gyroR{\dmu} = 0$ (see \eref{low-mudot}), $\gyroR{\bm{w}_\perp\dg\gkupot} = 0$ (see \eref{wperpdg}), and \mbox{$\bm{u}\cdot\partial F_{0s}\left/\partial\bm{R}_s\right. = 0$} (axisymmetry), we obtain
\begin{equation}
\label{avke1}
w_\parallel \Meanb \cdot\pd{F_{0s}}{\bm{R}_s} = \gyroR{\collop[F_{0s}]}.
\end{equation}
Multiplying this by $1+\ln F_{0s}$, integrating over all velocities and averaging over the flux surface, we find
\begin{equation}
\label{entropprodfs}
\fav{ \wint w_\parallel \Meanb \cdot\ddR{}\left( F_{0s}\ln F_{0s}\right)} = \fav{\wint  \ln F_{0s}\collop[F_{0s}]}.
\end{equation}
To lowest order, we can replace all instances of $\bm{R}_s$ on the left-hand side of this equation with $\bm{r}$ and write the velocity-space integral in terms of integrals over $\energy$, $\magmom$, $\gyr$ and $\sigma$ at constant $\bm{r}$ using
\begin{equation}
\wint  = \sum_\sigma\int \frac{\MeanMagB(\bm{r}) d\energy d\magmom d\gyr}{m_s^2 \left|w_\parallel\right|} + \Or(\gkeps\vth^3).
\end{equation}
This gives
\begin{equation}
\fl
\begin{eqalign}{
\fav{ \wint w_\parallel \Meanb \cdot\ddR{}\left( F_{0s}\ln F_{0s}\right)} &= \fav{ 2\pi\sum_{\sigma} \int \frac{ \MeanMagB d\energy d\magmom}{m_s^2 |w_\parallel|} w_\parallel\Meanb\dg\left( F_{0s}\ln F_{0s}\right)}\\
&= \fav{ 2\pi\sum_\sigma \sigma \int \frac{d\energy d\magmom}{m_s^2} \MeanB\dg\left(F_{0s}\ln F_{0s}\right)} \\
&= \fav{ \dv\left( \Meanb\wint w_\parallel F_{0s} \ln F_{0s} \right)}.
}\end{eqalign}
\label{flibble}
\end{equation}
By using \eref{favdiv} to express the action of the flux-surface average on a divergence and the fact that $\Meanb\dg\psi = 0$, we conclude that the above expression vanishes. Therefore, from~\eref{entropprodfs},
\begin{equation}
\label{entrop}
\fav{\wint  \ln F_{0s}\collop[F_{0s}]} = 0.
\end{equation}
By Boltzmann's $H$-Theorem, \eref{entrop} implies that $F_{0s}$ is a local Maxwellian~\cite{helander2002ctm}. We know that, by definition, this Maxwellian has density $n_s$, temperature $T_s$ and velocity $\bm{u}_s$, where $\bm{u}_s$ is given by \eref{omdef}.\footnote{Formally, \eref{entrop} implies that all species have the same temperature and mean velocity. We will show that they do indeed have the same mean velocity, but we will gratuitously retain the species dependence of the temperature. This will only be of importance when we come to discuss heat transport and so we defer the discussion of interspecies temperature differences to the end of \Sref{heattrans}.}
Thus, $F_{0s}$ can be written as
\begin{equation}
\fl
\label{f0tmp}
F_{0s} = n_s(\bm{r}) \left[\frac{m_s}{2\pi T_s(\bm{r})}\right]^{3/2} \exp\left\{-\frac{m_s\left[w^2 - 2 m_s w_\parallel \tu(\bm{r}) + \tu^2(\bm{r})\right]}{ 2T_s(\bm{r})}\right\},
\end{equation}
where
\begin{equation}
\tu = u_{\parallel s} - \frac{\angvel(\psi) I}{B}.
\end{equation}
This form does not contradict the condition \eref{f0indep} that $F_{0s}$ must be gyrophase-independent whilst holding $\bm{R}_s$, $\energy$ and $\magmom$ fixed because the difference between $\bm{R}_s$ and $\bm{r}$ is higher order. 
However, we still wish to have $F_{0s}$ expressed in the $(\bm{R}_s, \energy, \magmom, \sigma)$ variables. We accomplish this by using \eref{eloword} to express $m_s w^2$ in terms of $\energy$ and find
\begin{equation}
\fl
F_{0s} = {\NotN(\bm{R}_s)} \left[\frac{m_s}{2\pi T_s(\bm{R}_s)}\right]^{3/2} \exp\left[ - \frac{\energy}{T_s(\bm{R}_s)} - \frac{m_s w_\parallel \tu(\bm{R}_s)}{T_s(\bm{R}_s)}  \right] +\Or(\gkeps F_s),
	\label{eq87}
\end{equation}
where
\begin{equation}
	\NotN = n_s\exp\left[ -\frac{m_s \angvel^2(\psi) R^2}{2T_s} + \frac{Z_s e \pot_0}{T_s} + \frac{m_s\tu^2}{2T_s}\right],
		\label{ntwiddles}
\end{equation}
and we have used the fact that to lowest order the mean fields taken at the guiding-centre position $\bm{R}_s$ are the same as when taken
at the particle position $\bm{r}$.

Inserting $F_{0s}$, given by \eref{f0tmp}, back into \eref{avke1} and dividing through by $F_{0s}$ we get
\begin{equation}
\fl\begin{eqalign}{
	\label{badgermushroom}
	w_\parallel \Meanb\dg \left( \ln \NotN - \frac{3}{2} \ln T_s \right) + \frac{w_\parallel \energy}{T_s^2} \Meanb\dg T_s - w_\parallel \Meanb\dg\left(\frac{m_s w_\parallel \tu}{T_s}\right) = 0.
}\end{eqalign}
\end{equation}
This equation must hold for all velocities $\bm{w}$, so each term in this equation must vanish independently. Tackling the second term first, we see that the temperature $T_s$ must be a flux function to lowest order:
\begin{equation}
\Meanb\dg T_s = 0,
\end{equation}
so $T_s = T_s(\psi)$.
For the first term to vanish the same must be the case for $\NotN$:
\begin{equation}
\Meanb\dg\NotN = 0,
\end{equation}
so $\NotN = \NotN(\psi)$.
Turning now to the third term of \eref{badgermushroom}, we see that
\begin{equation}
2 w_\parallel^2 \Meanb\dg \tu + \tu\Meanb\dg w_\parallel^2  = 0,
\end{equation}
which can only be solved for arbitrary $w_\parallel$ by $\tu = 0$.
Thus, the background Maxwellian only depends on $\energy$ and so is
isotropic in $\bm{w}$ and the lowest-order (sonic) flow is a pure toroidal rotation: $\bm{u}_s = \bm{u} = \angvel(\psi)R^2\nabla\tor$.
As this flow is species-independent there are no currents in $F_{0s}$, since the plasma is quasineutral \eref{quasineutral0}. This is consistent with the lowest-order Amp\`ere's Law \eref{amp0}.
The vanishing of the parallel component of \eref{omdef} is in accord with the well known result~\cite{hintonwong1985nit,helander2002ctm,hintonrosenbluth} that poloidal flow is strongly damped on the ion-ion collision time, which, in our ordering is, indeed, $1 / \gkeps^2$ shorter than the timescale of the evolution of the bulk flow.

We can finally write the complete solution for $F_{0s}$ as
\begin{equation}
\label{F0R}
F_{0s} = \NotN(\psi(\bm{R}_s)) \left[\frac{m_s}{2\pi T_s(\psi(\bm{R}_s))}\right]^{3/2} e^{- \energy / T_s(\psi(\bm{R}_s))}.
\end{equation}
This form of the distribution function is manifestly gyrophase independent holding $\bm{R}_s$ and $\energy$ constant and so is consistent with \eref{f0indep}. Note that \eref{F0R} is now the definition of $F_{0s}$ to all orders in our expansion. Thus, any small terms neglected previously (in, e.g., \eref{eq87}), will be automatically included in $F_{1s}$ by virtue of using \eref{F0R} for $F_{0s}$ in the equations that determine $F_{1s}$.
\subsection{Gyrotropy of \texorpdfstring{$F_{1s}$}{F1}}
\label{SgyrF1}
Substituting \eref{F0R} back into the first-order kinetic equation \eref{ke:1}, we find
\begin{equation}
\label{ke:1:redo}
\cycfreq \pd{}{\gyr}\left( F_{1s} + \delta f_{1s} \right) = - \frac{Z_s e}{T_s}\bm{w}_\perp \dg \gkupot F_{0s}.
\end{equation}
Averaging this over the fluctuations gives
\begin{equation}
\cycfreq \left.\pd{F_{1s}}{\gyr}\right|_{\bm{R}_s,\magmom,\energy} = 0.
\label{F1indep}
\end{equation}
Thus, $F_{1s}$ is gyrophase independent.

\subsection{Poloidal Density Variation}
\label{poldens}
Rearranging \eref{ntwiddles} and using $\tu = 0$, we see that the physical density of a species is given by
\begin{equation}
\label{npol}
n_s = \NotN \left(\psi\right)\exp\left[ \frac{m_s \angvel^2(\psi) R^2}{2T_s} - \frac{Z_s e \pot_0}{T_s} \right] + \Or(\gkeps n_s),
\end{equation}
which is not a flux function, unless the rotation is subsonic (see \Sref{LowMach})~\cite{hintonwong1985nit,helander2002ctm}. We can insert this into the lowest-order quasineutrality condition \eref{quasineutral0} to determine $\pot_0$, subject to the constraint that $\fav{\pot_0} = 0$ (by definition, see \eref{potconstraint}):
\begin{equation}
\label{qn0}
\sum_s Z_s \NotN(\psi) \exp\left[ \frac{m_s \angvel^2(\psi) R^2}{2T_s} - \frac{Z_s e \pot_0}{T_s} \right]  = 0.
\end{equation}
Physically, this demonstrates that the poloidal variation of the density on a flux surface is governed by the balance between centrifugal forces which sweep heavy particles (i.e., ions) to the outboard side (higher $R$) and the electrostatic potential that is set up within the flux surface to maintain quasineutrality.
\subsection{The Boltzmann Response}
\label{boltzresp}
Taking the lowest-order fluctuating components of \eref{Edef} and \eref{Bdef}, we see that the fluctuating electric field in the toroidally rotating frame is electrostatic:
\begin{equation}
\label{emov}
\delE + \frac{1}{c}\bm{u}\times\delB = -\nabla_\perp \gkupot + \Or(\gkeps^2 \MeanE),
\end{equation}
where $\gkupot$ is defined in \eref{etaDef}.
Using \eref{F1indep}, the first-order kinetic equation \eref{ke:1:redo} is
\begin{equation}
\cycfreq \pd{\delta f_{1s}}{\gyr}  = -\frac{Z_s e}{T_s} \bm{w}_\perp \dg \gkupot F_{0s}.
\end{equation}
We can integrate this with respect to $\gyr$, using \eref{wperpdg1} for $\gkupot$:
\begin{equation}
\bm{w}_\perp\dg\gkupot = \cycfreq \left.\pd{\gkupot}{\gyr}\right|_{\bm{R}_s,\energy,\magmom},
\end{equation}
and obtain
\begin{equation}
\label{hindep}
\delta f_{1s} = -\frac{Z_s e\gkupot(\bm{r})}{T_s} F_{0s} + h_s\left(\bm{R}_s,\magmom,\energy,\sigma,t\right),
\end{equation}
where the constant of gyrophase integration $h_s$ is the gyrophase-independent distribution of Larmor rings, which will completely describe the kinetics of the small-scale turbulence.
The gyrophase-dependent part of $\delta f_{1s}$ is just the Boltzmann response in the moving frame to the electrostatic field \eref{emov}.

\subsection{Summary: The Lowest-Order Solution}
\label{s1sum}
Collating the results of the previous subsections we have the solution for $f_s$ to first order in $\gkeps$:
\begin{eqnarray}
\label{fSoln}
f_s &=& F_s + \delta f_s,\\
\label{FSoln}
F_s &=& F_{0s} \left( \psi(\bm{R}_s), \energy \right) + F_{1s}\left( \bm{R}_s,\energy,\magmom,\sigma\right) + \Or\left(\gkeps^2 f\right),\\
		 \label{hDef}
\delta f_s &=& - \frac{Z_s e}{T_s} \gkupot(\bm{r}) F_{0s} + h_s\left(\bm{R}_s,\energy,\magmom,\sigma\right) + \Or\left(\gkeps^2 f\right),\\
\label{F0def}
F_{0s} &=& \NotN(\psi(\bm{R}_s)) \left[\frac{m_s}{2\pi T_s(\psi(\bm{R}_s))}\right]^{3/2} e^{- \energy / T_s(\psi(\bm{R}_s))},
\end{eqnarray}
In the next section, we find equations that determine the gyrophase-independent functions $F_{1s}$ and $h_s$.
\section{Second Order \texorpdfstring{$\Or\left(\gkeps^2\cycfreq f_s\right)$}{O(2)}: Neoclassical Theory and Gyrokinetics}
\label{s2ord}
In order to completely determine the distribution function $f_s$ given by \eref{fSoln}, we need equations for $F_{1s}$ and $h_s$.
In order to find them, let us substitute the form of $f_s$ summarized in \Sref{s1sum} into the exact kinetic equation \eref{newvfp} and keep only terms up to 
second order in $\gkeps$:
\begin{equation}
\fl
\label{2ndord}
\begin{eqalign}{
\pd{h_{s}}{t} &+ \left({\dot{\bm{R}}_s}\cdot\pd{}{\bm{R}_s}+\dmu\pd{}{\magmom} + \denergy\pd{}{\energy}\right)\left( F_{0s} + F_{1s} + h_s \right)   =\\
&-\cycfreq \pd{}{\gyr}\left( F_{2s} + \delta f_{2s}\right) + \frac{d}{dt}\left(\frac{Z_s e\gkupot}{T_s} F_{0s}\right)+ \collop[F_{0s} + F_{1s} + h_s],
}\end{eqalign}
\end{equation}
where $d/dt$ is the full derivative along a particle orbit as in \eref{newvfp} and we have used the gyrophase independence of $F_{0s}$ \eref{f0indep}, $F_{1s}$ \eref{F1indep} and $h_s$ \eref{hindep}. 

\Eref{2ndord} contains $F_{2s}$ and $\delta f_{2s}$, about which we have no information. However, they only occur under a derivative with respect to $\gyr$ and so we gyroaverage \eref{2ndord} to find a closed equation that does not contain any second-order distribution functions:
\begin{equation}
\label{2complex}
\fl
\begin{eqalign}{
\pd{h_{s}}{t} &+ \gyroR{\dot{\bm{R}}_s}\cdot\pd{}{\bm{R}_s}\left( F_{0s} + F_{1s} + h_s \right) =\\
	&\gyroR{\denergy} \frac{F_{0s}}{T_s} +  \gyroR{\frac{d}{dt}\left(\frac{Z_s e\gkupot}{T_s} F_{0s}\right)} + \gyroR{\collop[F_{0s} + F_{1s} + h_s]},
}\end{eqalign}
\end{equation}
where we have used 
$F_{0s}$ given by \eref{F0def} and the fact, derived in \ref{apA}, that $\gyroR{\dmu} = \Or({\gkeps^2 \cycfreq \magmom})$ (see \eref{low-mudot}) and $\gyroR{\denergy} = \Or({\gkeps^2 \cycfreq T_s})$
(see \eref{edot-low-gyr}).
As we did for \eref{ke:1}, we can formally consider \eref{2ndord} to be an equation determining the $\gyr$-dependence of $F_{2s}$ and $\delta f_{2s}$. Under this interpretation, \eref{2complex} is the solubility constraint that must be satisfied if the $F_{2s}$ and $\delta f_{2s}$ found from \eref{2ndord} are to be single-valued in $\gyr$.

The gyroaverages appearing in \eref{2complex} are calculated in \ref{apA}: we use \eref{rdotg} for $\gyroR{\dot{\bm{R}}_s}$ and \eref{localboy} for the combination of gyroaverages appearing on the right-hand side:
\begin{equation}
\fl\begin{eqalign}{
\left[\pd{}{t}+\bm{u}(\bm{R}_s)\dgR{}\right] h_s + w_\parallel\Meanb\cdot\ddR{}\left(F_{1s}+h_s\right) + \left(\vdrift + \vchiR\right) \cdot\ddR{}\left( F_{0s} + h_s \right)\\
	=  \frac{Z_s e F_{0s}}{T_s}\left[\pd{}{t} +\bm{u}(\bm{R}_s)\dgR{}\right]\gyroR{\gkpot} - \frac{m_sF_{0s}}{T_s}\left[\frac{Iw_\parallel}{B} + \angvel(\psi) R^2\right]\frac{d\angvel}{d\psi}\vchiR\dg\psi\\
	\qquad-\frac{Z_s e}{T_s c} w_\parallel F_{0s}\pd{\MeanA}{t} \cdot\Meanb
	+\gyroR{\collop[F_{0s} + F_{1s} + h_s]},
}\end{eqalign}
\label{2ndorder}
\end{equation}
where we have defined the gyrokinetic potential
\begin{equation}
\gkpot = \delpot - \frac{1}{c}\bm{v}\cdot\delA = \gkupot - \frac{1}{c} \bm{w}\cdot\delA,
\label{chidef}
\end{equation}
the associated fluctuating velocity field\footnote{
This can be shown to consist, physically, of the fluctuating $\MeanE\times\MeanB$ drift in the rotating frame, the motion of guiding centres along fluctuating field lines and the fluctuating $\nabla B$ drift. This is proved in detail for gyrokinetics in a non-rotating slab in \cite{howes2006agb}. }
\begin{equation}
\vchi = \frac{c}{\MeanMagB}\Meanb\times\nabla\gkpot,\qquad \vchiR = \frac{c}{\MeanMagB} \Meanb\times\ddR{\gyroR{\gkpot}} + \Or(\gkeps^2\vth),
\label{vchi}
\end{equation}
and the guiding-centre drift velocity
\begin{equation}
\label{vdrift}
\begin{eqalign}{
\vdrift = \frac{\Meanb}{\cycfreq} \times &\left[ w_\parallel^2 \Meanb\dg\Meanb + \frac{1}{2}w_\perp^2\nabla \ln \MeanMagB\right.\\
	 &\left.- \angvel^2(\psi) R\nabla R - 2 w_\parallel \angvel(\psi) \Meanb\times\nabla z +\frac{Z_s e}{m_s} \nabla\pot_0\right],
}\end{eqalign}
\end{equation}
which consists of, in order of appearance in \eref{vdrift}, the curvature drift, the $\nabla B$ drift, the centrifugal drift, the Coriolis drift, and the mean first-order $\MeanE\times\MeanB$ drift.

The detailed derivation of \eref{2ndorder} is given in \ref{apA6}. In the rest of this section, we will use \eref{2ndorder} to derive a closed solution for $F_{1s}$ and $h_s$, accurate to first order in $\gkeps$, in terms of the unknown functions $\NotN(\psi)$, $T_s$, $\angvel(\psi)$, $I$ and $\psi$, which parametrise the equilibrium.
\Eref{2ndorder} can be split into a mean equation (see \eref{neoclass}), which will determine $F_{1s}$, and a fluctuating part (see \eref{gke}), which will determine $h_{s}$.
We will find that the mean equation is precisely the neoclassical drift-kinetic equation~\cite{catto1987ion} and the fluctuating equation is the gyrokinetic equation for a rotating plasma~\cite{sugama1998neg,brizardFlow}.
These two equations are closed by the mean and fluctuating Maxwell equations, which relate $\psi$ and $I$ to $F_{1s}$ (Sections \ref{ampmag} and \ref{magevolve}) and the fluctuating fields to $h_s$ (\Sref{Sfmag}).

\subsection{Neoclassical Distribution Function}
\label{nclasse}
Averaging \eref{2ndorder} over the fluctuations, we obtain
\begin{equation}
\fl
\label{neoclass}
w_\parallel \Meanb\cdot\ddR{F_{1s}} - \gyroR{\lincol[F_{1s}]} = - \vdrift\cdot\ddR{F_{0s}} -\frac{Z_s e}{T_s c} w_\parallel F_{0s} \pd{\MeanA}{t} \cdot \Meanb + \gyroR{\collop[F_{0s}]},
\end{equation}
where we have introduced the linearised collision operator $\lincol$, linearised about the Maxwellian part of $F_{0s}$ -- for more details on linearised collision operators see \cite{helander2002ctm}.
\Eref{neoclass} is just the usual neoclassical drift-kinetic equation~\cite{catto1987ion,helander2002ctm,hintonwong1985nit}.
Note that $\collop[F_{0s}]$ is only zero to lowest order, so in \eref{neoclass} it represents collisions acting on the first-order departures of $F_{0s}$, as defined formally by \eref{F0R}, from a pure Maxwellian 
(having to do with the absorption into $F_{0s}$ of some non-Maxwellian velocity dependence via $\psi(\bm{R}_s)$ and $\energy$; see \Sref{maxback} and \ref{apF}).

We now wish to solve \eref{neoclass} by first separating particular solutions corresponding to some of the source terms in its right-hand side. To deal with the first of these terms, we let
\begin{eqnarray}
\label{F1Hat}
F_{1s} &=& \FHat(\bm{R}_s, \energy, \magmom, \sigma) + \Fstar,\\
	\label{Fstardef}
\Fstar &=& \frac{m_s c}{Z_s e}\left[\frac{Iw_\parallel}{B} + \angvel(\psi) R^2\right] \pd{F_{0s}}{\psi},
\end{eqnarray}
where
\begin{equation}
\begin{eqalign}{
w_\parallel \Meanb\cdot\ddR{\Fstar} &= w_\parallel \Meanb\dgR{}\left[\frac{Iw_\parallel + \angvel(\psi)R^2 \MeanMagB}{\cycfreq}\right] \pd{F_{0s}}{\psi} \\
	&= - \left(\vdrift\dg\psi\right) \pd{F_{0s}}{\psi} = - \vdrift\dgR{F_{0s}},
}\end{eqalign}
\end{equation}
where we have used the well-known identity~\cite{hintonwong1985nit}
\begin{equation}
\label{VDpsi}
\vdrift\dg\psi = -w_\parallel \Meanb\left.\dg\right|_{\energy,\magmom,\gyr}\left[ \frac{Iw_\parallel}{\cycfreq} + \frac{\MeanMagB R^2}{\cycfreq} \angvel(\psi)\right].
\end{equation}
Therefore,
\begin{equation}
\fl
w_\parallel \Meanb\cdot\ddR{\FHat} - \gyroR{\lincol[\FHat]} = -\frac{Z_s e}{T_s c} w_\parallel F_{0s} \pd{\MeanA}{t} \cdot \Meanb + \gyroR{\collop[F_{0s} + \Fstar]}.
\label{FHatEq}
\end{equation}

We now wish to replace the term depending on $\MeanA$ with a source term that depends on $\fav{\MeanE\cdot\MeanB}$, which will later turn out to be opportune (see  Sections \ref{ampmag} and \ref{magevolve}).
To do this, we follow \cite{gabethesis,ericthesis,bernstein1974tia} and introduce the ansatz
\begin{eqnarray}
\fl
\widehat{F}_{1s} &=& \Fneo(\bm{R}_s,\energy,\magmom,\sigma) - \frac{Z_s e}{T_s} F_{0s} \int^l dl' \left( \MeanMagB\frac{\fav{\MeanE\cdot\MeanB}}{\fav{\MeanMagB^2}} + \frac{1}{c} \pd{\MeanA}{t}\cdot\Meanb\right),
\label{Fneodef}
\end{eqnarray}
where the integral is performed along a fixed field line and $l'$ is the distance along that field line. 
Note that the second term in \eref{Fneodef} carries no current.
Inserting \eref{Fneodef} into \eref{FHatEq}, we find
\begin{equation}
\label{neogeq}
\fl
w_\parallel \Meanb\cdot\ddR{\Fneo} - \gyroR{\lincol[\Fneo]} = 2\cycfreq \frac{w_\parallel c}{\vth[s]^2} \frac{\fav{\MeanE\cdot\MeanB}}{\fav{\MeanMagB^2}} F_{0s}+\gyroR{\collop[{F_{0s}} + \Fstar]},
\end{equation}
where the term arising from the insertion of the second term in \eref{Fneodef} into the collision operator is small, $\Or(\gkeps^3 \cycfreq F_{0s})$, and hence has been neglected.
Finally, we split
\begin{eqnarray}
\label{neog}
\Fneo &= \FE \frac{c}{\vth[s]} \frac{\fav{\MeanE\cdot\MeanB}}{\fav{\MeanMagB^2}} + \Fnct,
  \end{eqnarray}
where $\FE$ and $\Fnct$ satisfy
\begin{eqnarray}
	\label{neo1}
	w_\parallel \Meanb\cdot\ddR{\FE} - \gyroR{\lincol[\FE]} &=& 2\cycfreq \frac{w_\parallel}{\vth[s]} F_{0s},\\
	\label{neo2}
	w_\parallel \Meanb\cdot\ddR{\Fnct} - \gyroR{\lincol[\Fnct]} &=& \gyroR{\collop[F_{0s} + \Fstar]}.
\end{eqnarray}
This pair of equations can now be solved without knowing $\fav{\MeanE\cdot\MeanB}$ and then $\Fneo$ is given in terms of $\fav{\MeanE\cdot\MeanB}$ by \eref{neog}. This will be used in solving for the evolution of the mean magnetic field in \Sref{ampmag} and \Sref{magevolve} (where it is explained how $\fav{\MeanE\cdot\MeanB}$ is calculated).

\subsection{Amp\`ere's Law and the Magnetic Equilibrium}
\label{ampmag}
The average of Amp\`ere's Law \eref{ampere} over the fluctuations is
\begin{equation}
\label{avgampall}
\Meanj = \frac{c}{4\pi} \curl\MeanB,\qquad \Meanj = \sum_s Z_s e \wint \bm{w} F_s,
\end{equation}
where $\Meanj$ is first-order (to lowest order, $\Meanj = 0$; see \eref{amp0}).
Using the axisymmetric form for $\MeanB$ \eref{magfield}, we can express $\curl\MeanB$ in terms of derivatives of $I$ and $\psi$:
\begin{equation}
	\Meanj = \frac{c}{4\pi} \left[ \nabla I \times \nabla \tor - \left( \Delta^* \psi  \right)\nabla\tor\right],
\label{avgamp}
\end{equation}
where $\Delta^*$ is the Grad-Shafranov operator
\begin{equation}
\Delta^* \psi =\left( \pd{^2}{R^2} - \frac{1}{R} \pd{}{R} + \pd{^2}{z^2} \right) \psi.
\end{equation}

We can now use the complete solution for $F_s$ given by \eref{FSoln} to determine $\Meanj$. This is done in \ref{foflow} (see \eref{current}):
\begin{equation}
\fl
\label{main-current}
\begin{eqalign}{
\Meanj = c R^2 &  \sum_s n_s\left\{ T_s\frac{d \ln \NotN}{d\psi} + \left[Z_se \pot_0 - \frac{1}{2}{m_s\angvel^2(\psi) R^2} + T_s \right]\frac{d\ln T_s}{d\psi}\right\}\nabla\tor\\
	&+cR^2\left( \sum_s m_s n_s R^2 \right) \angvel(\psi) \frac{d\angvel}{d\psi} \nabla\tor + K(\psi) \MeanB,
}\end{eqalign}
\end{equation}
and we have defined
\begin{equation}
K(\psi) = \sum_s \frac{Z_s e }{\MeanMagB} \wint w_\parallel \Fneo,
	\label{KDef}
\end{equation}
with $\Fneo$ given by the neoclassical drift-kinetic equation \eref{neogeq}.
The fact that $K$ is a flux function follows directly from \eref{neogeq}, and is derived in \ref{foflow} (see \eref{Kfluxfn}). \Eref{main-current} is consistent with this, as can be seen by taking the divergence of both sides.
Note that the perpendicular part of \eref{main-current} is a statement of force balance for the mean quantities. Indeed, taking the cross product of it with $\MeanB$ and using $R^2 \nabla\tor\times\MeanB = \nabla\psi$, we get
\begin{equation}
\fl
\begin{eqalign}{
\frac{1}{c}\Meanj\times\MeanB &= \sum_s n_s\left\{ T_s\frac{d \ln \NotN}{d\psi} + \left[Z_s e\pot_0 - \frac{1}{2}m_s\angvel^2(\psi) R^2 + T_s \right]\frac{d\ln T_s}{d\psi} \right\}\nabla\psi\\
	&\qquad + \left( \sum_s m_s n_s R^2 \right) \angvel(\psi) \frac{d\angvel}{d\psi}\nabla\psi \\
		&= \sum_s \left( \nabla p_s + m_s n_s \bm{u}\dg\bm{u} \right),
}\end{eqalign}
\label{forcebalance}
\end{equation}
where we have defined the pressure $p_s = n_s T_s$ and used \eref{npol} for $\NotN(\psi)$ and \eref{torrot} for $\bm{u}$. This is just the balance between the Lorentz force, pressure and inertia.

We will now take three projections of \eref{avgamp}: onto $\nabla \psi$, onto $\nabla\psi\times\nabla\tor$, and onto $\nabla\tor$.
Taking the projection onto $\nabla \psi$ first, we have, from \eref{avgamp},
\begin{equation}
\Meanj\dg\psi = \frac{c}{4\pi}\left(\nabla I\times\nabla\tor\right)\dg\psi
\end{equation}
and from \eref{main-current}
\begin{equation}
\Meanj\dg\psi = 0,
\end{equation}
whence
\begin{equation}
\label{IofPsi}
\left(\nabla\psi\times\nabla\tor\right)\dg I = \MeanB\dg I = 0.
\end{equation}
Therefore, $I = I(\psi)$ is a flux function.

Taking the projection onto $\nabla\psi\times\nabla\tor$, we have from \eref{avgamp}
\begin{equation}
\Meanj\cdot\left(\nabla\psi\times\nabla\tor\right) = \frac{c}{4\pi R^2}\frac{d I}{d\psi} |\nabla\psi|^2
\end{equation}
and from \eref{main-current}
\begin{equation}
\Meanj\cdot \left(\nabla\psi\times\nabla\tor\right) = K(\psi) \MeanB\cdot\left( \nabla\psi\times\nabla\tor \right) = {K(\psi) |\nabla\psi|^2}{R^{-2}}.
\end{equation}
This gives us the following equation for $I(\psi)$
\begin{equation}
	\label{didpsi}
\frac{d I}{d\psi} = \frac{4\pi}{c} K(\psi).
\end{equation}

Finally, the toroidal component of \eref{avgamp} is just
\begin{equation}
\Meanj\dg\tor = -\frac{c}{4\pi R^2} \Delta^* \psi.
\end{equation}
Using the toroidal current from \eref{main-current} and expressing $K(\psi)$ via \eref{didpsi}, we obtain
\begin{equation}
\fl
\label{gradshaf}
\begin{eqalign}{
	\Delta^* \psi = &-4\pi R^2 \sum_s n_s\left\{ T_s\frac{d \ln \NotN}{d\psi} + \left[Z_s e\pot_0 - \frac{1}{2} {m_s\angvel^2(\psi) R^2} + T_s \right]\frac{d\ln T_s}{d\psi}\right\}\\
						 &- 4\pi R^2 \left(\sum_s m_s n_s R^2\right) \angvel(\psi) \frac{d\angvel}{d\psi}  - I(\psi)\frac{dI}{d\psi},
}\end{eqalign}
\end{equation}
which allows us to determine $\psi$ as a function of $R$ and $z$ if we know $\NotN$, $T_s$, $\angvel$ and $I$. This is the generalization of the Grad-Shafranov equation for a rotating plasma.

\subsection{Evolution of the Mean Magnetic Field}
\label{magevolve}

The results of the previous section complete the solution, including time dependence, for the magnetic field. We can solve \eref{neogeq} for $\Fneo$ to find $K(\psi)$ in terms of $\fav{\MeanE\cdot\MeanB}$, and then use \eref{didpsi} to find $\fav{\MeanE\cdot\MeanB}$ in terms of $I(\psi)$. We can then use \eref{IDef} to find $I(\psi)$ in terms of $\MeanA$ and so our expression for 
\begin{equation}
\fav{\MeanE\cdot\MeanB} = - \frac{1}{c} \fav{\pd{\MeanA}{t} \cdot\MeanB}
\end{equation}
is an evolution equation for $\MeanA$ (and thus $\MeanB$).
Indeed, taking the time derivative of \eref{IDef} divided by $R^2$, we find
\begin{equation}
\pd{}{t} \frac{I(\psi)}{R^2} = \pd{\MeanB}{t} \dg\tor = -c\dv\left(\MeanE\times\nabla\tor\right).
\end{equation}
Flux-surface averaging and using \eref{favdiv} and \eref{movingflux} we have
\begin{equation}
\label{qevolve}
\ddtpsi V'I(\psi)\fav{{R^{-2}}} =  c\pd{}{\psi} V'\fav{\MeanE\cdot\MeanB},
\end{equation}
where we have used \eref{magfield} to rewrite the right-hand side in terms of $\MeanB$.

In the study of tokamak plasmas, it is conventional to work with the safety factor $q(\psi)$ rather than $I(\psi)$.
The safety factor is defined in terms of the toroidal flux $\Psi$ and the poloidal flux $\psi$ by
\begin{equation}
q(\psi) = \frac{1}{2\pi} \frac{d \Psi}{d \psi},\qquad \Psi = \frac{1}{2\pi} \int_{D(\psi,t)} d^3\bm{r} \MeanB \dg \tor,
\label{qDef}
\end{equation}
where the integration is carried out over the volume interior to the flux surface labelled by $\psi$ (see \Sref{Sfluxav}). Geometrically, $q(\psi)$ is the number of times a field line on a given flux surface toroidally winds round the vertical symmetry axis for each poloidal transit around the magnetic axis.
From \eref{qDef}, the definition of the flux-surface average \eref{newfav}, and the axisymmetric form of the magnetic field \eref{magfield}, we can express $\Psi$ as
\begin{equation}
\fl
\Psi = \frac{1}{2\pi} \int_0^{\psi} d\psi' \int_{\partial D(\psi,t)} \frac{dS}{|\nabla\psi|} I(\psi')R^{-2} = \frac{1}{2\pi} \int^\psi_0 d\psi'V'(\psi') I(\psi') \fav{R^{-2}}(\psi'),
\end{equation}
where $V'$ is defined by \eref{Vprimdef}.
Therefore,
\begin{equation}
q(\psi) = \frac{1}{4\pi^2} V'I(\psi)\fav{R^{-2}}.
\label{qOfI}
\end{equation}

From \eref{qevolve} and \eref{qOfI}, we immediately find
\begin{equation}
\label{dqdt}
  \ddtpsi q = \frac{c}{4\pi^2} \pd{}{\psi} V'\fav{\MeanE\cdot\MeanB}.
\end{equation}
Thus, $q(\psi)$ evolves via \eref{dqdt}, $I(\psi)$ is determined by \eref{qOfI}, and $\psi(R,z)$ is then instantaneously determined from the generalized Grad-Shafranov equation \eref{gradshaf}.  This completes our solution for the evolution of the mean magnetic field.

It turns out that $q(\psi)$ can be shown to evolve on the resistive timescale, which is even slower than the transport timescale, i.e., $c\fav{\MeanE\cdot\MeanB}$ turns out to be of the order of $\vth[i] (m_e / m_i)^{1/2} \gkeps^2 \MeanMagB^2$. This result is proved in \ref{EXTP2-resistive} of \cite{flowtome2-electrons}, where we also relate the evolution of $q(\psi)$ to the conservation of magnetic flux. Indeed, the geometric interpretation of $q(\psi)$ is consistent with the notion that this longer timescale is the timescale on which the topology of the magnetic field changes, even if other properties of the field may change more rapidly.

\subsection{The Gyrokinetic Equation}
\label{Sgke}
\setcounter{footnote}{1}
Turning to the fluctuating component of the second-order kinetic equation \eref{2ndorder}, 
we find\footnote{
This equation is in agreement with the nonlinear gyrokinetic equation of \cite{sugama1998neg}, but not with that of \cite{artun1994nonlinear}. The discrepancy arises from a subtlety in the gyroaveraging of $\bm{u}\dg\gkpot$ that is explained in detail in \ref{apA6} (see \eref{tmp-fail-gyro} and \ref{Ap62}.)
}

\begin{equation}
\fl
\begin{eqalign}{
\left[\pd{}{t}+\bm{u}(\bm{R}_s)\dgR{}\right]h_s + \left( w_\parallel \Meanb + \vdrift + \vchiR \right)\cdot\ddR{h_s} - \gyroR{\lincol[h_s]} \\
\quad=\frac{Z_s e F_{0s}}{T_s}\left[\pd{}{t} + \bm{u}(\bm{R}_s)\dgR{}\right]\gyroR{ \gkpot}\\\qquad -
\left\{ \pd{F_{0s}}{\psi} + \frac{m_s F_{0s}}{T_s} \left[ \frac{I(\psi) w_\parallel}\MeanMagB + \angvel(\psi)R^2 \right]\frac{d\angvel}{d\psi}\right\} \vchiR\dg\psi
}\end{eqalign}
\label{gke}
\end{equation}
with $\vchi$ and $\vdrift$ given by \eref{vchi} and \eref{vdrift}, respectively. As in \Sref{nclasse}, $\lincol$ is the collision operator linearised about the Maxwellian part of $F_{0s}$.  Explicit forms for the collision term $\gyroR{\lincol[h_s]}$ are proposed in~\cite{catto1977col,abelcollisions,Tome} for various model collision operators, however, only the properties common to all such operators will be needed in this work so we leave this term in its general form.
Finally, we reiterate that this equation is written in terms of the variables $\bm{R}_s$, $\energy$, $\magmom$ and $\gyr$, so $w_\parallel$ is given by
\begin{equation}
  w_\parallel =  \sqrt{{\frac{2}{m_s}}\left[ \energy - \magmom \MeanMagB - Z_s e\pot_0 + \frac{1}{2} m_s \angvel^2(\psi) R^2 \right]} + \Or(\gkeps \vth).
\label{wparDef}
\end{equation}

It is the gyrokinetic equation \eref{gke} that is solved by the gyrokinetic simulation codes~\cite{dimits:969,jenko2000electron,dorlandETG,candy2003egm,gkwRef,linGTC,parkerGEM} and has formed the basis of a large body of theoretical analysis of tokamak microinstabilites and turbulence.
The turbulence is driven by the density and temperature gradients in the equilibrium distribution (through the $\left.\partial F_{0s}\right/\partial\psi$ term on the right-hand side of \eref{gke} \cite{dorlandETG,sagdeevITG,coppiITG}) and by the velocity gradient (through the $\left.d\angvel\right/d\psi$ term\footnote{This, potentially destabilizing~\cite{cattoPVG,newton2010understanding,schekochihin2010subcritical}, term comes from the gradient of the parallel part of the toroidal flow~\eref{torrot}. Gradients in the perpendicular part of the toroidal flow enter into \eref{gke} via the $\bm{u}(\bm{R_s})\cdot\partial/\partial\bm{R}_s$ terms and are stabilizing~\cite{BiglariDiamond,TerryRMPSuppress,newton2010understanding,schekochihin2010subcritical}.}).
It is important to note that this equation is only coupled to the neoclassical distribution function $\Fneo$ (\Sref{nclasse}) through the slow evolution of $\NotN$, $T_s$, $\angvel$ and $I$.

Although \eref{gke} is the most commonly used form of the gyrokinetic equation, two alternative formulations deserve discussion. Firstly, some derivations include the so-called ``parallel nonlinearity'' in the gyrokinetic equation, which is not present in \eref{gke}. The inclusion of this term is discussed in \ref{parallelNote}. Secondly, some formulations of the gyrokinetic equation explicitly emphasise polarisation effects. We discuss them in the context of our formulation in \ref{polarisationAp}.

\subsection{The Fluctuating Maxwell's Equations}
\label{Sfmag}
\setcounter{footnote}{1}
To solve \eref{gke}, we require a way of obtaining $\gyroR{\gkpot}$ from $h_s$. This is provided by the fluctuating parts of Maxwell's equations.

The fluctuating component of the quasineutrality condition \eref{qn} is
\begin{equation}
\label{fluct-qn}
 \sum_s \frac{Z_s^2 e^2 n_s\gkupot}{T_s}   = \sum_s Z_s e \wint \gyror{h_s},
\end{equation}
where $\gkupot$ is given by \eref{etaDef}. The integral over velocities is performed at constant $\bm{r}$ as discussed in \Sref{derivInt}. We have now made this explicit by introducing the gyroaverage at constant $\bm{r}$, $w_\parallel$, and $w_\perp$:
\begin{equation}
\fl
\gyror{h_s} = \frac{1}{2\pi} \oint d\gyr h_s(\bm{R}_s(\bm{r},w_\parallel,w_\perp,\gyr),\energy(\bm{r},w_\parallel,w_\perp,\gyr),\magmom(\bm{r},w_\perp),\sigma).
\end{equation}

The fluctuating component of Amp\`ere's Law \eref{ampere} is
\begin{equation}
\label{famp}
\curl \delB = -\nabla^2 \delA = \frac{4\pi}{c} \sum_s Z_s e \wint \gyror{\bm{w} h_s}.
\end{equation}
The parallel component of this equation gives us the following equation for $\delAp$:
\begin{equation}
\label{fluct-apar}
-\nabla_\perp^2 \delAp = \frac{4\pi}{c} \sum_s Z_s e \wint w_\parallel \gyror{h_s}.
\end{equation}

It turns out to be convenient for various calculations to work with $\delBp$ rather than $\delA_\perp$. Thus, we take the divergence of $\Meanb\,\,\times$
\eref{famp} to find
\begin{equation}
\dv\left[\Meanb\times\left(\curl\delB\right)\right] = \nabla_\perp^2 \delBp = \frac{4\pi}{c} \sum_s Z_s e\wint \nabla_\perp\cdot\gyror{\Meanb\times\bm{w}_\perp h_s}.
\end{equation}
We now use
\begin{equation}
\Meanb\times\bm{w} = -\pd{\bm{w}}{\gyr},
\end{equation}
integrate by parts inside the gyroaverage, and use (see \eref{wperpdg1})
\begin{equation}
\bm{w}_\perp\dg h_s = - \cycfreq\left.\pd{h_s}{\gyr}\right|_{\bm{r},\energy,\magmom} + \Or(\gkeps \cycfreq h_s)
\end{equation}
to obtain
\begin{equation}
\label{fluct-bpar}
\nabla_\perp^2 \frac{\delBp \MeanMagB}{4\pi} + \bm{\nabla}_\perp\bm{\nabla}_\perp\bm{:} \sum_s \wint \gyror{m_s\bm{w}_\perp\bm{w}_\perp h_s} = 0.
\end{equation}
Once again (cf.~\eref{forcebalance}), perpendicular Amp\`ere's law has become the equation for perpendicular force balance. This eliminates fluctuations such as the 
compressional Alfv\'en wave (see the appendix of \cite{microstab1}), which are not force-balanced.

This completes the description of the fluctuations, which are now determined by a closed set of equations: \eref{hDef}, \eref{gke}, \eref{fluct-qn}, \eref{fluct-apar} and \eref{fluct-bpar}.
\section{Third Order \texorpdfstring{$\Or(\gkeps^3\cycfreq f_s)$}{O(3)}: Transport Equations}
\label{transport}
So far we have derived equations for the fast evolution of the fluctuations, instantaneous relations between equilibrium quantities, and an equation for the slow evolution of the mean magnetic field. These hold for a given set of
three flux functions: $\NotN$, $\angvel$ and $T_s$. In this section, we derive equations for the slow (transport-timescale) evolution of these functions.
We refer to these as the transport equations.

To derive such equations, we will need to go to the next order in our expansion in $\gkeps$, namely $\Or(\gkeps^3\cycfreq f_s)$.
As we wish to maintain the physical interpretation of evolution equations for $n_s$, $\angvel$ and $T_s$  in terms of the transport of particles, momentum, and heat,  we return to the original $\bm{r}$ and $\bm{w}$ variables to begin our derivation\footnote{This is not the only way of deriving equations for $\NotN$, $T_s$, and $\angvel$. One could, instead, continue the procedure of the previous sections, expand \eref{newvfp} to $\Or(\gkeps^3\cycfreq f_s)$, and then derive the transport equations as solubility conditions for the resulting equation. The reader interested in this approach should see \cite{calvo2012long}, where such a calculation is performed to derive transport equations for density and temperature in the ``low-flow'' regime (i.e., when the lowest-order mean velocity is diamagnetic, $\bm{u}_s \sim \gkeps \vth$).}.
The averaged form of \eref{vfp}, written in these variables, is
\begin{equation}
\label{avke}
\fl
\begin{eqalign}{
\pd{F_s}{t} + \left(\bm{u}+\bm{w}\right)&\dg F_s + \left[\accel - \pd{\bm{u}}{t}-\left(\bm{u}+\bm{w}\right)\dg\bm{u}\right]\cdot \pd{F_s}{\bm{w}} + \ensav{\daccel \cdot \pd{\delta f_s}{\bm{w}}}\\
&= \collop[F_s] + \source,}\end{eqalign}
\end{equation}
where we have naturally split the particle acceleration into its mean part $\accel$ and its fluctuating part $\daccel$. The mean acceleration is
\begin{eqnarray}
\label{accel}
\fl
\hspace{3.5mm}\accel &= \frac{Z_s e}{m_s} \left[ \MeanE + \frac{1}{c} \left(\bm{u}+\bm{w}\right) \times \MeanB \right] = -\frac{Z_s e}{m_s} \left( \nabla\pot_0 + \frac{1}{c} \pd{\MeanA}{t}\right) + \cycfreq \bm{w}\times\Meanb,
\end{eqnarray}
where we have used the results of \Sref{constraints}.
The fluctuating acceleration is
\begin{equation}
\fl
\begin{eqalign}{
\daccel &= \frac{Z_s e}{m_s} \left[ \delE + \frac{1}{c} \left(\bm{u}+\bm{w}\right) \times \delB \right] \\
		  &= -\frac{Z_s e}{m_s} \left[ \nabla\gkpot + \frac{1}{c}\bm{w}\dg\delA + \frac{1}{c}\left(\pd{ }{t} + \bm{u}\dg\right)\delA\right]\\
		  &= -\frac{Z_s e}{m_s} \left[ \nabla\left(\gkupot - \frac{1}{c}\bm{w}\cdot\delA\right) + \frac{1}{c} \bm{w}_\perp \dg\delA\right] + \Or(\gkeps^2 \vth\cycfreq),
}\end{eqalign}
\label{daccel}
\end{equation}
where $\gkpot$ is defined by \eref{chidef} and $\gkupot$ by \eref{etaDef}.
\Eref{avke} will correctly describe the transport-timescale evolution of $F_s$ if we keep all terms up to order $\Or(\gkeps^3\cycfreq f_s)$.  This clearly requires knowledge of the distribution function
including $\Or(\gkeps^2 f_s)$ corrections.

From the previous two sections, the particle distribution function is
\begin{equation}
\label{fsol}
\fl
\begin{eqalign}{
f_s &= F_{0s}(\psi(\bm{R}_s),\energy) + F_{1s}(\bm{R}_s,\energy,\magmom,\sigma) + F_{2s}(\bm{r},\bm{v})\\
		\,&\qquad\qquad - \frac{Z_s e}{T_s} \gkupot(\bm{r}) F_{0s} + h_s(\bm{R}_s,\energy,\magmom,\sigma)  +\delta f_{2s}(\bm{r},\bm{v}) + \cdots,
}\end{eqalign}
\end{equation}
where the equilibrium distribution function $F_{0s}$ is given by \eref{F0R},
$F_{1s}$ and $h_s$ are obtained from the neoclassical drift-kinetic equation \eref{neoclass}, and the gyrokinetic equation \eref{gke}, respectively, and we have absorbed all (as yet unknown) higher-order terms into $F_{2s}$ and $\delta f_{2s}$. The exact electric and magnetic fields are
\begin{eqnarray}
\Efield &= \MeanE + \delE =  -\nabla \fpot - \nabla\pot_0 - \nabla \delpot - \frac{1}{c}\pd{\MeanA}{t} - \frac{1}{c}\pd{\delA}{t}
\end{eqnarray}
and
\begin{eqnarray}
\Bfield &= \MeanB + \delB = I(\psi) \nabla\tor + \nabla\psi \times\nabla\tor + \curl\delA.
\end{eqnarray}
In the expressions for the fields, $I(\psi)$ and $\psi(R,z)$ are determined as explained in Sections \ref{ampmag} and \ref{magevolve}, $\pot_0$ is obtained from the quasineutrality condition \eref{qn0}, and $\gkupot$ and $\delA$ are given by the fluctuating Maxwell's equations \eref{fluct-qn}, \eref{fluct-apar} and \eref{fluct-bpar}.

The solution \eref{fsol} for $f_s$ will need to be completed with information about $F_{2s}$ and $\delta f_{2s}$. We can obtain
the gyrophase-dependent part of $F_{s}$ to second order by expanding \eref{avke} to second order in $\gkeps$:
\begin{equation} \fl
\label{hoc-main}
\begin{eqalign}{
\cycfreq \pd{F_s}{\gyr} = -&\left(\bm{u}+\bm{w}\right)\dg F_s +\left[\frac{Z_s e}{m_s}\left( \nabla\pot_0+\frac{1}{c}\pd{\MeanA}{t}\right) + \left(\bm{u}+\bm{w}\right)\dg\bm{u}\right] \cdot\pd{F_s}{\bm{w}} \\
  & - \ensav{\daccel \cdot \pd{\delta f_s}{\bm{w}}} + \collop[F_s] + \Or(\gkeps^3 \cycfreq f_s),
}\end{eqalign}
\end{equation}
where we have used
\begin{equation}
\left(\bm{w}\times\Meanb\right) \cdot\left.\pd{ }{\bm{w}}\right|_{\bm{r}} = \left.\pd{ }{\gyr}\right|_{\bm{r},w_\parallel,w_\perp}.
\end{equation}
Similarly, the fluctuating part of \eref{vfp} can be expanded to obtain the gyrophase dependence of $\delta f_{s}$ to second order:
\begin{equation}
\fl
\label{fluct-main}
\begin{eqalign}{
\left(\bm{w}_\perp\dg + \cycfreq\pd{}{\gyr}\right) \delta f_s = 
-\left( \pd{}{t} + \bm{u}\dg\right) \delta f_s - w_\parallel \Meanb \dg \delta f_s- \daccel \cdot \pd{F_s}{\bm{w}}\\
\qquad+\left[\frac{Z_s e}{m_s}\left( \nabla\pot_0+\frac{1}{c}\pd{\MeanA}{t}\right) + \left(\bm{u}+\bm{w}\right)\dg\bm{u}\right]\cdot\pd{\delta f_s}{\bm{w}}-\daccel\cdot\pd{\delta f_s}{\bm{w}}\\
\qquad + \ensav{\daccel\cdot\pd{\delta f_s}{\bm{w}}} + \lincol[\delta f_s]+ \Or(\gkeps^3 \cycfreq f_s),
}\end{eqalign}
\end{equation}
where $\lincol$ is the collision operator linearised about the Maxwellian part of $F_{0s}$.
By examining the order of the terms on the right-hand side of \eref{hoc-main} and \eref{fluct-main}, we see that we only need the particle distribution function correct to $\Or(\gkeps f_s)$ in order to evaluate the left-hand side
correctly to order $\Or(\cycfreq\gkeps^2 f_s)$.

In the subsequent sections, we will take moments of \eref{avke} and show that \eref{hoc-main} and \eref{fluct-main} contain sufficient information about the second-order parts of the distribution function to close the resulting transport equations in terms of known quantities -- i.e., the gyrophase-independent parts of $F_{2s}$ and $\delta f_{2s}$ do not affect transport.

\subsection{Particle Transport}
\label{parttrans}
To derive the particle transport equation, we integrate \eref{avke} over all velocities to find
\begin{eqnarray}
\label{0moment}
\pd{n_s}{t} + \dv\left( n_s\bm{U}_s \right) = \Psource,
\end{eqnarray}
where 
\begin{eqnarray}
\label{psource}
\Psource &=& \wint \source,\\
\label{pflux}
n_s\bm{U}_s &=& \wint \bm{w} F_s.
\end{eqnarray}
\Eref{0moment} clearly describes all particle transport: $n_s$ is created or destroyed by the source $\Psource$ and advected by the mean flux $n_s\bm{U}_s$.
As the first and last terms in this equation are $\Or(\gkeps^3 \cycfreq n_s)$, we conclude that we require $n_s\bm{U}_s$ correct to order $\Or(\gkeps^2n_s\vth[s])$ to evaluate the divergence.
It is clear that the parallel part of the flux will depend on the gyrophase-independent part of $F_{2s}$, which we cannot easily find. In contrast, the perpendicular flux will only depend on the gyrophase-dependent piece of $F_{2s}$, which we can find via \eref{hoc-main}.

To reduce \eref{0moment} to an equation only containing the perpendicular flux, we apply the flux-surface average, defined by \eref{favdef}. This results in 
\begin{equation}
\label{dndt}
\frac{1}{V'}\left.\pd{}{t}\right|_\psi V'\fav{n_s} + \frac{1}{V'} \pd{}{\psi} V'\fav{ \ParticleFlux } = \fav{\Psource},
\end{equation}
where we have used \eref{favdiv} and \eref{movingflux} to simplify the flux-surface averages of the divergence and the time derivative.
We have combined the radial particle flux\footnote{Whilst this flux is not in the geometrically radial direction, but is in fact the cross-flux-surface flux, it is both convenient and conventional to refer to it as the radial flux.} and the terms due to the motion of the flux surface into
\begin{equation}
\label{GammaDef}
\ParticleFlux = n_s\bm{U}_s\dg\psi + n_s \pd{\psi}{t}. 
\end{equation}
We can write the first term of \eref{GammaDef} as
\begin{eqnarray}
\fl
\label{rpf}
n_s\bm{U}_s\dg\psi &= \wint \left(\bm{w}_\perp\dg\psi\right)F_s = \wint R^2 \MeanMagB \left(\bm{v}\dg\tor\right) \left.\pd{F_s}{\gyr}\right|_{\bm{r},w_\parallel,w_\perp},
\end{eqnarray}
where $\bm{v}$ is just shorthand for $\bm{u}+\bm{w}$ and we have used
\begin{equation}
\fl
\bm{w}_\perp \dg \psi = -R^2\MeanMagB \bm{w} \cdot \left(\Meanb\times \nabla\tor\right) = R^2\MeanMagB \left( \Meanb\times\bm{w} \right)\dg\tor = -R^2\MeanMagB \left.\pd{\bm{v}}{\gyr}\right|_{\bm{r},w_\parallel,w_\perp} \dg \tor
\label{magicGyrThing}
\end{equation}
and integrated by parts with respect to $\gyr$. So, in order to calculate the radial flux to second order, it is sufficient to know $\infrac{\partial F_s}{\partial\gyr}$ up to second order only.

In \ref{pfluxes}, we perform the explicit evaluation of $\fav{n_s\bm{U}_s\dg\psi}$ via the kinetic equation \eref{hoc-main}, resulting in \eref{pflux0}, which we substitute into \eref{GammaDef} to find:
\begin{equation}
\label{partflux}
\fl
\begin{eqalign}{
\fav{\ParticleFlux} = &\fav{\wint \left(\frac{\bm{w}\times\Meanb}{\cycfreq} \cdot\nabla\psi\right)\collop[F_{0s}]} + \fav{\wint \Fneo\vdrift \dg \psi} \\
&\qquad- \fav{n_s}I(\psi) \frac{\fav{\MeanE\cdot\MeanB}}{\fav{\MeanMagB^2}}  
  + \fav{\ensav{\wint \gyror{h_s\,\vchi}\dg\psi}}.
}\end{eqalign}
\end{equation}
The first term in the above expression is due to classical collisional transport with $F_{0s}$ given by \eref{F0R}\footnote{Note that $F_{0s}$ is only Maxwellian to lowest order, so $\collop[F_{0s}] = \Or(\gkeps^2\cycfreq F_{0s})$ is non-zero.}, the second and third terms are due to neoclassical transport with $\Fneo$ and $\fav{\MeanE\cdot\MeanB}$ calculated as explained in Sections \ref{nclasse}-\ref{magevolve}\footnote{
Explicit expressions for the collisional fluxes can be found in \cite{hinton1976tpt,braginskii1965tpp} or, for the case of non-zero $\angvel(\psi)$, in \cite{catto1987ion,hintonwong1985nit,BishopAndCowley}.}, and the final term gives the turbulent contribution to the particle flux in terms of $h_s$, which is the solution of the gyrokinetic equation \eref{gke}\footnote{Having determined the particle flux, we note that the classical, neoclassical and turbulent contributions are all independently ambipolar~\cite{sugama1998neg}. This is explicitly proved in \ref{ambipolar}.}.
We note that the composition of a flux-surface average (defined by \eref{favdef}) with the fluctuation average (defined by \eref{avdef}) is
\begin{equation}
\fav{\ensav{ g(\bm{r},t)}} = \frac{1}{V'} \frac{1}{T} \int\limits_{t-T/2}^{t+T/2} dt' \int\limits_{\Delta_\lambda} d^3 \bm{r}' V' g(\bm{r}',t'),
\end{equation}
where $\Delta_\lambda$ is the annulus bounded by two flux surfaces a distance $\lambda$ apart centered on $\psi(\bm{r})$. This is exactly the average over a flux tube that is implemented in current gyrokinetic linked-flux-tube codes~\cite{barnes2009trinity,candy2009tgyro}.

Thus, we now have an equation, \eref{dndt}, that determines $\fav{n_s}$, but we still need to use this to find $\NotN$ and hence $n_s$ via \eref{npol}. This is done by flux-surface averaging \eref{npol} to find 
\begin{equation}
\label{ntwiddle}
\NotN = \fav{n_s} \left/ \fav{\exp\left[ \frac{m_s \angvel^2(\psi) R^2}{2T_s} - \frac{Z_s e \pot_0}{T_s} \right]}\right.,
\end{equation}
which allows us to express $\NotN$ in terms of $\fav{n_s}$ and known quantities. In particular, we can solve \eref{ntwiddle} simultaneously with the quasineutrality condition \eref{qn0} to find $\NotN$ and $\pot_0$ from $\fav{n_s}$.

\subsection{Momentum Transport}
\label{momtrans}
To find an evolution equation for $\bm{u} = \angvel(\psi)R^2\nabla\tor$, we multiply \eref{avke} by $m_s \left(\bm{v}\cdot\nabla\tor\right) R^2$, where $\bm{v} = \bm{w} + \bm{u}$, integrate over $\bm{w}$, and sum over species to find
\begin{equation}
\fl
\begin{eqalign}{
  \pd{}{t} \sum_s m_s n_s R^2 \angvel(\psi) + \left[\dv\sum_s\left( \viscosity + m_s n_s\bm{U}_s\bm{u}+ m_s n_s \bm{u}\bm{U}_s\right) \right]\cdot\left(\nabla\tor\right)R^2\\
  \qquad- \sum_s {m_s \ensav{\wint  \left(\daccel\cdot\nabla\tor\right) R^2\delta f_s}} - \frac{1}{c}\left( \Meanj\times\MeanB \right)\cdot\left( \nabla\tor \right) R^2
  = {\Msource},
}\end{eqalign}
\label{2moment}
\end{equation}
where the viscous stress tensor $\viscosity$ is
\begin{equation}
\label{mflux}
\viscosity = \wint m_s \bm{w}\bm{w} F_s,
\end{equation}
the source of toroidal angular momentum is
\begin{equation}
\Msource = \sum_s\left[\wint m_s \left(\bm{w}\cdot\nabla\tor\right)R^2 \source + m_s \angvel(\psi)\Psource\right],
\label{msourceDef}
\end{equation}
where $\Psource$ is given by \eref{psource} and we have used the quasineutrality condition \eref{quasineutral0}. The second term in \eref{msourceDef} is due to injected particles joining the mean flow, while the first term contains also contributions from particle-conserving momentum injection mechanisms, e.g., momentum injection by waves~\cite{cmodrotation2009}.

We now wish to write all terms except the time derivative and the source as full divergences. The second term on the left-hand side of \eref{2moment} can be written so because,
  for any symmetric tensor field $\tensor{A}$,
\begin{equation}
\left(\dv\tensor{A}\right)\cdot\left(\nabla\tor\right) R^2 = \dv\left(R^2\tensor{A}\dg\tor\right).
  \label{divtor}
\end{equation}
The third term (due to the fluctuations) can also be written as a divergence:
\begin{equation}
\begin{eqalign}{
-\sum_s& {m_s \ensav{\wint  \left(\daccel\cdot\nabla\tor\right)R^2 \delta f_s}} \\
  &= -\frac{1}{c}\ensav{\delj\times\delB}\cdot\left( \nabla\tor \right)R^2 \\
  &= \dv\left[ \ensav{\frac{\MagDelB^2}{8\pi} \idmat - \frac{\delB\delB}{4\pi}}\cdot\left(\nabla\tor\right) R^2 \right],
}\end{eqalign}
\end{equation}
where $\idmat$ is the unit dyadic and we have used \eref{daccel} for $\daccel$, the fluctuating quasineutrality condition \eref{fluct-qn}, expressed the Lorentz force as the divergence of the Maxwell stress in the usual way and used \eref{divtor}.
The final term on the left-hand side of \eref{2moment} can similarly be written as the divergence of the Maxwell stress associated with the mean magnetic field:
\begin{equation}
-\frac{1}{c} \left( \Meanj\times\MeanB \right)\cdot\left(\nabla\tor\right)R^2 = \dv\left[\left(\frac{\MeanMagB^2}{8\pi}\idmat - \frac{\MeanB\MeanB}{4\pi}\right)\cdot\left( \nabla\tor \right)R^2\right].
\end{equation}

We now flux-surface average \eref{2moment}, using the identities \eref{favdiv} and \eref{movingflux} for the flux-surface average of a divergence and of a time derivative.
This gives
\begin{equation}
\begin{eqalign}{
\frac{1}{V'}\ddtpsi&{{V'\inertia \angvel(\psi)}} + \frac{1}{V'}\pd{}{\psi} V'\fav{\TotMomFlux}
= \fav{\Msource},
}\end{eqalign}
\label{domegadt}
\end{equation}
where 
\begin{equation}
\label{inertiaDef}
\inertia = \sum_s m_s \fav{n_s R^2}
\end{equation}
is the flux-surface-averaged moment of inertia and the total flux of angular momentum is given by
\begin{equation}
\begin{eqalign}{
\TotMomFlux = &\sum_s \left(\nabla\psi\right) \cdot \viscosity \cdot\left(\nabla\tor\right) R^2 + \sum_s m_s  \angvel(\psi) R^2\ParticleFlux\\
				  &- \frac{1}{4\pi} \left(\nabla\psi\right)\cdot\ensav{\delB \delB}\cdot\left(\nabla\tor\right)R^2,
}\end{eqalign}
\label{TMFdef}
\end{equation}
where $\ParticleFlux$ is given by \eref{GammaDef}.

Proceeding analogously to the derivation of the particle transport (see \eref{rpf}), we can write the radial angular momentum flux due to the particles of species $s$ as
\begin{equation}
\label{rmf}
\fl
\left(\nabla\psi\right) \cdot \viscosity \cdot\left(\nabla\tor\right) R^2 + m_s n_s\angvel(\psi) R^2 \bm{U}_s\dg\psi  = \wint  \MeanMagB m_s\frac{1}{2}\left(R^2\bm{v}\dg\tor\right)^2 \pd{F_s}{\gyr},
\end{equation}
where $\bm{v} = \bm{u}+\bm{w}$.
We evaluate this explicitly in \ref{mfluxes} to find
\begin{equation}
\fav{\TotMomFlux} = \sum_s \left[\,\fav{\MomentumFlux} + m_s \angvel(\psi) \fav{R^2\ParticleFlux}\right] + \fav{\EMViscosity},
\label{TMFresult}
\end{equation}
where $\MomentumFlux$ contains the classical, neoclassical and turbulent viscous stresses:
\begin{equation}
\label{momflux}
\fl
\begin{eqalign}{
\fav{\MomentumFlux} = \fav{\ClassMomFlux} + \fav{\NeoMomFlux} +\fav{\ensav{\wint m_sR^2\gyror{\left(\bm{w} \dg \tor\right)h_s\vchi}\dg \psi }}
}\end{eqalign}
\end{equation}
with
\begin{equation}
\fl\begin{eqalign}{
&\fav{\ClassMomFlux} = \\
	&\quad\frac{c}{Z_s e} \fav{\left(\wint \frac{m_s^2}{2} \bm{w}_\perp\bm{w}_\perp \collop[F_{0s}]\right)\bm{:}\left\{ R^4\left(\nabla\tor\right)\left(\nabla\tor\right) - \frac{1}{2}\left[R^2 - \frac{I^2(\psi)}{\MeanMagB^2}\right]\idmat\right\}}\\
	&\qquad+ \fav{\wint m_s \frac{I(\psi)w_\parallel}\MeanMagB\left( \frac{\bm{w}\times\Meanb}{\cycfreq} \dg\psi \right)\collop[F_{0s}]}
}\end{eqalign}
\label{classvisc}
\end{equation}
and
\begin{equation}
\fl
\fav{\NeoMomFlux} = -\frac{m_s c}{Z_s e}\fav{\wint \Fneo w_\parallel\Meanb\dg\left\{m_s{\frac{I^2(\psi) w^2_\parallel}{\MeanMagB^2}} + \magmom\frac{\left|\nabla\psi\right|^2}{\MeanMagB^2}\right\}},
\label{neovisc}
\end{equation}
$m_s \angvel(\psi)\fav{R^2\ParticleFlux}$ is the convective flux of angular momentum with $\fav{R^2\ParticleFlux}$ given by (cf. \eref{partflux}):
\begin{equation}
\fl
\begin{eqalign}{
\fav{R^2\ParticleFlux} =&
\fav{R^2\wint\left( \frac{\bm{w}\times\Meanb}{\cycfreq}\dg\psi \right)\collop[F_{0s}]}\\
	&+\fav{\wint \Fneo w_\parallel\Meanb\dg\left[ R^2\frac{I(\psi)w_\parallel}{\cycfreq} + R^2\frac{BR^2\angvel(\psi)}{\cycfreq} \right]} \\
	&- \fav{R^2n_s}I(\psi)\frac{\fav{\MeanE\cdot\MeanB}}{{\fav{\MeanMagB^2}}} + \fav{R^2\ensav{\wint \gyror{h_s \vchi}\dg \psi}},
}\end{eqalign}
\label{conMomflux}
\end{equation}
and the electromagnetic angular momentum flux $\EMViscosity$ is
\begin{equation}
\EMViscosity = -\left(\nabla\psi\right)\cdot\ensav{\frac{1}{4\pi} \delB\delB + \frac{1}{c} \delj\delA} \cdot \left(\nabla\tor\right) R^2.
\label{EMViscDef}
\end{equation}
The two terms in the last expression are, respectively, the Maxwell stress and the advection of electromagnetic momentum by the particles 
\begin{equation}
\frac{1}{c} \delj \delA = \sum_s\wint \gyror{h_s \bm{w}} \frac{Z_s e}{c} \delA
\end{equation}
(in the long-wavelength limit, this term is small as the fluctuating velocity is dominated by the $\MeanE\times\MeanB$ velocity, which carries no current).
\subsection{Heat Transport and Heating}
\label{heattrans}
Turning to heat transport next, we multiply \eref{avke} by $\infrac{m_s w^2}{2}$ and integrate over all velocities to obtain
\begin{equation}
\label{1moment}
\fl
\begin{eqalign}{
\frac{3}{2}\pd{}{t}n_s T_s +  &\dv \bm{Q}_s +  Z_s e\left( \nabla\pot_0  + \frac{1}{c} \pd{\MeanA}{t}\right) \cdot \left(n_s\bm{U}_s\right) + m_s n_s \bm{u}\cdot\left( \nabla\bm{u} \right)\cdot\bm{U}_s + \viscosity\bm{:}\nabla\bm{u}\\
	&- \ensav{\wint m_s\bm{w}\cdot \daccel \delta f_s} = \CollEnergy + \Esource,
}\end{eqalign}
\end{equation}
where we have defined the heat flux
\begin{equation}
\label{eflux}
\bm{Q}_s  = \wint \frac{1}{2} m_s w^2 \bm{w} F_s,
\end{equation}
the collisional energy transfer to (or from) species $s$
\begin{equation}
\CollEnergy   = \wint \frac{1}{2} m_s w^2 \collop[F_s],
	\label{CollEnergyDef}
\end{equation}
and the energy source 
\begin{equation}
\label{esource}
\Esource = \wint \frac{1}{2} m_s w^2 \source.
\end{equation}

Acting in the same vein as for the particle and momentum transport equations, we now flux-surface average \eref{1moment}. This flux-surface average is carried out in \ref{dpdtderiv}, where we obtain:
\begin{equation}
\label{dpdt}
\begin{eqalign}{
\frac{3}{2}&\frac{1}{V'}\ddtpsi V'\fav{n_s}T_s + \frac{1}{V'} \pd{}{\psi}  V' \fav{\HeatFlux} \\
	&= \ViscousHeat + \JouleHeat + \CompHeat + \PotEng+ \TurbPow 
+ \fav{\CollEnergy} + \fav{\Esource}.
}\end{eqalign}
\end{equation}
Let us detail the terms in this equation.

\subsubsection{The Heat Flux.}
\label{Sec831}
The ``heat flux'' $q_s$ in \eref{dpdt} is defined by the following formula, whose origin is explained in \ref{ApFavdpdpt}:
\begin{eqnarray}
\label{main-hflux-full}
\fl\nonumber
\fav{\HeatFlux} &= \fav{\left[ \bm{Q}_s +\left( Z_s e \pot_0 - \frac{1}{2} m_s \angvel^2(\psi) R^2 \right) n_s\bm{U}_s + Z_s e\ensav{\wint\gkupot \gyror{h_s\bm{w}}}\right]\dg\psi}\\
\fl
 &\qquad\qquad+\fav{n_s\left[\frac{5}{2} T_s + Z_s e\pot_0 - \frac{1}{2}m_s \angvel^2(\psi)R^2\right] \pd{\psi}{t}}.
\end{eqnarray}
This form of the heat flux is most useful for interpreting its constituent parts.
The first term in the first line of \eref{main-hflux-full} is the usual flux of thermal energy, the second term is the flux of the mean potential energy (electrostatic and centrifugal), and the third
is the flux of the fluctuating electrostatic potential energy. The second line of \eref{main-hflux-full} is minus the flow of heat and energy carried by the motion of the flux surfaces (remembering that $\vpsi\dg\psi = -\partial \psi/\partial t$ and observing that the expression multiplying $\partial \psi / \partial t$ is the enthalpy carried by the plasma\footnote{The first term is the enthalpy of the ideal gas of species $s$~\cite{landau-sp1} and the second and third terms are the potential energy contributions to the internal energy.}).
As we include the potential energy fluxes in \eref{main-hflux-full}, $\HeatFlux$ is not, strictly speaking, the heat flux in the usual sense. We will elaborate upon this in \Sref{potengsec} where we discuss the potential energy exchange term $\PotEng$. For simplicity, we will continue to refer to $\fav{\HeatFlux}$ as the radial heat flux.
This usage will be vindicated on physical grounds by the appearence of $\fav{\HeatFlux}$ multiplying the temperature gradient in the expression \eref{entropyProductionMain} for the entropy production.

Unfortunately, \eref{main-hflux-full} is not a useful form for actually evaluating the heat flux.
An explicit form for it is calculated in terms of known quantities in \ref{hfluxes}:
\begin{equation}
\fl
\label{heatflux}
\begin{eqalign}{
\fav{\HeatFlux} =& \fav{\wint\energy \left( \frac{\bm{w}\times\Meanb}{\cycfreq} \dg\psi \right) \collop[F_{0s}] } +\fav{\wint\energy \Fneo \vdrift\dg\psi}\\
&- \fav{n_s \left[\frac{5T_s}{2} + Z_s e \pot_0 - \frac{1}{2} m_s \angvel^2(\psi) R^2\right]}I(\psi)\frac{\fav{\MeanE\cdot\MeanB}}{\fav{\MeanMagB^2}} \\
&+ \fav{\ensav{\wint \energy \gyror{h_s \vchi}\dg\psi}}.
}\end{eqalign}
\end{equation}

Let us now examine the heating terms on the right-hand side of \eref{dpdt}.

\subsubsection{Viscous Heating.}
\label{VHsec}
This is (see \ref{C4ViscHeat})
\begin{equation}
  \label{VHDef}
\ViscousHeat = -  \left[ \,\fav{\MomentumFlux} + m_s\angvel(\psi)  \fav{R^2 \ParticleFlux}\right]\frac{d\angvel}{d\psi},
\end{equation}
where $\fav{\MomentumFlux}$ is given by \eref{momflux} and $\fav{R^2\ParticleFlux}$ by \eref{conMomflux}.
Usually, the heating due to the mass flow (the second term inside the square brackets in \eref{VHDef}) is not included in the viscous heating. We group them together here because the term in the square brackets in \eref{VHDef} is precisely the species-dependent term in the total momentum flux \eref{TMFresult}. We will come back to this point when interpreting the rotational part of $\PotEng$ in \Sref{potengsec}. The combination of momentum fluxes seen in \eref{VHDef} will appear multiplying $d\angvel/d\psi$ in the expression \eref{entropyProductionMain} for the entropy production, vindicating its interpretation as the total viscous heating of species $s$.

\subsubsection{Ohmic Heating.}
The mean Ohmic heating due to the induced parallel electric field is (see \ref{C4OhmHeat})
\begin{equation}
\JouleHeat = K_s(\psi) \fav{\MeanE\cdot\MeanB},
\label{OhmHeat}
\end{equation}
where 
\begin{equation}
{K}_s(\psi) = \frac{Z_s e}\MeanMagB \wint w_\parallel \FHat.
\label{KsDef}
\end{equation}

\subsubsection{Compressional Heating.}
This is due to the motion of the flux surfaces and is given by (see \ref{C4CompHeat})
\begin{equation}
\label{CompHeatDef}
\CompHeat = - T_s\fav{n_s  \dv\vpsi} = T_s\fav{n_s  \dv \left(\pd{\psi}{t} \frac{\nabla\psi}{|\nabla\psi|^2}\right)},
\end{equation}
where we have used the expression \eref{VpsiDef} for the velocity $\vpsi$ of a flux surface.

\subsubsection{Exchange between Potential and Thermal Energy.}
\label{potengsec}
The change in thermal energy due to exchange with the electrostatic and rotational potential energy is (see \ref{C4PotHeat})
\begin{equation}
\label{PotEngDef}
\fl
\begin{eqalign}{
\PotEng =  &- \fav{Z_s e \pot_0  \left(\pd{n_s}{t}+\vpsi\dg n_s\right)} - \fav{Z_s e n_s \pot_0\dv\vpsi} \\
 &\quad+ \frac{\angvel^2(\psi)}{2V'} \ddtpsi V'm_s \fav{R^2n_s} - \frac{1}{2} m_s \angvel^2(\psi)\fav{n_s \vpsi\dg R^2} \\
 &\quad+ \fav{\left(Z_s e\pot_0 - \frac{1}{2}m_s \angvel^2(\psi) R^2\right)\Psource}.
}\end{eqalign}
\end{equation}
The first line is the energy exchange with the electrostatic potential energy, the second the exchange with the rotational energy.
More transparently, we can group the terms in \eref{PotEngDef} involving $\pot_0$ together with the terms involving $\pot_0$ in our definition \eref{main-hflux-full} of the heat flux to write the total as 
\begin{equation}
\fl
\begin{eqalign}{
  &-\fav{Z_s e \pot_0  \left(\pd{n_s}{t}+\vpsi\dg n_s\right)} - \fav{Z_s e n_s \pot_0\dv\vpsi} +\fav{Z_s\pot_0\Psource}\\
	  &- \frac{1}{V'}\pd{ }{\psi} V'\fav{ Z_s e  \pot_0 n_s\bm{U}_s\dg\psi + Z_s e\pot_0 n_s \pd{\psi}{t}} \\
	&\quad\qquad = - Z_s e \fav{n_s \left(\bm{U}_s - \vpsi\right)\dg\pot_0},
}\end{eqalign}
\label{sighOhm}
\end{equation}
\setcounter{footnote}{1}
the work done by the mean electric field due to $\pot_0$.\footnote{$\vpsi$ appears here because we measure all velocities relative to $\vpsi$ and so, for a force $\bm{F}$, we consider $n_s \bm{F}\cdot\vpsi$ to be work done accelerating the plasma to $\vpsi$, not heating.}
Unfortunately, we do not know the small (i.e., of order $\gkeps^2\vth n_s$) poloidal component of $n_s \bm{U}_s$; the only way of calculating the right-hand side of \eref{sighOhm} is to calculate the left-hand side explicitly.
This was the reason for splitting the work done into the ``heating'' and ``flux'' terms and distributing them between \eref{PotEngDef} and \eref{main-hflux-full}, respectively.

Similarly, gathering the terms involving the rotation in \eref{PotEngDef} and \eref{main-hflux-full}, as well as the second term in \eref{VHDef} (see discussion in \Sref{VHsec}), we can rewrite them as
\begin{equation}
\fl
\begin{eqalign}{
  &\frac{\angvel^2(\psi)}{2V'} \ddtpsi V'm_s \fav{R^2n_s} - \frac{1}{2} m_s \angvel^2(\psi)\fav{n_s \vpsi\dg R^2} -\frac{1}{2}m_s\angvel^2(\psi)\fav{R^2\Psource}\\
	  &+\frac{1}{2V'} \pd{ }{\psi} V' m_s \angvel^2(\psi)\fav{n_s  R^2 \bm{U}_s\dg\psi -n_sR^2\pd{\psi}{t}} 
	 - m_s \angvel(\psi) \fav{R^2\ParticleFlux} \frac{d\angvel}{d\psi} \\
		 &\qquad= \frac{1}{2} m_s \angvel^2(\psi)\fav{n_s \left(\bm{U}_s - \vpsi\right)\dg R^2}.
}\end{eqalign}
\label{sighRot}
\end{equation}
which is the work done by the centrifugal force (the Coriolis force does no work).
Again, as with the work done by the electric field \eref{sighOhm}, we cannot explicitly evaluate the right-hand side of \eref{sighRot}. 

\subsubsection{Turbulent Heating.}
Finally, the energy exchange with the fluctuations is (see \ref{C4TurbHeat})
\begin{equation}
\begin{eqalign}{
\TurbPow &= Z_s e \fav{ \ensav{\wint \gyror{h_s\left( \pd{}{t} + \bm{u}\dg \right) \gkpot} }}\\
&\qquad- \frac{Z_s e}{c}\angvel(\psi) \fav{\ensav{\wint \gyror{h_s\bm{w}}\cdot \left(\delA\times\nabla z\right) }} \\
&= \TurbColl - \TurbInj.
}\end{eqalign}
\label{TurbPowDef}
\end{equation}
The last expression reflects the fact that $\TurbPow$ consists of two parts (see \eref{bing} in \ref{freeEnergyDeriv}): the turbulent heating  (dissipation of fluctuations on collisions):
\begin{equation}
\TurbColl = -\fav{\ensav{\wint \frac{T_s h_s}{F_{0s}} \lincol[h_s]}}
\label{TurbCollDef}
\end{equation}
and the energy injection into the fluctuations due to the mean density, temperature and velocity gradients:
\begin{equation}
\begin{eqalign}{
\TurbInj =& 
	  -\fav{\ensav{\wint \gyror{h_s \vchi}\dg\psi}} T_s \frac{d\ln \NotN}{d\psi} \\
	&\quad-\fav{\ensav{ \wint \left( \energy - \frac{3}{2} T_s \right) \gyror{h_s \vchi}\dg\psi }}\frac{d \ln T_s}{d\psi}  \\
&\quad+ \fav{\ensav{\wint m_s R^2\left(\nabla\tor\right) \cdot\gyror{\bm{v} h_s \vchi}\dg\psi}} \frac{d\angvel}{d\psi}.
}\end{eqalign}
\label{TurbInjDef}
\end{equation}
This is the energy borrowed from the internal energy of the equilibrium and transferred into turbulence by such mechanisms as the ITG, PVG, ETG, and other gradient-driven instabilities~\cite{dorlandETG,sagdeevITG,coppiITG,cattoPVG}. In \Sref{thermo}, we will show that all of this energy is eventually returned to the equilibrium via the turbulent heating term~\cite{waltzTurbulentHeating}: namely, after summation over species, net turbulent heating -- the difference between $\sum_s \TurbInj$ and $\sum_s \TurbColl$ -- is either viscous heating (conversion of rotational energy into heat at constant internal energy) or due to turbulent electromagnetic energy injected by antenna-like mechanisms, whose dissipation is the only piece of $\sum_s \TurbColl$ not cancelled by $\sum_s \TurbInj$.

\subsubsection{Collisional Heating.} Formally, the collisional heating term $\CollEnergy$ is both too large -- $\Or(\gkeps \cycfreq n_s T_s)$ or $\Or(\gkeps^2 \cycfreq n_s T_s)$ if all mean temperatures are equal, as $\collop[F_{0s}] \sim \gkeps \nu F_{0s}$ if all temperatures are equal -- and contains contributions from the gyrophase-independent part of $F_{2s}$, which we have not calculated. This problem has arisen from our decision to retain different temperatures for different species, a choice formally inconsistent with our orderings, as we acknowledged in the footnote preceding \eref{f0tmp}.
There are two na\"ive solutions to this problem. The first is to respect our ordering and assume that all temperatures are equal (which follows from \eref{entrop}). In this case we can sum the heat transport equation \eref{dpdt} over all species and, as collisions conserve energy, the collisional heating term will vanish
\begin{equation}
  \sum_s \CollEnergy = 0.
\end{equation}
Alternatively, if, completely formally, we order the interspecies collision frequency to be small $\nu_{s\,s'} \sim \gkeps^2 \nu_{s\,s}$, then this heating term will simply be the temperature equilibration between the Maxwellian equilibrium distributions of the species~\cite{helander2002ctm}:
\begin{equation}
  \CollEnergy = \sum_{s'} \nu^{\mathrm{(E)}}_{s\,s'} \left( T_{s'} - T_s \right),
  \label{TminusT}
\end{equation}
where
\begin{equation}
\nu^{\mathrm{(E)}}_{s\,s'} = 2^{5/2}\pi^{1/2} e^4 \frac{m_s^{1/2} m_{s'}^{1/2}n_s n_{s'} Z_s^2 Z_{s'}^2  \ln \Lambda_{s\,s'}}{\left(m_s T_{s'} + m_{s'} T_s\right)^{3/2}},
\label{TEquilRate}
\end{equation}
and $\ln \Lambda_{s\,s'}$ is the Coulomb logarithm.
However, this stretches the credibility of our ordering as there is no physical reason for $\nu_{s\,s'}$ to be smaller than $\nu_{s\,s}$ in general. Only in the special case of a very light species (electrons) or a very highly charged species (large-$Z_s$ impurities) is it reasonable to so order the temperature equilibration time. 
Thus, we elect to treat electrons separately but sum \eref{dpdt} over all other thermal species (ions), which should all have the same temperature.
\section{Energy Conservation in Multiscale Gyrokinetics}
\label{thermo}
In this section, we bring together the results of Sections \ref{s2ord} and \ref{transport} to determine how energy and entropy flow between mean and fluctuating quantities.

\setcounter{footnote}{1}
Firstly, in \Sref{energyNentropy}, we discuss energy conservation on the transport timescale. We show that the conserved quantity is the total energy of the plasma -- the sum of the thermal energy of each species, the rotational energy and the magnetic energy. We also show that the fluctuations can do no net work upon the plasma, i.e., there is no net turbulent heating (in the absence of direct injection of energy into the fluctuating scales -- as is, usually, the case in tokamaks).

Secondly, in \Sref{SFreeEnergy}, we derive the conservation laws that hold on the fluctuation timescale. We show that the conserved quantity is the free energy of the fluctuations. We then show that its conservation is local, i.e., neighbouring flux surfaces do not exchange free energy. We also use the steady-state version of this conservation law to interpret the absence of work done by the fluctuations (demonstrated in \Sref{energyNentropy}) as 
a precise balance between two energy flows: kinetic energy of the thermal particle motion and of the bulk plasma is converted into turbulent fluctuation energy via instabilities driven by mean-field gradients (e.g., ITG~\cite{sagdeevITG,coppiITG}, ETG~\cite{dorlandETG}, PVG~\cite{cattoPVG,newton2010understanding,AlexPVG}); it is then cascaded to small scales in phase space, where it is dissipated (converted back into thermal energy) by collisions.

Finally, in \Sref{specialMagField}, we discuss why the gradients of the magnetic field cannot drive fluctuations on their own and note the impossibility of anomalous resistivity in our formalism.

\subsection{Energy Conservation on the Transport Timescale}
\label{energyNentropy}
\subsubsection{Thermal Energy.}
\label{s9-1-1-thermal}
We begin by considering the evolution of the total thermal energy. In \ref{apEnergyConv}, we sum \eref{dpdt} over all species and simplify the result to obtain:
\begin{equation}
\fl
\begin{eqalign}{
\frac{3}{2}\frac{1}{V'}\hspace{-2pt}&\ddtpsi V'\sum_s \fav{n_s}T_s + \frac{1}{V'} \pd{}{\psi}V'\hspace{-2pt}\left[\sum_s \left(\fav{\HeatFlux}- T_s\fav{n_s\pd{\psi}{t}}\right) - \fav{\pot_0 \Meanj \dg\psi}\right] \\
	&=-\fav{\TotMomFlux}\frac{d\angvel}{d\psi} +  \frac{\angvel^2(\psi)}{2V'}\ddtpsi V'\inertia
+ \fav{\MeanE\cdot\Meanj} \\
	&\quad+ \fav{\EMViscosity}\frac{d\angvel}{d\psi}+ \sum_s \TurbPow
		+ \sum_s \fav{\Esource-\frac{1}{2}m_s\angvel^2(\psi)R^2\Psource},
}\end{eqalign}
\label{thermalenergy}
\end{equation}
where $\inertia$ is the plasma's moment of inertia, defined in \eref{inertiaDef}, $\TotMomFlux$ is defined in \eref{TMFdef}, $\EMViscosity$ in \eref{EMViscDef}, and $\TurbPow$ in \eref{TurbPowDef}.
Examining the terms on the right-hand side of the above equation, we identify the first and second as exchange between thermal and rotational energy, the third as the Ohmic heating associated with the mean fields, and the last term as the energy injected by the heat and particle sources.
We show in \ref{Sturbheat} (and, by a different route, in \Sref{S923}) that the fourth and fifth terms on the right-hand side of \eref{thermalenergy}, due to the fluctuations, cancel exactly (see \eref{TurbPowVisc}):
\begin{equation}
\sum_s \TurbPow + \fav{\EMViscosity} \frac{d\angvel}{d\psi} = 0.
\label{NoTHeat}
\end{equation}
Thus, the heating due to turbulence arises entirely out of the turbulent contribution to $\TotMomFlux$, which results in viscous heating in \eref{thermalenergy} (the first term on the right-hand side), i.e., the conversion of rotational energy into thermal energy or vice versa.
\subsubsection{Rotational Energy.}
The last point is made manifest by considering the evolution of the rotational kinetic energy. Multiplying \eref{domegadt} by $\angvel(\psi)$, we obtain
\begin{equation}
  \begin{eqalign}{
  \frac{1}{2} \frac{1}{V'}&\ddtpsi V' \inertia \angvel^2(\psi) + \frac{1}{V'}\pd{}{\psi} V'\angvel(\psi)\fav{\TotMomFlux} \\
	  &= \fav{\TotMomFlux} \frac{d\angvel}{d\psi} - \frac{\angvel^2(\psi)}{2V'}\ddtpsi V'\inertia  + \angvel(\psi)\fav{\Msource},
  }\end{eqalign}
\label{roteng}
\end{equation}
where we have written the left-hand side as an exact time derivative and an exact spatial derivative and collected the terms arising from this manipulation on the right-hand side so as to parallel \eref{thermalenergy}.
We now add \eref{roteng} to \eref{thermalenergy} to find
\begin{equation}
\label{energyconv}
\begin{eqalign}{
\frac{1}{V'}\ddtpsi V'{U}
	+& \frac{1}{V'} \pd{}{\psi} V'\fav{\JUflux}  = 
		\fav{\MeanE\cdot\Meanj} \\
	&+\sum_s \fav{\Esource + \angvel(\psi) \Msource -\frac{1}{2}m_s\angvel^2(\psi)R^2\Psource},
}\end{eqalign}
\end{equation}
where the total kinetic energy of the plasma (thermal energy and rotational kinetic energy) is
\begin{equation}
\label{UDef}
U = \sum_s \frac{3}{2}\fav{n_s}T_s +\frac{1}{2}\inertia \angvel^2(\psi),
\end{equation}
and its flux is
\begin{equation}
\label{JUflux}
\fl\begin{eqalign}{
&\fav{\JUflux} = \sum_s \left(\fav{\HeatFlux}-T_s\fav{n_s\pd{\psi}{t}}\right) + \angvel(\psi) \fav{\TotMomFlux} - \fav{\pot_0 \Meanj\dg\psi} \\
		&= \fav{\left\{ \sum_s \left[\bm{Q}_s + \frac{1}{2} m_s \angvel^2(\psi)R^2 n_s \bm{U}_s + \bm{u}\cdot\viscosity\right] + \frac{c}{4\pi}\ensav{\delE\times\delB} \right\}\dg\psi} \\
&\quad+ \fav{U\pd{\psi}{t}}.
}\end{eqalign}
\end{equation}
The two forms of $\fav{\JUflux}$ written above are shown to be equivalent in \ref{apJUSimples}. They serve two different purposes. The former is easier to evaluate for a specific practical implementation of the transport equations, while the latter is easier to interpret physically. 
In the second form of $\JUflux$, the terms are identified as: the heat flux (see footnote in \Sref{heattrans} before \eref{heatflux}), the flux of rotational energy, the energy flux due to the viscous stress,  the Poynting flux due to the fluctuations and a term due to the motion of flux surfaces.

\subsubsection{Magnetic Energy.}
\label{s9-1-3-magnetic}
The evolution equation \eref{energyconv} for $U$ has two sources on the right-hand side: Ohmic heating and energy injection due to heat and particle sources. 
Let us now express the Ohmic heating term (the first term on the right-hand side of \eref{energyconv}) by using Poynting's theorem for the mean fields (as Maxwell's equations are linear, Poynting's theorem holds separately for the mean and fluctuating fields):
\begin{equation}
\pd{ }{t} \frac{\MeanMagB^2}{8\pi} + \frac{c}{4\pi}\dv\left(\MeanE\times\MeanB\right) = - \MeanE\cdot\Meanj.
\label{avPoynt}
\end{equation}
Averaging this equation over a flux surface and adding it to \eref{energyconv}, we obtain
\begin{equation}
\label{TotalenergyTMP}
\label{Totalenergyconv}
\fl
\begin{eqalign}{
\frac{1}{V'}&\ddtpsi V'\left(U + \frac{\fav{\MeanMagB^2}}{8\pi}\right) \\
	&\quad+ \frac{1}{V'} \pd{}{\psi} V'\left(\fav{\JUflux} - \frac{c}{4\pi} I(\psi)\fav{\MeanE\cdot\MeanB} - \fav{\frac{B^2}{8\pi}\pd{\psi}{t}}\right)\\
		&= \sum_s \fav{\Esource + \angvel(\psi) \Msource -\frac{1}{2}m_s\angvel^2(\psi)R^2\Psource},
}\end{eqalign}
\end{equation}
where we have used the flux-surface-average identities \eref{favdiv} and \eref{movingflux} and $\MeanB\times\nabla\psi = -I(\psi)\MeanB + R^2\MeanMagB^2 \nabla\tor$ to rewrite the radial Poynting flux in the following way:
\begin{equation}
\fl
\left( \MeanE\times\MeanB \right)\dg\psi  = -I(\psi)\MeanE\cdot\MeanB + R^2\MeanMagB^2\MeanE\cdot\nabla\tor = -I(\psi) \MeanE\cdot\MeanB - \frac{\MeanMagB^2}{c} \pd{\psi}{t}.
\end{equation}
\Eref{Totalenergyconv} is the desired energy conservation law on the transport timescale: there are no sources save those explicitly contained in the kinetic equation.

Note that, as \eref{avPoynt} contains no contribution from the fluctuations, turbulence cannot cause any exchange between the total kinetic energy $U$ and the energy of the mean magnetic field.
These results, suprising at first glance, in fact follow directly from Poynting's theorem for the fluctuating fields:
\begin{equation}
\begin{eqalign}{
  \pd{}{t}  \frac{\MagDelB^2}{8\pi} + \frac{c}{4\pi}\dv\left(\delE\times\delB\right) = - \delE\cdot\delj,
}\end{eqalign}
\label{fluctPoynt}
\end{equation}
where we have dropped the energy in the electric field as it is negligible  for non-relativistic plasmas. Averaging \eref{fluctPoynt} over the fluctuations gives
\begin{equation}
\label{blah}
\frac{c}{4\pi}\dv\ensav{\delE\times\delB} = -\ensav{\delE\cdot\delj},
\end{equation}
as the time derivative of the averaged fluctuating magnetic energy is $\gkeps^2$ smaller than the other two terms. This clearly demonstrates that the work done on the particles by the fluctuating fields (the right-hand side) is balanced by the Poynting flux of fluctuating electromagnetic energy (the left-hand side), so the fluctuations cannot change the total kinetic energy $U$.
The Poynting flux associated with the fluctuations appears explicitly in the second line of \eref{JUflux}. If we wished to consider electromagnetic fluctuations driven by an external source, this would enter the energy-conservation equation not through a source term but through the fluctuating Poynting flux at the plasma boundary.
We will return to the physical interpretation of this result in \Sref{specialMagField}, after discussing the conservation of free energy in the next section.

\subsection{Free Energy Conservation and the Turbulent Cascade}
\label{SFreeEnergy}
In this section, we derive the fluctuating counterpart to the energy conservation law of the previous section. 
The conserved (cascading) quantity for kinetic turbulence is not the energy of the fluctuations, but is in fact
a combination of the entropy and the energy known as the free energy~\cite{Krommes2,hallatschek,Tome,schekcrete}.
In our notation, this quantity is
\begin{equation}
\label{FreeEnergyDef}
\FreeEnergy(\psi) = \sum_s\fav{\perpav{\wint \frac{T_s \delta f_s^2}{2 F_{0s}}}} + \fav{\perpav{\frac{\MagDelB^2}{8\pi}}},
\end{equation}
where we have averaged over an annular region of space centered on the flux surface labelled by $\psi$ (the composition of the flux-surface average and the perpendicular spatial average)\setcounter{footnote}{1}\footnote{For the purposes of this section, we can visualise a flux surface as having a finite width (comparable to the intermediate length scale $\lambda$ introduced in \Sref{ordering}) and so in our discussion we will drop the distinction between a flux surface and the annulus centered upon it.}, but {\it not} over time.  The first term in \eref{FreeEnergyDef} is $-\sum_s T_s \delEnt$, where $\delEnt$ is the part of the mean entropy density of species $s$ due to the fluctuations (note that this is not the same as the fluctuating part of the entropy density because $\delEnt$ has a nonzero average; see \Sref{subEntBal}). Thus, \eref{FreeEnergyDef} agrees with the usual thermodynamic definition of Helmholtz free energy.

In \ref{freeEnergyDeriv}, we derive an evolution equation for $W$ from the gyrokinetic equation \eref{gke}:
\begin{equation}
\fl
\begin{eqalign}{
&\ddtpsi W = \sum_s \fav{\perpav{\wint \gyror{\frac{T_sh_s }{F_{0s}} \lincol[h_s]}}}\\
	&- \sum_s \fav{\perpav{\wint \gyror{h_s \vchi}}\dg\psi} T_s\left( \frac{d\ln \NotN}{d\psi} -\frac{3}{2} \frac{d\ln T_s}{d\psi}\right)\\
	&-\sum_s \fav{ \perpav{\wint  \energy  \gyror{h_s \vchi}}\dg\psi }\frac{d  \ln T_s}{d\psi}\\
&- \fav{R^2\left(\nabla\tor\right) \cdot\left[\sum_s\perpav{ \wint \gyror{\bm{v}h_s \vchi}} - \perpav{\frac{1}{4\pi}\delB\delB + \frac{1}{c} \delA \delj}\right]\cdot \nabla\psi } \frac{d\angvel}{d\psi}.
}\end{eqalign}
\label{freeEnergyBalance}
\end{equation}
This is the free-energy balance on a given flux surface. 
\subsubsection{The Local Cascade Law.}
Before interpreting the sources and sinks in this equation, we would like to highlight the fact that there is no flux of free energy
-- the turbulent cascade of free energy is essentially local and does not couple neighbouring flux surfaces.
This means that turbulence excited on a given flux surface must dissipate on that surface and, moreover, that the turbulence is excited by the local gradients and cannot originate from turbulence propagating from neighbouring flux surfaces.
Succinctly: {\it that which is stirred up in the flux surface, stays in the flux surface} -- the local (in space) cascade law of tokamak turbulence.\footnote{This conclusion is at odds with the literature on turbulence spreading -- a process whereby the turbulence (characterised by the free energy $W$, or some other quadratic measure of the intensity of the turbulence) ``spreads'' out of the radial region where it is driven (e.g., by a linear instability) into a region where the fluctuations are damped. Observations from gyrokinetic simulations have been interpreted as evidence for turbulence spreading~\cite{lin:1099,linSizePRL}, and theoretical models have been proposed to describe this effect~\cite{hahmSpreading,garbetSpreading,naulin:122306}. The numerical evidence comes from global simulations, run with a finite (i.e., non-zero) value of $\gkeps = \rho_i / a$ -- i.e., simulations in which the scale separation between the fluctuations and the mean profiles is not explicitly enforced. This suggests that the appearance of turbulence spreading could be due to the lack of scale separation. Indeed, the evidence presented in \cite{linSizePRL} shows that, as the simulations become more asymptotic (smaller $\rho_i/a$), the observed spreading decreases. The fact that there is no spreading in our formulation corroborates this interpretation -- we have completely separated the spatial scales of the turbulence from those of the equilibrium and, consequently, there is no spreading.
	Mathematically, our equation \eref{freeEnergyBalance} for $W$ is the analog of (1) in \cite{hahmSpreading} or (2) in \cite{naulin:122306}. Comparing \eref{freeEnergyBalance} to these other equations, we observe that a key difference is that there are no spatial derivatives of $W$ present in \eref{freeEnergyBalance}. This, in turn, follows from the fact that the spatial/flux-surface averaging procedure eliminated all such terms (see \eref{NoSpreading} and \ref{NoSpreadProof}), again demonstrating that it is scale separation that prohibits turbulence spreading.
}
\subsubsection{Sources and Sinks of Free Energy.}
The first term on the right-hand side of \eref{freeEnergyBalance} is the sink of free energy due to collisional dissipation (the fact that it is a sink and not a source follows from the $H$-Theorem for the linearised collision operator $\lincol[h_s]$; see \Sref{Sboltequib}). The remaining source terms are the entropy generated by the turbulent fluxes transporting mean quantities along their gradients.  As the fluxes are generally dominated by the largest turbulent scales and the collisional dissipation is most important when acting on the smallest scales (in phase space), we can interpret \eref{freeEnergyBalance} in terms of a cascade in phase space analogous to the Kolmogorov cascade of energy in hydrodynamic turbulence~\cite{Tome,schekcrete}:
fluctuations are excited at large scales, then nonlinearly cascaded, whilst conserving $W$, to small scales, where they are destroyed by collisional dissipation.

It is important to note that only gradients in the thermodynamic and mechanical quantities (density, temperature and angular velocity) appear in the right-hand side of \eref{freeEnergyBalance}. Only gradients in those quantities represent deviations from the global thermodynamic equilibrium and can, therefore, drive fluctuations (as the presence of fluctuations implies non-zero $W$). Gradients in the magnetic field strength or magnetic curvature do not have this property; the relaxation of the mean magnetic field to a minimum energy state proceeds independently of the fluctuations (i.e., there is no fluctuating contribution to \eref{avPoynt}; see further discussion in \Sref{specialMagField}).

\subsubsection{Statistically Steady-State Turbulence.}
\label{S923}
Time averaging \eref{freeEnergyBalance}, we see that a balance between free-energy injection and dissipation is set up on a timescale much shorter than the intermediate time $T$ (introduced for the purposes of time averaging in \Sref{ordering}). This balance can be written as
\begin{equation}
\sum_s\TurbInj -\fav{\EMViscosity} \frac{d\angvel}{d\psi} = \sum_s \TurbColl,
	\label{steadystate}
\end{equation}
where $\TurbInj$ and $\TurbColl$ are defined by \eref{TurbInjDef} and \eref{TurbCollDef}, respectively, and $\EMViscosity$ by \eref{EMViscDef}.
This is another derivation of \eref{NoTHeat}.
This clearly demonstrates that the steady-state turbulence extracts all its energy from the equilibrium, but it also returns exactly the same amount of energy via collisions. This explains how the fluctuations can be continuously driven and dissipated without changing the total kinetic energy $U$. 
Turbulence can, however, move energy between various constituent parts of $U$, notably energy extracted from one species does not have to be returned to that species, and so fluctuations can redistribute energy between species (subject to the caveats about temperature equilibration at the end of \Sref{heattrans}). This can lead to systematic dependences of equilibrium temperatures on the nature of the species, e.g., in the case of effective heating of heavy minority ions~\cite{barnesMinority}.

Similarly, turbulence can convert rotational energy into thermal energy (or even vice versa, e.g., if there is an angular momentum pinch or intrinsic rotation~\cite{waltzCoriolis,waltzIntrinsicRotation,peeters2007toroidal,camenen2009intrinsic,jungpyo2013intrinsic}).

\Eref{freeEnergyBalance} implies that in order to sustain a non-zero steady-state flux of particles, heat or angular momentum, collisional dissipation is required. This is the reason, alluded to in \Sref{gkord}, why it was crucial to retain $\lincol[\delta f_s]$ at the same order as $\infrac{\partial \delta f_s}{\partial t}$. Similarly, this is why all simulations of the gyrokinetic equation \eref{gke} have to include some dissipation in order to reach a steady state \cite{Krommes2,watanabe2002kinetic,watanabe2004kinetic}.
\subsection{The Special Status of the Mean Magnetic Field}
\label{specialMagField}
At several points in the above discussion, we have noted that the mean magnetic field behaves very differently from the other mean fields. In this section, we collect these individual points together to further elucidate the special role that the mean magnetic field plays
in multiscale gyrokinetics.

The mean magnetic field is determined from the Grad-Shafranov equation \eref{gradshaf} and the evolution equation for the safety factor $q(\psi)$ \eref{dqdt} (see the detailed discussion in \Sref{magevolve}). These equations contain no direct contribution from the fluctuations. In this sense they are very different from the transport equations determining the mean density, velocity, and temperature profiles. 
This separation is also reflected in the energetics of the magnetic field. The fluctuations cannot directly convert magnetic energy into thermal energy (note the absence of any fluctuation-dependent term in \eref{avPoynt}). Correspondingly, the magnetic field is not a source of turbulent free energy (there are no source terms in \eref{freeEnergyBalance} arising from gradients in $\MeanB$), i.e., energy cannot be borrowed from the magnetic field to fuel fluctuations in the same way that it can be from the thermal or kinetic energy of the plasma.

Ultimately, these results imply that there is no turbulent contribution to the diffusion of mean magnetic flux. The fluctuations considered here cannot reconnect the mean magnetic field, i.e., they can change the topology of the total magnetic field $\Bfield = \MeanB + \delB$ but {\textit{only}} on small spatial and temporal scales, while the topology of $\MeanB$ is preserved -- indeed this is true automatically as we assumed axisymmetry of the equilibrium (see \Sref{maggeom}). Equivalently, we can say that there is no anomalous resistivity due to gyrokinetic turbulence (i.e., the evolution of $q(\psi)$, given by \eref{dqdt}, cannot be sped up by turbulence).

\section{Entropy Flows and the Thermodynamics of Multiscale Gyrokinetics}
\label{Sentropy}
In the preceeding section, we have established the nature of the energy flows, both between the equilibrium fields and the fluctuations, and between neighbouring flux surfaces.
In this section, we discuss these results in explicitly thermodynamic language.

In \Sref{subEntBal}, we derive an evolution equation for the mean entropy and show that this is not only consistent with the above results but also provides further evidence that flux surfaces interact only through particle, momentum and heat fluxes (as opposed to turbulence spreading). In \Sref{Sboltequib}, we relate the evolution of the mean entropy to notions of dissipation and equilibration through Boltzmann's $H$-Theorem. In \Sref{SapproachEquib}, we relate the
sources of mean entropy to sinks of free energy discussed previously.
We discover that dissipation increases mean entropy via three routes: collisional and turbulent temperature equilibration, Ohmic heating due to the induced electric field, and 
fluxes that strive to flatten gradients in the thermodynamic quantities $n_s$, $T_s$ and $\angvel(\psi)$. We also discuss the effects that can occur in systems where two or more equilibration mechanisms compete.
\subsection{Entropy Balance}
\label{subEntBal}
The (Boltzmann) entropy density of species $s$ is given by
\begin{equation}
\entropy[s] = - \wint f_s \ln \left(\frac{ 8\pi^3 \hbar^3 f_s}{m_s^3}\right), \qquad \entropy = \sum_s \entropy[s],
\label{EntropyDef}
\end{equation}
where $\hbar$ is Planck's constant (for an explanation of this suprise appearance of $\hbar$ in a classical formula see \S7 of \cite{landau-sp1}).
To lowest order in $\gkeps$, $f_s = F_{0s}$ and so the mean entropy density of the plasma is
\begin{equation}
\label{MeanEntropyDef}
\begin{eqalign}{
\MeanEntropy= \ensav{\entropy} 
	&= -\sum_s n_s \left[ \ln \left( \frac{n_s}{\quantconc}  \right) - \frac{3}{2} \right] + \Or(\gkeps\MeanEntropy),
}\end{eqalign}
\end{equation}
where $\quantconc = \binfrac{m_sT_s}{2\pi\hbar^2}^{3/2}$ is the quantum density of states for the ideal gas of species~$s$.
This is just the standard result for a mixture of ideal gases in local thermodynamic equilibrium~\cite{landau-sp1}, as expected. We are in local thermodynamic equilibrium: $f_s$ is Maxwellian to lowest order (the equilibrium is local, not global, as $n_s$ and $T_s$ are functions of space). Because we are considering a given flux surface, the presence of the rigid rotation $\angvel(\psi)$ does not contribute to the entropy density at this surface.
Expanding $f_s = F_s + \delta f_s$, we find an expression for the mean entropy density that retains 
terms up to order $\gkeps^2$:
\begin{equation}
\fl
\MeanEntropy = -\sum_s\wint F_s \ln\left( \frac{8\pi^3\hbar^3F_s}{m_s^3} \right) - \sum_s  \ensav{\wint\frac{\delta f_s^2}{2F_s}} + \Or(\gkeps^3 \MeanEntropy),
\label{deltaSdef}
\end{equation}
where the first term is the entropy of the mean distribution $F_s$ and the second term is the contribution from the fluctuations, $\delEnt$, which was introduced in the discussion of free-energy conservation (see \eref{FreeEnergyDef}; in \eref{deltaSdef}, $\delEnt$ is also time-averaged).

We now derive an evolution equation for $\MeanEntropy$. Multiplying the kinetic equation \eref{vfp} by $-\left[1 + \ln\binfrac{8\pi^3\hbar^3 f_s}{m_s^3}\right]$, integrating over all velocities, summing over species, averaging over the fluctuations and over a flux surface, we obtain
\begin{equation}
\frac{1}{V'}\ddtpsi V'\fav{\MeanEntropy} + \frac{1}{V'} \pd{}{\psi} V' \fav{\Rentropyflux} = \fav{\CollEntropy} + \fav{\EntropySource},
	\label{entropyConservation}
\label{dSdt}
\end{equation}
where we have used \eref{movingflux} to simplify the time derivative, \eref{favdiv} for the flux-surface average of a divergence, and defined the radial entropy flux
\begin{equation}
\label{RentDef}
\Rentropyflux = -\sum_s {\ensav{\wint \bm{v} f_s\ln \left(\frac{8\pi^3\hbar^3 f_s}{m_s^3}\right)}\dg\psi} +\MeanEntropy\pd{\psi}{t},
\end{equation}
the collisional entropy production
\begin{equation}
\CollEntropy = -\sum_s\ensav{\wint \ln\left( \frac{8\pi^3 \hbar^3 f_s}{m_s^3} \right) \collop[f_s]},
\end{equation}
and the explicit entropy source
\begin{equation}
\begin{eqalign}{
\EntropySource &= -\sum_s \ensav{\wint \left[ 1 + \ln \left(\frac{8\pi^3\hbar^3f_s}{m_s^3}\right) \right] \source} \\
					&= -\sum_s \left[ \ln \left( \frac{n_s}{\quantconc}\right) \Psource - \frac{3}{2} \frac{\Esource}{T_s} \right].
}\end{eqalign}
\end{equation}

In \ref{entropyDeriv}, we show that the entropy flux \eref{RentDef} can be written in terms of the particle and heat fluxes through a flux surface as follows
\begin{equation}
\begin{eqalign}{
\fav{\Rentropyflux}  
&= - \sum_s \left[\left( 1 + \frac{\chempot[s]}{T_s} \right) \fav{\ParticleFlux} + \frac{\fav{\HeatFlux}}{T_s}\right],
}\end{eqalign}
\label{RentFluxMain}
\end{equation}
where we have introduced the chemical potential of an ideal gas of particles of species~$s$:
\begin{equation}
\chempot(\psi) = T_s \ln\left(\frac{n_s}{\quantconc}\right) + Z_s e \pot_0 - \frac{1}{2} m_s \omega^2(\psi) R^2 = T_s \ln\left(\frac{\NotN[s]}{\quantconc}\right),
\label{ChemPotDef}
\end{equation}
(we used \eref{npol} to obtain the last equality; note that the last two terms of the first equality are the potential energy per particle).
Equation \eref{RentFluxMain} is analogous to the same result for a mixture of ideal gases~\cite{balescu1988tpp,deGrootMazur}. 
We have already seen, in \Sref{transport}, that each annulus interacts with neighbouring annuli via the particle, momentum, and heat fluxes. The expression \eref{RentFluxMain} for the entropy flux shows that the annuli exchange entropy only through the particle and heat fluxes (but not momentum, as we should expect, given that the mean entropy density does not depend upon $\angvel(\psi)$).

By expanding $f_s$ in terms of $F_{s}$ and $\delta f_s$, we can write the mean (collisional) entropy production term as
\begin{equation}
\begin{eqalign}{
\fav{{\EProd}} =&
	-\sum_s\wint \fav{\ln{\left( \frac{8\pi^3\hbar^3 F_{s}}{m_s^3} \right)} \collop[F_{s}]} \\
	&\quad- \sum_s\wint \fav{\ensav{\frac{h_s}{F_{0s}} \lincol[h_s]}}
		+ \Or(\gkeps^4 \cycfreq \MeanEntropy),
}\end{eqalign}
\label{eprodExp}
\end{equation}
where we have used \eref{hDef} for $\delta f_s$. 
In this expression, we can clearly identify the separate contributions from collisional dissipation of $F_s$ and collisional dissipation of $\delta f_s$. It is important to note that the second term in \eref{eprodExp} is minus the time average of the first term on the right hand side of \eref{freeEnergyBalance} -- the rate of entropy production due to the fluctuations is precisely the rate at which free energy is dissipated.

We determined in \Sref{thermo} that the energy budget for the fluctuations consists of borrowing some energy from the mean thermal or bulk kinetic energy, cascading it to small scales, and then returning it. We also showed that in this process no net work was done on the plasma. We seem to have arrived at a contradictory result: we know that, in the absence of viscous heating, there is no local heating, but \eref{eprodExp} shows that we have a monotonically increasing entropy (the right-hand  side of \eref{eprodExp} is positive-definite if we are away from thermodynamic equilibrium). This apparent contradiction is resolved in the next two sections, where we consider the precise nature of the thermodynamic equilibration of our system.
\subsection{Boltzmann's \texorpdfstring{$H$}{H}-Theorem and Global Equilibrium}
\label{Sboltequib}
Boltzmann's $H$-Theorem~\cite{boltzmann} states that for a homogenous dilute gas of hard spheres, the entropy $\entropy[s]$, as defined by \eref{EntropyDef}, is non-decreasing and the only steady-state distribution is a Maxwellian. Let us examine the corresponding theorem for the plasmas under consideration here.
The general form of the $H$-Theorem for a dilute plasma is:
if $f_s(\bm{r},\bm{v})$ is the distribution function for species $s$, then \cite{helander2002ctm}
\begin{equation}
\sum_s \int d^3{\bm{v}} \ln f_s \collop[f_s] \le 0,
\label{HTheorem}
\end{equation}
with equality achieved if and only if for each species $f_s$ is a Maxwellian distribution and all their temperatures and mean velocities are equal.
We have already called upon this result in \Sref{maxback} to prove that $F_{0s}$ is Maxwellian. We now examine its implications for the thermodynamics of the plasma.

First of all, we see that $\CollEntropy \ge 0$, i.e., collisions generate entropy. Moreover, the two terms in the expression \eref{eprodExp} for $\CollEntropy$ are separately non-negative. Thus, the relaxation of the equilibrium $F_s$ and the dissipation of the turbulent free energy are separate net sources of entropy. 

It is instructive to discuss the equilibrium towards which these relaxation process are driving the system. In the absence of sources or fluxes of entropy connecting the plasma to the outside world, the system is in a global equilibrium if $\CollEntropy$ vanishes when integrated over the plasma volume. However, as $\fav{\CollEntropy}$ is positive definite on each flux surface, for its volume integral to vanish, it must vanish locally on each surface. From the $H$-Theorem, we know that this is the case only if $f_s$ is a Maxwellian up to possible errors of $\Or(\gkeps^2 f_s)$ (these smaller deviations from a Maxwellian generate an evolution of the entropy on timescales longer than $\tau_E$ and so are negligible). By applying the $H$-theorem to the first term of \eref{eprodExp}, we see that $F_s$ must be Maxwellian for $\CollEntropy$ to vanish. From the explicit expression \eref{expF3} for $F_s$ (\ref{flowcurrent}), we conclude that for $F_s$ to be Maxwellian, all the gradients in $F_{0s}$ (i.e., those of $n_s$, $T_s$ and $\angvel$) must vanish and the neoclassical distribution function $\Fneo$ must also be Maxwellian (in the absence of gradients in $F_{0s}$, \eref{neogeq}, which determines $\Fneo$, admits purely Maxwellian solutions).\footnote{Even though we do not have an evolution equation for it, $\Fneo$ evolves in time due to the time evolution of the right-hand side of \eref{neogeq} and the time evolution of the collision frequency.}
Thus, collisions, which drove the distribution on each flux surface to be Maxwellian on shorter timescales, now strive to smooth out the gradients between the flux surfaces.

The $H$-theorem has determined the form of the background distribution function, the direction of the free-energy cascade {\it from} injection scales {\it to} collisional scales in phase space, and it has determined the ultimate equilibrium to which the system wishes, in some sense, to arrive at. There remains but one issue to clarify: how the relaxation towards this global equilibrium is achieved.
\subsection{Approach to Equilibrium and Multichannel Transport.}
\label{SapproachEquib}
The precise nature of the relaxation process is hidden inside the entropy production terms in \eref{eprodExp}.
In \ref{eprodDeriv}, we show that the collisional dissipation in \eref{eprodExp} can be expressed entirely in terms of fluxes through a flux surface and local relaxation terms\setcounter{footnote}{1}\footnote{We note that this is a somewhat arbitrary distinction. For example, the viscous heating (the last term in \eref{entropyProductionMain}) can be considered as a flux-gradient term for the flux of angular momentum, or a local heating term.}~\cite{sugama1998neg,sugama1997tpe}:
\begin{equation}
\fl
\begin{eqalign}{
&\fav{{\EProd}} = \sum_s \frac{1}{T_s}\left(  \fav{\CollEnergy} + {\TurbPow} + \JouleHeat\right)\\
	&\quad\,\,-\sum_s\left\{ \fav{\ParticleFlux}  \frac{d}{d\psi} \frac{\chempot}{T_s}\right.  + \frac{\fav{\HeatFlux}}{T_s}\frac{d \ln T_s}{d\psi} 
	\left. + \frac{1}{T_s}\left[\fav{\MomentumFlux} + m_s \angvel(\psi) \fav{R^2 \ParticleFlux}\right] \frac{d\angvel}{d\psi}\right\},
}\end{eqalign}
\label{entropyProductionMain}
\end{equation}
where $\CollEnergy$ is defined in \eref{CollEnergyDef}, $\TurbPow$ in \eref{TurbPowDef}, $\JouleHeat$ in \eref{OhmHeat},
$\fav{\ParticleFlux}$ in \eref{partflux}, $\chempot$ in \eref{ChemPotDef}, $\fav{\HeatFlux}$ in \eref{heatflux}, $\fav{\MomentumFlux}$ in \eref{momflux}, and $\fav{R^2\ParticleFlux}$ in \eref{conMomflux}. Note that
\begin{equation}
\frac{d}{d\psi} \frac{\chempot}{T_s} = \frac{d \ln\NotN}{d\psi} - \frac{3}{2}\frac{d\ln T_s}{d\psi}.
\end{equation}

We now interpret \eref{entropyProductionMain}, term by term.

The first term is just the collisional equilibration of mean temperatures; it vanishes if all temperatures are equal. 

The second term contains two effects. Firstly, if all temperatures are equal ($T_s=T$), then we can use \eref{NoTHeat} to write this term as a viscous heating term due to $\fav{\EMViscosity}$ and combine it with the last term in \eref{entropyProductionMain} to form the total viscous heating $-({1}/{T})\fav{\TotMomFlux} \infrac{d\angvel}{d\psi}$. Secondly, if there is no rotation ($\angvel=0$) then we find that, from \eref{NoTHeat}, $\sum_s \TurbPow = 0$. 
Therefore, in the absence of rotation, this term represents a change in entropy not due to net heating but due to
the {\textit{differential}} heating of different species by the turbulence. 
Hence, in the absence of rotation, this term is a source of entropy if it equilibrates temperatures but a sink of entropy if it drives them apart. That this term can generate entropy without net heating is part of the resolution of the apparent contradiction noted at the end of \Sref{subEntBal}.

The third term in \eref{entropyProductionMain} is the Ohmic heating due to the induced parallel electric field.

The remaining terms in \eref{entropyProductionMain} (the second sum) can be interpreted as fluxes multiplying their corresponding gradients. 

The first term contains the particle flux and the gradient of the local chemical potential.
The appearance of the chemical potential is natural as we can consider the flux surface to be a thermodynamic system that can gain or lose particles. 

The second term is just the heat flux (in the extended definition adopted in \Sref{Sec831}) multiplying the temperature gradient. 

Finally, the term multiplying the angular velocity gradient is not precisely the momentum flux as the electromagnetic momentum flux $\EMViscosity$ is hidden within $\TurbPow$ as discussed above. This ambiguity arises because, if the temperatures are unequal, it is impossible to separate uniquely the temperature equilibration due to turbulence driven by angular velocity shear and viscous heating due to the angular momentum flux driven by the turbulence.

Individually, these flux-gradient terms are positive for any flux that acts to flatten the corresponding gradient, and negative if the flux steepens the gradient.\setcounter{footnote}{1}\footnote{Take, for example, the first term in the sum, which is the entropy production arising from the particle flux $\fav{\ParticleFlux}$ and the density gradient $\infrac{d\ln \NotN}{d\psi}$; assume all other gradients are zero. If $\psi$ increases away from the magnetic axis, $\fav{\ParticleFlux}$ will be positive for a flux or particles that is going away from the magnetic axis and $\infrac{d\ln \NotN}{d\psi}$ will be negative for a density profile that is peaked on-axis. Thus, their contribution to the entropy production
$- \sum_s \fav{\ParticleFlux} \binfrac{d\ln\NotN}{d\psi}$
will be positive for a flux that relaxes the gradient, and negative for one that steepens it. This is independent of the definition of $\psi$ as it enters in both terms and the effect of flipping the sign of $\psi$ cancels between the two.
} The presence of turbulent contributions to the fluxes in these terms is the second part of the resolution of the apparent contradiction that was raised at the end of \Sref{subEntBal} -- the fluctuations can increase entropy without heating the plasma by flattening gradients.
From \eref{HTheorem}, we know that $\CollEntropy$ is positive. 
However, it is clearly possible to have some negative terms in \eref{entropyProductionMain} without violating this constraint -- up-gradient fluxes of particles, heat, or momentum are examples of just such negative terms.
It is the multi-channel nature of the transport that allows such phenomena: offsetting a pinch in one channel against a larger outward flux in another to observe a net increase in entropy.
 
Stating that the transport is multi-channel simply means that gradients in any one thermodynamic quantity ($n_s$, $T_s$, or $\angvel$) can drive fluxes of any other quantity. For example, trapped-electron-mode turbulence can be driven predominantly by the electron density gradient but produces both a particle flux and an electron heat flux~\cite{dannert:072309}. Combining this multi-channel property and the fact that the terms due to the fluctuations in \eref{entropyProductionMain} can have differing signs and are only constrained to be positive upon summation can result in suprising effects -- such as the spontaneous steepening of a gradient in $\angvel$ merely from temperature-gradient driven turbulence~\cite{waltzIntrinsicRotation,camenen2009intrinsic}.

Finally, we note that, as expected from the discussions in \Sref{specialMagField}, gradients of the mean magnetic field do not appear as sources of entropy.

\section{Multiscale Gyrokinetics at Low Mach Number}
\label{LowMach}
\setcounter{footnote}{1}
In Sections \ref{s0ord}--\ref{transport}, a multiscale hierarchy of equations is derived from the Fokker-Planck kinetic equation by a systematic expansion in $\gkeps = \rho / a$.
The full generality of these equations is, in fact, unnecessary for a large class of plasma conditions, namely those where the Mach number $M = \left. R\angvel(\psi)\right/ \vth$ of the toroidal rotation is low, i.e., $M \ll 1$. In this section, we investigate the low-Mach-number limit: formulating the expansion in $M\ll 1$ precisely in \Sref{SLowMachLimit} before presenting the results of this expansion in Sections~\ref{LM-Mean}--\ref{LM-Energy}.
It is not necessary to read the detailed derivations in Sections \ref{s0ord}--\ref{thermo} and their attendant appendices to use the results presented this section; however familiarity with the notation of Sections~\ref{kinetic-eq} and \ref{Sfluctmean} will be assumed. Because of the considerable simplifications that the low-Mach-number expansion brings, the exposition in this section can be read as an easier-to-grasp summary of the basic structure of multiscale gyrokinetics.
\subsection{The Low-Mach-Number Ordering}
\label{SLowMachLimit}
There are two distinct low-Mach-number regimes. In the extreme limit of $M \sim \gkeps$, all dependence on the angular velocity $\angvel$ drops out of the equations for both the mean and fluctuating fields, and \eref{domegadt} is no longer sufficient to calculate angular momentum transport.
This is the ``low-flow regime'' discussed in detail in \cite{parra2009via,parra2008lgt}.
In this paper, we deal instead with the intermediate limit where $\gkeps \ll M \ll 1$. This is the expected parameter regime for many current and future fusion experiments (see \Tref{TableInfo} in \Sref{sec_intro}~\footnote{Strictly speaking, the evidence in \Tref{TableInfo} only shows that $M$ is numerically larger than $\gkeps$ and smaller than $1$. However, as the rotation we are considering is {\textit{driven}} rotation, it seems plausible to assume that $M$ does not scale with $\gkeps$. This is in contrast to the phenomenon of {\emph intrinsic} rotation -- the spontaneous spin up of the plasma to mean velocities of a diamagnetic level~\cite{felixIntrinsic2}, i.e., first-order in $\gkeps$. Such rotataion is in the ``low-flow'' regime which we do not consider here (interested readers should consult \cite{parra2010momtrans,parra2011sources}.)}).

There are two physical effects of plasma rotation that survive in this limit. They are, firstly, the suppression of small-scale turbulence by the perpendicular flow shear~\cite{BiglariDiamond,TerryRMPSuppress}~\footnote{The physics of the suppression of turbulence by sheared flows was originally discussed in these papers in the context of poloidal flows. The crucial fact is not that the flow shear is poloidal, but that it is perpendicular to the magnetic field. Thus, the same physical mechanisms (shearing apart of eddies, tilting of eddies due to the flow shear) apply here, despite the fact that the perpendicular shear comes from a purely toroidal flow.}
	(the $\bm{u}\cdot\infrac{\partial h_s}{\partial \bm{R}_s}$ term in \eref{gke}) and, secondly, the instability drive due to the parallel flow shear that can itself excite turbulence~\cite{cattoPVG,newton2010understanding,AlexPVG} (the $d\angvel / d\psi$ term on the right-hand-side of \eref{gke}).
We retain the first of these terms by ordering the fluctuating frequency in the plasma frame
\begin{equation}
\left(\pd{}{t} + \bm{u}\dg\right) \sim \frac{\vth}{a}
\end{equation}
independently of the Mach number. The shear in the flow is hidden within the spatial dependence of the velocity. To make the presence of perpendicular flow shear more transparent, we assume that the domain of interest is shorter than the length scale of the flow and then expand the velocity around its value $\bm{u}_0$ at the centre of the domain $\bm{R}_{s0}$:
\begin{equation}
\fl
\begin{eqalign}{
\left( \pd{}{t} + \bm{u}(\bm{R}_s) \dgR{} \right) h_s = \\
		\qquad\qquad\left( \pd{ }{t} + \bm{u}_0\dgR{} \right) h_s + \left( \bm{R}_s - \bm{R}_{s0} \right)\cdot\left[ \nabla\bm{u}(\bm{R}_{s0}) \right]\dgR{h_s} \,+\, \cdots\,.
}\end{eqalign}
\end{equation}
The second term on the right-hand side of this expression is now explicitly the action of the perpendicular flow shear upon the distribution function $h_s$.
To retain only the part of this term due to shear in $\angvel$, we formally order the gradients of $\angvel(\psi)$ to be sharp: $a \nabla\ln \angvel(\psi) \sim M^{-1}$ so that $\infrac{d\angvel}{d\psi}$ is zeroth-order in $M$. For conciseness, we will continue to write the advection term unexpanded as $\bm{u}(\bm{R}_s)\cdot\infrac{\partial}{\partial\bm{R}_s}$, but the reader should keep in mind the idea of solving these equations in a small domain and Taylor-expanding the mean velocity field.
Note that, in the low-Mach-number limit, the toroidal rotation rate is proportional to the radial electric field not just to lowest
order in $\gkeps$ (see \eref{omdef}) but also to next order (see \ref{experimentalOmega}), with corrections of order $M^2$ (see \eref{LM-MeanE}).
This is perhaps why this shear is often referred to as ``$\bm{E}\times\bm{B}$ shear''~\cite{hahm1995flowshear,waltz1998shear}.

The $1/M$ ordering of the gradients of $\angvel$ is also precisely the ordering required to retain the parallel-velocity-gradient drive term on the right-hand side  of the gyrokinetic equation. 
We also order the timescale  of variation of $\angvel(\psi,t)$ as $M\tau_E$ rather than $\tau_E$, in order to retain the transport-time variation of the velocity.
With these orderings, we will be able to keep the two effects we wish to examine and neglect all other effects of rapid toroidal rotation, e.g., Coriolis and centrifugal forces.
In the following sections, we will apply this low-Mach-number expansion to the multiscale hierarchy derived in the preceding sections and present the ensuing equations to leading order in $M$.
\subsection{Mean Fields}
\label{LM-Mean}
First we derive the equations for the mean distribution function $F_s$ and the mean fields $\MeanE$ and $\MeanB$.

In the low-Mach-number limit, the mean distribution function is given by (see \Sref{vanishpot0} for details)
\begin{eqnarray}
\fl
F_s &=& F_{0s} \left( \psi(\bm{R}_s), \energy \right) + \FHat\left( \bm{R}_s,\energy,\magmom,\sigma\right) + \frac{I(\psi)w_\parallel}{\cycfreq} \pd{F_{0s}}{\psi} + \Or\left(\gkeps M f_s\right),\\
		\fl
F_{0s} &=& {n}_s(\psi(\bm{R}_s)) \left[\frac{m_s}{2\pi T_s(\psi(\bm{R}_s))}\right]^{3/2} e^{- \energy / T_s(\psi(\bm{R}_s))},
\end{eqnarray}
where $n_s$ and $T_s$ are the density and temperature of species $s$, the energy variable is now given by (cf. \eref{eloword})
\begin{equation}
\energy = \frac{1}{2} m_s w^2 + \Or(M^2 T_s),
\label{LM-energyvar}
\end{equation}
and $\bm{R}_s$ and $\magmom$ are still given by \eref{Rdef} and \eref{mudef}, respectively. 
The distinction between the particle velocity $\bm{v}$ and the peculiar velocity $\bm{w} = \bm{v} - \bm{u}$ can now be dropped, and so we use $\bm{w}$ as the particle velocity throughout this section.
The neoclassical part of the mean distribution function, $\FHat$, is 
found from \eref{Fneodef} and \eref{neog}--\eref{neo2}, as before.

The mean electric field is
\begin{equation}
\MeanE = - \nabla\fpot(\psi) + \Or\left(\gkeps M^2 \frac{\vth}{c} \MeanMagB\right),
	\label{LM-MeanE}
\end{equation}
where $\fpot$ is related to the angular velocity $\angvel(\psi)$ through \eref{omdef}. 
The mean magnetic field is given by the two functions $\psi(R,z)$ and $I(\psi)$ via \eref{magfield}. In the low-Mach-number limit, \eref{gradshaf} becomes the usual Grad-Shafranov equation:
\begin{equation}
\label{LM-GradShaf}
\begin{eqalign}{
	\Delta^* \psi = -4\pi R^2& \sum_s \frac{d p_s}{d\psi} - I(\psi)\frac{dI}{d\psi},
}\end{eqalign}
\end{equation}
where $p_s = n_s T_s$ is the pressure of species $s$, which is now a flux function (see \Sref{vanishpot0}). The second component of the magnetic field, $I(\psi)$, is determined by the same method as in the high-flow regime -- this is detailed in \Sref{magevolve}.
\subsection{Vanishing of Centrifugal Effects}
\label{vanishpot0}
As an example of how these results have been obtained, let us prove explicitly that $\pot_0$ and the poloidal variation of the density (see \Sref{poldens}) vanish in the low-Mach-number limit.
First, we expand the lowest-order quasineutrality condition \eref{qn0} in powers of $M$ to find
\begin{equation}
\label{lowMachpot}
\sum_s Z_s \NotN(\psi) \exp\left[ - \frac{Z_s e \pot_0}{T_s(\psi)}\right] = \Or \left( M^2 n_s\right).
\end{equation}
Since the only spatial dependence in \eref{lowMachpot} is via $\psi$, we know that the solution (for $\pot_0$, given $\NotN$) will be of the form $\pot_0 = \pot_0(\psi) + \Or(T_s M^2 / Z_s e)$. However, we have
defined $\pot_0$ so that it has no flux-surface average (see \eref{poteq} and \eref{potconstraint}). Thus, $\pot_0$ must vanish to lowest order in $M^2$\footnote{Furthermore, if we went to next order in $M^2$ to determine the small $\pot_0$, it would also be small in the ratio $a/R$ -- it is the variation of $R^2$ around the flux surface that drives $\pot_0$ and $R^2-\fav{R^2} \sim Ra$.}. 
Therefore, from \eref{npol}, we find $n_s = \NotN(\psi) + \Or(M^2 n_s)$. Thus, $n_s$ is a flux function to the required order in $M$ and we can drop the distinction between $\NotN$ and $n_s$. The quasineutrality constraint \eref{lowMachpot} then reduces to
\begin{equation}
\sum_s Z_s n_s(\psi) = 0.
\end{equation}
\subsection{Fluctuations}
\label{LM-Fluct}
In the low-Mach-number limit, the fluctuating distribution function is (cf. \eref{hDef})
\begin{equation}
\delta f_s = - \frac{Z_s e}{T_s} \delpot(\bm{r}) F_{0s} + h_s\left(\bm{R}_s,\energy,\magmom,\sigma\right) + \Or\left(\gkeps M f\right),
\end{equation}
where $\delpot$ is the fluctuating electrostatic potential, which determines the fluctuating electric field to the required order:
\begin{equation}
\delE = -\nabla\delpot + \Or\left(\frac{\vth[s]}{c} \gkeps M B\right).
\end{equation}
The gyrokinetic distribution function $h_s$ is given by the low-Mach-number gyrokinetic equation, which is the same as \eref{gke} but with centrifugal and Coriolis effects neglected:
\begin{equation}
\fl
\begin{eqalign}{
\left[\pd{}{t}+\bm{u}(\bm{R}_s)\dgR{}\right]h_s + \left( w_\parallel \Meanb+ \vdrift + \vchiR \right)\cdot\ddR{h_s} - \gyroR{\lincol[h_s]} \\
\quad=\frac{Z_s e F_{0s}}{T_s}\left[\pd{}{t} + \bm{u}(\bm{R}_s)\dgR{}\right]\gyroR{\gkpot}-
\left[ \pd{F_{0s}}{\psi} + \frac{m_s I(\psi) w_\parallel F_{0s}}{T_s \MeanMagB} \frac{d\angvel}{d\psi}\right] \vchiR\dg\psi,
}\end{eqalign}
\label{LM-gke}
\end{equation}
where the guiding-centre drift velocity is now (cf. \eref{vdrift})
\begin{equation}
\label{LM-vdrift}
\begin{eqalign}{
\vdrift = \frac{\Meanb}{\cycfreq} \times \left( w_\parallel^2 \Meanb\dg\Meanb+ \frac{1}{2}w_\perp^2\nabla \ln \MeanMagB\right) + \Or(\gkeps M \vth[s]).
}\end{eqalign}
\end{equation}
The definition of $\gkpot$ is now
\begin{equation}
\gkpot = \delpot - \frac{1}{c}\bm{w}\cdot\delA,
\end{equation}
because $\gkupot\approx \delpot$ in the low-Mach-number limit.
The definitions \eref{vchi} of $\vchi$ and \eref{gyroaverage} of the gyroaverage remain unchanged.

The equation for $h_s$ is closed through constitutive relations for the fluctuating fields.
The fluctutating potential $\delpot$ obeys the quasineutrality condition (cf. \eref{fluct-qn}):
\begin{equation}
\label{LM-fluct-qn}
 \sum_s \frac{Z_s^2 e^2 n_s\delpot}{T_s}   = \sum_s Z_s e \wint  \gyror{h_s}
\end{equation}
and the fluctuating magnetic field is
$\delB = \delBp \Meanb + \Meanb \times \nabla \delAp$,
with $\delAp$ and $\delBp$ determined from the parallel \eref{fluct-apar} and perpendicular \eref{fluct-bpar} components of Amp\`ere's law \eref{fluct-bpar}, respectively.

\subsection{Transport}
\label{LM-Transp}
The long-time evolution of the densities $n_s$ and temperatures $T_s$ are given by the low-Mach-number limits of the transport equations \eref{dndt}, \eref{domegadt}, and \eref{dpdt} of \Sref{transport}:\footnote{With the transport-time variation of $\omega(\psi)$ ordered as $M^{-1}$ as discussed in \Sref{SLowMachLimit}, all terms in \eref{domegadt} are the same order.}
\begin{equation}
\fl
\phantom{\frac{3}{2}}\frac{1}{V'}\left.\pd{}{t}\right|_\psi V'{n_s} + \frac{1}{V'} \pd{}{\psi} V'\fav{ \ParticleFlux } = \fav{\Psource},
\label{dndt-lowmach}
\end{equation}
\begin{equation}
\fl
\begin{eqalign}{
\phantom{\frac{3}{2}}\frac{1}{V'}\ddtpsi&{{V'\inertia \angvel}} + \frac{1}{V'}\pd{}{\psi} V'\fav{\TotMomFlux}
= \fav{\Msource},
}\end{eqalign}
\end{equation}
\begin{equation}
\fl\begin{eqalign}{
\frac{3}{2}\frac{1}{V'}\ddtpsi V'p_s &+ \frac{1}{V'} \pd{}{\psi}  V' \fav{\HeatFlux} = \\
&\ViscousHeat + \JouleHeat + \CompHeat + \TurbPow 
+ \fav{\CollEnergy}+  \fav{\Esource}.
}\end{eqalign}
\label{dpdt-lowmach}
\end{equation}
The expressions \eref{partflux} for the particle flux $\fav{\ParticleFlux}$, \eref{psource} for the particle source $\Psource$, \eref{inertiaDef} for the moment of inertia $\inertia$, and \eref{esource} for the energy source $\Esource$ are all unchanged in the low-Mach-number limit.
The momentum flux \eref{TMFresult}, source \eref{msourceDef}, and the heat flux \eref{heatflux} lose some terms that are higher-order in $M$:
\begin{equation}
\fav{\TotMomFlux} = \sum_s \fav{\MomentumFlux} + \fav{\EMViscosity},
\end{equation}
where $\fav{\MomentumFlux}$ and $\EMViscosity$ are given by \eref{momflux} and \eref{EMViscDef}, respectively,
\begin{eqnarray}
\fl
\Msource &= \sum_s \wint m_s \left(\bm{w}\dg\tor\right) R^2 \source,\\
\label{msource-lowmach}
\fl
\label{hflux-lowmach}
\fav{\HeatFlux} &= \fav{\wint\energy \left( \frac{\bm{w}\times\Meanb}{\cycfreq} \dg\psi \right) \collop[F_{0s}] } +\fav{\wint\energy \Fneo \vdrift\dg\psi}\\
		\fl\nonumber
&\qquad
- \frac{5}{2}p_s(\psi)I(\psi)\frac{\fav{\MeanE\cdot\MeanB}}{\fav{\MeanMagB^2}}
+ \fav{\ensav{\wint \energy \gyror{h_s \vchi}\dg\psi}}.
\end{eqnarray}
The Ohmic heating $\JouleHeat$ is given by \eref{OhmHeat}, the compressional heating $\CompHeat$ by \eref{CompHeatDef}; they are unchanged in the low-Mach-number limit.
The other two heating terms, viz., viscous \eref{VHDef} and turbulent \eref{TurbPowDef}, are simplified:
\begin{equation}
\ViscousHeat = -  \fav{\MomentumFlux}\frac{d\angvel}{d\psi},
\end{equation}
\begin{equation}
\begin{eqalign}{
\TurbPow &= Z_s e \fav{ \ensav{\wint \gyror{h_s\left( \pd{}{t} + \bm{u}\dg \right) \gkpot} }}\\
&= \TurbColl - \TurbInj,
}\end{eqalign}
\label{LowMachTurbHeat}
\end{equation}
where $\fav{\MomentumFlux}$ is given by \eref{momflux} and $\TurbColl$ and $\TurbInj$ are given by \eref{TurbCollDef} and \eref{TurbInjDef}, respectively.

The system of equations \eref{dndt-lowmach}--\eref{LowMachTurbHeat} with the solutions for the neoclassical (\Sref{LM-Mean}) and fluctuating (\Sref{LM-Fluct}) distribution functions and the mean magnetic field (\Sref{LM-Mean}) is a closed system for coupled turbulence and transport in the low-Mach-number limit.
\subsection{Energy and Entropy}
\label{LM-Energy}
In the low-Mach-number limit, we are able to neglect the rotational energy compared to the thermal energy of the system (despite the viscous heating appearing in the temperature evolution equation \eref{dpdt-lowmach} -- because we allow the turbulence to generate a large momentum flux from the sharp gradients in $\omega(\psi)$). Thus, the energy conservation law \eref{Totalenergyconv} reads
\begin{equation}
\label{Totalenergyconv-lowmach}
\fl
\begin{eqalign}{
\frac{1}{V'}&\ddtpsi V'\left(\sum_s \frac{3}{2} n_s T_s + \frac{\fav{\MeanMagB^2}}{8\pi}\right)\\
	&+ \frac{1}{V'} \pd{}{\psi} V'\left(\fav{\JUflux} - \frac{c}{4\pi} I(\psi)\fav{\MeanE\cdot\MeanB} + \fav{\frac{3B^2}{8\pi}\pd{\psi}{t}}\right)
		=\sum_s \fav{\Esource},
}\end{eqalign}
\end{equation}
where the flux of kinetic energy (in the low-Mach-number limit, this is purely thermal energy), previously given by \eref{JUflux}, is
\begin{equation}
\label{JUflux-lowmach}
\fl\begin{eqalign}{
\fav{\JUflux} = \sum_s \fav{\HeatFlux} 
		= &\fav{ \left( \sum_s \bm{Q}_s + \frac{c}{4\pi}\ensav{\delE\times\delB}  \right)\dg\psi} \\
		  &\qquad+ \sum_s \frac{3}{2} n_s T_s \fav{\pd{\psi}{t}}.
}\end{eqalign}
\end{equation}
Here $\bm{Q}_s$ is the usual heat flux defined by \eref{eflux} and $\fav{\HeatFlux}$ is given by \eref{hflux-lowmach}.

The free energy \eref{FreeEnergyDef} of the fluctuations continues to satisfy \eref{freeEnergyBalance} in the low-Mach-number limit, but with all instances of $\NotN$ replaced by $n_s$.
 Note that because the gradients of $\omega(\psi)$ have been retained in the low-Mach-number ordering, they contribute to the free-energy balance, and hence can drive turbulence.
Equations \eref{Totalenergyconv-lowmach} and \eref{freeEnergyBalance} should be considered as constraints on the low-Mach-number system of Sections~\ref{LM-Mean}--\ref{LM-Transp}: any solution of these equations should identically satisfy these constraints.

The entropy balance of \Sref{Sentropy} is unaffected by the low-Mach-number expansion (up to changes in the chemical potential \eref{ChemPotDef} and the transport fluxes). Importantly, the entropy generation and source of free energy due to the angular velocity gradient remains. This is understandable as we have constructed our ordering so that the momentum flux is finite and the gradient in the flow is finite, and thus the entropy generation due to viscous heating remains finite and comparable to all other forms of entropy generation.
\section{Summary}
\label{SConc}
In this paper, we have presented a complete and self-consistent derivation of multiscale gyrokinetics from first principles, drawing on and in some cases correcting previous work~\cite{sugama1998neg,sugama1997tpe,sugama1996tpa,gabethesis,ericthesis,michaelthesis,artun1994nonlinear}, as well as incorporating the evolution of the equilibrium magnetic field 
into this theoretical framework. We have shown how, in this framework, the gyrokinetic equation for fluctuations can be derived~\cite{sugama1997tpe,frieman1982nge,artun1994nonlinear,PeetersGKE} (\Sref{Sgke}), in parallel with the neoclassical drift kinetic equation~(\Sref{nclasse}) more usually derived in a different setting~\cite{catto1987ion}. 
This system of equations describes fast, small-scale and slow, large-scale perturbations of the local Maxwellian equilibrium driven by the gradients in the local equilibrium. 
We close this system on the long (transport) time scale by deriving evolution equations for the equilibrium magnetic field (\Sref{magevolve}) and equations governing the transport of particles (\Sref{parttrans}), momentum (\Sref{momtrans}) and energy (\Sref{heattrans}).

We are now in a position to draw together all of these results to form a clear picture of the evolution of a rotating turbulent plasma in a tokamak. The summary of our work can be reduced to three key points.

Firstly, flux surfaces form effectively isolated systems on timescales shorter than the transport time. They rotate toroidally as rigid bodies \eref{torrot}, they are in local thermodynamic equilibrium \eref{F0R} and the turbulence within them is established via a local balance between energy injection and dissipation \eref{steadystate}. Thus, we can conceptually think of a tokamak plasma as being made up of annular regions with a radial width given by some intermediate length \eref{LambdaScale} -- these regions are independent on the intermediate time scale \eref{TScale}. It is this decomposition that underlies analytical and numerical considerations of the gyrokinetic system \eref{gke}, \eref{fluct-qn}, \eref{fluct-apar}, and \eref{fluct-bpar} in any one such annular region, the so-called ``local approximation''~\cite{candy2004local}.

Secondly, on long timescales \eref{TauEScale}, these independent annular regions are coupled by the transport equations of \Sref{transport}. These are local equations for the plasma density \eref{dndt}, angular velocity \eref{domegadt}, and temperature \eref{dpdt}, linking each annular region to its neighbours via transport fluxes that can be computed from averages over local solutions to the small-scale equations. Despite the dramatic simplification that has occured in going from the full kinetic equation \eref{vfp} to evolving only a few flux functions, we have retained all the critical information about the plasma. This linked-flux-surface approach to transport provides both a transparent theoretical framework and a basis for practially feasible numerical simulations of global tokamak dynamics on confinement timescales~\cite{michaelthesis,barnes2009trinity,candy2009tgyro}.

Thirdly, the (global) mean fields and the (local) fluctuations are also linked via energy and entropy flows (and so are individual flux surfaces). 
\Sref{thermo} is dedicated to the flows of energy, which express the basic conservation properties of the system.
The results of this section are consequences of the multiscale representation of the plasma dynamics that is derived in the preceding sections. No extra information was used in their derivation. Thus, the energy balance \eref{Totalenergyconv} is a statement that if the mean quantities are evolved according to the transport equations of \Sref{transport}, then the mean energy of the system is conserved.
Similarly, if the gyrokinetic equation \eref{gke} and the field equations \eref{fluct-qn}, \eref{fluct-apar}, and \eref{fluct-bpar} are solved to determine the fluctuating distribution function and fluctuating electromagnetic fields, then the free energy \eref{FreeEnergyDef} is the conserved quantity in the sense that \eref{freeEnergyBalance} is satisfied by the solutions.
These two statements are what is meant by energy conservation in our multiscale system.

A direct consequence of the multiscale plasma dynamics is the evolution of our multiscale system towards thermodynamic equilibrium (\Sref{Sentropy}) -- the entropy of the system evolves according to \eref{entropyConservation}.
As a consequence of writing the entropy production in a convenient form \eref{entropyProductionMain}, we are able to diagnose how our system proceeds towards a global thermodynamic equilibrium and what that equilibrium is. 
As the system approaches a global equilibrium, the collisional and (relative) turbulent heating equilibrate the temperatures and the Ohmic heating dissipates the parallel induced electric field; the fluxes flatten gradients.
Therefore, the global equilibrium is a state with no temperature differences, no parallel induced  electric field, and no spatial gradients in $n_s$, $T_s$ or $\angvel$ (as a consequence of this, there are no fluctuations).
However, this simple diagnosis is complicated by the multichannel nature of the transport in our system: far from equilibrium, it may be thermodynamically favourable for the system initially to {\textit{steepen}} gradients in
some of $n_s$, $T_s$ or $\angvel$ in order to flatten other gradients more rapidly.

In the course of our presentation of this general framework, we have elucidated a number of issues that have the potential to be confusing and indeed have been so in the past. It is useful to itemize some of them here:
\begin{itemize}
\item Turbulence is local to a given flux surface and does not spread -- \Sref{SFreeEnergy}.
\item There is no turbulent heating -- \Sref{s9-1-1-thermal} (except in the sense of turbulent redistribution of energy between species).
\item The evolution of the mean magnetic field is determined only by the mean profiles, and not directly by the turbulence -- \Sref{magevolve}.
\item Gradients in the magnetic field do not drive fluctuations -- \Sref{specialMagField}
\item The absence of the parallel nonlinearity in the gyrokinetic equation \eref{gke} does not upset the energy or entropy conservation properties of the system -- see \ref{parallelNote}.
\end{itemize}

Let us now discuss some further directions and natural extensions following from this work.

We have accomplished the construction of multiscale gyrokinetics by exploiting the scale separations inherent to plasma turbulence: between the fluctuations and the equilibrium and between the particle motion and the fluctuations (\Sref{Sfluctmean}). However, these are not the only scale separations available to us. In \cite{flowtome2-electrons}, this framework is extended by exploiting the scale separation between ion- and electron-scale fluctuations, where this scale separation is found to have profound implications for the structure of the fluctuating magnetic field.

In this paper, we have restricted ourselves to situations where the equilibrium distribution functions for all species are Maxwellian. In a burning plasma, we expect there to be at least one non-Maxwellian species -- the fusion-produced $\alpha$-particles. In \cite{flowtome3-alphas}, we will consider precisely this situation -- demonstrating how 
such a species can be incorporated into the theoretical framework developed here.

We have also restricted ourselves to the so-called ``high-flow ordering,'' where the mean rotation velocity is ordered to be comparable to the sound speed. The ``low-flow'' ordering, where the mean velocity is comparable to the drift velocities, requires a much more involved calculation to determine the transport of toroidal angular momentum correctly~\cite{parra2008lgt,parra2009via,parra2010momtrans}. It is precisely this regime that must be considered if a detailed and complete study of the so-called ``intrinsic'' rotation is to be undertaken~\cite{felixIntrinsic2}. This alternative ordering may hold the key to how a slowly-moving or stationary plasma can transition into a rotating one~\cite{parra2011sources} or how L-H transitions occur in the edge of a slowly-rotating plasma \cite{gohil2008dependence}.

Finally, it is impossible to present a set of equations such as the ones found in this paper without discussing how to solve them numerically. Currently, linked-flux-tube codes~\cite{barnes2009trinity,candy2009tgyro} solve the low-Mach-number system of equations presented in \Sref{LowMach}. It is anticipated that
 \verb#TRINITY#~\cite{barnes2009trinity} will be extended to solve the complete set of equations derived here. With the advent of ever faster supercomputers, the combination of multiscale theoretical approaches and advanced numerical algorithms may allow us, for the first time, to simulate a burning plasma, from the small scales of kinetic turbulence, through the scales associated with energetic fusion products, to the plasma equilibrium scales and even the timescales of the resistive evolution of the plasma. As a tool to study potential reactor designs, this would be invaluable. Hitherto, experiments have been optimised for MHD stability. Henceforth, it is possible to envision a time when they will be optimised not only for macro-stability but also for micro-stability and 
designed to achieve dramatically reduced turbulent transport (initial forays into this exciting possiblity include, e.g., \cite{highcock2012zero} and \cite{mynick2011reducing}).

\ack
We wish to thank G. Hammett, E. Highcock and F. I. Parra for many useful discussions. We also wish to thank L. F. van Wyk, M. F. J. Fox, and the anonymous referees for their careful reading of the paper, which has greatly improved its content.
We are grateful to the Leverhulme Trust Academic Network for Magnetized Plasma Turbulence for travel support
(GGP, EW, MAB and WD) and to the Isaac Newton Institute, Cambridge, for its hospitality during the programme ``Gyrokinetics for Laboratory and Astrophysical Plasmas''.
IGA was supported by a CASE EPSRC studentship jointly with the EURATOM/CCFE Fusion Association and by a Junior Research Fellowship at Merton College, Oxford; GGP and WD were supported by the US DOE Center for Multiscale Plasma Dynamics; MB was supported by the Oxford-Culham Fusion Research Fellowship and a DoE Fusion Energy Sciences Postdoctoral Fellowship; AAS was supported in part by an STFC Advanced Fellowship and an STFC Astronomy Grant. 
Although this work was carried out within the framework of the European Fusion Development Agreement, the views and opinions expressed herein do not necessarily reflect those of the European Commission.

\appendix

\section{Derivation of \texorpdfstring{$\dot{\bm{R}}_s$, $\dmu$, $\denergy$ and $\dot{\gyr}$}{time derivatives of the gyrokinetic variables}}
\label{apA}
Here we calculate the time derivatives of the variables $\bm{R}_s$, $\magmom$, $\energy$ and $\gyr$ along the particle trajectory, which allows the derivation of the second order gyrokinetic equation \eref{2ndorder} from the general kinetic equation \eref{newvfp}.
Note that we only require these derivatives correct to first order in $\gkeps$, because the higher-order (transport-timescale) effects are treated separately in \Sref{transport}.
\subsection{Some useful properties of the gyroaverage}
In performing gyroaverages in this Appendix, we will need to replace temporal and spatial derivatives at constant $\bm{v}$ with derivatives with respect to $\bm{R}_s$, $\energy$, $\magmom$ and $\gyr$. For any quantity $g(\bm{r},\bm{v})$ that has small-scale spatial structure (i.e., spatial variation on scales $\sim \rho_s$),
\begin{equation}
\begin{eqalign}{
\left.\nabla\right|_{\bm{v}} g &= \left[ \left( \nabla\bm{R}_s \right) \cdot \ddR{} + \left( \nabla\energy \right) \pd{}{\energy}+ \left( \nabla\magmom \right)\pd{}{\magmom}+ \left( \nabla\gyr \right)\pd{}{\gyr}\right]g\\
	&= \left.\ddR{}\right|_{\energy,\magmom,\gyr}  g + \Or(\gkeps k_\perp g).
}\end{eqalign}
\end{equation}
Similarly, for time derivatives,
\begin{equation}
\left.\pd{}{t} \right|_{\bm{r},\bm{v}}  g = \left.\pd{}{t}\right|_{\bm{R}_s,\energy,\magmom,\gyr}  g + \Or(\gkeps^2 \cycfreq g).
\end{equation}
We will also use the fact that the gyroaverage $\gyroR{\cdot}$ commutes with derivatives in the $\bm{R}_s$, $\energy$ and $\magmom$ variables. 

Another general result we will use is that
\begin{equation}
\label{wperpdg1}
\begin{eqalign}{
\cycfreq {\left.\pd{g}{\gyr}\right|_{\bm{R}_s,\energy,\magmom}} &= \cycfreq \left( \left.\pd{\bm{r}}{\gyr} \right|_{\bm{R}_s,\energy,\magmom} \dg  g + \left. \pd{\bm{v}}{\gyr}\right|_{\bm{R}_s,\energy,\magmom} \cdot\pd{ g}{\bm{v}} \right)\\
&= \bm{w}_\perp\dg g + \cycfreq \left(\bm{w}\times\Meanb\right)\cdot\pd{g}{\bm{w}} + \Or(\gkeps\cycfreq g) \\
&= \bm{w}_\perp\dg g + \cycfreq \left.\pd{g}{\gyr}\right|_{\bm{r},w_\parallel,w_\perp} + \Or(\gkeps \cycfreq g),
}\end{eqalign}
\end{equation}
where we have used the definitions \eref{Rdef}, \eref{energydef}, \eref{mudef} and \eref{wDef} of $\bm{R}_s$, $\energy$, $\magmom$ and $\bm{w}$, respectively.
Gyroaveraging the identity \eref{wperpdg1} at constant $\bm{R}_s$ or at constant $\bm{r}$, we find that
\begin{equation}
\fl
\label{wperpdg}
\gyroR{\bm{w}_\perp \dg g} = - \cycfreq \gyroR{\left.\pd{g}{\gyr}\right|_{\bm{r},w_\parallel,w_\perp}},\qquad \gyror{\bm{w}_\perp\dg g} = \cycfreq \gyror{\left.\pd{g}{\gyr}\right|_{\bm{R}_s,\energy,\magmom}}.
\end{equation} 

Finally, we will need the following identities in order to evaluate gyroaverages explicitly:
\begin{equation}
  \gyroR{\bm{w}_\perp} = \Or(\gkeps\vth),\qquad\gyroR{\bm{w}_\perp \bm{w}_\perp} = \frac{w_\perp^2}{2} \left(\idmat-\Meanb\Meanb\right) + \Or(\gkeps \vth[s]^2),
\label{gyrww}
\end{equation}
where we have used \eref{wDef} to carry out the integration over $\gyr$.
Similarly, we can use the definition \eref{Rdef} of $\bm{R}_s$ in terms of $\bm{r}$ to find, for any mean quantity $g(\bm{r},\energy,\magmom)$,
\begin{equation}
\begin{eqalign}{
\gyroR{g(\bm{r},\energy,\magmom)} &= \gyroR{g(\bm{R}_s,\energy,\magmom)} + \gyroR{\frac{\Meanb\times\bm{w}}{\cycfreq} \dg g} + \Or(\gkeps^2)\\
	&= g(\bm{R}_s,\energy,\magmom) + \Or(\gkeps^2).
}\end{eqalign}
\label{toR}
\end{equation}

\subsection{Derivation of \texorpdfstring{$\dot{\bm{R}}_s$}{R-dot}}
\label{derivRdot}
 ${\bm{R}}_s$ is defined by \eref{Rdef}. Taking the full time derivative along the particle trajectory, we get
\begin{equation}
	\dot{\bm{R}}_s = \frac{d}{dt} \left( \bm{r} - \frac{\Meanb\times\bm{w}}{\cycfreq} \right) = \dot{\bm{r}} - \frac{\Meanb\times\dot{\bm{w}}}{\cycfreq}  - \bm{v}\left.\dg\right|_{\bm{v}}\left( \frac{\Meanb\times\bm{w}}{\cycfreq} \right),
\end{equation}
where we have neglected terms of order $\Or(\gkeps^3\vth)$ (time derivatives of the mean quantities $\cycfreq$ and $\Meanb$) and the spatial derivatives are taken at constant $\bm{v}$. We now use $\dot{\bm{r}} = \bm{v}$ and 
\begin{equation}
\dot{\bm{w}} = \dot{\bm{v}} - \bm{v}\dg\bm{u} = \accel +\daccel - \bm{v}\dg\bm{u},
\label{wDot}
\end{equation} where the accelerations are given by \eref{accel} and \eref{daccel}, to find
\begin{equation}
\fl
\begin{eqalign}{
\label{rdot}
\dot{\bm{R}}_s &= w_\parallel \Meanb + \bm{u} + \frac{c}\MeanMagB \Meanb\times\nabla \left(\pot_0 + \gkpot\right) +\frac{1}\MeanMagB\Meanb\times \left(\bm{w}_\perp\dg\delA\right) \\
&+ \frac{1}{\cycfreq} \left[ \Meanb\times\left( \bm{u}\dg\bm{u} \right) +\Meanb\times\left( \bm{w}\dg\bm{u} \right) - \left(\bm{u}\dg\Meanb\right)\times \bm{w}- \left(\bm{w}\dg\Meanb\right)\times \bm{w}\right.\\
&\left.\qquad+ \left(\Meanb\times\bm{w}\right)\bm{w}\dg\ln \MeanMagB \right] + \Or(\gkeps^2\vth),
}\end{eqalign}
\end{equation}
where we have used $\bm{u}\dg\ln \MeanMagB =0$ (axisymmetry). The lowest-order component of this expression is
\begin{equation}
\label{low-rdot}
\dot{\bm{R}}_s = w_\parallel \Meanb +\bm{u} + \Or\left(\gkeps \vth[s]\right).
\end{equation}

We now proceed to gyroaverage \eref{rdot} term by term.  We use \eref{toR} on the first two terms to find\setcounter{footnote}{1}\footnote{In this expression $w_\parallel$ is considered to be a function of $\bm{R}_s$, $\energy$ and $\magmom$ via \eref{wparDef} with all functions of $\bm{r}$ evaluated at $\bm{R}_s$ in accordance with \eref{toR}.}
\begin{equation}
  \gyroR{w_\parallel \Meanb+\bm{u}} = w_\parallel \Meanb(\bm{R}_s)+\bm{u}(\bm{R}_s) + \Or(\gkeps^2\vth).
\end{equation}
This is the combination of parallel streaming along the field line and convection by the mean flow (but it is now the guiding centres, not the particles, that stream along the field line and are advected by the flow).
The third term in \eref{rdot} gyroaverages to
\begin{equation}
  \fl
  \gyroR{\frac{c}\MeanMagB\Meanb\times\nabla\left(\pot_0 + \gkpot\right)} = \frac{c}\MeanMagB\Meanb\times\nabla\left(\pot_0 + \gyroR{\gkpot}\right) = \frac{c}\MeanMagB\Meanb\times\nabla\pot_0 + \vchiR,
\end{equation}
where $\vchi$ is defined by \eref{vchi}. These terms are the $\MeanE\times\MeanB$ drifts due to $\pot_0$ and due to the averaged fluctuating potential $\gyroR{\gkpot}$ seen by the particle ($\gkpot$ is just the electrostatic potential in the particle frame). The fourth term of \eref{rdot} vanishes upon gyroaveraging:
\begin{equation}
  \gyroR{\frac{1}\MeanMagB \Meanb\times\left( \bm{w}_\perp\dg\delA \right)} = \frac{1}\MeanMagB \Meanb\times\gyroR{\bm{w}_\perp\dg\delA} = 0,
\end{equation}
where we have used \eref{wperpdg}.
The fifth term in \eref{rdot} becomes the centrifugal drift:
\begin{equation}
  \gyroR{\frac{1}{\cycfreq}\Meanb\times\left( \bm{u}\dg\bm{u} \right) } = -\frac{1}{\cycfreq} \Meanb \times \left[ \angvel^2(\psi)R\nabla R \right]+\Or(\gkeps^2\vth),
\end{equation}
where we have used $\bm{u}=\angvel(\psi)R^2\nabla\tor$ and expressed $\nabla\bm{u}$ as
\begin{equation}
\begin{eqalign}{
\nabla \bm{u} &= \frac{d\angvel}{d\psi} R^2\left(\nabla\psi\right)\left(\nabla\tor\right)+\angvel(\psi) R\left[\left(\nabla R\right)\left( \nabla\tor\right) - \left(\nabla \tor\right)\left(\nabla R\right)\right].
}\end{eqalign}
\label{nablaU}
\end{equation}
Combining the gyroaverages of the sixth and seventh terms of \eref{rdot}, we find that they give rise to the Coriolis drift:
\begin{equation}
  \begin{eqalign}{
  \gyroR{\frac{1}{\cycfreq} \left[ \Meanb\times\left(\bm{w}\dg\bm{u}\right) - \left( \bm{u}\dg\Meanb\right)\times\bm{w}  \right]}	\\
  \qquad= \frac{2w_\parallel}{\cycfreq}\Meanb\times\left( \Meanb\dg\bm{u} \right) =-\frac{1}{\cycfreq} \Meanb\times\left[ 2 w_\parallel \angvel(\psi) \Meanb\times\nabla z \right], 
  }\end{eqalign}
  \label{coriolistmp}
\end{equation}
where we have used $\gyroR{\bm{w}} = w_\parallel \Meanb$, the fact that $\bm{u}\dg\Meanb = \Meanb\dg\bm{u}$, which follows from
\begin{equation}
\label{udgb}
B\left(\bm{u}\dg\Meanb - \Meanb\dg\bm{u}\right) = \curl\left(\MeanB\times\bm{u}\right) = -\curl\left[ \angvel(\psi) \nabla\psi \right] = 0,
\end{equation}
and also the fact that, for any $\bm{a}$,
\begin{equation}
R\left[\left(\nabla R\right)\left( \nabla\tor\right) - \left(\nabla \tor\right)\left(\nabla R\right)\right] \cdot \bm{a} = \bm{a}\times\nabla z.
\label{aCrossZ}
\end{equation}
The penultimate term in \eref{rdot} is gyroaveraged as follows:
\begin{equation}
\fl
  -\gyroR{\frac{1}{\cycfreq} \left[ \left( \bm{w}\dg\Meanb \right)\times\bm{w}\right]} = \frac{1}{\cycfreq} \left\{ w_\parallel^2\Meanb\times\left( \Meanb\dg\Meanb \right) + \frac{w_\perp^2}{2} \left[ \curl\Meanb + \left( \Meanb\dg\Meanb \right)\times\Meanb \right] \right\},
\end{equation}
where we have used \eref{gyrww}. The first and second terms here are, respectively, the curvature drift and the so-called Ba\~nos drift. We can use the vector identity
\begin{equation}
\Meanb\dg\Meanb = -\Meanb\times\left( \curl\Meanb \right)
\end{equation}
to write the Ba\~nos drift as 
\begin{equation}
\fl
\frac{w_\perp^2}{2\cycfreq} \left[ \curl\Meanb + \Meanb\Meanb\cdot \left( \curl\Meanb \right) - \curl\Meanb\right] = \frac{w_\perp^2}{2\cycfreq}\left[\Meanb\cdot\left( \curl\Meanb \right) \right] \Meanb.
\end{equation}
Thus, this drift is purely along the field line and one order smaller than parallel streaming. Therefore, as parallel derivatives of fluctuating quantities are small and $\Meanb\dg F_{0s} = 0$, this drift will not appear in the second-order gyrokinetic equation. Finally, gyroaverging the last term of \eref{rdot} using \eref{gyrww} we obtain
\begin{equation}
\gyroR{\frac{1}{\cycfreq} \left[\left(\Meanb\times\bm{w}\right) \bm{w}\dg\ln \MeanMagB\right]} = \frac{w_\perp^2}{2\cycfreq} \Meanb\times\nabla\ln \MeanMagB,
\end{equation}
which is the $\nabla \MeanMagB$ drift.

Assembling these results, we arrive at
\begin{equation}
\begin{eqalign}{
\gyroR{\dot{\bm{R}}_s} &=   \Meanb\left(\bm{R}_s\right)\left[ w_\parallel +\frac{w_\perp^2}{2\cycfreq}\Meanb\cdot\left( \curl\Meanb \right)\right]+ \bm{u}\left(\bm{R}_s\right) \\
	&\qquad+ \vchiR + \vdrift  + \Or(\gkeps^2 \vth[s]),
}\end{eqalign}
\label{rdotg}
\end{equation}
where $\vchi$ and $\vdrift$ are defined in \eref{vchi} and \eref{vdrift}, respectively (all the drifts associated with the mean magnetic, electric and velocity fields have been combined into
$\vdrift$).
\subsection{Derivation of \texorpdfstring{$\dmu$}{mu-dot}}
The magnetic moment $\magmom$ is defined by \eref{mudef}.
Taking the time derivative along a particle trajectory, we find
\begin{equation}
\fl
\dmu = \frac{m_s}{2\MeanMagB}\left\{ -w_\perp^2 \bm{w}\dg\ln \MeanMagB + 2\left[ \bm{v}\dg\left( \Meanb\times\bm{w} \right) \right]\cdot\left( \Meanb\times\bm{w} \right) + 2 \bm{w}_\perp \cdot\dot{\bm{w}} \right\} + \Or(\gkeps^3\cycfreq \magmom),
\end{equation}
where we have used $w_\perp^2 = \left(\Meanb\times\bm{w}\right)\cdot\left( \Meanb\times\bm{w} \right)$ and $\bm{u}\dg \MeanMagB =0$ (axisymmetry) . Substituting \eref{wDot} for $\dot{\bm{w}}$ and rearranging terms gives
\begin{equation}
\begin{eqalign}{
\label{mudot}
\dmu =& -\magmom \bm{w}\cdot\nabla\ln \MeanMagB - \frac{m_s}\MeanMagB w_\parallel \left(\bm{v}\cdot \nabla\Meanb \right)\cdot \bm{w}_\perp \\
		&+\frac{Z_s e}\MeanMagB \bm{w}_\perp \cdot \left(-\nabla\pot_0-\nabla\gkpot - \frac{1}{c}\bm{w}_\perp\dg \delA\right)\\
		&- \frac{m_s}\MeanMagB  \left(\bm{v}\cdot\nabla\bm{u}\right)\cdot\bm{w}_\perp + \Or(\gkeps^2\cycfreq \magmom).
}\end{eqalign}
\end{equation}

Gyroaveraging the first two terms in \eref{mudot}, we have
\begin{equation}
\begin{eqalign}{
-\magmom\gyroR{\bm{w}\dg\ln \MeanMagB} - \frac{m_s}\MeanMagB w_\parallel\gyroR{\left( \bm{v}\dg\Meanb \right)\cdot\bm{w}_\perp} \\
 \qquad=-\magmom w_\parallel \left(\Meanb\dg\ln \MeanMagB + \dv\Meanb\right) = -\magmom \frac{w_\parallel}{\MeanMagB} \dv\MeanB = 0,
}\end{eqalign}
\end{equation}
where we have used \eref{wperpdg} and \eref{gyrww}.
Proceeding to the third term of \eref{mudot}, we have $\gyroR{\bm{w}_\perp\dg\pot_0} = 0$, and, substituting  $\gkpot = \delpot - \bm{v}\cdot\delA/c$, we find
\begin{equation}
\begin{eqalign}{
\gyroR{\bm{w}_\perp\left.\dg\right|_{\bm{v}} \gkpot }+\frac{1}{c} \gyroR{\left(\bm{w}_\perp\dg\delA_\perp\right) \cdot\bm{w}_\perp} \\
	\qquad\qquad= \gyroR{\bm{w}_\perp \dg\delpot} - \frac{1}{c} \gyroR{\bm{w}_\perp\dg\delA}\cdot\left(w_\parallel\Meanb+\bm{u}\right) = 0,
}\end{eqalign}
\label{wperpchiA}
\end{equation}
where we have used \eref{wperpdg} on each term on the right hand side. 
Finally, gyroaveraging the last term in \eref{mudot} we find
\begin{equation}
\gyroR{\frac{m_s}\MeanMagB \left(\bm{v}\dg\bm{u}\right)\cdot\bm{w}_\perp} = \frac{m_s}\MeanMagB \gyroR{\bm{w}_\perp\bm{w}_\perp} \bm{:} \nabla\bm{u} = 0,
\end{equation}
where we have used \eref{gyrww} and then \eref{nablaU}.

Combining all these results, we have 
	\begin{eqnarray}
	\label{low-mudot}
	\gyroR{\dmu} =& \Or(\gkeps^2\cycfreq \magmom),
	\end{eqnarray}
and so $\magmom$ is conserved to second order in $\gkeps$.
\subsection{Derivation of \texorpdfstring{$\denergy$}{energy-dot}}
\label{energydot}
The energy variable $\energy$ is defined by \eref{energydef}. Taking the time derivative along a particle trajectory, we find
\begin{equation}
	\fl
	\label{destart}
\begin{eqalign}{
\denergy &= m_s\bm{v}\cdot\dot{\bm{v}} +Z_s e\left(\pd{}{t} + \bm{v}\dg\right)\left[\fpot(\psi) + \pot_0\right]- Z_s e \left.\frac{d\fpot}{d\psi}\right|_{\psi=\psistar} \dot{\psistar} \\
				  &= Z_s e\bm{v}\cdot\left( \MeanE + \delE\right)+ Z_s e \left(\pd{}{t} + \bm{v}\dg\right) \left[\fpot(\psi) + \pot_0\right] \\
				  &\qquad\quad- Z_s e \left[\frac{d\fpot}{d\psi} + \frac{m_s c}{Z_s e} R^2 \left( \bm{v}\dg\tor \right) \frac{d^2\fpot}{d\psi^2}\right]\dot{\psistar} + \Or(\gkeps^3\cycfreq T_s),
}\end{eqalign}
\end{equation}
where we have used \eref{accel} and \eref{daccel} for $\dot{\bm{v}} = \accel + \daccel$, and \eref{psistardef} to expand $\psistar$ around $\psi$ in the argument of $\infrac{d\fpot}{d\psi}$.
In the above expression, 
\begin{equation}
\dot{\psistar} = \pd{\psi}{t} + \bm{v}\dg\psi + \frac{m_s c}{Z_s e}R^2\left(\accel+\daccel\right)\dg\tor  = \frac{m_s c}{Z_s e} R^2 \daccel\dg\tor,
\end{equation}
where we have used \eref{accel}, axisymmetry, the axisymmetric form of the mean magnetic field \eref{magfield} and $\psi = R^2\MeanA\dg\tor$.
Expanding the electric field in terms of potentials and using $\infrac{d\fpot}{d\psi} = \angvel(\psi)/c$ (see \eref{omdef}), we can write \eref{destart} as
\begin{equation}
\begin{eqalign}{
\denergy = -\frac{Z_s e}{c} \left( {\bm{w}}\cdot\pd{\MeanA}{t} + {\bm{v}}\cdot \pd{\delA}{t} \right) -Z_s e \bm{v} \dg \delpot\\
		\qquad\qquad- {m_s} R^2\left[\angvel(\psi)  + \frac{m_sc}{Z_s e} R^2 \left( \bm{v}\dg\tor  \right)\frac{d\angvel}{d\psi} \right] \daccel\dg\tor,
}\end{eqalign}
\label{tmpA32}
\end{equation}
where we have used \eref{psiDef} and \eref{torrot} to show that $\bm{u}\cdot\infrac{\partial \MeanA}{\partial t} = c \infrac{\partial\fpot}{\partial t}$, so it cancelled with the corresponding term in \eref{destart}.
Finally, we rearrange \eref{tmpA32} into
\begin{equation}
\fl
\begin{eqalign}{
\label{edot}
\denergy &= -\frac{Z_s e}{c} \left(\pd{}{t} + \bm{u}\dg\right) \left(\bm{w}\cdot\delA\right) - \frac{Z_s e}{c} \bm{w}\cdot \pd{\MeanA}{t}\\
				&- Z_s e\bm{w}\dg\gkupot- \frac{Z_s e}{c} \left(\bm{w} \cdot \nabla\bm{u} \right)\cdot\delA
				   - \frac{m_s^2c}{Z_s e} R^4\frac{d\angvel}{d\psi} \left(\bm{v}\cdot\nabla\tor\right) \daccel\dg\tor,
}\end{eqalign}
\end{equation}
where we have used $\bm{u}=\angvel(\psi)R^2\nabla\tor$ and \eref{daccel}.

The lowest-order contribution to $\denergy$ is, therefore,
\begin{equation}
\label{low-edot}
\denergy = -Z_s e \bm{w}_\perp \dg\gkupot + \Or\left(\gkeps^2 \cycfreq T_s\right),
\end{equation}
and so, by using \eref{wperpdg}, we conclude that
\begin{equation}
\label{edot-low-gyr}
\gyroR{\denergy} = \Or\left(\gkeps^2 \cycfreq T_s\right).
\end{equation}
\subsection{Derivation of \texorpdfstring{$\dot{\gyr}$}{theta-dot}}
The gyrophase $\gyr$ is defined implicitly by \eref{wDef}, viz.,
\begin{equation}
\label{apwdef}
\bm{w} = \bm{v} - \bm{u}= w_\parallel \Meanb + w_\perp \left(\cos \gyr\,\bm{e}_2 - \sin\gyr\,\bm{e}_1 \right).
\end{equation}
Taking the time derivative of this equation along a particle orbit gives
\begin{equation}
\label{bikkit}
\fl
\bm{a}_s + \delta\bm{a}_s - \bm{v}\dg\bm{u} = \frac{d w_\parallel}{dt} \Meanb + w_\parallel \bm{v}\dg\Meanb + \frac{\bm{w}_\perp}{w_\perp} \frac{d w_\perp}{dt} 
- w_\perp \left( \cos\gyr\,\bm{e}_1 + \sin\gyr\,\bm{e}_2\right) \dot{\gyr}.
\end{equation}
Since $ w_\perp \left( \cos\gyr\,\bm{e}_1 + \sin\gyr\,\bm{e}_2\right) = \Meanb\times\bm{w}_\perp$,  we take the inner product of \eref{bikkit} with $\Meanb\times\bm{w_\perp}$ to find
\begin{equation}
\label{gyrdotlong}
\dot{\gyr} = -\frac{1}{w_\perp^2} \left( \bm{a}_s +\delta\bm{a}_s - \bm{v}\dg\bm{u} - w_\parallel \bm{v}\dg\Meanb \right) \cdot\left(\Meanb\times\bm{w}_\perp\right).
\end{equation}
Expanding $\bm{a}_s$ according to \eref{accel} and taking only the leading-order contribution, we find
\begin{equation}
\label{gyrdot}
\dot{\gyr} = \cycfreq + \Or(\gkeps \cycfreq).
\end{equation}
\subsection{Derivation of \eref{2ndorder}}
\label{apA6}
To derive \eref{2ndorder} from \eref{2complex}, we first note that, from \eref{rdotg}, 
\begin{equation}
\begin{eqalign}{
\gyroR{\dot{\bm{R}}_s} \dgR{}&\left( F_{0s} + F_{1s} + h_s \right) = w_\parallel \Meanb\dgR{ } \left(F_{1s} + h_s \right) \\
	&+\bm{u}(\bm{R}_s) \dgR{h_s} + \left( \vchiR + \vdrift \right) \dgR{ } \left( F_{0s} + h_s\right)
}\end{eqalign}
\end{equation}
because $\Meanb\cdot\infrac{\partial F_{0s}}{\partial\bm{R}_s} = 0$, and $F_{0s}$ and $F_{1s}$ are axisymmetric. Note that the Ba\~nos drift (second term in \eref{rdotg}) is negligible, as explained in \ref{derivRdot}.

Now we must calculate the gyroaverages in the right-hand side of \eref{2complex}:
\begin{equation}
\gyroR{\denergy} \frac{F_{0s}}{T_s} + \gyroR{\frac{d}{dt} \left( \frac{Z_s e \gkupot}{T_s} F_{0s}\right)}.
\label{startBleh}
\end{equation}
This quantity is $\Or(\gkeps^2\cycfreq F_s)$.
Tackling the Boltzmann response (the second term) first, we find
\begin{equation}
\fl
\begin{eqalign}{
\gyroR{\frac{d}{dt} \left( \frac{Z_s e \gkupot}{T_s} F_{0s}\right)} = \gyroR{ \left(\pd{}{t} + \bm{v}\dg\right)\gkupot } \frac{Z_s e}{T_s} F_{0s} \\
	\qquad+ \gyroR{\frac{Z_s e\gkupot}{T_s}\dot{\bm{R}}_s \cdot\nabla\psi}\pd{F_{0s}}{\psi} -  \gyroR{\frac{Z_s e\gkupot}{T_s} \denergy} \frac{F_{0s}}{T_{s}} + \Or(\gkeps^3\cycfreq F_{0s}),
}\end{eqalign}
\label{tmpA42}
\end{equation}
where we have used \eref{F0def} for $F_{0s}$ and the fact that $\infrac{\partial F_{0s}}{\partial t} = \Or(\gkeps^3\cycfreq F_{0s})$. Using the lowest-order expressions \eref{low-rdot} for $\dot{\bm{R}}_s$ and \eref{low-edot} for $\denergy$, 
and inferring from \eref{wperpdg} that $Z_s^2 e^2 \gyroR{\bm{w}_\perp\dg\gkupot^2} = \Or(\gkeps^3 \cycfreq T_s^2)$,
we conclude that the second and third terms in \eref{tmpA42} are $\Or(\gkeps^3 \cycfreq F_{0s})$ and so can be neglected.
Thus, we are left with
\begin{equation}
\fl\label{gkudt}
\gyroR{\frac{d}{dt} \left( \frac{Z_s e \gkupot}{T_s} F_{0s}\right)} = \gyroR{ \left(\pd{}{t} + \bm{v}\dg\right)\gkupot } \frac{Z_s e}{T_s} F_{0s} + \Or(\gkeps^3\cycfreq F_{0s}).
\end{equation}
Factoring out $F_{0s} / T_s$ in \eref{startBleh} and using \eref{edot} to express $\denergy$, we find, therefore, that calculating \eref{startBleh} reduces to calculating the following gyroaverage:
\begin{equation}
\fl
\begin{eqalign}{
\label{edots}
\gyroR{\denergy+ Z_s e\left(\pd{}{t} + \bm{v}\dg\right)\gkupot} = \\
	\qquad\quad Z_s e\gyroR{ \left(\pd{}{t} + \bm{u}\dg\right) \gkpot} - \frac{Z_s e}{c} w_\parallel \pd{\MeanA}{t}\cdot\Meanb
		 - \frac{Z_s e}{c} \gyroR{\left(\bm{w}\cdot \nabla\bm{u} \right)\cdot \delA} \\
		 \qquad\quad - \frac{m_s^2c}{Z_s e} R^4 \frac{d\angvel}{d\psi} \left(\nabla\tor\right)\cdot \gyroR{\bm{v} \daccel }\cdot\nabla\tor,
}\end{eqalign}
\end{equation}
where $\gkpot = \gkupot - \bm{w}\cdot\delA / c$.

Tackling the last term in \eref{edots} first, we use the second line of \eref{daccel} for $\daccel$ to find (neglecting $\Or(\gkeps^2 \vth \cycfreq)$ contributions and so keeping only the first two terms)
\begin{equation}
\label{badger2}
\fl
\begin{eqalign}{
- &\frac{m_s^2c}{Z_s e} R^4 \frac{d\angvel}{d\psi} \left(\nabla\tor\right)\cdot \gyroR{\bm{v} \daccel }\cdot\nabla\tor  \\
	&= m_sc R^4 \frac{d\angvel}{d\psi} \left(\nabla\tor\right)\cdot \gyroR{\bm{v} \nabla\gkpot}\cdot\nabla\tor  + m_s R^4\frac{d\angvel}{d\psi} \left(\nabla\tor\right)\cdot\gyroR{\bm{w}\left( \bm{w}_\perp\dg\delA \right)} \cdot\nabla\tor.
}\end{eqalign}
\end{equation}
In the first of these terms, we split $\bm{v} = \left(\bm{u}+w_\parallel\Meanb\right) + \bm{w}_\perp$ and use $\left(\nabla\tor\right)_\perp = \Meanb\times\nabla\psi / \MeanMagB R^2$ (which follows from the axisymmetric form of the magnetic field, \eref{magfield});
in the third term, we use \eref{wperpdg1}, integrate by parts with respect to $\gyr$, and again use $\left(\nabla\tor\right)_\perp = \Meanb\times\nabla\psi / \MeanMagB R^2$. The result is
\begin{equation}
\fl
\begin{eqalign}{
&- \frac{m_s^2c}{Z_s e} R^4 \frac{d\angvel}{d\psi} \left(\nabla\tor\right)\cdot \gyroR{\bm{v} \daccel }\cdot\nabla\tor = - m_s \vchiR\cdot\bm{W} \\
		&\quad-\frac{m_sc}\MeanMagB R^2\frac{d\angvel}{d\psi} \left(\nabla\psi\right)\cdot \gyroR{ \left(\Meanb\times\bm{w}_\perp\right) \nabla\gkpot}\cdot\nabla\tor
		+ \frac{Z_s e}{c} R^2\frac{d\angvel}{d\psi} \left(\nabla\psi\right)\cdot\gyroR{\bm{w}_\perp\delA}\cdot\nabla\tor,
}\end{eqalign}
\label{newA46}
\end{equation}
where $\vchiR$ is defined by \eref{vchi} and we have abbreviated
\begin{equation}
\fl
\begin{eqalign}{
\bm{W} &= \left[R^2\left(w_\parallel\Meanb + \bm{u}\right)\dg\tor\right]\frac{d\angvel}{d\psi}\nabla\psi = \left[ \frac{Iw_\parallel}\MeanMagB + \angvel(\psi) R^2 \right] \frac{d\angvel}{d\psi} \nabla \psi \\
	&= \left(w_\parallel \Meanb + \bm{u}\right)\cdot\nabla\bm{u} + \left( \nabla\bm{u} \right)\cdot\left(w_\parallel \Meanb+\bm{u}\right).
}\end{eqalign}
\label{Wsimple}
\end{equation}

We wish to write the second term on the right-hand side of \eref{newA46} in terms of $\nabla \bm{u}$. 
Using \eref{nablaU}, we observe that
\begin{equation}
\fl
\begin{eqalign}{
\gyroR{\left( \Meanb\times\bm{w}_\perp \right)\cdot\left( \nabla\bm{u} \right)\cdot\nabla\gkpot} = R^2\frac{d\angvel}{d\psi} \left( \nabla\psi \right)\cdot\gyroR{\left( \Meanb\times\bm{w}_\perp \right)\nabla_\perp\gkpot}\cdot\nabla\tor \\
		\qquad\qquad+ R\angvel(\psi)\gyroR{\left( \Meanb\times\bm{w}_\perp \right)\cdot\left[ \left( \nabla R \right)\left( \nabla\tor \right)-\left( \nabla\tor \right)\left( \nabla R \right) \right]\cdot\nabla_\perp\gkpot}.
}\end{eqalign}
\end{equation}
Rearranging this equation and using \eref{aCrossZ}, we find
\begin{equation}
\fl
\begin{eqalign}{
R^2\frac{d\angvel}{d\psi} &\left( \nabla\psi \right)\cdot\gyroR{\left( \Meanb\times\bm{w}_\perp \right)\nabla_\perp\gkpot}\cdot\nabla\tor \\
	&= \gyroR{\left( \Meanb\times\bm{w}_\perp \right)\cdot\left( \nabla\bm{u} \right)\cdot\nabla\gkpot} - \angvel(\psi)\gyroR{\left( \Meanb\times\bm{w}_\perp \right)\cdot\left(\nabla_\perp\gkpot\times\nabla z\right)} \\
&=\gyroR{\left( \Meanb\times\bm{w}_\perp \right)\cdot\left( \nabla\bm{u} \right)\cdot\nabla\gkpot}+\angvel(\psi)\left(\Meanb\dg z\right)\gyroR{\bm{w}_\perp\dg\gkpot}.
}\end{eqalign}
\label{tmpTmpBadger}
\end{equation}
In \ref{derivA61}, we show that $\gyroR{\bm{w}_\perp\dg\gkpot} = 0$ and so the second term on the right-hand side of \eref{tmpTmpBadger} vanishes.
Substituting the remainder into \eref{newA46}, we obtain
\begin{equation}
\label{badger2.5}
\fl
\begin{eqalign}{
- \frac{m_s^2c}{Z_s e} R^4 \frac{d\angvel}{d\psi} \left(\nabla\tor\right)\cdot \gyroR{\bm{v} \daccel }\cdot\nabla\tor 
		= - m_s \vchiR\cdot\bm{W}\\
		\quad -Z_s e \gyroR{\left( \frac{\Meanb\times\bm{w}}{\cycfreq} \right)\cdot\left( \nabla\bm{u} \right)\cdot \nabla \gkpot}+ \frac{Z_s e}{c} R^2\frac{d\angvel}{d\psi} \left(\nabla\psi\right)\cdot\gyroR{\bm{w}_\perp\delA}\cdot\nabla\tor.
}\end{eqalign}
\end{equation}

Collecting our results, we substitute \eref{badger2.5} back into \eref{edots}
\begin{equation}
\fl
\begin{eqalign}{
\label{edots2}
\gyroR{\denergy+ Z_s e\left(\pd{}{t} + \bm{v}\dg\right)\gkupot} =\\
	\qquad Z_s e\gyroR{\left[\pd{}{t} + \bm{u}(\bm{R}_s)\dg\right] \gkpot} - \frac{Z_s e}{c} w_\parallel\Meanb\cdot \pd{\MeanA}{t} - \frac{Z_s e}{c} \gyroR{\bm{w}\cdot\left(\nabla\bm{u}\right)\cdot \delA} \\
	\qquad- m_s \vchiR\cdot\bm{W}+ \frac{Z_s e}{c} R^2 \frac{d\angvel}{d\psi} \left(\nabla\psi\right)\cdot\gyroR{\bm{w}_\perp \delA }\cdot\left(\nabla\tor \right)
,
}\end{eqalign}
\end{equation}
where we have the fact that $\bm{r} - \bm{R}_s = \Meanb\times\bm{w} / \cycfreq$ to absorb the second term in the right-hand side of \eref{badger2.5} into the first term in the right-hand side of \eref{edots2} (this is valid up to corrections of order $\Or(\gkeps^3\cycfreq T_s)$, which we neglect).

Turning now to the third term in \eref{edots2}, we use \eref{nablaU} and \eref{aCrossZ} to find
\begin{equation}
\label{tmp}
\fl
\begin{eqalign}{
\gyroR{\bm{w}\cdot\left(\nabla\bm{u}\right)\cdot\delA} 
   %&= R^2 \frac{d\angvel}{d\psi} \left(\nabla\psi\right)\cdot\gyroR{\bm{w}_\perp \delA_\perp }\cdot\left(\nabla\tor \right) +\angvel(\psi)\gyroR{\bm{w}_\perp\cdot\left(\delA_\perp\times\nabla z\right)}\\
	%&\qquad+ \gyroR{w_\parallel \Meanb \cdot\left(\nabla\bm{u}\right)\cdot \delA_\perp} + \gyroR{\bm{w}_\perp \cdot\left(\nabla\bm{u}\right)\cdot \Meanb \delAp} \\
	&=R^2 \frac{d\angvel}{d\psi} \left(\nabla\psi\right)\cdot\gyroR{\bm{w}_\perp \delA }\cdot\left(\nabla\tor \right)
	+\angvel(\psi)\gyroR{\left(\bm{w}_\perp\times\delA_\perp\right)\cdot\nabla z} \\
	&\qquad+ \left(\Meanb \cdot\nabla\bm{u}\right)\cdot\gyroR{w_\parallel \delA_\perp - \delAp \bm{w}_\perp}.
}\end{eqalign}
\end{equation}
In \ref{derivA61}, we show that the second term on the right-hand side of \eref{tmp} vanishes (see \eref{aperpvanish} with $\bm{a} = \nabla z$).
Substituting the remainder of \eref{tmp} into \eref{edots2}, and noting the cancellation between the last term in \eref{edots2} and the first term in \eref{tmp}, we obtain
\begin{equation}
\fl
\label{edotnotsofinal}
\begin{eqalign}{
\gyroR{\denergy+ Z_s e\left(\pd{}{t} + \bm{v}\dg\right)\gkupot} = Z_s e\gyroR{\left[\pd{}{t} + \bm{u}(\bm{R}_s)\dg\right] \gkpot} - \frac{Z_s e}{c} w_\parallel\pd{\MeanA}{t}\cdot\Meanb \\
\quad-m_s\vchiR\cdot\bm{W}
+ \frac{Z_s e}{c}\left(\Meanb\cdot\nabla\bm{u}\right)\cdot\gyroR{\delAp \bm{w}_\perp - w_\parallel \delA_\perp}.
}\end{eqalign}
\end{equation}

It merely remains to write the first term in this expression in terms of $\gyroR{\gkpot}$. 
This is done in \ref{Ap62}. We find that
\begin{equation}
\fl
\gyroR{\bm{u}(\bm{R}_s) \dg \gkpot} + \Meanb\cdot\left(\nabla\bm{u}\right)\cdot\gyroR{\bm{w}_\perp \delAp-w_\parallel\delA_\perp } = \bm{u}(\bm{R}_s)\cdot \ddR{\gyroR{\gkpot}}.
\label{tmp-fail-gyro}
\end{equation}
Substituting \eref{tmp-fail-gyro} into \eref{edotnotsofinal}, using the second expression in \eref{Wsimple} to expand $\bm{W}$, and using \eref{gkudt}, we obtain our final result:
\begin{equation}
\fl
\begin{eqalign}{
\gyroR{\denergy} \frac{F_{0s}}{T_s}& + \gyroR{\frac{d}{dt} \left(\frac{Z_s e \gkupot}{T_s} F_{0s}\right)} =
\frac{Z_s eF_{0s}}{T_s}\left[\pd{}{t} + \bm{u}(\bm{R}_s)\dgR{}\right] \gyroR{\gkpot}\\
& - \frac{m_s F_{0s}}{T_s}\left[\frac{Iw_\parallel}\MeanMagB + \angvel(\psi) R^2\right] \frac{d\angvel}{d\psi}\vchiR\dg\psi
- \frac{Z_s e}{T_sc} w_\parallel F_{0s}\pd{\MeanA}{t}\cdot\Meanb.
}\end{eqalign}
\label{localboy}
\end{equation}
\subsubsection{Gyroaverages involving \texorpdfstring{$\delA_\perp$}{delta A perp}:}
\label{derivA61}
As $\nabla_\perp\cdot\delA_\perp =0$ (the Coulomb gauge condition to lowest order in $\gkeps$), we can always find a scalar function $\zeta(\bm{r},t)$ such that 
\begin{equation}
\delA_\perp = \Meanb\times\nabla_\perp\zeta.
\label{AperpZeta}
\end{equation}
This expression allows us to gyroaverage explicitly various quantities involving $\delA_\perp$.

Firstly, for any vector $\bm{a}$,
\begin{equation}
{\left( \bm{w}_\perp\times\delA_\perp \right) \cdot \bm{a}} = \Meanb\cdot\bm{a} \left( {\bm{w}_\perp \dg \zeta} \right).
\end{equation}
Gyroaveraging both sides of this equation, we find that the right-hand side vanishes upon use of \eref{wperpdg}.
Thus, to lowest order in $\gkeps$,
\begin{equation}
\gyroR{\left( \bm{w}_\perp\times\delA_\perp \right) \cdot \bm{a}} = 0,
\label{aperpvanish}
\end{equation}
for any vector $\bm{a}$.

We now use \eref{aperpvanish} to evaluate another gyroaverage involving $\delA_\perp$: again using \eref{wperpdg}, we find for $\gkpot$, defined in \eref{chidef}, that
\begin{equation}
\fl
\begin{eqalign}{
\gyroR{\bm{w}_\perp\dg\gkpot} &= -\cycfreq\gyroR{\left.\pd{ }{\gyr}\right|_{\bm{r},\energy,\magmom}\gkpot}
= \frac{\cycfreq}{c}\gyroR{\delA_\perp\cdot\left( \Meanb\times\bm{w}_\perp \right)} = 0.
}\end{eqalign}
\label{wperpdgchi}
\end{equation}
The last equality follows from \eref{aperpvanish}.
\subsubsection{Derivation of \eref{tmp-fail-gyro}:}
\label{Ap62}

We start by transforming the derivative \mbox{$\bm{u}(\bm{R}_s)\dg\gkpot$} into a derivative with respect to $\bm{r}$ at constant $\energy$, $\magmom$ and $\gyr$. 
As $Z_s e \bm{u}\dg\chi \sim \Or(\gkeps \cycfreq T_s)$, we keep track of first-order corrections.\footnote{This is where we disagree with the derivation in \cite{artun1994nonlinear}, where the difference between $\bm{u}(\bm{R}_s)$ and $\bm{u}(\bm{r})$ is correctly retained, but the difference between $\bm{u}(\bm{R}_s) \cdot \infrac{\partial\chi}{\partial \bm{R}_s}$ and $\bm{u}(\bm{R}_s)\dg \chi$ is incorrectly neglected.}
We begin from
\begin{equation}
\fl
\label{toenergy}
\begin{eqalign}{
\bm{u}(\bm{R}_s)\dg\gkpot &= \bm{u}(\bm{R}_s)\cdot\left.\nabla\right|_{\energy,\magmom,\gyr}\gkpot \\
	&\qquad+\bm{u}\cdot\left[ \left( \nabla\energy \right) \pd{}{\energy}+ \left( \nabla\magmom \right)\pd{}{\magmom}+ \left( \nabla\gyr \right)\pd{}{\gyr}\right] \gkpot + \Or(\gkeps^2 \cycfreq \gkpot),
}\end{eqalign}
\end{equation}
where, as the second term is $\gkeps$ smaller than the first, we are able to drop the distinction between $\bm{r}$ and $\bm{R}_s$ in the argument of $\bm{u}$.
Using axisymmetry and the definitions \eref{energydef}, \eref{mudef}, and \eref{wDef} of $\energy$, $\magmom$, and $\gyr$, we have (correct to lowest order in~$\gkeps$)
\begin{eqnarray}
\label{udgEnergy}
\bm{u}\dg\energy &= 0,\\
\label{udgMu}
\bm{u}\dg\magmom &= -\frac{m_s}\MeanMagB w_\parallel \left(\bm{u}\cdot\nabla\Meanb\right)\cdot\bm{w}_\perp,\\
\label{udgGyr}
\bm{u}\dg\gyr &= \left(\bm{u}\cdot\nabla\bm{e}_1\right) \cdot\bm{e}_2 + \frac{w_\parallel}{w_\perp^2} \left(\bm{u}\dg\Meanb\right)\cdot\left(\Meanb\times\bm{w}_\perp\right).
\end{eqnarray}
Differentiating \eref{chidef}, we find
\begin{eqnarray}
\left.\pd{}{\magmom}\right|_{\bm{r},\energy,\gyr} \gkpot &= \frac\MeanMagB{cm_s w_\parallel}\delAp  -\frac{1}{2 c \magmom} \delA_\perp \cdot\bm{w}_\perp,\\
\left.\pd{}{\gyr}\right|_{\bm{r},\energy,\magmom} \gkpot &= \frac{1}{c} \delA_\perp \cdot\left(\Meanb\times\bm{w}_\perp\right).
\end{eqnarray}
Inserting these results into \eref{toenergy} and gyroaveraging at constant $\bm{R}_s$, we find
\begin{equation}
\fl
\begin{eqalign}{
\gyroR{\bm{u}(\bm{R}_s)\dg\gkpot} = \gyroR{\bm{u}(\bm{R}_s)\cdot\left.\nabla\right|_{\energy,\magmom,\gyr} \gkpot} \\
\qquad	+ \frac{w_\parallel}{cw_\perp^2}\left(\bm{u}\cdot\nabla\Meanb\right)\cdot\gyroR{\bm{w}_\perp \bm{w}_\perp \cdot\delA_\perp + \Meanb\times\bm{w}_\perp \left(\Meanb\times\bm{w}_\perp \right)\cdot\delA_\perp} \\
\qquad - \left(\bm{u}\cdot\nabla\Meanb\right) \cdot\gyroR{\bm{w}_\perp \delAp} \\
	\quad = \gyroR{\bm{u}(\bm{R}_s)\cdot\left.\nabla\right|_{\energy,\magmom,\gyr} \gkpot} + \left( \Meanb\dg\bm{u} \right)\cdot\gyroR{w_\parallel\delA_\perp - \bm{w}_\perp \delAp},
}\end{eqalign}
\label{tmpA44}
\end{equation}
where we have used \eref{udgb} and the fact that $\bm{w}_\perp \bm{w}_\perp + \left( \Meanb\times\bm{w}_\perp \right) \left( \Meanb\times\bm{w}_\perp \right) = w_\perp^2 \left(\idmat - \Meanb\Meanb\right)$.
Since $\bm{R}_s = \bm{r} - \infrac{\Meanb\times\bm{w}}{\cycfreq}$, the spatial derivative with respect to $\bm{r}$ at constant $\energy$, $\magmom$ and $\gyr$ is converted to a derivative with respect to $\bm{R}_s$ as follows:
\begin{equation}
\fl
\begin{eqalign}{
\gyroR{\bm{u}(\bm{R}_s)\left.\dg\right|_{\energy,\magmom,\gyr} \gkpot} 
&= \gyroR{\bm{u}(\bm{R}_s)\cdot\left( \left.\nabla\right|_{\energy,\magmom,\gyr} \bm{R}_s\right) \cdot\ddR{\gkpot}} \\
%&= \gyroR{\bm{u}(\bm{R}_s)\cdot\ddR{\gkpot} + \frac{\bm{u}}{\cycfreq} \left.\dg\right|_{\energy,\magmom,\gyr} \left(\Meanb\times\bm{w}\right) \dg\gkpot }\\
&= \bm{u}(\bm{R}_s)\cdot\ddR{\gyroR{\gkpot}} - \gyroR{\frac{\bm{u}}{\cycfreq} \cdot\left[ \left.\nabla\right|_{\energy,\magmom,\gyr} \left(\Meanb\times\bm{w}\right) \right] \dg\gkpot }.
}\end{eqalign}
\label{tmpA46}
\end{equation}
The last term in \eref{tmpA46} vanishes:
\begin{equation}
\begin{eqalign}{
&\gyroR{\frac{\bm{u}}{\cycfreq}\left.\cdot\left[\nabla\right|_{\energy,\magmom,\gyr}\left(\Meanb\times\bm{w}\right)\right] \dg\gkpot}\\
	&\qquad= \frac{w_\perp}{\cycfreq}\gyroR{\left[\left(\bm{u}\cdot\nabla\bm{e}_1\right) \cos\gyr + \left(\bm{u}\cdot\nabla\bm{e}_2\right) \sin \gyr\right]\dg_\perp\gkpot}\\
		&\qquad= \frac{1}{\cycfreq}\left(\bm{u}\cdot\nabla\bm{e}_1\right)\cdot\bm{e}_2 \gyroR{\bm{w}_\perp\dg\gkpot} = 0,
}\end{eqalign}
\label{tmpA47}
\end{equation}
where we have used the definition \eref{wDef} of $\bm{w}$, the identity $\left( \nabla \bm{e}_1 \right) \cdot\bm{e}_2 = - \left( \nabla \bm{e}_2 \right) \cdot\bm{e}_1$, and finally \eref{wperpdgchi}.
Substituting the remainder of \eref{tmpA46} into \eref{tmpA44}, we arrive at the final result \eref{tmp-fail-gyro}.

\section{Alternative Formulations of the Gyrokinetic Equation}
\label{altGKap}
There are many ways to derive the gyrokinetic equation \eref{gke}. In this Appendix, we discuss two ways in which the results of these derivations may differ from the one presented in this paper.
\subsection{Parallel Nonlinearity}
\label{parallelNote}
In the gyrokinetic literature, there is much discussion of the so-called ``parallel nonlinearity''~\cite{ParallelNonlinearity,hinton:102301}, which is proportional to
$\delta E_\parallel\binfrac{\partial\delta f_s}{\partial{v}_\parallel}$.
Some derivations of gyrokinetics include this term in the gyrokinetic equation, claiming that it has beneficial numerical properties and an important role in turbulent heating and conservation of energy.
The gyrokinetic equation \eref{gke} derived here does not contain the parallel nonlinearity because the latter is higher order than the terms we retain and only appears in the derivation of the transport equations.   We have, in \Sref{thermo}, derived conservation laws that prove that our equations nevertheless conserve, in the mean, energy and satisfy Boltzmann's $H$-theorem. It is true that in our formalism there is no formal {\it energy} conservation law for gyrokinetics on the fluctuating timescale (in contrast to versions of gyrokinetics derived from Hamiltonians; see, e.g., \cite{dubin1983nge}). This is because energy is exchanged between the fluctuations and the mean fields. We showed in \Sref{SFreeEnergy} that, on fluctuating timescales, our gyrokinetic equation conserves the {\textit{free energy}}, which is the relevant conserved quantity for kinetic turbulence~\cite{Krommes2,hallatschek,Tome,schekcrete} and, moreover, the conservation of which is crucial for the mean energy to be conserved on the transport timescale.
Note that numerical simulations~\cite{ParallelNonlinearity} have shown that the inclusion of the parallel nonlinearity in the gyrokinetic equation does not affect the solutions, which is how it should be for a term that is ordered smaller than all the other terms in \eref{gke}. This result implies that, even though small scales in velocity space may be generated in the solutions of \eref{gke}, they are not small enough to violate our formal ordering of $\infrac{\partial \delta f_s }{\partial\bm{v}} \sim \delta f_s / \vth$.

\subsection{Gyrokinetic Polarisation}
\label{polarisationAp}
The second difference we wish to discuss is the emphasis that some versions of the gyrokinetic formalism place upon the effects of the polarisation drift.
Through $\vchi$ and $\vdrift$, \eref{gke} contains the explicit action of all the lowest-order drifts, but we have yet to mention the polarisation drift. This drift, given in the long-wavelength, $k_\perp\rho_i \ll 1$, limit by 
\begin{equation}
\poldrift =  \frac{c}{\MeanMagB \cycfreq}\pd{\delta\bm{E}}{t},
\label{poldriftDef}
\end{equation}
 is $\Or(\gkeps^2\vth)$ and, therefore, smaller than $\vchi$ or $\vdrift$. Interestingly, despite the fact that $\poldrift$ is second order in $\gkeps$, its effects couple to the gyrokinetic equation through the quasineutrality condition and Amp\`ere's law~\cite{dubin1983nge}. This coupling arises because the polarisation drift causes an accumulation of charge density and because it gives rise to a perpendicular current.
	
	Unfortunately, due to our choice of decomposition \eref{hDef} for $\delta f_s$, this is not apparent from the formulation \eref{fluct-qn} of the fluctuating component of the quasineutrality condition. If, instead of using $h_s$, we work with $g_s = \gyroR{\delta f_s} = h_s - Z_s e F_{0s} \gyroR{\gkupot}/T_s$~\cite{lee1983gyrokinetic}, then \eref{fluct-qn} becomes
\begin{equation}
 \sum_s \frac{Z_s^2 e^2 n_s(1-\Gamma_{0s})\gkupot}{T_s}   = \sum_s Z_s e \wint \gyror{g_s},
 \label{qn-with-g}
\end{equation}
where $\Gamma_{0s}$ is an operator defined by $\Gamma_{0s} \gkupot = (1/n_s)\wint \gyror{\gyroR{\gkupot}} F_{0s}$. In this formulation, the left-hand side of \eref{qn-with-g} can be interpreted as the density of polarisation charge (i.e. the density that accumulates because $\dv\poldrift \ne 0$). 

Let us demonstrate this in the long-wavelength limit. In this limit, the polarisation density $\npol$ is defined by the following continuity equation:
\begin{equation}
\pd{\npol}{t} + \dv\left(n_s \poldrift\right) = 0,
\end{equation}
where $\poldrift$ is given by \eref{poldriftDef}.
Integrating over time, we obtain
\begin{equation}
\npol = -\dv\left(\frac{ n_s c}{\MeanMagB \cycfreq} \delta \bm{E}\right) = \frac{c}{B\cycfreq} \nabla_\perp^2 \delpot.
\label{npol-lw}
\end{equation}
Now, taking \eref{qn-with-g} in the long-wavelength limit, we approximate $1-\Gamma_0 \approx (1/2)\rho_i^2 \nabla_\perp^2$ and obtain
\begin{equation}
\left( \sum_s \frac{Z_s^2 e^2 n_s \rho_i^2}{2T_s} \right) \nabla_\perp^2 \gkupot = \sum_s Z_s e \wint \gyror{g_s}.
\label{qn-lw}
\end{equation}
Comparing this with \eref{npol-lw}, we see that the left-hand side is precisely $\sum_s Z_s e \npol$ -- the polarisation-charge density.
In this limit, another interpretation of \eref{qn-lw} also becomes apparent: the quasineutrality condition has the form of Poisson's equation for $\gkupot$, with an enhanced permittivity (known as ``the dielectric permittivity of the gyrokinetic vacuum''~\cite{krommesGyroReview}). 
	
The difference between the two formulations is purely interpretative; we interpret the gyrokinetic equation \eref{gke} and the field equations in \Sref{Sfmag} as describing the dynamics of physically extended rings of charge moving in a vacuum, whereas the approach emphasising the polarisation drift interprets gyrokinetics as describing a gas of point-particle-like gyrocenters (with the distribution function $g_s = \gyroR{\delta f_s}$) in a polarisable vacuum, with the quasineutrality condition in the form \eref{qn-with-g} playing the role of Poisson's equation. For more discussion of this second interpretation, see \cite{krommesGyroReview} and references therein.

\section{Flows within a Flux Surface and Moments of \texorpdfstring{$F_s$}{F}}
\label{flowcurrent}
Throughout this paper we need to take moments of the mean distribution function $F_s$. We need these moments in
\Sref{ampmag} to calculate the lowest-order mean current, in \Sref{heattrans} to evaluate the mean flow and thence the mean Ohmic heating \eref{OhmHeat} and in \ref{fluxes} to prove that certain parts of the radial transport are negligible.

In order to integrate $F_s$ over $\bm{w}$ we need to express it in terms of $\bm{r}$ and $\bm{w}$ rather than $\bm{R}_s$, $\energy$ and $\magmom$.
Assembling together the form \eref{FSoln} of $F_s$ and the expressions \eref{F1Hat}, \eref{Fstardef} and \eref{Fneodef} for $F_{1s}$ in terms of $\Fneo$ and $\Fstar$, we find 
\begin{equation}
\begin{eqalign}{
F_s = &\left[ 1 - \frac{Z_s e}{T_s}\int^ldl' \left( \MeanMagB\frac{\fav{\MeanE\cdot\MeanB}}{\fav{\MeanMagB^2}}+\frac{1}{c}\pd{\MeanA}{t}\cdot\Meanb\right)\right] F_{0s}(\psi(\bm{R}_s),\energy)\\
		&+ \frac{m_s c}{Z_s e} R^2\left( w_\parallel \Meanb + \bm{u} \right)\cdot\left( \nabla\tor \right) \pd{F_{0s}}{\psi}  + \Fneo(\bm{R}_s,\energy,\magmom,\sigma),
}\end{eqalign}
	 \label{expF}
\end{equation}
where the penultimate term is a convenient form for $\Fstar$ equivalent to \eref{Fstardef}.
We now expand $\psi(\bm{R}_s)$ around $\psi = \psi(\bm{r})$ in the first argument of $F_{0s}$ to find
\begin{equation}
\fl\begin{eqalign}{
F_{0s}(\psi(\bm{R}_s),\energy) &= F_{0s}(\psi,\energy) + \left(\bm{R}_s - \bm{r}\right)\cdot\left(\nabla\psi\right) \pd{F_{0s}}{\psi} + \Or(\gkeps^2 F_s) \\
 &= F_{0s}(\psi,\energy) + \frac{c m_s}{Z_s e} R^2\left( \bm{w}_\perp\dg\tor  \right)\pd{F_{0s}}{\psi}.
}\end{eqalign}
\label{F0sExp}
\end{equation}
Substituting this into \eref{expF} and combining the second term in this equation with the penultimate term in \eref{expF}, we find
\begin{equation}
\begin{eqalign}{
F_s = &\left[ 1 - \frac{Z_s e}{T_s}\int^ldl' \left(\MeanMagB\frac{\fav{\MeanE\cdot\MeanB}}{\fav{\MeanMagB^2}}+\frac{1}{c}\pd{\MeanA}{t}\cdot\Meanb\right)\right] F_{0s}(\psi,\energy)\\
		&\quad+ \frac{cm_s}{Z_s e}R^2 \left(\bm{v}\dg\tor\right) \pd{F_{0s}}{\psi} + \Fneo(\bm{r},\energy,\magmom,\sigma) + \Or(\gkeps^2 F_s),
}\end{eqalign}
\label{expF2}
\end{equation}
where we have neglected the difference between $\bm{R}_s$ and $\bm{r}$ in the argument of $\Fneo$ and abbreviated $\bm{v} = \bm{w} + \bm{u}$.
We now
expand $\fpot(\psistar)$ around $\fpot(\psi)$ in the definition \eref{energydef} of $\energy$ to obtain
\begin{equation}
\fl
\begin{eqalign}{
\energy &= \frac{1}{2} m_sw^2 - \frac{1}{2} m_s \angvel^2(\psi) R^2 + Z_s e \pot_0 - \frac{m_s\MeanMagB}{2 \cycfreq} R^4 \left( \bm{v}\dg\tor \right) ^2 \frac{d\angvel}{d\psi} + \Or(\gkeps^2 T_s) \\
	&= \energy_0- \frac{m_s\MeanMagB}{2 \cycfreq} R^4 \left( \bm{v}\dg\tor \right) ^2 \frac{d\angvel}{d\psi} + \Or(\gkeps^2 T_s),
}\end{eqalign}
\label{energyBong}
\end{equation}
where the lowest-order energy is
\begin{equation}
\energy_0 = \frac{1}{2} m_sw^2 + \Xi_s  = \energy + \Or(\gkeps T_s),
\end{equation}
which consists of the kinetic energy of the particle motion in the rotating frame and 
the zeroth-order potential energy of the particle (centrifugal plus electrostatic)
\begin{equation}
\Xi_s = - \frac{1}{2} m_s \angvel^2(\psi) R^2 + Z_s e \pot_0.
\label{XiDef}
\end{equation}
Substituting \eref{energyBong} into \eref{expF2} and using $\infrac{\partial F_{0s}}{\partial\energy} = - F_{0s} / T_s$ gives 
\begin{equation}
\fl\begin{eqalign}{
F_s = &\left[ 1 - \frac{Z_s e}{T_s}\int^ldl' \left(\MeanMagB\frac{\fav{\MeanE\cdot\MeanB}}{\fav{\MeanMagB^2}}+\frac{1}{c}\pd{\MeanA}{t}\cdot\Meanb\right)\right] F_{0s}(\psi,\energy_0)\\
		&+ \frac{cm_s}{Z_s e} R^2\left(\bm{v}\dg\tor\right) \left[ \frac{d\ln \NotN}{d\psi} + \left(\frac{\energy_0}{T_s} - \frac{3}{2}\right)\frac{d\ln T_s}{d\psi}\right] F_{0s}(\psi,\energy_0)\\
&+\frac{m_s\MeanMagB}{2 \cycfreq T_s}  R^4 \left( \bm{v}\dg\tor \right) ^2 \frac{d\angvel}{d\psi} F_{0s}(\psi,\energy_0)
 + \Fneo(\bm{r},\energy_0,\magmom,\sigma) + \Or(\gkeps^2 F_s),
}\end{eqalign}
\label{expF3}
\end{equation}
where (see \eref{F0def} and \eref{npol})
\begin{equation}
\fl
F_{0s}(\psi,\energy_0) = \frac{\NotN}{\pi^{3/2} \vth^3} \exp\left( -\frac{\Xi_s}{T_s} - \frac{m_s w^2}{2T_s} \right) = \frac{n_s}{\pi^{3/2}\vth^3} \exp\left( -\frac{m_s w^2}{2T_s} \right).
\label{F00}
\end{equation}
In the next two sections, we will use \eref{expF3} and \eref{F00} to explicitly evaluate moments of $F_s$ correct to first order.

\subsection{Flows within a Flux Surface}
\label{foflow}
To calculate the first-order mean flow, we multiply \eref{expF3} by $\bm{w}$ and integrate over all velocities to obtain
\begin{equation}
\fl
\begin{eqalign}{
n_s \bm{U}_s &= \wint \bm{w} F_s \\
	&= \wint \bm{w} \left\{
\frac{cm_s}{Z_s e} R^2\left(\bm{v}\dg\tor\right) \left[ \frac{d\ln \NotN}{d\psi} + \left(\frac{\energy_0}{T_s} - \frac{3}{2}\right)\frac{d\ln T_s}{d\psi}\right] F_{0s}(\psi,\energy_0)\right.\\
&\quad\left.
+\frac{m_s\MeanMagB}{2 \cycfreq T_s}  R^4 \left( \bm{v}\dg\tor \right) ^2 \frac{d\angvel}{d\psi} F_{0s}(\psi,\energy_0)
 + \Fneo(\bm{r},\energy_0,\magmom,\sigma) \right\} + \Or(\gkeps^2n_s\vth).
}\end{eqalign}
\label{flowlow}
\end{equation}
Using \eref{F00} for $F_{0s}(\psi,\energy_0)$, we can perform the first two integrals in \eref{flowlow} explicitly:
\begin{equation}
\fl
\begin{eqalign}{
\wint \bm{w}& \frac{cm_s}{Z_s e}R^2 \left(\bm{v}\dg\tor\right) \left[ \frac{d\ln \NotN}{d\psi} + \left(\frac{\energy_0}{T_s} - \frac{3}{2}\right)\frac{d\ln T_s}{d\psi}\right] F_{0s}(\psi,\energy_0)\\ 
		&= \frac{n_s}{Z_s e} c R^2  \left[ T_s\frac{d \ln \NotN}{d\psi} + \left(  { \Xi_s }+{T_s} \right) \frac{d\ln T_s}{d\psi} \right]\nabla\tor
}\end{eqalign}
\end{equation}
and
\begin{equation}
\fl
\wint \bm{w} \left[ \frac{m_s\MeanMagB}{2 \cycfreq T_s}  R^4 \left( \bm{v}\dg\tor \right) ^2 \frac{d\angvel}{d\psi}{F_{0s}}\right] 
		= \frac{n_s}{Z_s e} m_s c R^4 \angvel(\psi) \frac{d\angvel}{d\psi} \nabla\tor.
\end{equation}
Substituting these into \eref{flowlow} we find the expression for the first-order flow:
\begin{eqnarray}
\label{flow}
\fl
n_s \bm{U}_s &=& \frac{n_s}{Z_s e} c R^2 \left[ T_s\frac{d \ln \NotN}{d\psi} + \left( { \Xi_s }+{T_s}  \right) \frac{d\ln T_s}{d\psi} + m_s  R^2 \angvel(\psi) \frac{d\angvel}{d\psi}\right]\nabla\tor
		 + \frac{1}{Z_s e} {K}_s \MeanB,\\
\fl
\quad {K}_s &=& \frac{Z_s e}\MeanMagB \wint w_\parallel \Fneo,
\label{Khatdef}
\end{eqnarray}
where we have used the fact that $\wint \bm{w}_\perp \Fneo = 0$ to lowest order because $\Fneo$ is gyrophase-independent.

Let us now prove that $K_s$ is a flux function.
Integrating the drift-kinetic equation \eref{neogeq}, which determines $\Fneo$, over all velocities gives
\begin{equation}
\wint w_\parallel \Meanb\cdot\ddR{\Fneo} = 0,
\end{equation}
where we have used the fact that $\wint w_\parallel F_{0s} = 0$ and $\wint \collop[F_{0s}+\Fstar + \Fneo] =0$.
By exactly the same method as used in \eref{flibble} to derive the lowest-order $H$-theorem, we can rewrite this constraint as
\begin{equation}
\fl
\begin{eqalign}{
{ \wint w_\parallel \Meanb \cdot\ddR{\Fneo}} &= { 2\pi\sum_{\sigma} \int \frac{ \MeanMagB d\energy d\magmom}{m_s^2 |w_\parallel|} w_\parallel\Meanb\dg\left( \Fneo\right)}\\
&= { 2\pi\sum_\sigma \sigma \int \frac{d\energy d\magmom}{m_s^2} \MeanB\dg\Fneo} \\
&= \frac{1}{Z_s e}\MeanB\dg{K}_s = 0.
}\end{eqalign}
\label{Kfluxfn}
\end{equation}
 Thus, ${K}_s = K_s(\psi)$. This immediately implies that the first-order mean velocity is divergence-free (the divergence of the first term of \eref{flow} vanishes by axisymmetry), i.e.,
 \begin{equation}
 \label{lowdiv}
 \dv\left(n_s\bm{U}_s\right) = \Or(\gkeps^3 n_s \cycfreq).
 \end{equation}
This result will be useful in several places in \ref{fluxes}.

Finally, multiplying \eref{flow} by $Z_s e$ and summing over species we find the mean current
\begin{equation}
\fl
\label{current}
\begin{eqalign}{
\Meanj = c& R^2 \sum_s n_s\left\{ T_s\frac{d \ln \NotN}{d\psi} + \left[Z_s e\pot_0 - \frac{1}{2}{m_s\angvel^2(\psi) R^2} + T_s\right]\frac{d \ln T_s}{d\psi}\right\}\nabla\tor\\
			&+ cR^2\left( \sum_s n_s m_s R^2 \right) \angvel(\psi) \frac{d\angvel}{d\psi} \nabla\tor + K(\psi) \MeanB,
}\end{eqalign}
\end{equation}
where 
\begin{equation}
K(\psi) = \sum_s {K}_s(\psi) = \sum_s \frac{Z_s e}\MeanMagB\wint w_\parallel \Fneo.
\end{equation}
This is the result \eref{main-current} of \Sref{ampmag}.
\subsection{Moments of \texorpdfstring{$F_s$}{F}}
\label{lohflux}
An important corollary of \eref{flow} is that the first-order flows stay within the flux surface:
\begin{equation}
\label{flowinsurface}
n_s \bm{U}_s \dg\psi = \Or(\gkeps^2 n_s \vth |\nabla\psi|),
\end{equation}
which is consistent with our orderings for the timescale on which particles are transported.
We now derive a general result that will show that toroidal momentum transport and heat transport are also slow across flux surfaces.
We could calculate the lowest-order viscous stresses and heat fluxes by explicitly taking moments of \eref{expF3} as we did for the flows in \ref{foflow}, but we do not require these explicit expressions, only the fact that their cross-flux-surface components are small. This result will be useful in \ref{fluxes}.

First we prove that,
for any gyrophase-independent single-valued function $g$ and any positive integer $n$,
\begin{equation}
\wint g(\bm{r},w_\parallel,w_\perp,\sigma)\left(R^2\bm{w}\dg\tor\right)^n \bm{w}_\perp\dg\psi = 0,
	\label{MagicIdentity}
\end{equation}
Indeed, using the identity \eref{magicGyrThing}, we find
\begin{equation}
\fl\begin{eqalign}{
\wint &g(\bm{r},w_\parallel,w_\perp,\sigma) \left(R^2\bm{w}\dg\tor\right)^n \bm{w}_\perp\dg\psi\\
	&= -\MeanMagB\wint g(\bm{r},w_\parallel,w_\perp,\sigma) \left(R^2\bm{w}\dg\tor\right)^n \pd{}{\gyr} \left(R^2\bm{w}\dg\tor\right) \\
	&= - \frac\MeanMagB{n+1}\wint \pd{}{\gyr}\left[ g(\bm{r},w_\parallel,w_\perp,\sigma) \left( R^2\bm{w}\dg\tor \right)^{n+1}\right] = 0
}\end{eqalign}
\end{equation}
because single-valued functions must be periodic in $\gyr$.

We can now prove that
\begin{equation}
\fl
\wint g(\bm{r},w_\parallel,w_\perp,\sigma) F_s \left(R^2\bm{w}\dg\tor\right)^n \bm{w}_\perp\dg\psi = \Or(\gkeps^2  n_s g R^n \vth^{n+1} |\nabla\psi|).
\label{MagicTheorem}
\end{equation}
This follows from expanding $F_s$ via \eref{expF3} and applying \eref{MagicIdentity} term by term (since $\bm{w}\dg\tor$ differs from $\bm{v}\dg\tor$ by a gyrophase-independent function, we can replace the former by the latter in the statement of \eref{MagicIdentity} where necessary).

The radial heat flux, from \eref{eflux}, is
\begin{equation}
\bm{Q}_s\dg\psi = \wint \frac{1}{2} m_s w^2F_s\bm{w}_\perp\dg\psi.
\end{equation}
Letting $g= \binfrac{1}{2} m_s w^2$ and $n=0$, we can apply \eref{MagicTheorem} to find
\begin{equation}
\bm{Q}_s\dg\psi = \Or(\gkeps^2 n_s T_s \vth |\nabla\psi|).
\end{equation}
We can show in a similar fashion that the radial transport of toroidal angular momentum (see \eref{mflux}) is also small:
\begin{equation}
\label{lowviscous}
\left(\nabla\psi\right)\cdot\viscosity\cdot\left(\nabla\tor\right)R^2 = \Or(\gkeps^2 n_s T_s R|\nabla\psi|).
\end{equation}

\subsection{Relating \texorpdfstring{$\angvel(\psi)$}{omega} to physical plasma flows}
\label{experimentalOmega}
As mentioned at the end of \Sref{constraints}, the fact that the flow of a given species $\bm{u}_s$ is composed both of the toroidal flow \eref{torrot} and the corrections $\bm{U}_s$ given by \eref{flow} complicates the process of relating $\angvel(\psi)$ to experimental measurements.
If the mean flow $\bm{u}$ is sonic, i.e., the Mach number $M = \left. R\angvel(\psi)\right/ \vth$ is close to unity, then 
any differences between $\angvel(\psi)$ and the measured rotation rate that arise from simply neglecting $\bm{U}_s$ entirely will be small. 
However, experimentally, the Mach number is often relatively small (even though it is asymptotically larger than $\gkeps$ and so does not violate the gyrokinetic ordering).
In this case, the difference in magnitude between $R\angvel(\psi)$ and $\bm{U}_s$ may not be very large and we would like a better estimate both for $\angvel(\psi)$ and for the error arising in various approximations of it.

Formally, we may write the quantity $\angvel(\psi)$ as
\begin{equation}
\angvel(\psi) = \angvel^{(0)} + \angvel^{(1)} + \Or\left(\gkeps^2 \frac{\vth}{R}\right),
\label{approxAngVel}
\end{equation}
where $\angvel^{(1)} \sim \gkeps\angvel^{(0)}$ and the $\angvel^{(i)}$ are expressed in terms of experimentally measured quantities.
Examining \eref{omdef}, we see that there are two ways to infer $\angvel(\psi)$ from experimental data: either from a measurement of the mean flow $\bm{u}_s$ or from a measurement of the mean radial electric field (and thence $d\fpot/d\psi$). In analysing this problem, we will assume that all measurements are suitably averaged so as to remove any fluctuating component.

Suppose first that we have a measurement of the radial mean electric field $\bm{E}\dg\psi$~\cite{bieniosek1980hibp,melnikov2011potential}. Then, from \eref{electrostatic} and \eref{poteq}, we have
\begin{equation}
\bm{E}\dg\psi = -\frac{d\fpot}{d\psi}|\nabla\psi|^2 - (\nabla\pot_0)\dg\psi + \Or\left( \gkeps^2 \frac{\vth}{c} \MeanMagB |\nabla\psi| \right).
\end{equation}
Using \eref{omdef}, we can write this as an equation for $\angvel(\psi)$:
\begin{equation}
\angvel(\psi) = -\frac{c\MeanE\dg\psi}{|\nabla\psi|^2} - \frac{c(\nabla\pot_0)\dg\psi }{|\nabla\psi|^2} + \Or\left(\gkeps^2 \frac{\vth}{a}\right).
\label{angvelFromEr}
\end{equation}
To lowest order,
\begin{equation}
\angvel^{(0)} = \frac{c\MeanE\dg\psi}{|\nabla\psi|^2}.
\label{lowOrdEradialMeas}
\end{equation}
From the second term on the right-hand side of \eref{angvelFromEr}, we estimate the relative difference between this approximation and the quantity $\angvel(\psi)$ defined by \eref{omdef} to be
\begin{equation}
\frac{\angvel-\angvel^{(0)}}{\angvel^{(0)}} \sim \frac{c |\nabla\pot_0|}{|\nabla\psi|\angvel^{(0)}} \sim M \frac{\rho_i}{a},
\label{errEradial}
\end{equation}
where we have used the low-Mach-number estimate $\pot_0 \sim (a/R) M^2 T_s / Z_s e$ (see \Sref{vanishpot0}).
From this expression, we see that a measurement of $\angvel(\psi)$ via the electric field is correct to lowest order both in $\gkeps = \rho_i/a$ and in the Mach number.

Now let us instead suppose that we have a measurement of the flow velocity $\bm{u}_s$ of a single
species $s$ (as is indeed commonly the case; see, e.g., \cite{fonck1984cxrs}). 
In terms of the quantities that appear in the theory,
\begin{equation}
\bm{u}_s = \angvel(\psi) R^2 \nabla\tor + \bm{U}_s[\angvel(\psi),\NotN,T_s] + \Or(\gkeps^2 \vth),
	\label{flowSpecies}
\end{equation}
where $\bm{U}_s$ is given by \eref{flow}, and we have explicitly emphasised its dependence upon $\angvel(\psi)$ and other equilibrium parameters.
Keeping only the lowest-order term, we have
\begin{equation}
\angvel^{(0)} = \bm{u}_s\cdot\nabla\tor.
\label{lowOrdVelMeas}
\end{equation}
If $M \lesssim 1$,
\begin{equation}
\frac{\angvel- \angvel^{(0)}}{\angvel^{(0)}} = -\frac{\bm{U}_s\dg\tor}{\angvel^{(0)}} \sim \frac{1}{M} \frac{\rho_s}{a} \frac{R}{L_{p_s}},
\label{errFlowMeas}
\end{equation}
where $M$ is the Mach number of the flow estimated from \eref{lowOrdVelMeas}, $R/L_{p_s} = \max\left\{ R|\nabla \ln \NotN|,R|\nabla \ln T_s|\right\}$  and we have used \eref{flow} and $\psi \sim B a^2$. We can see that if $M\lesssim 1$ and $R/L_{p_s} \gtrsim 1$, as is usually the case in current experiments, then the difference between $\angvel(\psi)$ and the measured rotation rate \eref{lowOrdVelMeas} can be far larger than the simple $\Or(\rho_s/a)$ error estimate would suggest. 
If measurements of $\NotN$ and $T_s$ are available, then we can calculate $\angvel^{(1)}$ by returning to \eref{flowSpecies} and using \eref{lowOrdVelMeas} for $\angvel(\psi)$ in the argument of $\bm{U}_s$:
\begin{equation}
\angvel^{(1)} = -\bm{U}_s[\angvel^{(0)}(\psi),\NotN,T_s]\dg\tor.
	\label{highOrdVelMeas}
\end{equation}
In order to determine $\bm{U}_s\dg\tor$ in this expression, we need to solve for $K_s(\psi)$ -- the neoclassical parallel flow of species $s$ given by \eref{Khatdef}. 
This can, in principle, be done by solving the neoclassical equation \eref{neogeq} for $\Fneo$ (see \ref{nclasse}), but a far more straightforward way is afforded us if we know the poloidal component of $\bm{u}_s$: indeed, then $K_s = Z_s e R^2 n_s \bm{u}_s \cdot \left( \nabla\psi\times\nabla\tor \right) /  |\nabla\psi|^2$. With this in hand, we infer from \eref{flow}, to second order in $\gkeps$,
\begin{equation}
\fl
\begin{eqalign}{
\angvel(\psi) &\approx \angvel^{(0)} + \angvel^{(1)} \\
	&= \angvel^{(0)} - \frac{c}{Z_s e}  \left[ T_s\frac{d \ln \NotN}{d\psi} + \left( { \Xi_s }+{T_s}  \right) \frac{d\ln T_s}{d\psi} + m_s  R^2 \angvel^{(0)} \frac{d\angvel^{(0)}}{d\psi}\right] \\
		&\qquad- \bm{u}_s\cdot\left( \nabla\psi\times\nabla\tor \right) \frac{I(\psi)}{|\nabla\psi|^2},
		\label{omegaFromExp}
}\end{eqalign}
\end{equation}
where $\Xi_s = Z_s e \pot_0 + (1/2)m_s (\angvel^{(0)})^2$ and $\pot_0$ is determined by \eref{qn0} with $\angvel = \angvel^{(0)}$. 
Note that this requires a density and temperature measurement for all species. The situation is considerably simplified in the low-Mach number limit: retaining in \eref{omegaFromExp} only terms that are lowest-order in $M$, we have 
\begin{equation}
\begin{eqalign}{
\angvel(\psi) = \angvel^{(0)} - \frac{c}{Z_s e n_s} \frac{d}{d\psi} \left( n_s T_s \right) - \bm{u}_s\cdot\left( \nabla\psi\times\nabla\tor \right) \frac{I(\psi)}{|\nabla\psi|^2}.
		\label{LM-omegaFromExp}
}\end{eqalign}
\end{equation}
This expression is equivalent to the usual radial force balance relation used to find $\MeanE\dg\psi$ (see, e.g., \cite{meister2001measurement}).
Indeed, as we showed at the beginning of this appendix, in the low-Mach-number limit, $\angvel(\psi) = -c (\MeanE\dg\psi) / |\nabla\psi|^2 + \Or(\gkeps M^2 \vth / R)$ and so \eref{LM-omegaFromExp} is the radial force balance multiplied by $c/|\nabla\psi|$, with errors of order $\Or(\gkeps M^2 + \gkeps^2)$.
It is because of this connection between $d \angvel / d \psi$ and $\MeanE\dg\psi$, satisfied particularly accurately at low $M$, that the toroidal flow shear is often referred to as the $\bm{E}\times\bm{B}$ shear~\cite{hahm1995flowshear,waltz1998shear}.

\section{Derivations for \Sref{transport}}
\label{fluxes}
In this Appendix, we derive the explicit expressions \eref{partflux}, \eref{momflux}, and \eref{heatflux} for the mean fluxes in the transport equations of \Sref{transport}. 
We also provide detailed derivations of other results of that section that were stated there without proof.

Equations \eref{rpf}, \eref{rmf}, and \eref{rhf} express the radial fluxes as moments of
$
\infrac{\partial F_s}{\partial\gyr}
$.
This is calculated in terms of known quantities via \eref{hoc-main}:
\begin{equation}
\fl
\label{hoc}
\begin{eqalign}{
\cycfreq \pd{F_s}{\gyr} = -&\left(\bm{u}+\bm{w}\right)\dg F_s +\left[\frac{Z_s e}{m_s}\left( \nabla\pot_0+\frac{1}{c}\pd{\MeanA}{t}\right) + \left(\bm{u}+\bm{w}\right)\dg\bm{u}\right] \cdot\pd{F_s}{\bm{w}} \\
	& - \ensav{\daccel \cdot \pd{\delta f_s}{\bm{w}}} + \collop[F_s] + \Or(\gkeps^3 \cycfreq f_s).
}\end{eqalign}
\end{equation}
In what follows we will use this equation to find the explicit form of the fluxes in terms of $h_s$, $F_{0s}$ and $\Fneo$.
Note that we will use $\bm{v}$ as a shorthand for $\bm{u} + \bm{w}$, but derivatives are always taken at constant $\bm{w}$.
\subsection{The Particle Flux}
\label{pfluxes}
The radial particle flux is given by \eref{rpf}.
We multiply \eref{hoc} by $\binfrac{m_s c}{Z_s e}R^2\bm{v}\dg\tor$ and integrate over velocities to obtain
\begin{equation}
\label{pfluxtmp}
\fl
\begin{eqalign}{
 n_s \bm{U}_s&\dg\psi =  -\frac{c}{Z_s e}\dv\left[\left( \viscosity + m_s n_s\bm{U}_s\bm{u} + m_s n_s\bm{u}\bm{U}_s + m_s n_s \bm{u}\bm{u} \right)\cdot\left( \nabla\tor \right) R^2\right] \\
		&- n_s \pd{\psi}{t}
	+ \frac{c}{Z_s e}\left(\bm{F}_s\dg\tor\right) R^2 +\ensav{\wint{\frac{m_s c}{Z_s e} R^2\left(\daccel\dg\tor\right) \delta f_s}},
}\end{eqalign}
\end{equation}
where we have used \eref{lowdiv} and the identity \eref{divtor} to write the first term as a full divergence, used $\psi = R^2\MeanA\dg\tor$ to simplify the second term and used $\bm{u}\dg\pot_0 = 0$ (axisymmetry). The viscous stress tensor $\viscosity$ is defined by \eref{mflux} and 
\begin{equation}
\bm{F}_s = \wint m_s\bm{v} \collop[F_s]
\label{FrictionDef}
\end{equation}
is the collisional friction force.
Interpreting \eref{pfluxtmp}, we see that it relates the radial component of the particle flow velocity to the toroidal forces, so we can view the particle flux as composed of effective second-order drifts, driven by these forces. However, these drifts will only cause net transport of particles if they do not vanish when averaged over a flux surface, as indeed is obvious from the fact that we were able to average the particle transport equation over a flux surface.

So we average \eref{pfluxtmp} over a flux surface. 
The first term becomes
\begin{equation}
	\fl
\begin{eqalign}{
\frac{c}{Z_s e}&\fav{\dv\left[\left(\viscosity +m_s n_s \bm{U}_s \bm{u} + m_s n_s \bm{u}\bm{U}_s\right)\cdot\left(\nabla\tor\right)R^2\right]} =\\
&\qquad\qquad\frac{c}{Z_s e} \frac{1}{V'} \pd{}{\psi} \left[ V'\fav{\left(\nabla\psi\right)\cdot\viscosity\cdot\left(\nabla\tor\right) R^2 + m_s n_s \angvel(\psi)R^2\bm{U}_s\dg\psi}\right],
}\end{eqalign}
\label{blahtmpflux}
\end{equation}
where we have used \eref{favdiv} to express the flux-surface average of this divergence (recall also that $\bm{u} = \angvel(\psi)R^2\nabla\tor$).
From \eref{flow}, $m_s n_s \angvel(\psi)R^2\bm{U}_s\dg\psi = \Or(\gkeps^2 n_s T_s R |\nabla\psi|)$. From \eref{lowviscous}, the same is true for the off-diagonal viscous stress.
Thus, the right-hand side of \eref{blahtmpflux} is $\Or(\gkeps^3 n_s\vth |\nabla\psi|)$ and so the first term in the flux-surface average of \eref{pfluxtmp} can be dropped.
The second term does not simplify upon averaging.
In \ref{apF1}, we show that we can write the third term entirely in terms of
$F_{0s}$ and $\Fneo$ (see \eref{collpflux}) :
\begin{equation}
\label{tmpcollpflux}
\begin{eqalign}{
\frac{c}{Z_s e}\fav{ \left(\bm{F}_s\dg\tor\right)R^2} = \fav{\wint \left( \frac{\bm{w}\times\Meanb}{\cycfreq}\dg\psi \right) \collop[F_{0s}]}\\ 
	\qquad+ \fav{\wint \Fneo\vdrift\dg\psi }-\fav{n_s} I(\psi)  \frac{\fav{\MeanE\cdot\MeanB}}{\fav{\MeanMagB^2}}.
}\end{eqalign}
\end{equation}

The fourth and final (turbulent) term can be written as follows
\begin{equation}
\begin{eqalign}{
&\fav{\ensav{\wint{\frac{m_s c}{Z_s e} R^2\left( \daccel\dg\tor \right) \delta f_s}}}=\\
&\qquad- \fav{\frac{m_s c}{T_s}\ensav{\wint R^2\left(\daccel\dg\tor\right) \gkupot F_{0s}}}\\
&\qquad+\fav{\ensav{\wint{\frac{m_s c}{Z_s e} R^2\gyror{\left(\daccel\dg\tor\right) h_s}}}},
}\end{eqalign}
\label{D6}
\end{equation}
where we have split $\delta f_s$ into the Boltzmann response and the gyrokinetic distribution $h_s$ according to \eref{hDef}. Substituting from \eref{daccel} for $\daccel$, the first term in \eref{D6} becomes
\begin{equation}
\fl\begin{eqalign}{
&\frac{m_s c}{T_s}\ensav{\wint R^2\left(\daccel\dg\tor\right) \gkupot F_{0s}} \\
	&\quad\hspace{-3pt}= -\frac{Z_s e c}{T_s}R^2\left(\nabla\tor\right)\cdot\ensav{\wint \left[\nabla\left(\gkupot - \frac{1}{c}\delA\cdot\bm{w}\right)+ \frac{1}{c}\left(\bm{w}_\perp\dg\right)\delA\right] \gkupot F_{0s}} \\
		&\quad\hspace{-3pt}= -\frac{Z_s e c}{2T_s}R^2\left(\nabla\tor\right)\cdot\nabla\ensav{\gkupot^2} n_{s} = \Or(\gkeps^3 n_s\vth |\nabla\psi| ),
}\end{eqalign}
\label{boltzVanP}
\end{equation}
where we have used the fact that $\wint \bm{w} F_{0s} = 0$ to lowest order. The term in \eref{D6} containing $h_s$ can be manipulated as follows (again using \eref{daccel} for $\daccel$)
\begin{equation}
\fl
\begin{eqalign}{
&\fav{\ensav{\wint{\frac{m_s c}{Z_s e} R^2\gyror{\left(\daccel\dg\tor\right) h_s}}}}\\
	&\quad= -c\fav{\ensav{\wint{ R^2 \left(\nabla\tor\right)\cdot\gyror{\left[\nabla\gkpot + \frac{1}{c}\left(\bm{w}_\perp\dg\right)\delA\right] h_s}}}}\\
	&\quad= \fav{\frac{c}\MeanMagB\ensav{\wint{\gyror{h_s \left(\Meanb\times\nabla\gkpot\right)}} \dg\psi}} = \fav{\ensav{\wint \gyror{h_s\vchi}\dg\psi}},
}\end{eqalign}
\label{gyroavtmp}
\end{equation}
where we have used the axisymmetric form of the magnetic field \eref{magfield} to write $R^2 \left(\nabla\tor\right)\dg\gkpot = \left(\Meanb\times\nabla\gkpot\right)\cdot\left.\left(\nabla\psi\right)\right/\MeanMagB$. The term involving $\delA$ vanished because
\begin{equation}
\fl\begin{eqalign}{
\ensav{\gyror{h_s \left(\bm{w}_\perp\dg\right)\delA }} = \ensav{\gyror{ \dv\left(\bm{w}_\perp h_s \delA\right)}} - \ensav{\delA\gyror{ \left(\bm{w}_\perp\dg\right)h_s}}\\
\qquad= \dv\ensav{\gyror{\bm{w}_\perp h_s \delA}} + \ensav{\delA\gyror{\cycfreq \left.\pd{h}{\gyr}\right|_{\bm{r},w_\parallel,w_\perp}}} = \Or(\gkeps\cycfreq h_s \delA),
}\end{eqalign}
\label{interchange}
\end{equation}
where we have used the fact that $\delA$ is independent of gyrophase at constant $\bm{r}$ and also used the identity \eref{wperpdg1} in conjunction with the fact that $h_s$ is independent of gyrophase at constant $\bm{R}_s$.

Combining the above results we can write the flux-surface average of \eref{pfluxtmp} as
\begin{equation}
\fl
\begin{eqalign}{
\fav{n_s\bm{U}_s\dg\psi} = -&\fav{n_s\pd{\psi}{t}}+ \fav{\wint \left(\frac{\bm{w}\times\Meanb}{\cycfreq}\cdot \nabla\psi \right) \collop[F_{0s}]}\\
&+ \fav{\wint \Fneo\vdrift\dg\psi } - \fav{n_s}I(\psi) \frac{\fav{\MeanE\cdot\MeanB}}{\fav{\MeanMagB^2}}\\
	 &+\fav{\ensav{\wint \gyror{h_s \vchi}\dg \psi}}.
}\end{eqalign}
\label{pflux0}
\end{equation}
The first four terms are known from classical and neoclassical theory of collisional transport (compare with equation 37 of \cite{catto1987ion}) and the final term is the turbulent particle flux.
After substitution of \eref{pflux0} into \eref{GammaDef}, the terms containing $\infrac{\partial \psi}{\partial t}$ cancel and we obtain \eref{partflux}.
\subsection{Ambipolarity of The Particle Flux}
\label{ambipolar}
It is well known that in axisymmetric tokamaks the radial particle fluxes are ambipolar, i.e. there is no net radial current~\cite{kovrizhnikhAmbipolar,rutherfordAmbipolar,parra2009via,calvo2012long}. This
follows immediately from the $\nabla\psi$ component of the mean Amp\`ere's Law \eref{avgampall} averaged over a flux surface:
\begin{equation}
\fav{\Meanj \dg\psi} = \frac{c}{4\pi} \fav{\left( \curl\MeanB \right)\dg\psi} = \frac{c}{4\pi} \fav{\dv\left( \MeanB\times\nabla\psi \right)} = 0,
	\label{ambipolarcon}
\end{equation}
where we have used the axisymmetric form of the magnetic field, \eref{magfield}.

Since $\Meanj=\sum_s Z_s e n_s \bm{U}_s$, to prove that our fluxes are consistent with the constraint \eref{ambipolarcon}, we multiply \eref{pfluxtmp} by $Z_s e$ and sum over species (neglecting the first term because of \eref{blahtmpflux}):
\begin{equation}
\label{pfluxtmp2}
\begin{eqalign}{
\fav{\Meanj\dg\psi} =&
	\sum_s Z_s e\fav{n_s\pd{\psi}{t}} + \sum_s \fav{c \left(\bm{F}_s\dg\tor\right)R^2} \\
	&+ \sum_s m_s c\fav{\ensav{\wint R^2\left(\daccel\dg\tor\right)\delta f_s}}.
}\end{eqalign}
\end{equation}
The first term in this equation vanishes due to quasineutrality of the plasma and the second term also vanishes because collisions conserve momentum and hence the species-summed friction force must vanish. The third term is
\begin{equation}
\fl
\begin{eqalign}{
\sum_s m_s c\fav{\ensav{\wint R^2 \left(\daccel\dg\tor\right)\delta f_s}} =\\
\qquad\qquad	\sum_s Z_s ec\fav{R^2\ensav{ \delta n_s\delE\dg\tor}}
	+ \fav{R^2\ensav{\delj \times\delB}\dg\tor},
}\end{eqalign}
\label{ambitmp}
\end{equation}
where we have used the first line of the expression \eref{daccel} for $\daccel$. The first term in \eref{ambitmp} vanishes by quasineutrality and so the only contribution to the radial current is the fluctuating Lorentz torque. But this can be written as the divergence of the Maxwell stress:
\begin{equation}
\fl
\fav{R^2\ensav{\delj\times\delB}\cdot\left(\nabla\tor\right) } = \frac{1}{4\pi}\fav{\ensav{ \dv\left[R^2\left( \frac{1}{2}\MagDelB^2\idmat - \delB\delB \right)\dg\tor \right]}},
\end{equation}
whereupon we can interchange the divergence with the average over the fluctuations and the right-hand side becomes $\Or(\gkeps^3 e n_s \vth)$.

Thus, the flux-surface-averaged total radial current vanishes to second order, $\fav{\Meanj\dg\psi} \sim \Or(\gkeps^3 e n_s \vth|\nabla\psi|)$. To the next order, the ambipolarity condition becomes equivalent to the conservation of toroidal angular momentum: for a more detailed discussion of how these two effects are connected see \cite{parra2009via}.

\subsection{Toroidal Angular Momentum Flux}
\label{mfluxes}
The part of the radial angular momentum flux that we need to calculate is given by \eref{rmf}. Multiplying \eref{hoc} by $\binfrac{m_s^2 c}{2 Z_s e} \left(R^2\bm{v}\dg \tor\right)^2$ and integrating over $\bm{w}$ gives
\begin{equation}
\label{torflux}
\fl
\begin{eqalign}{
	&\left( \nabla\psi \right)\cdot\viscosity\cdot\left( \nabla\tor\right)R^2 + m_s n_s\angvel(\psi) R^2 \bm{U}_s\dg\psi =\\
		&-\dv\left[\wint \frac{m_s^2c}{2Z_s e} \left(R^2\bm{v}\dg\tor\right)^2 \bm{v} F_s\right] - m_sn_s\angvel(\psi)R^2 \pd{\psi}{t}\\
		&+\wint \frac{m_s^2c}{2Z_s e} \left(R^2\bm{v}\dg\tor\right)^2 \collop[F_s] + \ensav{\wint \frac{m_s^2c}{Z_s e} R^4 \left(\bm{v}\dg\tor\right) \left(\daccel\dg\tor\right)\delta f_s },
}\end{eqalign}
\end{equation}
where we have used axisymmetry and integrated by parts where opportune.

Completely analogously to the calculation of the particle flux in \ref{pfluxes}, we take the flux-surface average of \eref{torflux}. The first term becomes
\begin{equation}
\begin{eqalign}{
\fav{\dv\left[\wint \frac{cm_s^2}{2Z_s e} \left(R^2\bm{v}\dg\tor\right)^2 \bm{v} F_s\right]} \\
	\qquad\quad= \frac{1}{V'}\pd{}{\psi}\left[V'\fav{\wint \frac{cm_s^2}{2Z_s e} \left(R^2\bm{v}\dg\tor\right)^2 F_s\bm{w}_\perp\dg\psi }\right],
}\end{eqalign}
\end{equation}
where we have used \eref{favdiv} to simplify the flux-surface average of the divergence. From \eref{MagicTheorem}, we can see that we can neglect this term as it is $\Or(\gkeps^3 n_s T_s R |\nabla\psi|)$.
The second term in \eref{torflux} has to do with the motion of the flux surface and will be absorbed into the left-hand side via the definition \eref{GammaDef} of $\ParticleFlux$. 
We split the third (collisional) term as follows
\begin{equation}
  \begin{eqalign}{
\frac{c}{Z_s e} &\fav{\wint \frac{m_s^2}{2}  \left( R^2\bm{v}\dg\tor\right)^2 \collop[F_s]} = \\
	&\frac{c}{Z_s e} \fav{ \wint \frac{m_s^2}{2} \left( R^2\bm{w}\dg\tor\right)^2 \collop[F_s]} \\
&+\frac{c}{Z_s e} \fav{ \wint m_s^2 R^4 \angvel(\psi) \left( \bm{v}\dg\tor\right) \collop[F_s]},
}\end{eqalign}
\label{tmpE13}
\end{equation}
where we have used the fact that the full Landau collision operator conserves particles and so $\wint \left(R^2\bm{u}\dg\tor\right)^2 \collop[F_s] = 0$.
The integrals in the right hand side of \eref{tmpE13} are calculated in \ref{apF}: using \eref{collvstress} and \eref{collconmflux}, we find
\begin{equation}
\fl\begin{eqalign}{
\frac{c}{Z_s e} &\fav{\wint \frac{1}{2} m_s^2 \left( R^2\bm{v}\dg\tor\right)^2 \collop[F_s]} = \fav{\ClassMomFlux}+ \fav{\NeoMomFlux}\\
	&+ m_s\angvel(\psi)\fav{R^2\wint \left(\frac{\bm{w}\times\Meanb}{\cycfreq}\cdot\nabla\psi\right) \collop[F_{0s}]}\\
	&+ m_s\angvel(\psi)\wint \fav{w_\parallel \Fneo \Meanb\dg\left[ R^2\frac{I(\psi)w_\parallel}{\cycfreq} +  R^2\frac{\MeanMagB R^2 \angvel(\psi)}{\cycfreq}\right]} \\
&- m_s\angvel(\psi)\fav{R^2n_s}I(\psi)\frac{\fav{\MeanE\cdot\MeanB}}{\fav{\MeanMagB^2}}.
}\end{eqalign}
\label{apDcollmflux}
\end{equation}
where $\fav{\ClassMomFlux}$ and $\fav{\NeoMomFlux}$ are given by \eref{classvisc} and \eref{neovisc} respectively.

Finally, we split the term involving the fluctuations as follows
\begin{equation}
\fl
\begin{eqalign}{
\ensav{\wint \frac{c m_s^2}{Z_s e} R^4 \left(\bm{v}\dg\tor\right) \left(\daccel\dg\tor\right) \delta f_s } =\\
	\quad\,-\ensav{\wint \frac{cm_s^2}{T_s} \left( \bm{v}\dg\tor \right)\left(\daccel\dg\tor\right)\gkupot F_{0s}}\\
	\quad\,-\ensav{\wint cm_s R^4 \gyror{\left(\bm{v}\dg\tor\right)\left[ \nabla\gkpot + \frac{1}{c}\left( \bm{w}_\perp\dg \right)\delA \right]\cdot\left(\nabla\tor\right) h_s} },
}\end{eqalign}
\label{tmpD17}
\end{equation}
where we have used \eref{hDef} for $\delta f_s$ and \eref{daccel} for $\daccel$.
The first term in \eref{tmpD17}, due to the Boltzmann response, becomes
\begin{equation}
\fl
\begin{eqalign}{
-\ensav{\wint \frac{cm_s^2}{T_s} \left( \bm{v}\dg\tor \right)\left(\daccel\dg\tor\right)\gkupot F_{0s}} =\\
		\qquad \ensav{\wint \frac{Z_s e c m_s}{T_s} \left( \bm{v} \dg\tor\right) \left( \nabla\tor \right)\cdot\left( \nabla \gkupot\right) \gkupot F_{0s}}\\
		\qquad +\ensav{\wint \frac{Z_s e m_s}{T_s} \left( \bm{v}\dg\tor \right)\left[ \left( \nabla\delA \right)\cdot\bm{w} - \bm{w}\cdot\nabla\delA \right]\cdot\left( \nabla\tor \right)\gkupot F_{0s}},
}\end{eqalign}
\end{equation}
where we have used the third line of \eref{daccel}.
The first term in this expression is small by the same logic as in \eref{boltzVanP} and the second term vanishes because to lowest order, $\wint\,\bm{w} F_{0s} = 0$ and $\wint \,m_s\bm{w}\bm{w} F_{0s} = n_s T_s \idmat$.
As in the calculation of the particle flux, the term in \eref{tmpD17} involving $h_s$ and $\nabla\gkpot$ can be written in terms of $\vchi$:
\begin{equation}
\begin{eqalign}{
-&\ensav{\wint cm_s R^4 \gyror{\left(\bm{v}\dg\tor\right)\left( \nabla\gkpot \right)\cdot\left(\nabla\tor\right) h_s} } = \\
&\qquad {\ensav{\wint m_s R^2 \gyror{ \left(\bm{w}\dg\tor\right) h_s\vchi}\dg \psi }}\\
&\qquad\qquad+m_s\angvel(\psi){R^2\ensav{\wint  \gyror{h_s\vchi}\dg \psi }}.
}\end{eqalign}
\end{equation}
The final term in \eref{tmpD17} (involving $\delA$) simplifies to
\begin{equation}
\fl\begin{eqalign}{
-\ensav{\wint{m_s R^4\gyror{\left(\bm{v}\dg\tor\right) \left(\bm{w}_\perp\dg\delA\right)\cdot\left(\nabla\tor\right) h_s}}}=\\
		\quad -\frac{Z_s e}{c}\ensav{\wint  \MeanMagB R^4\left(\delA\dg\tor\right)\gyror{\left(\bm{v}\dg\tor\right)\left.\pd{h_s}{\gyr}\right|_{\bm{r},w_\parallel,w_\perp}}}=\\
		\quad -\frac{Z_s e}{c}\ensav{\wint R^2 \left(\delA\cdot\nabla\tor\right)\gyror{h_s\bm{w}_\perp}}\dg\psi,
}\end{eqalign}
\label{bypartsbaby}
\end{equation}
where the first equality followed from a manipulation analogous to that in \eref{interchange} and the second equality from integrating by parts with respect to $\gyr$ and using  $\MeanMagB R^2\left( \infrac{\partial \bm{v}}{\partial\gyr} \right)\dg\tor =-\bm{w}_\perp \dg\psi$ (which follows from \eref{magfield} and \eref{wDef}).

Collecting the above results, the flux-surface-average of \eref{torflux} becomes
\begin{equation}
\fl
\label{pifinalap}
\begin{eqalign}{
&\fav{\left( \nabla\psi \right)\cdot\viscosity\cdot\left( \nabla\tor \right)R^2 + m_s\angvel(\psi) R^2\ParticleFlux} = \fav{\ClassMomFlux} + \fav{\NeoMomFlux}\\
&\quad + m_s\angvel(\psi)\fav{R^2\wint \left(\frac{\bm{w}\times\Meanb}{\cycfreq}\cdot\nabla\psi\right) \collop[F_{0s}]} \\
&\quad + m_s\angvel(\psi)\wint \fav{w_\parallel \Fneo \Meanb\dg\left[ R^2\frac{I(\psi)w_\parallel}{\cycfreq} + R^2\frac{\MeanMagB R^2 \angvel(\psi)}{\cycfreq}  \right]}\\
&\quad- m_s\angvel(\psi)\fav{R^2n_s}I(\psi)\frac{\fav{\MeanE\cdot\MeanB}}{\fav{\MeanMagB^2}}\\
&\quad- \fav{\ensav{\wint \frac{Z_s e}{c}R^2 \left(\delA\dg\tor\right)\gyror{h_s\bm{w}}\dg\psi }}\\
&\quad+\fav{\ensav{\wint m_s R^2 \gyror{ \left(\bm{w}\dg\tor\right) h_s\vchi}\dg \psi }}\\
&\quad+m_s\angvel(\psi)\fav{R^2\ensav{\wint  \gyror{h_s\vchi}\dg \psi }}.
}\end{eqalign}
\end{equation}
This will lead to the final result of \Sref{momtrans} if we split the above expression into the viscous stress $\MomentumFlux$, as defined by \eref{momflux}, and the convective momentum flux $m_s \angvel(\psi) \fav{R^2\ParticleFlux}$ with $\fav{R^2 \ParticleFlux}$ given by \eref{conMomflux}:
\begin{equation}
\fl\begin{eqalign}{
\fav{\left( \nabla\psi \right)\cdot\viscosity\cdot\left( \nabla\tor \right)R^2 + m_s\angvel(\psi) R^2\ParticleFlux} = 
	\fav{\MomentumFlux} + m_s \angvel(\psi) \fav{R^2\ParticleFlux} \\
\quad- \fav{\left( \nabla\psi\right)\cdot\ensav{\wint  \gyror{h_s\bm{w}}\frac{Z_s e}{c}\delA }\cdot\left( \nabla\tor\right)R^2}.
}\end{eqalign}
\label{pipsifinalap}
\end{equation}
Inserting this into \eref{TMFdef} then gives \eref{TMFresult}, where the third term in \eref{pipsifinalap} has been grouped together with the Maxwell stress to give the electromagnetic stresses $\EMViscosity$, defined by \eref{EMViscDef}.

\subsection{Derivation of the Pressure Evolution Equation \eref{dpdt}}
\label{dpdtderiv}
In this Appendix, we simplify \eref{1moment} and take its flux-surface average to derive \eref{dpdt}.
\Eref{1moment} reads
\begin{equation}
\label{ap1moment}
\fl
\begin{eqalign}{
	\frac{3}{2}\pd{}{t}n_s T_s +  &\dv \bm{Q}_s +  Z_s e\left(\nabla\pot_0+\frac{1}{c} \pd{\MeanA}{t}\right) \cdot \left(n_s\bm{U}_s\right) + m_s n_s \bm{u}\cdot\left( \nabla\bm{u} \right)\cdot\bm{U}_s \\
		&+ \viscosity\bm{:}\nabla\bm{u}- \ensav{\wint m_s\bm{w}\cdot \daccel \delta f_s} = \CollEnergy + \Esource.
}\end{eqalign}
\end{equation}
First, denoting the potential energy of a particle $\Xi_s = Z_s e \pot_0 - m_s \angvel^2(\psi)R^2 / 2$, we observe that
\begin{equation}
\fl
\begin{eqalign}{
\dv\left(\bm{Q}_s + \Xi_s n_s \bm{U}_s\right) = \dv\bm{Q}_s + n_s\bm{U}_s\dg\Xi_s - \Xi_s\left( \pd{n_s}{t} -\Psource \right)  \\
	= \dv\bm{Q}_s + n_s\bm{U}_s\cdot\left( Z_s e \nabla\pot_0 + m_s\bm{u}\dg\bm{u}\right) - R^2\angvel(\psi) \left( n_s\bm{U}_s\dg\psi \right) \frac{d\angvel}{d\psi}\\
	\qquad -\Xi_s\left( \pd{n_s}{t} -\Psource \right),
}\end{eqalign}
\end{equation}
where we have used the continuity equation \eref{0moment} to substitute for $\dv\left(n_s\bm{U}_s\right)$. 
Adding this to \eref{ap1moment} and cancelling terms where possible gives
\begin{equation}
\fl
\begin{eqalign}{
&\frac{3}{2}\pd{}{t}n_s T_s +  \dv\left( \bm{Q}_s +\Xi_s n_s\bm{U}_s\right) =- \left[R^2\left( \nabla\tor \right)\cdot\viscosity + m_s R^2\angvel(\psi) n_s\bm{U}_s\right]\cdot\left( \nabla\psi \right) \frac{d\angvel}{d\psi}   \\
&\quad- \frac{Z_s e}{c} \pd{\MeanA}{t} \cdot \left( n_s\bm{U}_s \right)-\Xi_s \left( \pd{n_s}{t} - \Psource \right)+ \CollEnergy+ \Esource\\
&\quad+ \ensav{\wint m_s\bm{w}\cdot\daccel \delta f_s}.
}\end{eqalign}
\end{equation}
Before we work on simplifying the right-hand side, it is convenient to subtract $\dv\left[\left(n_s T_s + Z_s e \pot_0\right)\vpsi\right]$, where $\vpsi$ is the velocity of the flux surfaces given by \eref{VpsiDef}, from both sides of this equation.
Then our pressure evolution equation becomes
\begin{equation}
\label{1momap}
\begin{eqalign}{
&\frac{3}{2}\pd{}{t}n_s T_s +  \dv\left[ \bm{Q}_s +\Xi_s n_s\bm{U}_s -\left( n_s T_s + Z_s e\pot_0 \right)\vpsi\right] = \\
&\quad- \left[R^2\left( \nabla\tor \right)\cdot\viscosity + m_s R^2\angvel(\psi) n_s\bm{U}_s\right]\cdot\left( \nabla\psi \right) \frac{d\angvel}{d\psi}   \\
&\quad- \frac{Z_s e}{c} \pd{\MeanA}{t} \cdot \left( n_s\bm{U}_s \right)-\Xi_s \left( \pd{n_s}{t} - \Psource \right)+ \CollEnergy+ \Esource\\
&\quad- \dv\left[\left(n_s T_s + Z_s e \pot_0\right)\vpsi\right] + \ensav{\wint m_s\bm{w}\cdot\daccel \delta f_s}.
}\end{eqalign}
\end{equation}

\subsubsection{Turbulent Heating.}
\label{C4TurbHeat}
Let us first work out the turbulent contribution to \eref{1momap} -- the last term on the right-hand side.
Using the third line of \eref{daccel} to express $\daccel$, we can write it as follows
\begin{equation}
\label{fluct-heat-tmp-1}
\fl
\begin{eqalign}{
\ensav{\wint m_s \bm{w}\cdot\daccel \delta f_s} = 
	- Z_s e\ensav{\wint \left(\bm{w}\dg\gkupot\right) \delta f_s} \\
	\qquad\qquad
	- \frac{Z_s e}{c}\ensav{\wint \delta f_s\left(\bm{w}\cdot\nabla\bm{u}\right)\cdot\delA + \wint\delta f_s\left( \pd{}{t} + \bm{u}\dg \right) \bm{w}\cdot\delA}.
}\end{eqalign}
\end{equation}
The first term of \eref{fluct-heat-tmp-1} can be rewritten by observing that
\begin{equation}
\fl
	\label{tmp-2}
	\begin{eqalign}{
&\ensav{\wint \left(\bm{w}\cdot   \nabla \gkupot \right)\delta f_s}=\\
&\qquad \ensav{\wint\dv \left( \bm{w} \gkupot \delta f_s \right) - \wint \gkupot w_\parallel \Meanb\dg \delta f_s -\wint  \gkupot \bm{w}_\perp \dg \delta f_s}.
}\end{eqalign}
\end{equation}
Estimating the size of terms in this expression, we notice that we require $\delta f_s$ correct to $\Or(\gkeps^2 f_s)$ only in the last term, as the first term is an exact divergence and the second contains a small parallel derivative so in them we only need the first-order part of $\delta f_s$ given by \eref{hDef}.
We evaluate the last term by using \eref{fluct-main}: multiplying it by $\gkupot$ and integrating over all velocities, we find
\begin{equation}
\fl\begin{eqalign}{
	\wint \gkupot \bm{w}_\perp\dg\delta f_s = - \wint \gkupot \left( \pd{ }{t} + \bm{u}\dg \right) \delta f_s - \wint \gkupot w_\parallel \Meanb\dg\delta f_s,
}\end{eqalign}
\label{tmpD29}
\end{equation}
where on the right-hand side we now only require $\delta f_s$ correct to $\Or(\gkeps f_s)$ to calculate the left-hand side to $\Or(\gkeps^3 \cycfreq n_s T_s / e)$. Substituting \eref{tmpD29} into \eref{tmp-2} and then \eref{tmp-2} into \eref{fluct-heat-tmp-1} gives
\begin{equation}
\fl
\begin{eqalign}{
&\ensav{\wint m_s \bm{w}\cdot\daccel \delta f_s} = \\
&\quad -Z_s e \ensav{\wint \dv\left( \bm{w} \gkupot \delta f_s \right) + \frac{1}{c}\delta f_s \bm{w}\cdot\left( \nabla\bm{u} \right)\cdot\delA - \delta f_s\left( \pd{}{t} + \bm{u}\dg \right) \gkpot},
}\end{eqalign}
\label{tmpD31}
\end{equation}
where we have used $\gkpot = \gkupot - \bm{w}\cdot\delA/c$ and the fact that $\left(\infrac{\partial}{\partial t} + \bm{u}\dg \right) \ensav{ \gkpot \delta f_s } = \Or(\gkeps^3 \cycfreq \gkpot \delta f_s )$, so it can be neglected as it is the convective derivative of a mean quantity. We expand $\delta f_s$ in the right-hand side of \eref{tmpD31} by using the decomposition \eref{hDef} of $\delta f_s$ into the Boltzmann response and the gyrokinetic distribution function:
\begin{equation}
\begin{eqalign}{
\ensav{\wint m_s \bm{w}\cdot\daccel \delta f_s} = \\
	\quad -Z_s e \ensav{\wint \dv\left( \gkupot\gyror{\bm{w}h_s} \right) + \frac{1}{c}\gyror{h_s\bm{w}}\cdot\left( \nabla\bm{u} \right)\cdot\delA}\\
	\quad + Z_s e\ensav{\wint \gyror{h_s\left( \pd{}{t} + \bm{u}\dg \right) \gkpot}},
}\end{eqalign}
\label{tmpD32}
\end{equation}
where we have used $\wint \bm{w}F_{0s} = 0$ and the fact that $\left(\infrac{\partial}{\partial t} + \bm{u}\dg\right) \ensav{\gkupot^2} =\Or(\gkeps^3 \cycfreq \gkupot^2)$ to eliminate the terms arising from the Boltzmann response.
Next, we use \eref{nablaU} and \eref{aCrossZ} to work out the term in \eref{tmpD32} involving $\nabla\bm{u}$ and find the following expression for the turbulent contribution to \eref{1momap}:
\begin{equation}
\fl
\begin{eqalign}{
\label{fluct-heat-result}
&\ensav{\wint m_s \left(\bm{w}\cdot\daccel\right)\delta f_s} =
	-Z_s e \dv\ensav{\wint \left( \gkupot\gyror{\bm{w}h_s} \right)}\\
&	\quad- \frac{Z_s e}{c}\angvel(\psi)\ensav{\wint\gyror{h_s\bm{w}}\cdot\left( \delA\times\nabla z\right)}\\
&	\quad- \frac{Z_s e}{c}{\ensav{\wint R^2 \left(\delA\dg\tor\right)\gyror{h_s\bm{w}}\dg\psi }}\frac{d\angvel}{d\psi}\\
&	\quad+ Z_s e\ensav{\wint \gyror{h_s\left( \pd{}{t} + \bm{u}\dg \right) \gkpot}}.
}\end{eqalign}
\end{equation}
Averaging this over a flux surface, we have
\begin{equation}
\begin{eqalign}{
\label{fav-fluct-heat-result}
&\fav{\ensav{\wint m_s \left(\bm{w}\cdot\daccel\right)\delta f_s}} =\\
	&- \frac{1}{V'}\pd{ }{\psi} V'\fav{\ensav{\wint Z_s e\gkupot\gyror{h_s\bm{w}}\dg\psi}} + \TurbPow\\
&	\quad- \frac{Z_s e}{c}\fav{\ensav{\wint R^2 \left(\delA\dg\tor\right)\gyror{h_s\bm{w}}\dg\psi }}\frac{d\angvel}{d\psi},\\
}\end{eqalign}
\end{equation}
where we have used \eref{favdiv} for the flux-surface average of a divergence and the definition \eref{TurbPowDef} of $\TurbPow$.
When we average \eref{1momap} over a flux surface in \ref{ApFavdpdpt} we will collect the first term in \eref{fav-fluct-heat-result} together with all other divergences to form the heat flux and the final term will cancel with part of the turbulent viscous heating, as we will show momentarily.

%When we substitute this back into \eref{1momap}, in \ref{ApFavdpdpt}, we will move the first term to the left-hand-side of the equation and group it with all the other divergences in the expression for the heat flux \eref{main-hflux-full}, the second and fourth terms comprise the turbulent heting, and the third term will be grouped with all other terms proportional to ${d\angvel}/{d\psi}$ as the turbulent part of the viscous heating \eref{VHDef}.

\subsubsection{Viscous Heating.}
\label{C4ViscHeat}
Next, let us handle the viscous heating -- the first term on the right-hand side of \eref{1momap}. Flux-surface averaging this term, we have
\begin{equation}
\fl
\begin{eqalign}{
- \fav{\left[R^2\left( \nabla\tor \right)\cdot\viscosity + m_s R^2\angvel(\psi) n_s\bm{U}_s\right]\cdot\left( \nabla\psi \right) }\frac{d\angvel}{d\psi} \\
\qquad= -\left[\fav{\left( \nabla\psi \right)\cdot\viscosity\cdot\left( \nabla\tor \right)R^2} + m_s\angvel(\psi)\fav{R^2 \ParticleFlux}\right] \frac{d\angvel}{d\psi}\\
\qquad\quad+ m_s\angvel(\psi)\fav{R^2 n_s \pd{\psi}{t}}\frac{d\angvel}{d\psi}\\
\qquad= \ViscousHeat + \fav{\left( \nabla\psi \right)\cdot\ensav{\wint\gyror{h_s\bm{w}} \frac{Z_se}{c} \delA}\cdot\left( \nabla\tor \right) R^2}\frac{d\angvel}{d\psi}\\
\qquad\quad+ m_s\angvel(\psi)\fav{R^2 n_s \pd{\psi}{t}}\frac{d\angvel}{d\psi},\\
}\end{eqalign}
\label{fav-vischeat-result}
\end{equation}
where we have used the definition \eref{GammaDef} of $\ParticleFlux$, then substituted for the first two terms inside the brackets from \eref{pipsifinalap}, and finally introduced $\ViscousHeat$ as defined by \eref{VHDef}.
The term involving $\delA$ is precisely the same as the last term of \eref{fav-fluct-heat-result} with the opposite sign, as promised at the end of \ref{C4TurbHeat}. The last term in \eref{fav-vischeat-result} will cancel a similar term appearing in the expression for the potential energy exchange in \ref{C4PotHeat}.
\subsubsection{Ohmic Heating.}
\label{C4OhmHeat}
We now deal with the term in \eref{1momap} involving ${\partial\bm{A}}/{\partial t}$. Using \eref{flow} to express $n_s \bm{U}_s$ and $\bm{A}\dg\tor = \psi / R^2$ (see \eref{psiDef}), we have
\begin{equation}
\begin{eqalign}{
- \frac{Z_s e}{c} &\pd{\MeanA}{t} \cdot \left( n_s\bm{U}_s \right) = \\
		&- \pd{\psi}{t} n_s \left[ T_s \frac{d\ln \NotN}{d\psi} + \left(\Xi_s + T_s\right)\frac{d\ln T_s}{d\psi} + m_s R^2 \angvel(\psi)\frac{d\angvel}{d\psi}\right]\ \\
		&\quad + K_s(\psi) \left(\MeanE+\nabla\pot\right)\cdot\MeanB,
}\end{eqalign}
\end{equation}
where we have used the definition of $\MeanE$ in terms of $\bm{A}$ and $\pot$.
Taking the flux-surface average of this equation, we find that
\begin{equation}
\fl
\begin{eqalign}{
- \frac{Z_s e}{c} &\fav{\pd{\MeanA}{t} \cdot \left( n_s\bm{U}_s \right)} = \\
		&- \fav{\pd{\psi}{t} n_s \left[ T_s \frac{d\ln \NotN}{d\psi} + \left(\Xi_s + T_s\right)\frac{d\ln T_s}{d\psi} + m_s R^2 \angvel(\psi)\frac{d\angvel}{d\psi}\right]}
		+ \JouleHeat,
}\end{eqalign}
\label{ohm-heat-result}
\end{equation}
where we have used \eref{annhilator} to show that $\fav{\bm{B}\dg\pot} =0$ and $\JouleHeat$ is defined by \eref{OhmHeat}.
We will show, momentarily, that the terms involving $\partial\psi/\partial t$ cancel with identical terms in the expression for the compressional heating.

\subsubsection{Compressional Heating Due to Motion of Flux Surfaces.}
\label{C4CompHeat}
We simplify the compressional heating due to $\vpsi$, i.e., the $\dv[( n_s T_s + Z_s e\pot_0)\vpsi]$ term in \eref{1momap}, as follows. Using \eref{npol} to express $n_s$, we take the divergence explicitly to find
\begin{equation}
\begin{eqalign}{
-\dv&{\left[\left( n_s T_s + Z_s e n_s \pot_0 \right)\vpsi\right]} = -{n_s T_s\dv\vpsi} \\
		&\quad+{\pd{\psi}{t}n_s \left[ T_s \frac{d\ln \NotN}{d\psi} + \left(\Xi_s + T_s\right)\frac{d\ln T_s}{d\psi} + m_s R^2 \angvel(\psi)\frac{d\angvel}{d\psi}\right] }\\
		&\quad-{ Z_s e \pot_0 \left(\vpsi\dg n_s + n_s\dv\vpsi\right) - \frac{1}{2} m_s n_s \angvel^2(\psi) \vpsi\dg R^2}.
}\end{eqalign}
\end{equation}
where we have used $\vpsi\dg\psi = -\partial \psi / \partial t$.
We now average this over a flux surface and use the definition \eref{CompHeatDef} of $\CompHeat$ to obtain
\begin{equation}
\fl
\begin{eqalign}{
-&\fav{\dv\left[\left( n_s T_s + Z_s e n_s \pot_0 \right)\vpsi\right]} = \CompHeat \\
		&\quad+\fav{\pd{\psi}{t}n_s \left[ T_s \frac{d\ln \NotN}{d\psi} + \left(\Xi_s + T_s\right)\frac{d\ln T_s}{d\psi} + m_s R^2 \angvel(\psi)\frac{d\angvel}{d\psi}\right] }\\
		&\quad-\fav{ Z_s e \pot_0 \left(\vpsi\dg n_s + n_s\dv\vpsi\right)} - \frac{1}{2} m_s \angvel^2(\psi)\fav{n_s \vpsi\dg R^2}.
}\end{eqalign}
\label{comp-heat-result}
\end{equation}
When this is substituted back into \eref{1momap} in \ref{ApFavdpdpt}, the second line will cancel exactly with the similar term in \eref{ohm-heat-result} and the last line will cancel with terms from $\PotEng$ -- as we shall see in the next section.

\subsubsection{Heating Due to Exchange between Potential and Thermal Energy.}
\label{C4PotHeat}
We now handle the terms on the right-hand side of \eref{1momap} involving $\Xi_s$. Averaging them over a flux surface, we have
\begin{equation}
\fl
-\fav{\Xi_s\left( \pd{n_s}{t} - \Psource \right)} =
-\fav{\left( Z_s e\pot_0 - \frac{1}{2} m_s \angvel^2(\psi) R^2 \right)\left(\pd{n_s}{t} - \Psource\right)}.
\label{C4PotHeat-1}
\end{equation}
We wish to relate this to $\PotEng$ defined by \eref{PotEngDef}. Using \eref{movingflux}, we can show that
\begin{equation}
\fl\begin{eqalign}{
&\fav{\frac{1}{2}m_s \angvel^2(\psi)R^2\pd{n_s}{t}} \\
	&\quad= \frac{1}{2}m_s \angvel^2(\psi) \left(\frac{1}{V'}\ddtpsi V'\fav{R^2 n_s} + \frac{1}{V'}\pd{}{\psi} V' \fav{R^2 n_s\pd{\psi}{t}}\right)\\ 
&\quad= \frac{\angvel^2(\psi)}{2V'} \ddtpsi V'm_s \fav{R^2 n_s} + \frac{1}{V'} \pd{ }{\psi} V'\frac{1}{2} m_s \angvel^2(\psi) \fav{R^2 n_s \pd{\psi}{t}}\\
&\qquad- m_s\angvel(\psi) \fav{R^2n_s  \pd{\psi}{t} }\frac{d\angvel}{d\psi}.
}
\end{eqalign}
\label{tmptrick}
\end{equation}
Substituting \eref{tmptrick} back into \eref{C4PotHeat-1}, we have
\begin{equation}
\fl
\begin{eqalign}{
-\fav{\Xi_s\left( \pd{n_s}{t} - \Psource \right)} =
-\fav{ Z_s e\pot_0 \pd{n_s}{t} } + \frac{\angvel^2(\psi)}{2V'} \ddtpsi V'm_s\fav{R^2 n_s}\\
		\qquad+ \frac{1}{V'} \pd{ }{\psi} V'\frac{1}{2}m_s \angvel^2(\psi)\fav{R^2 n_s \pd{\psi}{t}} 
		-  m_s \angvel(\psi)\fav{ R^2 n_s \pd{\psi}{t}}\frac{d\angvel}{d\psi}\\
		\qquad +\fav{\left(Z_s e \pot_0- \frac{1}{2} m_s \angvel^2(\psi)R^2 \right)\Psource}.
}\end{eqalign}
\label{C4PotHeat-2}
\end{equation}
Using \eref{PotEngDef} for $\PotEng$, we obtain
\begin{equation}
\fl
\begin{eqalign}{
-\fav{\Xi_s\left( \pd{n_s}{t} - \Psource \right)} =
\PotEng + \frac{1}{V'} \pd{ }{\psi} V'\frac{1}{2}m_s \angvel^2(\psi)\fav{R^2 n_s \pd{\psi}{t}} \\
\qquad+\fav{ Z_s e \pot_0 \left(\vpsi\dg n_s + n_s\dv\vpsi\right)} + \frac{1}{2}m_s \angvel^2(\psi)\fav{n_s  \vpsi\dg R^2}\\
\qquad-  m_s \angvel(\psi)\fav{ R^2 n_s \pd{\psi}{t}}\frac{d\angvel}{d\psi}.
}\end{eqalign}
\label{C4PotResult}
\end{equation}
When we substitute this into the flux-surface average of \eref{1momap}, the last term will cancel with the corresponding term in \eref{fav-vischeat-result} and the entire second line of \eref{C4PotResult} will cancel with the third line of \eref{comp-heat-result}.
The second term \eref{C4PotResult} will be absorbed into the definition \eref{main-hflux-full} of the heat flux.
\subsubsection{The Flux-Surface-Averaged Pressure Evolution Equation.}
\label{ApFavdpdpt}
Taking the flux-surface average of \eref{1momap} and using \eref{favdiv}, \eref{movingflux}, \eref{fav-fluct-heat-result}, \eref{fav-vischeat-result}, \eref{ohm-heat-result}, \eref{comp-heat-result} and \eref{C4PotResult}, we obtain \eref{dpdt}. In \eref{dpdt}, we have grouped the divergences from \eref{1momap} together with the first term in \eref{fav-fluct-heat-result} and the second term in \eref{C4PotResult} to get the divergence of the total heat flux as defined by \eref{main-hflux-full}.
The explicit expression \eref{heatflux} for the heat flux is calculated in the next Appendix.

\subsection{Derivation of \eref{heatflux}: The Heat Flux}
\label{hfluxes}
The part of the heat flux that remains to be calculated is the first term in \eref{main-hflux-full}, ``the heat flux proper''.
By exactly the same procedure as the one employed to derive \eref{rpf} and \eref{rmf} for the particle and momentum fluxes, we can write
\begin{eqnarray}
\label{rhf}
\bm{Q}_s\dg\psi  &= \wint \frac{1}{2} m_sw^2 R^2 B\left(\bm{v}\dg\tor\right) \pd{F_s}{\gyr}.
\end{eqnarray}
In order calculate this, we multiply \eref{hoc} by $\binfrac{m_s^2 c}{2 Z_s e} w^2 R^2 \bm{v}\dg\tor$, integrate over all velocities, and find
\begin{equation}
\fl
\label{hfluxtmp}
\begin{eqalign}{
  \bm{Q}_s&\dg\psi = -\frac{m_s c}{Z_s e}\left[\dv\left(\tensor{H}_s+ \bm{u}\bm{Q}_s\right)\right]\cdot\left(\nabla\tor\right) R^2 - \frac{m_s c}{Z_s e}\angvel(\psi)R^2\dv\bm{Q}_s \\
	 &-{m_s c}R^2\left(\nabla\tor\right)\cdot\left(\viscosity+m_s n_s\bm{u} \bm{U}_s\right)\cdot\nabla\pot_0- \frac{5}{2} {m_s n_s T_s} \pd{\psi}{t}\\
	 &-\frac{m_s c}{Z_s e}R^2\left(\nabla\tor\right)\cdot\left(\viscosity+m_s n_s\bm{u} \bm{U}_s\right)\cdot\left(\bm{u}\dg{\bm{u}}  \right)
	 -\frac{m_s c}{Z_s e}\left(\bm{Q}_s\cdot\nabla\bm{u}\right)\cdot\left(\nabla\tor\right)R^2\\
	 &+\frac{c}{Z_s e}\left(\bm{G}_s\cdot\nabla\tor\right) R^2 
	 + \ensav{\wint \frac{m_s^2 c}{2Z_s e}  w^2R^2 \left(\daccel\dg\tor\right)  \delta f_s} \\
	 &+ \ensav{\wint \frac{m_s^2 c}{Z_s e} R^2\left(\bm{v}\dg\tor\right)\bm{w}\cdot\daccel \delta f_s },
}\end{eqalign}
\end{equation}
where 
\begin{equation}
\tensor{H}_s = \wint \frac{m_s w^2}{2} \bm{w}\bm{w} F_s
\end{equation}
is the ``kinetic-energy-weighted stress tensor'' and 
\begin{equation}
\bm{G}_s = \frac{1}{2}\wint m_s^2w^2 \bm{v} \collop[F_s]
\label{collHeatFricDef}
\end{equation}
is the ``collisional heat friction.''
In \eref{hfluxtmp}, we have dropped the following term:
\begin{equation}
\begin{eqalign}{
-\wint &\frac{m_s^2c}{Z_s e} R^2\left(\bm{v}\dg\tor\right) \left(\bm{w}\cdot\nabla\bm{u}\right)\cdot\bm{w} F_s \\
	 &=-\frac{m_s^2 c}{Z_s e} R^4 \frac{d\angvel}{d\psi} \wint \left(\bm{w}\dg\tor\right)\left(\bm{v}\dg\tor\right) \left(\bm{w}_\perp\dg\psi\right) F_s,
}\end{eqalign}
\end{equation}
which is $\Or(\gkeps^3 n_s \vth T_s |\nabla\psi|)$ by virtue of \eref{MagicTheorem} and, therefore, negligible.

Since, from \eref{daccel}, it follows that
$\binfrac{m_s}{Z_s e}\bm{w}\cdot\daccel = - \bm{w}_\perp\dg\gkupot + \Or\left(\binfrac\MeanMagB{c}\gkeps^2 \vth^2\right)$
the last term of \eref{hfluxtmp} is
\begin{equation}
\begin{eqalign}{
\frac{m_s^2 c}{Z_s e}& R^2\ensav{\wint \left(\bm{v}\dg\tor\right)\bm{w}\cdot\daccel \delta f_s}  \\
 &= -m_s c R^2\ensav{\wint  \gyror{\left(\bm{v}\dg\tor\right)\left(\bm{w}_\perp\cdot\nabla\gkupot\right) h_s}} \\
 &\quad +\frac{Z_s e m_s c}{2T_s} R^2 \wint\left(\bm{v}\dg\tor\right)\bm{w}_\perp\dg\ensav{\gkupot^2}F_{0s} \\\
 &= -Z_se \ensav{\wint \gkupot \gyror{h_s \bm{w}_\perp}}\dg\psi + \Or(\gkeps^3 n_s T_s\vth |\nabla\psi|),
}\end{eqalign}
\end{equation}
where we have used the decomposition \eref{hDef} of $\delta f_s$ and the second equality is derived in the same way as \eref{bypartsbaby}.
Using this result and adding the mean potential energy flux, $\Xi_s= Z_s e \pot_0 - m_s \angvel^2(\psi)R^2/2$ times \eref{pfluxtmp}, to \eref{hfluxtmp}, we get
\begin{equation}
\fl
\begin{eqalign}{
 &\left(\bm{Q}_s + \Xi_s n_s\bm{U}_s+ {Z_s e}\ensav{\wint \gkupot \gyror{h_s \bm{w}}} \right)\dg\psi =\\
	 &-\frac{c}{Z_s e}\dv\left[ m_s\left( \tensor{H}_s\dg\tor \right) R^2 + \Xi_s \left( \viscosity\dg\tor \right) R^2 + \angvel(\psi)R^2\Xi_s n_s \bm{U}_s + \angvel(\psi) R^2\bm{Q}_s\right]\\ 
	 &- \left( \frac{5}{2} T_s + \Xi_s \right)n_s \pd{\psi}{t}
	  + \frac{c}{Z_s e}\left( \bm{G}_s + \Xi_s \bm{F}_s\right)\cdot\left(\nabla\tor\right) R^2\\
	 &+ \frac{c}{Z_s e}\ensav{\wint \energy R^2\left(\daccel\dg\tor\right) \delta f_s} ,
}\end{eqalign}
\label{tmpD41}
\end{equation}
where $\bm{F}_s$ is the collisional friction force defined by \eref{FrictionDef}.
Taking the flux-surface average of \eref{tmpD41}, we obtain 
\begin{equation}
\begin{eqalign}{
&\fav{\left(\bm{Q}_s + \Xi_s n_s \bm{U}_s + Z_s e\ensav{\wint \gkupot \gyror{h_s \bm{w}}}\right)\dg\psi} =\\
& \quad- \fav{ \left(\frac{5}{2} T_s + \Xi_s \right)n_s \pd{\psi}{t} } + \frac{c}{Z_s e}\fav{\left(\bm{G}_s + \Xi_s \bm{F}_s\right)\cdot\left(\nabla\tor\right)R^2}  \\
&\qquad  + \fav{\ensav{\wint \energy  \gyror{h_s\vchi}\dg\psi}},
}\end{eqalign}
\label{qfinalap}
\end{equation}
where we have used \eref{lowviscous} and \eref{MagicTheorem} to demonstrate that the flux-surface average of a divergence (the first term on the right-hand side of \eref{tmpD41}) is negligible (a similar argument to the one that led to the neglect of \eref{blahtmpflux}) and by exactly the same procedure as for the particle flux 
written the fluctuating heat flux in terms of $h_s$ and $\vchi$ (see
\eref{boltzVanP} and \eref{gyroavtmp}).
The collisional terms are calculated in \ref{apF1} (see \eref{collhflux}):
\begin{equation}
\fl
\begin{eqalign}{
 \frac{c}{Z_s e} \fav{\left(\bm{G}_s+\Xi_s \bm{F}_s\right)\cdot\left(\nabla\tor\right)R^2}  = \\
 \qquad\quad \fav{\wint\energy \left( \frac{\bm{w}\times\Meanb}{\cycfreq}\dg\psi \right) \collop[F_{0s}]}\\
	 \qquad\qquad+\fav{\wint \energy \Fneo\vdrift \dg\psi} - I(\psi)\fav{n_s \left(\frac{5T_s}{2} + \Xi_s\right)}\frac{\fav{\MeanE\cdot\MeanB}}{\fav{\MeanMagB^2}}.
}\end{eqalign}
\label{colltmphflux}
\end{equation}

Finally, we substitute \eref{colltmphflux} into \eref{qfinalap} and then \eref{qfinalap} into \eref{main-hflux-full}
to find the desired result \eref{heatflux}.
\section{Collisional Transport}
\label{apF}
In \ref{fluxes}, we derived explicit forms for the particle, momentum and heat fluxes. A term due to collisions on the mean distribution function $F_s$ appeared in each of these fluxes and had to be expressed in terms of $F_{0s}$, $\Fneo$, and $\fav{\MeanE\cdot\MeanB}$ (see \eref{tmpcollpflux}, \eref{apDcollmflux}, and \eref{colltmphflux}). In this Appendix, we 
detail the relevant derivations.

In these derivations, we will need to be able to replace $\gyror{\collop[F_s]}$ by $\gyror{\gyroR{\collop[F_s]}}$. Let us prove that this is legitimate for
a general function $g$.
Taking the perpendicular spatial average of \eref{wperpdg1}, we find that the first term on the right-hand side vanishes to lowest order because it is a divergence, and we get
\begin{equation}
\perpav{\left.\pd{ g}{\gyr}\right|_{\bm{R}_s,\energy,\magmom}} = \perpav{\left.\pd{ g}{\gyr}\right|_{\bm{r},w_\parallel,w_\perp}} + \Or(\gkeps g),
\end{equation}
which we can gyroaverage at constant $\bm{r}$ to find
\begin{equation}
\gyror{\perpav{\left.\pd{ g}{\gyr}\right|_{\bm{R}_s,\energy,\magmom}}} = \Or(\gkeps g).
\label{tmpE2}
\end{equation}
Since	any function $g(\bm{r},\bm{w})$ can be written as $g=  \gyroR{g} + \infrac{\partial G}{\partial \gyr}|_{\bm{R}_s}$ for some $G$, we have,
using \eref{tmpE2},
\begin{equation}
\gyror{\perpav{\gyroR{g}}} = \gyror{\perpav{g}} - \gyror{\perpav{\left.\pd{G}{\gyr}\right|_{\bm{R}_s}}} = \gyror{\perpav{g}} + \Or(\gkeps g).
\label{interchangeRrgyr}
\end{equation}
If $g$ is a mean quantity, then the perpendicular averages have no effect ($\perpav{g} = g$) and we obtain
\begin{equation}
\gyror{\gyroR{g}} = \gyror{g} + \Or(\gkeps g).
\label{interchangeMeanRr}
\end{equation}
Therefore,
\begin{equation}
\gyror{\collop[F_s]} = \gyror{\gyroR{\collop[F_s]}}
\label{gyroCFs}
\end{equation}
to lowest order in $\gkeps$.
\subsection{Collisional Fluxes}
\label{apF1}
There are three collisional fluxes to calculate, the particle flux, the convective angular momentum flux and the heat flux resulting in the collisional contributions to \eref{partflux}, \eref{conMomflux} and \eref{heatflux} or, within \ref{fluxes}, \eref{tmpcollpflux}, \eref{tmpE13}, and \eref{colltmphflux}; we handle the collisional viscous stress, the first term in \eref{tmpE13}, separately in \ref{apF2}. As can be seen from \eref{pfluxtmp}, \eref{tmpE13}, and \eref{hfluxtmp}, all these fluxes can be written in the form
\begin{equation}
  \frac{m_s c}{Z_s e}\wint g(\bm{r},\energy) \left(\bm{v}\dg\tor\right) R^2 \collop[F_s],
  \label{generalFlux}
\end{equation}
where $g = 1$ for the particle flux, $g = m_s R^2\angvel(\psi)$ for the convective angular momentum flux, and $g = \energy$ for the heat flux.
The calculation proceeds almost identically for each of these, and so we shall perform the calculation for the particle flux, highlighting the one point where the calculations diverge, and state the results for the other two fluxes at the end of this section.

We start from the expression \eref{FrictionDef} for the collisional friction force,
\begin{equation}
\label{fricexp}
\fl
\begin{eqalign}{
\bm{F}_s = \wint m_s\bm{v}\collop[F_s] = \wint m_s \left(w_\parallel \Meanb+\bm{u}\right) \gyror{\collop[F_s]} + \wint m_s \gyror{\bm{w}_\perp \collop[F_s]},
}\end{eqalign}
\end{equation}
where we have split $\bm{v}$ into its gyroaverage, $w_\parallel\Meanb + \bm{u}$, and its gyrophase-dependent part $\bm{w}_\perp$ (the term involving $\bm{u}$ is kept in order to maintain the symmetry between the three flux calculations despite the fact that it vanishes identically for the particle flux).
The second term, the perpendicular friction, gives rise to the so-called ``classical'' collisional fluxes, whilst the first term gives rise to the ``neoclassical'' ones~\cite{helander2002ctm}.
In the perpendicular friction, we can expand $F_s$ via \eref{FSoln}, \eref{F1Hat},  \eref{Fstardef} and \eref{Fneodef} to find
\begin{equation}
\fl \begin{eqalign}{
\wint m_s \gyror{\bm{w}_\perp \collop[F_s]} &= \wint m_s \gyror{\bm{w}_\perp \left( \collop[F_{0s}] + \lincol[\Fstar + \Fneo] \right)}\\
		&= \wint m_s \gyror{\bm{w}_\perp \collop[F_{0s}]},
}\end{eqalign}
\label{tmpE8}
\end{equation}
where we have used the fact that the linearised collision operator acting on a gyrophase-independent function is also gyrophase-independent (the neoclassical distribution functions only depend on the gyrophase via their slow dependence on $\bm{R}_s$, which can be replaced by $\bm{r}$ in \eref{tmpE8} to the order we require).
Thus, we are left only with the perpendicular friction associated with the equilibrium distribution function \eref{F0def}. We remind the reader that $F_{0s}$ is only a Maxwellian to lowest order, so $\collop[F_{0s}]$ will be $\Or(\gkeps^2\cycfreq F_{0s})$\footnote{Here we are assuming, as in the discussion about collisional energy exchange at the end of \Sref{heattrans}, that either all temperatures are equal or that collisional temperature equilibration is a slow process. Thus, there is no contribution to $\collop[F_{0s}]$ from terms like $\nu^{(E)}_{s\,s'}\left(T_s - T_{s'}\right)F_{0s}/ T_{s} \sim \gkeps \cycfreq F_{0s}$.}.

Considering now the neoclassical friction force, we have
\begin{equation}
\fl
\begin{eqalign}{
\wint m_s  (w_\parallel \Meanb + \bm{u}) \gyror{\collop[F_s]}  \\
	\quad= \wint m_s \left(w_\parallel \Meanb + \bm{u}\right)\gyror{ \gyroR{\collop[{F_{0s}} + \Fstar]} +\gyroR{\lincol[\Fneo]}} + \Or(\gkeps^3 n_s m_s \vth\cycfreq),
}\end{eqalign}
\label{donkey}
\end{equation}
where we have again substituted for $F_s$ via \eref{FSoln}, \eref{F1Hat}, \eref{Fstardef} and \eref{Fneodef} and used \eref{gyroCFs}.
The collision operators in \eref{donkey} are exactly those operators that appear in \eref{neogeq}, so we use \eref{neogeq} to substitute for them and find
\begin{equation}
\fl
\begin{eqalign}{
\wint m_s (w_\parallel \Meanb + \bm{u}) \gyror{\collop[F_{s}]} \\
	\qquad\qquad= \wint m_s \left( w_\parallel \Meanb + \bm{u} \right) \left( w_\parallel \Meanb\cdot\ddR{\Fneo} - w_\parallel \cycfreq \frac{m_sc}{T_s}  \frac{\fav{\MeanE\cdot\MeanB}}{\fav{\MeanMagB^2}}F_{0s}\right),
}\end{eqalign}
\label{tmpE10}
\end{equation}
where we have used the fact that the integrand on the right-hand side is gyrophase-independent at constant $\bm{r}$.

Substituting \eref{tmpE8} and \eref{tmpE10} into the expression for the collisional particle flux (the third term on the right hand side of \eref{pfluxtmp}) and carrying out the velocity integration involving $F_{0s}$ (the value of this integral will be different for the calculation of the heat flux, but the procedure is identical), we find
\begin{equation}
\fl
\begin{eqalign}{
 \frac{c}{Z_s e}\fav{\left(\bm{F}_s\cdot\nabla\tor\right)R^2}  = \fav{\wint \left(\frac{\bm{w}\times\Meanb}{\cycfreq}\cdot\nabla\psi\right) \collop[F_{0s}]}\\
 \qquad+\frac{m_sc}{Z_s e} \wint \fav{R^2\left(\nabla\tor\right)\cdot\left(w_\parallel \Meanb + \bm{u}\right) w_\parallel \Meanb\cdot\ddR{\Fneo}} 
		- \fav{n_s}I(\psi)\frac{\fav{\MeanE\cdot\MeanB}}{\fav{\MeanMagB^2}} ,
}\end{eqalign}
\label{tmpE12}
\end{equation}
where we have used the axisymmetric form of the magnetic field \eref{magfield} to rewrite the classical flux.

For the convective flux of angular momentum, $g=m_s R^2 \angvel(\psi)$, and this is as far as we can proceed. Thus, the collisional contribution to the convective flux of angular momentum (the second term in \eref{tmpE13}) is
\begin{equation}
\fl
\begin{eqalign}{
\frac{c}{Z_s e}& \fav{ \wint m_s^2 R^4 \angvel(\psi) \left( \bm{v}\dg\tor\right) \collop[F_s]} 
= \\
 &  m_s\angvel(\psi)\fav{R^2\wint \left(\frac{\bm{w}\times\Meanb}{\cycfreq}\cdot\nabla\psi\right) \collop[F_{0s}]}\\
 &+m_s\angvel(\psi)\wint \fav{w_\parallel \Fneo \Meanb\dg\left[ R^2 \frac{I(\psi)w_\parallel}{\cycfreq} + R^2\frac{\MeanMagB R^2\angvel^2(\psi)}{\cycfreq} \right]}\\
		&- m_s\angvel(\psi)\fav{R^2n_s}I(\psi)\frac{\fav{\MeanE\cdot\MeanB}}{\fav{\MeanMagB^2}}.
}\end{eqalign}
\label{collconmflux}
\end{equation}
This expression is used in \eref{apDcollmflux} and then in \eref{conMomflux}.

For the particle and heat fluxes ($g = 1$ and $g=\energy$, respectively) we can make one further simplification.
Using $\wint = \sum_\sigma\int d\energy d\magmom d\gyr (\MeanMagB/m_s^2 {w_\parallel})$ and integrating by parts inside the flux-surface average, we can show that:
\begin{equation}
\fl\begin{eqalign}{
\frac{m_s c}{Z_s e}&\wint \fav{R^2\left(\nabla\tor\right)\cdot\left(w_\parallel \Meanb + \bm{u}\right) w_\parallel \Meanb\cdot\ddR{\Fneo}}\\
&= -\wint \fav{ w_\parallel \Fneo \Meanb\dg\left[ \frac{I(\psi)w_\parallel + \MeanMagB R^2\angvel(\psi)}{\cycfreq} \right]}\\
&=\fav{\wint \Fneo\vdrift\dg\psi},
}\end{eqalign}
\label{tmpE1}
\end{equation}
where we have used the identity \eref{VDpsi} obtain the last expression.
Substituting this into \eref{tmpE12}, we obtain the final form of the collisional particle flux
\begin{equation}
\label{collpflux}
\begin{eqalign}{
\frac{c}{Z_s e}\fav{\left(\bm{F}_s\cdot\nabla\tor\right)R^2} =
	 \fav{\wint \left( \frac{\bm{w}\times\Meanb}{\cycfreq}\dg\psi  \right)\collop[F_{0s}]}\\
	 \qquad+\fav{\wint \Fneo\vdrift\dg\psi } - \fav{n_s} I(\psi) \frac{\fav{\MeanE\cdot\MeanB}}{\fav{\MeanMagB^2}},
}\end{eqalign}
\end{equation}
which is \eref{tmpcollpflux} and contains the collisional terms appearing in \eref{partflux}.

As all the derivatives in \eref{tmpE1} are taken at constant $\energy$, the derivation presented for the particle flux ($g=1$) can be carried out identically for the heat flux ($g=\energy)$ (the fact that $\Meanb\dg \left[m_sR^2\angvel(\psi) \right] \ne 0$ is what prevented us from using \eref{tmpE1} to simplify the convective angular momentum flux). Thus, the heat flux is
\begin{equation}
\label{collhflux}
\fl
\begin{eqalign}{
\frac{c}{Z_s e} \fav{\left( \bm{G}_s+\Xi_s \bm{F}_s\right)\cdot\left(\nabla\tor\right)R^2} = \frac{m_s c}{Z_s e} \wint \energy \left( \bm{v}\dg\tor \right)R^2 \collop[F_s] \\
		\qquad\quad=\fav{\wint\energy \left( \frac{\bm{w}\times\Meanb}{\cycfreq}\dg\psi \right) \collop[F_{0s}]}\\
		\qquad\qquad+ \fav{\wint \energy \Fneo\vdrift \dg\psi} - I(\psi)\fav{n_s \left(\frac{5T_s}{2} + \Xi_s\right)}\,\frac{\fav{\MeanE\cdot\MeanB}}{\fav{\MeanMagB^2}},
}\end{eqalign}
\end{equation}
which is \eref{colltmphflux} as required.

\subsection{Collisional Viscous Stress}
\label{apF2}
We now calculate the collisional contribution to the viscous stress, the first term in \eref{tmpE13}.
Splitting $\left(R^2 \bm{w}\dg\tor \right)^2$ into its gyroaverage and a gyrophase-dependent part, we find
\begin{equation}
\fl
\begin{eqalign}{
\frac{c}{Z_s e} &\fav{\wint \frac{m_s^2}{2}  \left( R^2\bm{w}\dg\tor\right)^2 \collop[F_s]}=\\
	&\frac{c}{Z_s e} \fav{\wint \frac{m_s^2}{2} \gyror{ \left\{ \left(R^2\bm{w}_\perp \dg \tor\right)^2 - \frac{w_\perp^2}{2\MeanMagB^2} \left[\MeanMagB^2R^2 - I^2(\psi)\right]\right\} \collop[F_{0s}]}}\\
	&+ \frac{c}{Z_s e} \fav{\wint m_s^2   \frac{I(\psi)w_\parallel}\MeanMagB R^2\gyror{\left( \bm{w}_\perp\dg\tor \right)\collop[F_{0s}]}}\\
	&+\frac{c}{Z_s e} \fav{\wint \frac{m_s^2}{2\MeanMagB^2}  \left\{ I^2(\psi)w_\parallel^2 + \frac{w_\perp^2}{2} \left[\MeanMagB^2R^2 - I^2(\psi)\right]\right\} \gyror{\gyroR{\collop[F_s]}}},
}\end{eqalign}
\label{collvstresstmp}
\end{equation}	
where we have used $\gyror{(R^2\bm{w}\dg\tor)^2} = \left[I^2(\psi)/\MeanMagB^2\right] w_\parallel^2 +  \left[R^2 - I^2(\psi)/\MeanMagB^2\right]w_\perp^2/2$ and \eref{gyroCFs}. 
The first two terms in \eref{collvstresstmp} are the classical contribution to the collisional viscous stress. They can be rewritten using the axisymmetric form of the magnetic field \eref{magfield} and thus shown to be equal to $\fav{\ClassMomFlux}$ as given by \eref{classvisc}.
The last of the terms in \eref{collvstresstmp} is the neoclassical viscous stress. In the same way as for the neoclassical particle flux, we use \eref{neogeq} to substitute for $\gyroR{\collop[F_s]}$ and integrate by parts under the flux-surface average to find that it is equal to $\fav{\NeoMomFlux}$ as given by \eref{neovisc}.

Thus, the collisional viscous stress is
\begin{equation}
\begin{eqalign}{
\frac{c}{Z_s e} \fav{\wint \frac{m_s^2}{2}  \left( R^2\bm{w}\dg\tor\right)^2 \collop[F_s]} = \fav{\ClassMomFlux}+ \fav{\NeoMomFlux},
}\end{eqalign}
\label{collvstress}
\end{equation}
where the classical and neoclassical parts are given by \eref{classvisc} and \eref{neovisc}, respectively.
\section{Derivations for Sections~\ref{thermo} and \ref{Sentropy}}
\label{thermoDeriv}
In this Appendix we provide detailed derivations of equations that are stated without proof in Sections~\ref{thermo}\ and\ \ref{Sentropy}.
\subsection{Derivation of \eref{thermalenergy}: Evolution of Thermal Energy}
\label{apEnergyConv}
Summing \eref{dpdt} over all species we find
\begin{equation}
\fl
\begin{eqalign}{
\frac{3}{2} \frac{1}{V'} &\ddtpsi V'\sum_s \fav{n_s} T_s + \frac{1}{V'} \pd{ }{\psi}V'\sum_s \fav{\HeatFlux} \\
	&=\sum_s \ViscousHeat + \sum_s \JouleHeat +\sum_s \CompHeat + \sum_s \PotEng + \sum_s \TurbPow + \sum_s \fav{\Esource},
}\end{eqalign}
\label{initialSumDp}
\end{equation}
where we have used the fact that collisions conserve energy and so $\sum_s \CollEnergy = 0$.
We now proceed to simplify the source terms on the right-hand side of this equation. 

Starting with the viscous-heating term, defined in \eref{VHDef}, we find
\begin{equation}
\begin{eqalign}{
\sum_s \ViscousHeat %&=- \sum_s \left[\fav{\MomentumFlux} + m_s \angvel(\psi) \fav{R^2 \ParticleFlux}\right] \frac{d\angvel}{d\psi}\\
							 &= -\left( \fav{\TotMomFlux}- \fav{\EMViscosity}\right)\frac{d\angvel}{d\psi},
}\end{eqalign}
	\label{VHTmp}
\end{equation}
where we have used \eref{TMFresult} to substitute for the momentum flux in terms of the total flux and its electromagnetic part.

We handle the Ohmic-heating and compressional-heating terms together. Adding \eref{ohm-heat-result} to \eref{comp-heat-result} and summing over species, we obtain
\begin{equation}
\fl
\begin{eqalign}{
	-\sum_s \frac{Z_s e}{c}& \fav{\pd{\bm{A}}{t} \cdot \left( n_s \bm{U}_s \right)} + \fav{\dv\left(\sum_s n_sT_s \pd{\psi}{t}\right)} \\
		&= \sum_s\JouleHeat + \sum_s\CompHeat-\sum_s\fav{ \frac{1}{2}m_s n_s \angvel^2(\psi)\vpsi\dg R^2 },
}\end{eqalign}
\end{equation}
where we have used \eref{VpsiDef} for $\vpsi$ and used quasineutrality to eliminate all terms involving~$\pot_0$.
Rearranging this equation, we have
\begin{equation}
\fl
\begin{eqalign}{
\sum_s\left(\JouleHeat + \CompHeat\right) = &
	\fav{\left( \MeanE+\nabla\pot \right)\cdot\bm{j}} + \frac{1}{V'}\pd{ }{\psi} V'\sum_s T_s\fav{n_s\pd{\psi}{t}} \\
		&\qquad+\sum_s \fav{\frac{1}{2}m_s n_s \angvel^2(\psi)\vpsi\dg R^2 },
}\end{eqalign}
\label{OhmTMP}
\end{equation}
where we have used \eref{favdiv} for the flux-surface average of a divergence and used $-(1/c) \partial \bm{A} / \partial t = \bm{E} + \nabla\pot$.
The first term on the right-hand side of \eref{OhmTMP} can be written as
\begin{equation}
\fl
\begin{eqalign}{
						\fav{\left(\MeanE + \nabla \pot \right)\cdot\Meanj} &= \fav{\MeanE\cdot\Meanj} + \fav{\dv\left(\pot \Meanj\right)} =\fav{\MeanE\cdot\Meanj} + \frac{1}{V'} \pd{ }{\psi} V'\fav{\pot_0 \Meanj\dg\psi},
}\end{eqalign}
\label{OhmHeatTmp}
\end{equation}
where we have used $\dv\Meanj = 0$, the identity \eref{favdiv} again,  $\pot = \fpot(\psi) + \pot_0$, and $\fav{\Meanj\dg\psi} = 0$ (see \eref{ambipolarcon}).

Finally, turning to the potential-energy term, we collect the second term in the last line of \eref{OhmTMP} together with the potential-energy-exchange term, and substitute the definition \eref{PotEngDef} to find:
\begin{equation}
\begin{eqalign}{
\sum_s &\left(\PotEng + \frac{1}{2}m_s\angvel^2(\psi)\vpsi\dg R^2\right)  \\
	&= \frac{\angvel^2(\psi)}{2V'} \ddtpsi V' \inertia - \frac{1}{2}\angvel^2(\psi)\fav{ R^2\sum_s m_s\Psource},
}\end{eqalign}
	\label{PotEngTmp}
\end{equation}
where the moment of inertia $\inertia$ is defined in \eref{inertiaDef} and the terms involving $\pot_0$ vanished by quasineutrality:
$\sum_s Z_s e n_s= 0$, $\sum_s Z_s e \Psource = 0$.

Substituting \eref{VHTmp}, \eref{OhmTMP}, \eref{OhmHeatTmp}, and \eref{PotEngTmp} back into \eref{initialSumDp}, we find the desired result \eref{thermalenergy},
where we have grouped the second term in \eref{OhmTMP} and the last term in \eref{OhmHeatTmp} with the divergence of the heat flux on the left-hand side of \eref{initialSumDp} and also grouped the particle source term from \eref{PotEngTmp} with the heat sources in \eref{initialSumDp}. 
We postpone the discussion of the turbulent-heating term $\sum_s \TurbPow$ until \ref{Sturbheat}, as it is particularly involved.
\subsection{Derivation of \eref{NoTHeat}: Turbulent Heating}
\label{Sturbheat}
Here we calculate the contribution to the evolution of the thermal energy due to fluctuations, i.e., the source term $\sum_s \TurbPow$ on the right-hand side of \eref{thermalenergy}.
Starting from the definition \eref{TurbPowDef} of $\TurbPow$ we have
\begin{equation}
\fl\begin{eqalign}{
\sum_s \TurbPow =& \sum_s Z_s e \fav{ \ensav{\wint \gyror{h_s\left( \pd{}{t} + \bm{u}\dg \right) \gkpot} }}\\
&-\sum_s  \frac{Z_s e}{c}\angvel(\psi) \fav{\ensav{\wint \gyror{h_s\bm{w}}\cdot \left(\delA\times\nabla z\right) }}.
}\end{eqalign}
\label{turbheat}
\end{equation}
When we come to derive the conservation law for free energy in \ref{freeEnergyDeriv}, we will need an expression for the right-hand side of \eref{turbheat} except without the time average that is implicit in the average over fluctuations (see the definition \eref{avdef} of the fluctuation average). We will therefore first carry out our calculation of the right-hand side of \eref{turbheat} without this time average and then apply the time average at the last step to obtain $\sum_s \TurbPow$. As in \Sref{SFreeEnergy}, we refer to the composition of the perpendicular spatial average with the flux-surface average as the annulus average.

Thus, averaged only over an annulus but not over time, the first term on the right-hand side of \eref{turbheat} becomes (using the perpendicular spatial average as defined by \eref{PerpAvDef})
\begin{equation}
\fl
\begin{eqalign}{
\sum_s Z_s e& \fav{\perpav{\wint\gyror{ h_s \left(\pd{}{t} +\bm{u}\dg\right)\gkpot } }} \\
&= -\frac{1}{c}\fav{\perpav{\delj\cdot \left(\pd{}{t} + \bm{u}\dg\right)\delA}} + \sum_s\fav{ \pd{}{t} \perpav{\frac{Z_s^2 e^2 \gkupot^2 n_s}{2 T_s}}}  \\
&= \fav{\perpav{\left(\delE+\nabla\gkupot\right)\cdot\delj+\frac{1}{c}\left(\bm{u}\times\delB\right)\cdot\delj +\frac{1}{c}\left(\delj\cdot \nabla\bm{u} \right)\cdot\delA}}\\
&\qquad\qquad +  \sum_s\fav{\pd{}{t} \perpav{\frac{Z_s^2 e^2 \gkupot^2 n_s}{2 T_s}}},}\end{eqalign}
\label{tmptmpF7}
\end{equation}
where we have used $h_s = \delta f_s + Z_s e \gkupot F_{0s} / T_s$, $\gkpot = \gkupot - \bm{w}\cdot\delA / c$, the quasineutrality constraint \eref{fluct-qn}, and
the fact that $\fav{\bm{u}\dg\perpav{\cdots}} = 0$, which follows from $\dv\bm{u}=0$, $\bm{u}\dg\psi =0$ and the formula \eref{favdiv} for the flux-surface average of a divergence.
The last line in \eref{tmptmpF7} is obtained by using $\infrac{\partial \delA}{\partial t} = -c\left( \delE+\nabla\delpot \right) {= -c\left( \delE + \nabla\gkupot \right)-\dv\left(\bm{u}\cdot\delA \right)}$.

We now find an expression for $\perpav{\left( \delE+\nabla\gkupot \right)\cdot\delj}$. We start from the perpendicular spatial average of Poynting's theorem for the fluctuating fields \eref{fluctPoynt}:
\begin{equation}
\label{avgPoynt}
  \pd{}{t} \perpav{\frac{\MagDelB^2}{8\pi}} + \frac{c}{4\pi}\dv\perpav{\delE\times\delB} = -\perpav{\delE\cdot\delj}.
\end{equation}
Taking the cross product of \eref{emov} with $\delB$, we have the following expression for the Poynting flux:
\begin{equation}
\frac{c}{4\pi} \delE\times\delB = \frac{1}{4\pi} \bm{u}\cdot \left(\MagDelB^2\idmat-\delB\delB \right) - \frac{c}{4\pi} \left(\nabla\gkupot\right)\times\delB.
\label{tmpPoynFl}
\end{equation}
Using \eref{tmpPoynFl}, we obtain
\begin{equation}
\label{udgAvgPoynt}
  \fl
  \left( \pd{}{t} + \bm{u}\dg \right) \perpav{\frac{\MagDelB^2}{8\pi}} - \frac{1}{4\pi}\dv\left( \perpav{\delB\delB}\cdot\bm{u} \right) = -\perpav{\left( \delE + \nabla\gkupot \right)\cdot\delj}.
\end{equation}
Using this to substitute for $\perpav{\left( \delE + \nabla\gkupot \right)\cdot\delj}$ in \eref{tmptmpF7}, we find
\begin{equation}
\fl\begin{eqalign}{
\sum_s Z_s e& \fav{\perpav{\wint\gyror{h_s \left(\pd{}{t} +\bm{u}\dg\right)\gkpot}}}\\
&= \fav{\perpav{\frac{1}{4\pi}\delB\delB\bm{:}\nabla\bm{u} + \frac{1}{c}\left( \delj\dg\bm{u} \right)\cdot\delA }}\\
&\qquad\quad+ \sum_s\fav{\pd{}{t} \perpav{\frac{Z_s^2 e^2 \gkupot^2 n_s}{2 T_s}}}- \fav{\pd{}{t} \perpav{\frac{\MagDelB^2}{8\pi}}},
}\end{eqalign}
\label{heatingbong}
\end{equation}
where we have used Amp\`ere's law to write $\binfrac{1}{c}\perpav{\left( \bm{u}\times\delB \right)\cdot\delj} = -\binfrac{1}{4\pi} \perpav{\delB\cdot\left( \nabla\delB \right)\cdot\bm{u}}$.
Using \eref{nablaU} and \eref{aCrossZ} to express $\nabla\bm{u}$, we obtain
\begin{equation}
\fl\begin{eqalign}{
\sum_s &Z_s e \fav{\perpav{\wint \gyror{h_s \left(\pd{}{t} +\bm{u}\dg\right)\gkpot }}}\\
&= \fav{\left(\nabla\psi\right)\cdot\perpav{\frac{1}{4\pi} \delB\delB + \frac{1}{c} \delj\delA}\cdot\left( \nabla\tor \right)R^2} \frac{d\angvel}{d\psi}\\
&\qquad+ \frac{1}{c}  \fav{\perpav{ \delj\cdot \left( \delA\times\nabla z\right)}}\angvel(\psi)\\
&\qquad+ \sum_s\fav{\pd{}{t} \perpav{\frac{Z_s^2 e^2 \gkupot^2 n_s}{2 T_s}}}- \fav{\pd{}{t} \perpav{\frac{\MagDelB^2}{8\pi}}}.
}\end{eqalign}
\end{equation}
Using $\delj = \sum_s Z_s e\wint \gyror{h_s \bm{w}}$ in the second term on the right-hand side of this equation, we arrive at the expression for the non-time-averaged turbulent heating:
\begin{equation}
\fl\begin{eqalign}{
&\sum_s Z_s e  \fav{\perpav{\wint\gyror{h_s \left( \pd{}{t} + \bm{u}\dg\right)\gkpot}}} \\
	&\qquad- \frac{Z_s e}{c} \angvel(\psi)\fav{\perpav{\wint \gyror{h_s\bm{w}}\cdot\left(  \delA\times\nabla z\right)}} \\
&\quad= \fav{\left(\nabla\psi\right)\cdot\perpav{\frac{1}{4\pi} \delB\delB + \frac{1}{c} \delj\delA}\cdot\left( \nabla\tor \right)R^2} \frac{d\angvel}{d\psi} \\
&\quad\qquad+ \sum_s\pd{}{t} \fav{\perpav{\frac{Z_s^2e^2 n_s \gkupot^2}{2T_s} }} - \pd{}{t} \fav{\perpav{\frac{\delta \MeanMagB^2}{8\pi}}},
}\end{eqalign}
\label{binky}
\end{equation}
where we have moved the term involving $\delA\times\nabla z$ to the left-hand side to make the comparison to \eref{turbheat} more transparent.

Finally, taking the time average of \eref{binky}, we obtain from \eref{turbheat}
\begin{equation}
\sum_s \TurbPow = - \fav{\EMViscosity} \frac{d\angvel}{d\psi},
\label{TurbPowVisc}
\end{equation}
where we have used the definition \eref{EMViscDef} of $\EMViscosity$ and the fact that time derivatives of averaged quantities are $\gkeps^2$ smaller than the other terms and can be neglected. This is precisely \eref{NoTHeat}, as required.
\subsection{Derivation of \eref{JUflux}: A Simple Form of the Kinetic Energy Flux \texorpdfstring{$\JUflux$}{}}
\label{apJUSimples}
Starting from the first expression for $\JUflux$ of \eref{JUflux}, we wish to derive the second.
Using \eref{main-hflux-full} to express $\fav{\HeatFlux}$ in the first line of \eref{JUflux}, we obtain
\begin{equation}
\fl\begin{eqalign}{
\fav{\JUflux} = \sum_s &\fav{\left[ {\bm{Q}_s} - \frac{1}{2} m_s \angvel^2(\psi) R^2 n_s\bm{U}_s + Z_s e\ensav{\wint\gkupot \gyror{h_s\bm{w}}}\right]\dg\psi}\\
	&+ \sum_s \fav{n_s\left[\frac{3}{2} T_s - \frac{1}{2} m_s \angvel^2(\psi) R^2 \right] \pd{\psi}{t}} + \angvel(\psi)\fav{\TotMomFlux},
}\end{eqalign}
\label{tmpF15}
\end{equation}
where we have used $\Meanj = \sum_s Z_s e n_s \bm{U}_s$ and so cancelled the last term in the first line of \eref{JUflux} with corresponding term in \eref{main-hflux-full}.
Next, we expand $\TotMomFlux$ in \eref{tmpF15} by using its definition \eref{TMFdef} and obtain
\begin{equation}
\fl\begin{eqalign}{
&\fav{\JUflux} = \sum_s \fav{\left[ {\bm{Q}_s} - \frac{1}{2} m_s \angvel^2(\psi)R^2 n_s\bm{U}_s + Z_s e\ensav{\wint\gkupot \gyror{h_s\bm{w}}}\right]\dg\psi}\\
	&\quad+ \sum_s \fav{n_s\left[\frac{3}{2} T_s - \frac{1}{2} m_s \angvel^2(\psi) R^2 \right] \pd{\psi}{t}} \\
	&\quad+ \sum_s \fav{\bm{u}\cdot\viscosity\dg\psi}
	+ \sum_s \angvel^2(\psi) m_s \fav{R^2\ParticleFlux} \\
	&\quad- \frac{1}{4\pi}\fav{\bm{u}\cdot\ensav{\delB \delB}\dg\psi},
}\end{eqalign}
\label{tmpF16}
\end{equation}
where we have used $\bm{u} = \angvel(\psi)R^2\nabla\tor$ and the symmetry of $\viscosity$.
We now use \eref{GammaDef} to express $\ParticleFlux$ in terms of $n_s\bm{U}_s$ and $\infrac{\partial \psi}{\partial t}$ and find that
\begin{equation}
\fl\begin{eqalign}{
\fav{\JUflux} = \sum_s& \fav{\left[ {\bm{Q}_s} + \frac{1}{2} m_s \angvel^2(\psi) R^2n_s\bm{U}_s + Z_s e\ensav{\wint\gkupot \gyror{h_s\bm{w}}}\right]\dg\psi}\\
	&+ \fav{U \pd{\psi}{t}}+ \sum_s \fav{\bm{u}\cdot\viscosity\dg\psi} - \frac{1}{4\pi}\fav{\bm{u}\cdot\ensav{\delB \delB}\dg\psi},
}\end{eqalign}
\label{tmpF17}
\end{equation}
where we have collected all terms proportional to $n_s\bm{U}_s$ and $\infrac{\partial \psi}{\partial t}$ together and used the definition \eref{UDef} of the total kinetic energy $U$.

To simplify the terms involving the fluctuations, we need to relate $\bm{u}\cdot\ensav{\delB\delB}$ to the Poynting flux.
Averaging \eref{tmpPoynFl} over the fluctuations, we obtain
\begin{equation}
\fl
\frac{c}{4\pi}\ensav{\delE\times\delB}\dg\psi = -\frac{1}{4\pi}\bm{u}\cdot\ensav{\delB\delB}\dg\psi - \frac{c}{4\pi}\ensav{\left(\nabla\gkupot\right)\times\delB}\dg\psi.
\label{tmpFiiick}
\end{equation}
Using the fluctuating part of Amp\`ere's law \eref{famp}, we rewrite the last term as
\begin{equation}
\fl
\begin{eqalign}{
- \frac{c}{4\pi}\ensav{\left(\nabla\gkupot\right)\times\delB}\dg\psi &= \sum_s Z_s e \ensav{\wint \gkupot \gyror{h_s \bm{w}}}\dg\psi\\
	 &\qquad- \frac{c}{4\pi}\dv\ensav{\gkupot\delB\times\nabla\psi}.
}\end{eqalign}
\label{tmpFmoreick}
	\end{equation}
The second term in the right-hand side of \eref{tmpFmoreick} is order $\gkeps$ smaller than the first and so we can neglect it.
Thus, substituting \eref{tmpFmoreick} into \eref{tmpFiiick}, we find that the sum of the two fluctuation terms in \eref{tmpF17} is equal to
$\binfrac{c}{4\pi}\ensav{\delE\times\delB}\dg\psi$. This gives us the second line of \eref{JUflux}, as required.

\subsection{Derivation of \eref{freeEnergyBalance}: Free Energy Balance}
\label{freeEnergyDeriv}
In this Appendix, we derive the free-energy evolution equation \eref{freeEnergyBalance} from the gyrokinetic equation \eref{gke}.
The free energy is defined in \eref{FreeEnergyDef}.
We start by using the expression \eref{hDef} for $\delta f_s$ in terms of $h_s$ to rewrite $W$ as
\begin{equation}
\fl
\begin{eqalign}{
W(\psi) %&= \sum_s \wint \fav{\perpav{\frac{T_s \gyror{h_s^2}}{2F_{0s}}  - Z_s e \gyror{h_s} \gkupot + \frac{Z_s^2 e^2 \gkupot^2}{2T_s} F_{0s}}}\\
		%	  &\qquad\qquad+ \fav{\perpav{\frac{\MagDelB^2}{8\pi}}} \\
		  &= \sum_s \fav{\perpav{\wint \frac{T_s \gyror{h_s^2}}{2F_{0s}}}}  - \sum_s \fav{\perpav{\frac{Z_s^2 e^2 n_s \gkupot^2}{2T_s}}} \\
		  &\qquad\qquad+ \fav{\perpav{\frac{\MagDelB^2}{8\pi}}},
}\end{eqalign}
\label{tmpFrEng}
\end{equation}
where we have used the fluctuating quasineutrality constraint \eref{fluct-qn}.

We now find an evolution equation for the first of the terms in \eref{tmpFrEng}. Multiplying the gyrokinetic equation \eref{gke} by $T_s h_s / F_{0s}$, we obtain
\begin{equation}
\label{hsquared}
\fl
\begin{eqalign}{
\left[\pd{}{t}+\bm{u}(\bm{R}_s)\dgR{}\right]\frac{T_s h_s^2}{2 F_{0s}} + \left(w_\parallel\Meanb + \vdrift + \vchiR\right)\dgR{}\frac{T_s h_s^2}{2 F_{0s}}- {\frac{ T_s h_s}{F_{0s}} \gyroR{\lincol[h_s]}}\\
\quad= Z_s e {h_s \left[ \pd{}{t} + \bm{u}\left(\bm{R}_s\right)\cdot\ddR{}\right] \gyroR{\gkpot}}  \\
\qquad-\left\{T_s \pd{\ln F_{0s}}{\psi} + {m_s} \left[ \frac{I(\psi) w_\parallel}\MeanMagB + \angvel(\psi)R^2 \right]\frac{d\angvel}{d\psi}\right\} h_s\vchiR\dg\psi.
}\end{eqalign}
\end{equation}
We now average this equation over an annulus and integrate over all velocities.

We start by processing the left-hand side of \eref{hsquared}.
Using the identity \eref{movingflux}, the time derivative becomes
\begin{equation}
\fl
\begin{eqalign}{
&\fav{\perpav{\wint\pd{}{t} \frac{T_s \gyror{h_s^2}}{2 F_{0s}}}} = \ddtpsi \fav{\perpav{\wint\frac{T_s \gyror{h_s^2}}{2F_{0s}}}},
}\end{eqalign}
\label{mehDdt}
\end{equation}
where have dropped the term arising from the motion of the flux surfaces because it is $\gkeps^2$ smaller than the rapid variation of $h_s^2$. We have neglected the time variation of $V'$ on the same grounds.
In \ref{NoSpreadProof}, we show that
\begin{equation}
\fl
\label{NoSpreading}
\fav{\perpav{\wint\gyror{ \left[ \bm{u}(\bm{R}_s) + w_\parallel\Meanb + \vdrift + \vchiR \right]\dgR{} \frac{T_s h_s^2}{2F_{0s}}}}} = \Or(\gkeps^2 \cycfreq W).
\end{equation}
Thus, the various advection terms in \eref{hsquared} do not contribute to the evolution of free energy. Note in particular the vanishing of the term involving $\vchiR$ -- a reflection of free energy's fundamental status as a cascaded invariant with respect to the gyrokinetic nonlinearity.

Consider now the right-hand side of \eref{hsquared}.
The first term can be interpreted as the source of $h_s^2$ (equivalently the perturbed entropy) due to turbulent heating. In \ref{THeatFEngSourceProof}, we show that the average of this term is
\begin{equation}
\fl
\label{THeatFEngSource}
\begin{eqalign}{
&\fav{\perpav{\wint Z_s e\gyror{ h_s \left[ \pd{}{t} + \bm{u}(\bm{R}_s)\dgR{}\right] \gyroR{\gkpot}}}}\\ 
	&\qquad=Z_s e \fav{\perpav{\wint \gyror{h_s \left(\pd{}{t} + \bm{u}\dg\right) \gkpot}}} \\
	&\qquad\qquad- \frac{Z_s e}{c} \angvel(\psi) \fav{\perpav{\wint \gyror{h_s \bm{w}}\cdot\left(\delA\times\nabla z\right)}} \\
   &\qquad\qquad- Z_s e \fav{\perpav{\wint \gyror{ h_s \left(\frac{\Meanb\times\bm{w}}{\cycfreq}  \dg\bm{u}\right) \dg \chi }}}.
}\end{eqalign}
\end{equation}
Finally, we turn to the second term on the right-hand side of \eref{hsquared} and calculate the source of $h_s^2$ due to background gradients (equivalently, free-energy injection due to instabilities).
In \ref{FluxFENG}, we show that
\begin{equation}
\fl
\begin{eqalign}{
&-\fav{\perpav{\wint \left\{T_s \pd{\ln F_{0s}}{\psi} + {m_s} \left[ \frac{I(\psi) w_\parallel}\MeanMagB + \angvel(\psi)R^2 \right]\frac{d\angvel}{d\psi}\right\} \gyror{h_s\vchiR}\dg\psi}}\\
&\qquad=- \fav{\perpav{\wint  \gyror{h_s \vchi}\dg\psi}} T_s\left(\frac{d\ln \NotN}{d\psi} - \frac{3}{2}\frac{d\ln T_s}{d\psi} \right)\\
	&\qquad\quad-\fav{ \perpav{\wint  \energy  \gyror{h_s \vchi}\dg\psi }}\frac{d \ln T_s}{d\psi}\\
	&\qquad\quad- \fav{\perpav{\wint R^2\left(\nabla\tor\right) \cdot\gyror{\bm{v}h_s \vchi}\dg\psi}} \frac{d\angvel}{d\psi}\\
	&\qquad\quad	+ Z_se\fav{ \perpav{\wint\gyror{h_s\left( \frac{\Meanb\times\bm{w}}{\cycfreq}  \cdot \nabla\bm{u} \right) \dg\gkpot }}}.
}\end{eqalign}
\label{FluxesFEngSource}
\end{equation}

Collecting \eref{mehDdt}, \eref{THeatFEngSource}, and \eref{FluxesFEngSource} together, we find that the annulus average of \eref{hsquared} integrated over all velocities is
\begin{equation}
\fl
\begin{eqalign}{
 \ddtpsi &\fav{\perpav{\wint\frac{T_s \gyror{h_s^2}}{2F_{0s}}}} - \fav{\perpav{\wint \frac{T_s}{F_{0s}}\gyror{h_s \lincol[h_s]}}}= \\
	 &Z_s e  \fav{\perpav{\wint \gyror{h_s \left( \pd{}{t} + \bm{u}\dg\right)\gkpot}}} \\ 
		& - \frac{Z_s e}{c} \angvel(\psi)\fav{\perpav{\wint \gyror{h_s\bm{w}}\cdot\left( \delA\times\nabla z \right)}} \\
	  &- \fav{\perpav{\wint \gyror{h_s \vchi}\dg\psi}} T_s\left( \frac{d\ln \NotN}{d\psi} -\frac{3}{2}\frac{d\ln T_s}{d\psi}\right) \\
	&-\fav{\gyror{ \wint  \energy\gyror{h_s \vchi}\dg\psi }}\frac{d \ln T_s}{d\psi} \\
	&- \fav{\perpav{\wint R^2\left(\nabla\tor\right) \cdot\gyror{\bm{v} h_s \vchi}\dg\psi}} \frac{d\angvel}{d\psi},
}\end{eqalign}
\label{ap-deltaSdt}
\end{equation}
where the last term in \eref{THeatFEngSource} and the last term in \eref{FluxesFEngSource} have cancelled.
Summing \eref{ap-deltaSdt} over all species, using \eref{binky} for the first two terms on the right-hand side, and collecting together the time derivatives, we obtain \eref{freeEnergyBalance}.

\subsubsection{Derivation of \eref{TurbPowDef}.} Averaging \eref{ap-deltaSdt} over the intermediate timescale $T$ and using the fact that $\ensav{g} = \left<\perpav{g}\right>_T$, we find
\begin{equation}
	\fl
\begin{eqalign}{
\TurbColl &= Z_s e  \fav{\ensav{\wint\gyror{h_s \left( \pd{}{t} + \bm{u}\dg\right)\gkpot}}} \\
				&\qquad\qquad- \frac{Z_s e}{c} \angvel(\psi)\fav{ \ensav{\wint\gyror{h_s\bm{w}}\cdot\left(  \delA\times\nabla z\right)}} + \TurbInj.
}\end{eqalign}
\label{bing}
\end{equation}
where $\TurbColl$ and $\TurbInj$ are defined by \eref{TurbCollDef} and \eref{TurbInjDef}, respectively. Rearranging terms and using the first line of \eref{TurbPowDef} (the definition of $\TurbPow$) gives the second line of \eref{TurbPowDef} immediately.

\subsubsection{Derivation of \eref{NoSpreading}.}
\label{NoSpreadProof}
Here we prove that the advection terms in \eref{hsquared} do not contribute to the evolution of the free energy.
Starting with the mean rotation, we have
\begin{equation}
\fl
\begin{eqalign}{
\perpav{\wint\gyror{\bm{u}(\bm{R}_s) \dgR{} \frac{T_s h_s^2}{2 F_{0s}}}} = \perpav{\wint\gyror{\bm{u}(\bm{R}_s)\left.\dgR{}\right|_{\bm{w}}\frac{T_s {h_s^2}}{2F_{0s}}}} \\
	\qquad- \perpav{\wint\gyror{\left[\left( \bm{u}\dg\magmom \right)\pd{ }{\magmom} + \left( \bm{u}\dg\gyr \right)\pd{ }{\gyr} \right]\frac{T_s h_s^2}{2F_{0s}}}} + \Or(\gkeps^2 \cycfreq \FreeEnergy),
}\end{eqalign}
\label{tmpE30}
\end{equation}
where we have used \eref{toenergy} and \eref{udgEnergy} to change the spatial derivative from one at constant $\energy$, $\magmom$ and $\gyr$ to one at constant $\bm{w}$.
Substituting for $\bm{u}\dg\magmom$ and $\bm{u}\dg\gyr$ from \eref{udgMu} and \eref{udgGyr}, respectively, carrying out the gyroaverage, and integrating by parts with respect to $\gyr$ where necessary,\footnote{Note that we are allowed to interchange the $\gyror{ }$ and $\perpav{ }$ averages.} we find that the $\infrac{\partial}{\partial\magmom}$ and $\infrac{\partial}{\partial\gyr}$ terms in \eref{tmpE30} vanish and we are left with
\begin{equation}
\fl
\begin{eqalign}{
\perpav{\wint{\gyror{\bm{u}(\bm{R}_s) \dgR{} \frac{T_s h_s^2}{2 F_{0s}}}}} = 
\perpav{\wint{\bm{u}(\bm{R}_s)\left.\dgR{}\right|_{\bm{w}} \frac{T_s \gyror{h_s^2}}{2 F_{0s}}}}.
}\end{eqalign}
\end{equation}
We now need to rewrite the derivative with respect to $\bm{R}_s$ as a derivative with respect to $\bm{r}$ in order to interchange the derivative and the perpendicular spatial average.
Using 
\begin{equation}
\fl
\bm{u} \cdot \pd{ }{\bm{R}_s} = \bm{u}\cdot\nabla + \bm{u}\cdot\pd{\bm{r}}{\bm{R}_s} \dgR{} = \bm{u}\dg + \left[\frac{\bm{u}}{\cycfreq} \cdot\left(\nabla\bm{b}\right)\times\bm{w}\right] \dg + \Or(\gkeps^2 \cycfreq)
\end{equation} 
and the definition \eref{wDef} of $\bm{w}$ to do this, we obtain
\begin{equation}
\fl
\begin{eqalign}{
\perpav{\wint{\gyror{\bm{u}(\bm{R}_s) \dgR{} \frac{T_s h_s^2}{2 F_{0s}}}}} = \\
\quad\hspace{.5em}
\perpav{\wint{\gyror{\bm{u}(\bm{R}_s)\dg \frac{T_s h_s^2}{2 F_{0s}}}}}
- \left( \bm{u}\dg\bm{e}_1 \right)\cdot\bm{e}_2 \wint \gyror{\frac{\bm{w}_\perp}{\cycfreq}\dg\perpav{\frac{T_s h_s^2}{2F_{0s}}}}.
}\end{eqalign}
\label{tmpE26}
\end{equation}
Note that, as we have been able to interchange the derivative and the average, the first term in \eref{tmpE26} is now one power of $\gkeps$ larger than second term, and so we can drop the latter.
Similarly, we can drop the distinction between $\bm{R}_s$ and $\bm{r}$ in the argument of $\bm{u}$ and flux-surface average \eref{tmpE26} to obtain
\begin{equation}
\begin{eqalign}{
&\fav{\perpav{\wint\gyror{\bm{u}(\bm{R}_s) \dgR{} \frac{T_s h_s^2}{2 F_{0s}}}}} \\
		&\qquad\qquad= \fav{\perpav{\wint\bm{u}\dg\frac{T_s \gyror{h_s^2}}{2F_{0s}}}} = \Or(\gkeps^2 \cycfreq W),
}\end{eqalign}
\label{thankGoodness}
\end{equation}
where we have used $\dv\bm{u}=0$, \eref{favdiv} for the flux-surface average of a divergence, and $\bm{u}\dg\psi = 0$.

Turning to rest of the advection terms in \eref{hsquared}, we find that they can be rewritten in divergence form:
\begin{equation}
\label{ick}
\fl
\begin{eqalign}{
& \fav{\perpav{\wint\gyror{\left(w_\parallel\Meanb + \vdrift + \vchiR\right)\dgR{}\frac{T_s h_s^2}{2 F_{0s}}}}}\\
	&\quad= \fav{ \dv\perpav{\wint\gyror{\left( w_\parallel \Meanb + \vdrift + \vchiR\right)\frac{T_s h_s^2}{2 F_{0s}}}}}
	+ \Or(\gkeps^2\cycfreq W),
}\end{eqalign}
\end{equation}
where we have used the fact that derivatives with respect to $\bm{R}_s$ and with respect to $\bm{r}$ are equivalent to lowest order, used an analogous calculaiton to \eref{flibble} to interchange the divergence and the velocity integral of $w_\parallel \bm{b}\dg$, and used $\dv\vchi = \Or(\gkeps^2 \cycfreq)$.
Using \eref{favdiv}, we see that the term involving $w_\parallel\Meanb$ vanishes.  As the spatial derivatives of a perpendicularly averaged quantity are small, the remaining terms in the right-hand side of \eref{ick} are $\Or(\gkeps^2 \cycfreq W)$ and are thus negligible.
Combining this with \eref{thankGoodness}, we obtain \eref{NoSpreading} as required.
\subsubsection{Derivation of \eref{THeatFEngSource}.}
\label{THeatFEngSourceProof}
Using \eref{tmp-fail-gyro} to substitute for $\bm{u}(\bm{R}_s) \cdot\binfrac{\partial\gyroR{\gkpot}}{\partial \bm{R}_s}$, we have
\begin{equation}
\label{blahick}
\begin{eqalign}{
Z_s e&\fav{\perpav{\wint \gyror{h_s \left[\pd{ }{t} + \bm{u}(\bm{R}_s) \dgR{ }\right] \gyroR{\gkpot}}}} = \\
& Z_s e\fav{\perpav{\wint \gyror{h_s \left(\pd{ }{t} + \bm{u} \dg\right) \gkpot}}}\\
&\quad- \frac{c}{m_s}\fav{\frac{1}{\MeanMagB}\perpav{\wint \gyror{h_s\left( \Meanb\times\bm{w} \right) \cdot\left( \nabla\bm{u} \right) \dg\gkpot }}} \\
&\quad + \frac{Z_s e}{c} \fav{\left(\Meanb\cdot \nabla\bm{u} \right) \cdot \perpav{\wint\gyror{ h_s \left( \bm{w}_\perp \delAp - w_\parallel \delA_\perp \right)}}},
}\end{eqalign}
\end{equation}
where we have expanded $\bm{u}(\bm{R}_s)$ about $\bm{u}(\bm{r})$ and used \eref{interchangeRrgyr} to change from gyroaverages at constant $\bm{R}_s$ to gyroaverages at constant $\bm{r}$.
The last term in \eref{blahick} can be written as
\begin{equation}
\begin{eqalign}{
\frac{Z_s e}{c} &\fav{\left(\Meanb\cdot\nabla\bm{u}\right) \cdot \perpav{\wint \gyror{h_s \left( \bm{w}_\perp \delAp - w_\parallel \delA_\perp \right)}}}\\
	&= -\frac{Z_s e}{c} \angvel(\psi)\fav{ \wint\perpav{ \gyror{h_s\bm{w}}\cdot\left( \delA \times\nabla z \right) }} \\
	&\qquad +\frac{Z_s e}{c} \angvel(\psi)\fav{\left(\Meanb\dg z\right) \wint\perpav{\gyror{ h_s \Meanb\cdot\left( \bm{w}_\perp\times\delA_\perp\right) }}},
}\end{eqalign}
\label{godNo}
\end{equation}
where we have used \eref{nablaU} and \eref{aCrossZ} to expand $\Meanb\dg\bm{u}$. Using \eref{AperpZeta} to express $\delA_\perp$, we see that 
\begin{equation}
\fl
\begin{eqalign}{
\perpav{\gyror{ h_s \Meanb\cdot\left( \bm{w}_\perp\times\delA_\perp\right) }} &= \perpav{\gyror{h_s \bm{w}_\perp \dg \zeta}} = \dv\perpav{\gyror{\bm{w}_\perp h_s \zeta}} \\
		&= \Or(\gkeps\vth h_s\delA_\perp),
}\end{eqalign}
\end{equation}
where we have used \eref{wperpdg} and the fact that $h_s$ is independent of $\gyr$ at constant $\bm{R}_s$. Thus, the second term on the right-hand side of \eref{godNo} can be neglected.
Substituting the remaining term back into \eref{blahick} gives \eref{THeatFEngSource} as required.	
\subsubsection{Derivation of \eref{FluxesFEngSource}.}
\label{FluxFENG}
By using \eref{interchangeRrgyr}, we see that
\begin{equation}
\gyror{\perpav{h_s \vchiR}} = \perpav{\gyror{h_s\vchi}}.
\end{equation}
Therefore, 
\begin{eqnarray}
\begin{eqalign}{
&\fav{\perpav{\wint{ T_s \pd{\ln F_{0s}}{\psi} \gyror{{h_s \vchiR}}\dg\psi}}} = \\
&\qquad- \fav{\perpav{\wint T_s {\gyror{h_s \vchi}}\dg\psi}}\left( \frac{d\ln \NotN}{d\psi} -\frac{3}{2}\frac{d\ln T_s}{d\psi}\right) \\
	&\qquad-\fav{ \perpav{\wint \energy {\gyror{h_s \vchi}}\dg\psi }}\frac{d \ln T_s}{d\psi},
}\end{eqalign}
\label{tmpLnF0}
\end{eqnarray}
where we have used \eref{F0def} for $F_{0s}$.
Turning to the term in \eref{FluxesFEngSource} involving $\infrac{d\angvel}{d\psi}$, we subtract \eref{badger2.5} from \eref{badger2} and rearrange terms to find
\begin{equation}
\begin{eqalign}{
m_s c R^4 &\frac{d\angvel}{d\psi} \left(\nabla\tor\right)\cdot\gyroR{\bm{v}\nabla\chi}\cdot \nabla\tor = \\
	&m_s \vchiR \cdot \bm{W} + Z_s e\gyroR{\left(\frac{\Meanb\times\bm{w}}{\cycfreq} \cdot\nabla\bm{u}\right)\cdot\nabla\gkpot},
}\end{eqalign}
\end{equation}
where $\bm{W}$ is defined by \eref{Wsimple}.
Therefore,
\begin{equation}
\begin{eqalign}{
-{m_s}&\fav{\perpav{\wint{ \left[ \frac{I(\psi) w_\parallel}\MeanMagB + \angvel(\psi)R^2 \right]\frac{d\angvel}{d\psi} \gyror{h_s\vchiR}\dg\psi}}} \\
	&= - \fav{\perpav{\wint \gyror{h_s m_s \vchiR\cdot\bm{W}}}} \\
		&= - \fav{\perpav{\wint R^2\left(\nabla\tor\right) \cdot\gyror{\bm{v}{ h_s \vchi}}\dg\psi}} \frac{d\angvel}{d\psi}\\
		&\qquad+ 
\fav{\frac{m_s c}\MeanMagB\perpav{\wint \gyror{h_s\left( \Meanb\times\bm{w} \right) \cdot\left( \nabla\bm{u} \right) \dg\gkpot }}},
}\end{eqalign}
\label{RHS-F2}
\end{equation}
where we have used \eref{Wsimple}, $\Meanb\times\nabla\psi = \MeanMagB R^2\nabla\tor$ and \eref{interchangeRrgyr}.
Adding \eref{RHS-F2} to \eref{tmpLnF0}, we obtain \eref{FluxesFEngSource} as required.
\subsection{Derivation of \eref{RentFluxMain}: The Entropy Flux}
\label{entropyDeriv}
Here we derive an expression for the radial entropy flux in terms of known quantities. 
Letting $f_s= F_{s} + \delta f_s$ in the definition \eref{RentDef} of this flux
and then expanding $F_s = F_{0s} + F_{1s} + \cdots$ inside the logarithm, we obtain, keeping terms up to $\Or(\gkeps^2)$,
\begin{equation}
\fl
\label{tmpE45-2}
\begin{eqalign}{
\fav{\Rentropyflux} 
&= -\sum_s \left<\wint \left[ F_{s}\ln\left(\frac{8\pi^3\hbar^3 F_{0s}}{m_s^3}\right) + F_{1s} + F_{2s} + \frac{F_{1s}^2}{2F_{0s}} \right.\right.\\
	&\qquad\left.\left.- \frac{\ensav{\delta f_s}^2}{2F_{0s}}\right] \bm{w}_\perp\dg\psi\right>_\psi  + \fav{\MeanEntropy \pd{\psi}{t}} \\
&= -\sum_s \fav{\wint F_s\left\{1 + \ln \left[ \NotN\left(\frac{2\pi \hbar^2}{m_s T_s} \right)^{3/2}\right] - \frac{\energy}{T_s}\right\} \bm{w}_\perp\dg\psi} \\
&\qquad +\sum_s\frac{Z_s e}{T_s}\fav{\ensav{\wint \gkupot\gyror{h_s \bm{w}_\perp}}\dg\psi}+ \fav{\MeanEntropy \pd{\psi}{t}}\\
&\qquad-\sum_s\fav{\wint F_{0s}\bm{w}_\perp \dg\psi}.
}\end{eqalign}
\end{equation}
In deriving the last expression, we are helped by the fact that only the gyrophase-dependent parts of the distribution function give non-vanishing contributions -- so we have  used the gyrophase-independence of $F_{1s}$ (see \Sref{nclasse}), \eref{hDef} and \eref{interchangeRrgyr} to simplify the term involving $\delta f_s$, eliminated $F_{2s}$ by writing $F_{1s} +F_{2s} = F_{s} - F_{0s}$,
and used \eref{expF3} and \eref{MagicTheorem} to ascertain that we do not need to keep $\Or(\gkeps F_s)$ corrections to $F_{0s}$ when calculating $\ln F_{0s}$.
Expanding $F_{0s}(\psi(\bm{R}_s),\energy)$ about a local Maxwellian, as in \eref{F0sExp}, and retaining terms up to $\Or(\gkeps^2 F_{0s})$, we can apply \eref{MagicIdentity} multiple times to show that
\begin{equation}
\fav{\wint F_{0s}\bm{w}_\perp \dg\psi} = \Or(\gkeps^3\vth n_s |\nabla\psi|)
\label{vanihsInt}
\end{equation}
and so the final term in \eref{tmpE45-2} can be dropped.
Using \eref{npol} inside the logarithm in the lowest-order expression for the mean entropy density \eref{MeanEntropyDef}, we find
\begin{equation}
\MeanEntropy = -\sum_s n_s \left\{ \ln \left[ \NotN\left( \frac{2\pi\hbar^2}{m_s T_s} \right)^{3/2}   \right] + \frac{m_s \angvel^2 R^2}{2T_s} - \frac{Z_s e \pot_0}{T_s} - \frac{3}{2} \right\},
\label{logNotN}
\end{equation}
which we substitute into the last term of \eref{tmpE45-2}.
Now, using \eref{GammaDef} to write ${\wint F_s\bm{w}_\perp\dg\psi}$ in terms of $\ParticleFlux$, \eref{energyBong} to expand $\energy$, and working to the lowest non-zero order in $\gkeps$, we have
\begin{equation}
\fl
\begin{eqalign}{
\fav{\Rentropyflux} = &-\sum_s \left\{ 1 + \ln\left[\NotN \left(\frac{2\pi\hbar^2}{m_s T_s}\right)^{3/2}\right]\right\} \fav{\ParticleFlux} \\
&\quad-\sum_s \frac{1}{T_s}\fav{\wint  F_s\left( \frac{1}{2} m_s w^2 + Z_s e \pot_0 - \frac{1}{2}m_s\angvel^2R^2\right) \bm{w}\dg\psi}\\
&\quad+ \sum_s\frac{Z_s e}{T_s}\fav{\ensav{\wint \gkupot\gyror{h_s \bm{w}}}\dg\psi} \\
&\quad- \sum_s\fav{\left(\frac{m_s\angvel^2R^2}{2T_s} - \frac{Z_s e\pot_0}{T_s} - \frac{5}{2}\right)n_s\pd{\psi}{t}}.
}\end{eqalign}
\label{tmpEntFlux}
\end{equation}
Using the definitions \eref{eflux} of $\bm{Q}_s$ and \eref{pflux} of $n_s\bm{U}_s$ in the definition \eref{main-hflux-full} of $\HeatFlux$, we 
see that we can rewrite the second and third lines of \eref{tmpEntFlux} in terms of $\HeatFlux$ to arrive at
\begin{equation}
\fl
\fav{\Rentropyflux}  = - \sum_s \left\{ 1 + \ln\left[\NotN \left(\frac{2\pi \hbar^2}{m_s T_s}\right)^{3/2}\right] \right\} \fav{\ParticleFlux} + \sum_s \frac{\fav{\HeatFlux}}{T_s}.
\end{equation}
Using \eref{ChemPotDef} for the chemical potential $\chempot$ results in \eref{RentFluxMain} as required.
\subsection{Derivation of \eref{entropyProductionMain}: Collisional Entropy Production}
\label{eprodDeriv}
In this Appendix, we derive \eref{entropyProductionMain}, expressing \eref{eprodExp} in terms of known quantities. In particular, we wish to 
write as many terms as possible in terms of the transport fluxes derived in \ref{fluxes}.
To this end, we start by splitting the fluxes into collisional and turbulent contributions.

Using \eref{partflux} and \eref{tmpcollpflux}, we can write the particle flux in terms of its collisional and turbulent contributions as follows
\begin{equation}
\fl
\fav{\ParticleFlux} = \frac{c}{Z_s e} \fav{\left(\bm{F}_s\dg\tor\right)R^2} + \fav{\ensav{\wint\gyror{h_s\vchi}\dg\psi}}.
\label{PfluxEntrop}
\end{equation}
Similarly, from \eref{heatflux} and \eref{colltmphflux}, we have for the heat flux
\begin{equation}
\fl
\fav{\HeatFlux} = \frac{c}{Z_s e}\fav{\left(\bm{G}_s + \Xi_s \bm{F}_s\right)\cdot\left(\nabla\tor\right)R^2} + \fav{\ensav{\wint \energy\gyror{h_s\vchi}\dg\psi}}.
\label{HFluxentrop}
\end{equation}
For the momentum flux, we first use \eref{apDcollmflux} in \eref{pifinalap} to obtain 
\begin{equation}
\label{tmpMfluxEnt}
\begin{eqalign}{
&\fav{\left( \nabla\psi \right)\cdot\viscosity\cdot\left( \nabla\tor \right)R^2 + m_s\angvel(\psi) R^2\ParticleFlux} =\\
&\quad\frac{c}{Z_s e}\fav{ \wint \frac{1}{2} m_s^2\left(R^2\bm{v}\dg\tor\right)^2 \collop[F_s]}\\
&\quad- \fav{\ensav{\wint \frac{Z_s e}{c}R^2 \left(\delA\dg\tor\right)\gyror{h_s\bm{w}}\dg\psi }}\\
&\quad+\fav{\ensav{\wint m_s R^2 \gyror{ \left(\bm{v}\dg\tor\right) h_s\vchi}\dg \psi }},\\
}\end{eqalign}
\end{equation}
where we have used $\bm{v}=\bm{u}+\bm{w}$ to group the last two lines of \eref{pifinalap} together. Now, after we use \eref{pipsifinalap} to rewrite the left-hand side of \eref{tmpMfluxEnt} in terms of $\MomentumFlux$ and $\ParticleFlux$, the term involving $\delA$ cancels and we find
\begin{equation}
\fl
\begin{eqalign}{
\fav{\MomentumFlux + m_s \omega R^2 \ParticleFlux} = &\frac{c}{Z_s e}\fav{ \wint \frac{1}{2} m_s^2\left(R^2\bm{v}\dg\tor\right)^2 \collop[F_s]}\\
		&+ \fav{\ensav{\wint m_sR^2\left(\nabla\tor\right)\cdot \gyror{\bm{v}h_s\vchi}\dg\psi}}.
}\end{eqalign}
\label{MfluxEntrop}
\end{equation}

We will now proceed to show that the first term in \eref{eprodExp} can be expressed via the collisional terms in \eref{PfluxEntrop}, \eref{HFluxentrop}, and \eref{MfluxEntrop},
and that the second term in \eref{eprodExp} can be expressed via the turbulent terms in \eref{PfluxEntrop}, \eref{HFluxentrop}, and \eref{MfluxEntrop}, with precisely the same coefficients.

Using \eref{expF3} to expand $F_s$ inside the logarithm in the first term in \eref{eprodExp}, we obtain
\begin{equation}
\fl\begin{eqalign}{
&\wint {\fav{\ln\left( \frac{8\pi^3\hbar^3 F_s}{m_s^3} \right) \collop[F_s]}} =
		\wint \fav{\ln \left[\frac{8\pi^3\hbar^3F_{0s} (\psi,\energy_0)}{m_s^3}\right] \collop[F_s]}\\
&\quad+ \frac{cm_s}{Z_s e} \fav{\wint\left[ \frac{d\ln \NotN}{d\psi} + \left(\frac{\energy_0}{T_s} - \frac{3}{2}\right)\frac{d\ln T_s}{d\psi}\right]  R^2\left(\bm{v}\dg\tor\right) \collop[F_s]}\\
&\quad+\frac{m_s\MeanMagB}{2 \cycfreq T_s}  \frac{d\angvel}{d\psi} \fav{\wint R^4\left( \bm{v}\dg\tor \right) ^2  \collop[F_s]}
 + \fav{\wint \frac{\Fneo}{F_{0s}}\collop[F_s]} \\
	 &\quad+ \Or(\gkeps^4 \cycfreq \MeanEntropy),
}\end{eqalign}
\label{tmpF44}
\end{equation}
where we have used the fact that collisions conserve particles (this eliminates the $\infrac{\partial\MeanA}{\partial t}$ term arising in \eref{expF3}) and the fact that $\collop[F_s] \sim \gkeps^2\cycfreq F_s$ because $\nu \sim \gkeps\cycfreq$ and $F_s$ is Maxwellian to lowest order.
Multiplying the neoclassical kinetic equation \eref{neogeq} by $\Fneo / F_{0s}$, integrating over all velocities, and flux-surface averaging, we obtain
\begin{equation}
\fav{\wint \frac{\Fneo}{F_{0s}}\collop[F_s]} = - \frac{K_s(\psi)}{T_s} \fav{\MeanE\cdot\MeanB} = -\frac{\JouleHeat}{T_s},
\label{tmpF50}
\end{equation}
where $K_s$ is defined in \eref{KsDef} and we have used a manipulation analogous to the one in \eref{flibble} to eliminate the first term in the left-hand side of \eref{neogeq}.
Substituting \eref{tmpF50} into \eref{tmpF44}, and making use of \eref{F00} for $F_{0s}(\psi,\energy_0)$ and the definitions \eref{CollEnergyDef} of the collisional energy exchange $\CollEnergy$, \eref{FrictionDef} of the friction force $\bm{F}_s$, and \eref{collHeatFricDef} of the ``collisional heat friction'' $\bm{G}_s$, we find
 
\begin{equation}
\fl\begin{eqalign}{
&\fav{\wint {\ln\left( \frac{8\pi^3\hbar^3 F_s}{m_s^3} \right) \collop[F_s]}} = -\frac{1}{T_s} \fav{\CollEnergy} - \frac{1}{T_s} \JouleHeat\\
&\quad+ \frac{c}{Z_s e}  \fav{\left(\bm{F}_s\cdot\nabla\tor\right)R^2}\left( \frac{d\ln \NotN}{d\psi} - \frac{3}{2}\frac{d\ln T_s}{d\psi}\right)\\
&\quad+ \frac{c}{Z_s eT_s} \fav{\left(\bm{G}_s + \Xi_s \bm{F}_s\right)\cdot\left( \nabla\tor \right)R^2}\frac{d\ln T_s}{d\psi}\\
&\quad+\frac{c}{Z_s T_s}   \fav{\wint \frac{1}{2}m_s^2 \left(R^2 \bm{v}\dg\tor \right) ^2  \collop[F_s]}\frac{d\angvel}{d\psi} + \Or(\gkeps^4 \cycfreq \MeanEntropy),
}\end{eqalign}
\label{tmpF46}
\end{equation}
where $\Xi_s = Z_s e\pot_0 - \binfrac{1}{2} m_s \angvel^2 R^2$ is the potential energy of a particle as in \eref{XiDef}. We see that the moments of the collision operator in the last three lines of \eref{tmpF46} are precisely those in the expressions for the fluxes derived above.

We now complete the derivation by obtaining the fluctuating part of the fluxes from the second term in \eref{eprodExp}. We have, by definition \eref{TurbCollDef}, that
\begin{equation}
\fl
\begin{eqalign}{
-&\fav{\ensav{\wint \frac{h_s}{F_{0s}} \lincol[h_s]}} = \frac{1}{T_s} \TurbColl = \frac{1}{T_s} \left( \TurbPow + \TurbInj \right) \\
&\qquad=\frac{1}{T_s} \TurbPow
-\fav{\ensav{\wint \gyror{h_s \vchi}\dg\psi}} \left( \frac{d\ln \NotN}{d\psi} - \frac{3}{2}\frac{d\ln T_s}{d\psi}\right) \\
	&\qquad\quad-\frac{1}{T_s}\fav{\ensav{ \wint \energy \gyror{h_s \vchi}\dg\psi }}\frac{d \ln T_s}{d\psi}  \\
&\qquad\quad+ \frac{1}{T_s}\fav{\ensav{\wint m_s R^2\left(\nabla\tor\right) \cdot\gyror{\bm{v} h_s \vchi}\dg\psi}} \frac{d\angvel}{d\psi},
}\end{eqalign}
\label{tmpF47}
\end{equation}
where we have used the second line of \eref{TurbPowDef}
and then \eref{TurbInjDef} to write $\TurbInj$ in terms of the gradients of $n_s$, $T_s$, and $\angvel$. We see that, again, the turbulent fluxes in \eref{tmpF47} match up precisely with the turbulent fluxes
in \eref{PfluxEntrop}, \eref{HFluxentrop}, and \eref{MfluxEntrop}.

Substituting \eref{tmpF46} and \eref{tmpF47} into \eref{eprodExp} and using \eref{PfluxEntrop}, \eref{HFluxentrop} and \eref{MfluxEntrop}, we arrive at \eref{entropyProductionMain} as required.

\section*{References}
\addcontentsline{toc}{section}{References}
\bibliographystyle{unsrt}
\bibliography{references/references.bib}
\end{document}